\documentclass[final,5p,times,twocolumn]{elsarticle}
\usepackage{xspace}
\usepackage{graphicx}
\usepackage{amssymb}

\journal{Surface science reports}

\begin{document}
\begin{frontmatter}

\title{Electron counting in quantum dots}

\author[eth]{S.~Gustavsson}
\author[eth]{R.~Leturcq}
\author[eth]{M.~Studer}
\author[eth]{I.~Shorubalko}
\author[eth]{T.~Ihn}
\author[eth]{K.~Ensslin}
\author[ucsb]{D.~C.~Driscoll}
\author[ucsb]{A.~C.~Gossard} 

\address[eth]{Solid State Physics Laboratory, ETH Zurich, CH-8093 Zurich, Switzerland}
\address[ucsb]{Materials Departement, University of California, Santa Barbara, CA-93106, USA}

\begin{abstract}
We use time-resolved charge detection techniques to investigate single-electron tunneling in semiconductor quantum dots. The ability to detect individual charges in real-time makes it possible to count electrons one-by-one as they pass through the structure. The setup can thus be used as a high-precision current meter for measuring ultra-low currents, with resolution several orders of magnitude better than that of conventional current meters.
In addition to measuring the average current, the counting procedure also makes it possible to investigate correlations between charge carriers. Electron correlations are conventionally probed in noise measurements, which are technically challenging due to the difficulty to exclude the influence of external noise sources in the experimental setup. Using real-time charge detection techniques, we circumvent the problem by studying the electron correlation directly from the counting statistics of the tunneling electrons. In quantum dots, we find that the strong Coulomb interaction makes electrons try to avoid each other. This leads to electron anti-bunching, giving stronger correlations and reduced noise compared to a current carried by statistically independent electrons. 

The charge detector is implemented by monitoring changes in conductance in a near-by capacitively coupled quantum point contact. We find that the quantum point contact not only serves as a detector but also causes a back-action onto the measured device. Electron scattering in the quantum point contact leads to emission of microwave radiation. The radiation is found to induce an electronic transition between two quantum dots, similar to the absorption of light in real atoms and molecules. Using a charge detector to probe the electron transitions, we can relate a single-electron tunneling event to the absorption of a single photon. Moreover, since the energy levels of the double quantum dot can be tuned by external gate voltages, we use the device as a frequency-selective single-photon detector operating at microwave energies. The ability to put an on-chip microwave detector close to a quantum conductor opens up the possibility to investigate radiation emitted from mesoscopic structures and give a deeper understanding of the role of electron-photon interactions in quantum conductors. 

A central concept of quantum mechanics is the wave-particle duality;
matter exhibits both wave- and particle-like properties and can not
be described by either formalism alone. To investigate the wave
properties of the electrons, we perform experiments on a structure
containing a double quantum dot embedded in the Aharonov-Bohm ring
interferometer. Aharonov-Bohm rings are traditionally used to study
interference of electron waves traversing different arms of the
ring, in a similar way to the double-slit setup used for
investigating interference of light waves.  In our case, we use the
time-resolved charge detection techniques to detect electrons
one-by-one as they pass through the interferometer. We find that the
individual particles indeed self-interfere and give rise to a strong
interference pattern as a function of external magnetic field. The
high level of control in the system together with the ability to
detect single electrons enables us to make direct observations of
non-intuitive fundamental quantum phenomena like single-particle
interference or time-energy uncertainty relations.
\end{abstract}


\end{frontmatter}

%
\def\imat  {i}
\def\be{\begin{equation}}
\def\ee{\end{equation}}
\def\bea{\begin{eqnarray}}
\def\eea{\end{eqnarray}}
\def\la{\langle}
\def\ra{\rangle}
\def\l{\langle\!\langle}
\def\r{\rangle\!\rangle}

\def\Gin{\Gamma_\mathrm{in}}
\def\Gout{\Gamma_\mathrm{out}}
\def\Gs{\Gamma_\mathrm{S}}
\def\Gd{\Gamma_\mathrm{D}}
\def\Gc{\Gamma_\mathrm{C}}
\def\Grel{\Gamma_\mathrm{rel}}
\def\Gabs{\Gamma_\mathrm{abs}}
\def\Gem{\Gamma_\mathrm{em}}
\def\Gdet{\Gamma_\mathrm{det}}
\def\tdet{\tau_\mathrm{det}}
\def\Vtdeg{V_\mathrm{2DEG}}
\def\Iqpc{I_\mathrm{QPC}}
\def\Gqpc{G_\mathrm{QPC}}
\def\Vsd{V_\mathrm{SD}}
\def\Vgl{V_\mathrm{G1}}
\def\Vgr{V_\mathrm{G2}}
\def\aGl{\alpha_\mathrm{G1}}
\def\aGr{\alpha_\mathrm{G2}}
\def\Idqd{I_\mathrm{DQD}}
\def\tc{t_\mathrm{C}}
\def\mul{\mu_\mathrm{1}}
\def\mur{\mu_\mathrm{2}}
\def\mus{\mu_\mathrm{S}}
\def\mud{\mu_\mathrm{D}}
\def\Ec{E_\mathrm{C}}
\def\Ecm{E_\mathrm{Cm}}
\def\Ecl{E_\mathrm{C1}}
\def\Ecr{E_\mathrm{C2}}
\def\Vqpc{V_\mathrm{QPC}}
\def\Vqsd{V_\mathrm{QPC-SD}}
\def\Vdsd{V_\mathrm{DQD-SD}}
\def\Rs{R_\mathrm{S}}
\def\Rf{R_\mathrm{F}}
\def\Cc{C_\mathrm{C}}
\def\Cf{C_\mathrm{F}}
\def\Vout{V_\mathrm{out}}
\def\Vn{V_\mathrm{n}}
\def\Vn{I_\mathrm{n}}
\def\fBW{f_\mathrm{BW}}


\newcommand*{\eg}{e.\,g.\xspace}
\newcommand*{\ie}{i.\,e.\xspace}
\newcommand*{\cf}{c.\,f.\,,\xspace}
\newcommand*{\etc}{etc.\@\xspace}
\newcommand*{\etal}{\textsl{et al.\@\xspace}}

\newcommand*{\e}{\mathrm{e}}       
\newcommand*{\ci}{\mathrm{i}}
\newcommand*{\kb}{\ensuremath{\,k_\mathrm{B}}\xspace}
\newcommand*{\abs}[1]{\mid#1\mid}


\newcommand*{\um}{\ensuremath{\,\mu\mathrm{m}}\xspace}
\newcommand*{\nm}{\ensuremath{\,\mathrm{nm}}\xspace}
\newcommand*{\mm}{\ensuremath{\,\mathrm{mm}}\xspace}
\newcommand*{\m}{\ensuremath{\,\mathrm{m}}\xspace}
\newcommand*{\sqm}{\ensuremath{\,\mathrm{m}^2}\xspace}
\newcommand*{\sqmm}{\ensuremath{\,\mathrm{mm}^2}\xspace}
\newcommand*{\squm}{\ensuremath{\,\mu\mathrm{m}^2}\xspace}
\newcommand*{\psqm}{\ensuremath{\,\mathrm{m}^{-2}}\xspace}
\newcommand*{\psqmV}{\ensuremath{\,\mathrm{m}^{-2}\mathrm{V}^{-1}}\xspace}
\newcommand*{\cm}{\ensuremath{\,\mathrm{cm}}\xspace}

\newcommand*{\nF}{\ensuremath{\,\mathrm{nF}}\xspace}
\newcommand*{\pF}{\ensuremath{\,\mathrm{pF}}\xspace}

\newcommand*{\emob}{\ensuremath{\,\mathrm{m}^2/\mathrm{V}\mathrm{s}}\xspace}
\newcommand*{\edos}{\ensuremath{\,\mu\mathrm{C}/\mathrm{cm}^2}\xspace}
\newcommand*{\mbar}{\ensuremath{\,\mathrm{mbar}}\xspace}

\newcommand*{\A}{\ensuremath{\,\mathrm{A}}\xspace}
\newcommand*{\nA}{\ensuremath{\,\mathrm{nA}}\xspace}
\newcommand*{\pA}{\ensuremath{\,\mathrm{pA}}\xspace}
\newcommand*{\fA}{\ensuremath{\,\mathrm{fA}}\xspace}
\newcommand*{\uA}{\ensuremath{\,\mu\mathrm{A}}\xspace}

\newcommand*{\Ohm}{\ensuremath{\,\Omega}\xspace}
\newcommand*{\kOhm}{\ensuremath{\,\mathrm{k}\Omega}\xspace}
\newcommand*{\MOhm}{\ensuremath{\,\mathrm{M}\Omega}\xspace}
\newcommand*{\GOhm}{\ensuremath{\,\mathrm{G}\Omega}\xspace}

\newcommand*{\Hz}{\ensuremath{\,\mathrm{Hz}}\xspace}
\newcommand*{\kHz}{\ensuremath{\,\mathrm{kHz}}\xspace}
\newcommand*{\MHz}{\ensuremath{\,\mathrm{MHz}}\xspace}
\newcommand*{\GHz}{\ensuremath{\,\mathrm{GHz}}\xspace}
\newcommand*{\THz}{\ensuremath{\,\mathrm{THz}}\xspace}

\newcommand*{\K}{\ensuremath{\,\mathrm{K}}\xspace}
\newcommand*{\mK}{\ensuremath{\,\mathrm{mK}}\xspace}

\newcommand*{\kV}{\ensuremath{\,\mathrm{kV}}\xspace}
\newcommand*{\V}{\ensuremath{\,\mathrm{V}}\xspace}
\newcommand*{\mV}{\ensuremath{\,\mathrm{mV}}\xspace}
\newcommand*{\uV}{\ensuremath{\,\mu\mathrm{V}}\xspace}
\newcommand*{\nV}{\ensuremath{\,\mathrm{nV}}\xspace}

\newcommand*{\eV}{\ensuremath{\,\mathrm{eV}}\xspace}
\newcommand*{\meV}{\ensuremath{\,\mathrm{meV}}\xspace}
\newcommand*{\ueV}{\ensuremath{\,\mu\mathrm{eV}}\xspace}

\newcommand*{\T}{\ensuremath{\,\mathrm{T}}\xspace}
\newcommand*{\mT}{\ensuremath{\,\mathrm{mT}}\xspace}
\newcommand*{\uT}{\ensuremath{\,\mu\mathrm{T}}\xspace}

\newcommand*{\ms}{\ensuremath{\,\mathrm{ms}}\xspace}
\newcommand*{\s}{\ensuremath{\,\mathrm{s}}\xspace}
\newcommand*{\us}{\ensuremath{\,\mathrm{\mu s}}\xspace}
\newcommand*{\rpm}{\ensuremath{\,\mathrm{rpm}}\xspace}
\newcommand*{\minute}{\ensuremath{\,\mathrm{min}}\xspace}
\newcommand*{\degree}{\ensuremath{\,^\circ\mathrm{C}}\xspace}

\newcommand*{\AppRef}[1]{Appendix$\,$\ref{#1}}
\newcommand*{\SecRef}[1]{Sec.$\,$\ref{#1}}
\newcommand*{\TabRef}[1]{Tab.$\,$\ref{#1}}
\newcommand*{\ChaRef}[1]{chapter$\,$\ref{#1}}

\newcommand*{\Cite}[1]{Ref.$\,$\citealp{#1}}
\newcommand*{\citei}[2]{\cite{#1}$\,$(#2)}
\newcommand*{\Refi}[2]{\ref{#1}$\,$(#2)}
\newcommand*{\EqRef}[1]{Eq.~(\ref{#1})}
\newcommand*{\FigRef}[1]{Fig.~\ref{#1}}
\newcommand*{\FigRefi}[2]{\FigRef{#1}$\,$(#2)}
\newcommand*{\SampRef}[2]{\AppRef{#1}$\,$(#2)}

\tableofcontents


\section{The quantum point contact as a charge detector}
The quantum point contact (QPC) is the electron analogue of a photon
waveguide. Since the width of the constriction is of the order of
the electron wavelength, constructive and destructive interference
only allow electron wavefunctions corresponding to standing waves in
the directions of the confinement. Due to the Fermionic nature of
electrons, each mode within the QPC carries a fixed conductance of
$G_0=e^2/h$. The conductance of a QPC with $N$ available modes is
thus equal to \cite{vanWees:1998, wharam:1998}
\begin{equation}
\label{eq:QP_QuantC } G = N \, G_0,
\end{equation}
with $N$ integer. The effect is called \emph{conductance
quantization}. If the measurement is performed in the absence of
magnetic fields, the electron spin states are degenerate and the
conductance quantization appears in units of $2e^2/h$ instead of
$e^2/h$.

The conductance of the QPC depends strongly on its electrostatic surroundings. This may be
utilized to detect charge fluctuations in a quantum dot (QD) close to the constriction
with single-electron resolution. In this section, we show how to
operate the quantum point contact as a charge detector and
investigate how to optimize the device to obtain the best charge
sensitivity

We are concerned with quantum point contacts formed in a
two-dimen\-sional electron gas (2DEG). For such structures, the
confinement in growth direction is usually much stronger than in the
lateral direction. In the following, we assume the part of the
electron wavefunction in the growth direction to be in its ground
state and consider additional modes only in the lateral direction.
Quantum point contacts may be fabricated using a variety of methods,
for example by depleting the 2DEG by applying negative voltages to
metallic gates put on the heterostructure surface
\cite{vanWees:1998,wharam:1998}. Here we investigate structures
formed by etching or by local oxidation of the heterostructure
surface.

\begin{figure}[tb]
\centering
\includegraphics[width=\linewidth]{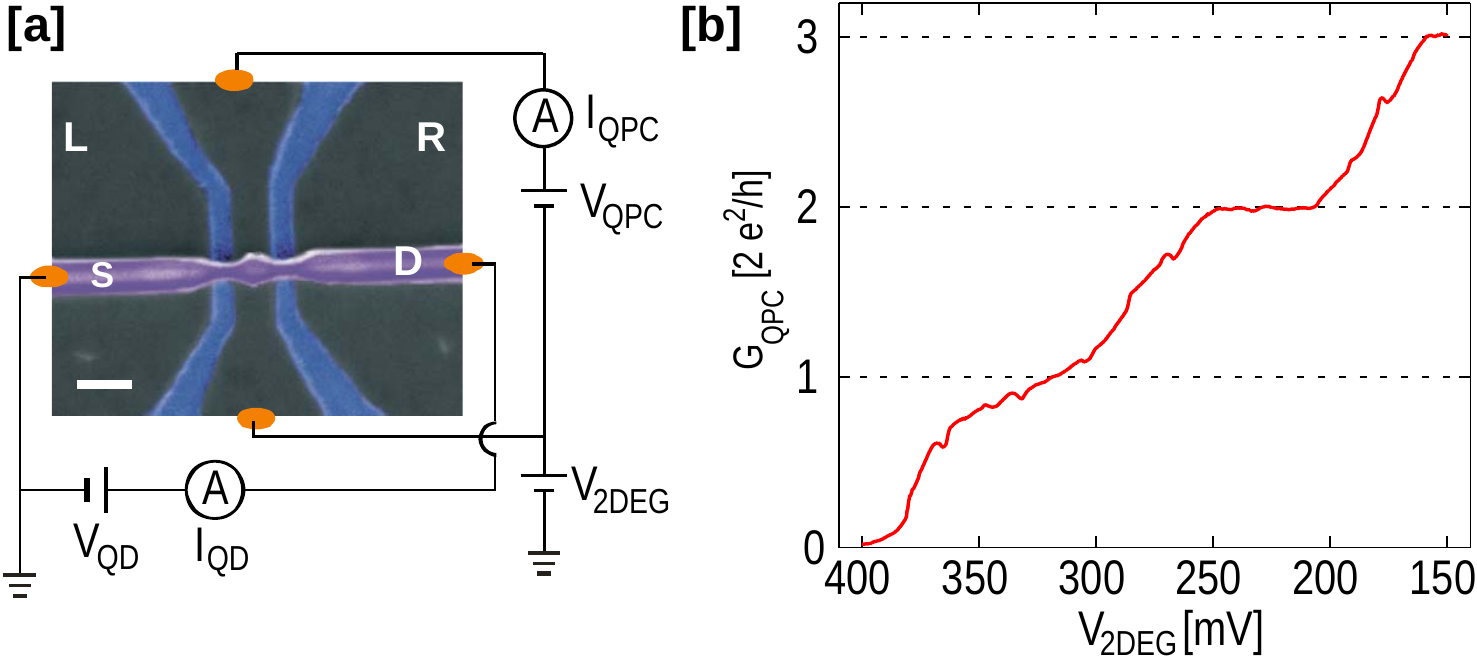}
\caption{(a) Quantum point contact, defined by etching trenches
(marked in blue in the figure) in a GaAs heterostructure containing
a 2DEG 37 nm below the surface. A quantum dot defined in an InAs
nanowire (purple) is lying on top of the structure.
 (b) Conductance of the QPC versus voltage applied to the 2DEG.
 The measurement was performed in a two-terminal setup, with $V_\mathrm{QPC-SD}=200\uV$ applied across the QPC.
 A series resistance of $4~\mathrm{k\Omega}$ was subtracted because of the ohmic contact resistance.
 The measurement was performed at a temperature of $T=1.7\K$. }
\label{fig:QP_QuantizedC}
\end{figure}

Figure~\ref{fig:QP_QuantizedC}(a) shows a scanning electron microscope
(SEM) image of the device used in the experiments in this section. An InAs
nanowire is deposited on top of a shallow (37 nm below the surface) AlGaAs/GaAs
heterostructure based two-dimensional electron gas. The QPC
is defined by etched trenches, which separate the QPC from the rest
of the 2DEG. The parts of the 2DEG marked by L and R are used as
in-plane gates. The horizontal object in the figure is the nanowire
lying on top of the surface, electrically isolated from the QPC. The
QD in the nanowire and the QPC in the underlying 2DEG are defined in
a single etching step using patterned electron beam resist as an
etch mask. The technique ensures perfect alignment between the two
devices. Details of the fabrication procedure can be found in
Ref.~\cite{shorubalko:2007}. The QD charging energy is around
$10\meV$, due to the small size of the structure. 

In Fig.~\ref{fig:QP_QuantizedC}(b), we plot the conductance of the
QPC, measured when shifting the voltage on the part of the 2DEG
connected to the QPC ($\Vtdeg$) and keeping the other contacts
grounded. Making $\Vtdeg$ more positive has the same effect as
making the surrounding gates more negative, leading to pinch-off of
the QPC. As $\Vtdeg$ is lowered, the constriction opens up to allow
the first electron mode to populate the QPC. Further lowering
$\Vtdeg$ makes more modes available and the conductance increases
stepwise.

\subsection{Charge detection}
\begin{figure}[tb]
\centering
\includegraphics[width=\linewidth]{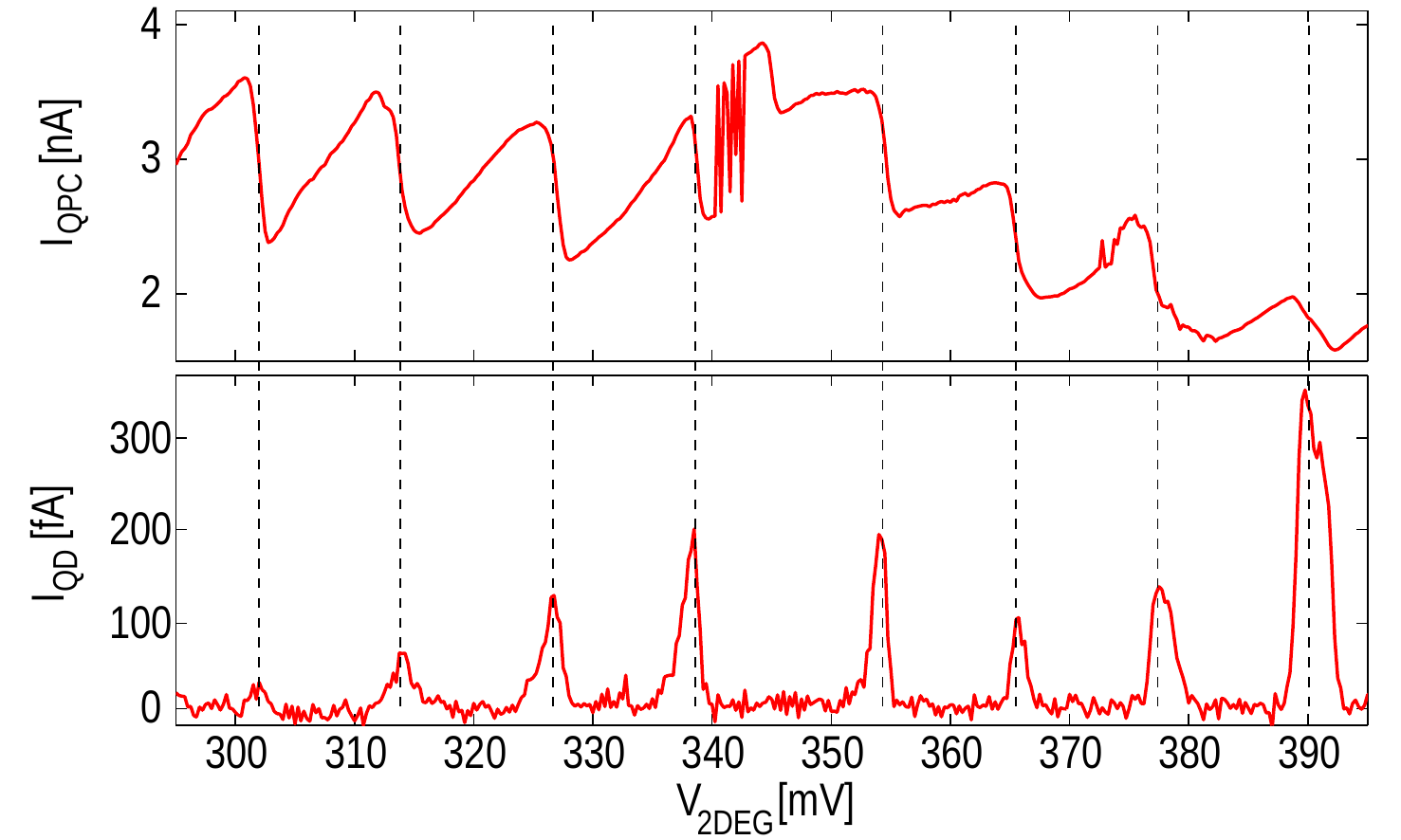}
\caption{Current in the QPC and the QD for the structure shown in
\FigRef{fig:QP_QuantizedC}(a), measured vs voltage on the 2DEG. A
second gate is tuned simultaneously to keep the potential of the QPC roughly
constant during the sweep. As the voltage of the 2DEG is lowered,
electrons are unloaded from the QD. At each transition there is a
corresponding increase in the QPC conductance. At the same gate
voltages, sequential tunneling gives rise to peaks in the QD
current.} \label{fig:QP_QPCQDcurr}
\end{figure}

Next, we investigate the electrostatic interactions between the QPC and the QD in the nanowire. Figure~\ref{fig:QP_QPCQDcurr} displays simultaneous measurements of
QPC and QD currents for the structure of
Fig.~\ref{fig:QP_QuantizedC}(a). As the gate voltage is lowered,
electrons are unloaded from the QD and the QD current shows clear
Coulomb peaks at each charge transition. At the same time, the QPC
conductance changes in steps at the positions of the Coulomb peaks \cite{field:1993}.
The QPC was voltage biased with $V_\mathrm{QPC-SD}=200\uV$ and
operated between pinch-off and the first plateau. The QPC
conductance is kept roughly constant during the sweep by applying a
compensation voltage to the side gate marked by L in
\FigRef{fig:QP_QuantizedC}(a). 

The left-most peak in the QD current in \FigRef{fig:QP_QPCQDcurr} is barely measurable due to weak tunnel coupling between the QD and its leads. However, the charge transition is still clearly visible in the QPC signal. This demonstrates one of the advantages of the charge detection method compared to a
standard current measurement. A conventional current meter has a
resolution of $\sim \! 10~\fA/\sqrt{\Hz}$, meaning that the
tunneling rates of the QD must be kept larger than $\Gamma > 10\fA/ e
\sim 60\kHz$ for reasonable integration times. Moreover, in order to
measure current through the QD it needs to be hooked up to two
leads. On the other hand, a charge detector can measure electron
tunneling which occurs on much slower timescales as well as detect
equilibrium fluctuations between a QD and a single lead.

\begin{figure}[b]
\centering
 \includegraphics[width=\columnwidth]{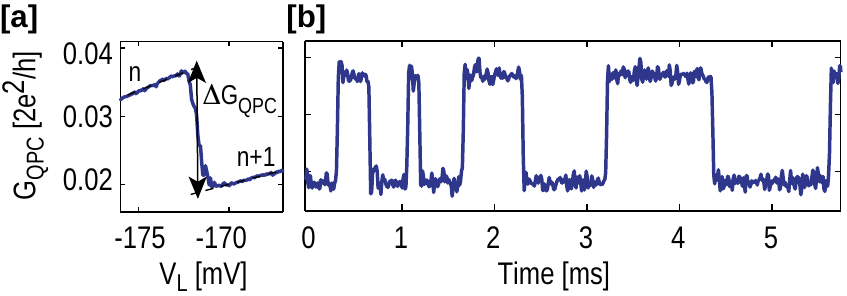}
 \caption{(a) QPC conductance measured versus voltage on gate L.
 At $V_\mathrm{L} = -172\mV$ an electron is added to the QD, leading to a decrease of $\Gqpc$.
 (b) Time trace of the QPC
 conductance measured at $V_\mathrm{L} = -172\mV$, showing a few electrons tunneling into and out of the
 QD. The upper level corresponds to a situation with $n$ electrons
 on the QD. Adapted from Ref.~\cite{gustavssonNWPRB:2008}.
 }
\label{fig:QP_timeResolved}
\end{figure}

Figure~\ref{fig:QP_timeResolved}(a) shows a measurement of the QPC conductance for a small region around one charge transition in the QD. The measurement was performed without any bias voltage
applied to the QD and with the drain lead of the QD pinched off. At
$V_\mathrm{L} = -172\mV$, the electrochemical potential of the QD
shifts below the Fermi levels of the source lead and an electron may
tunnel onto the QD. This gives a decrease $\Delta \Gqpc$ of the QPC
conductance corresponding to the change $\Delta q = \mathrm{e}$ of
the charge population on the QD. The curve in
Fig.~\ref{fig:QP_timeResolved}(a) shows the average QPC conductance, which gives
the time-averaged QD population. In Fig.~\ref{fig:QP_timeResolved}(b), we perform a time-resolved measurement of the QPC conductance at $V_\mathrm{L} = -172\mV$. The QPC
conductance fluctuates between the two levels corresponding to $(n)$
and $(n+1)$ electrons on the QD. Transitions between the levels occur
on a millisecond timescale, which provides a direct measurement of
the tunnel coupling between the QD and the source lead
\cite{schleser:2004}. 

\subsection{Time-resolved operation}
\begin{figure}[tb]
\centering
\includegraphics[width=\linewidth]{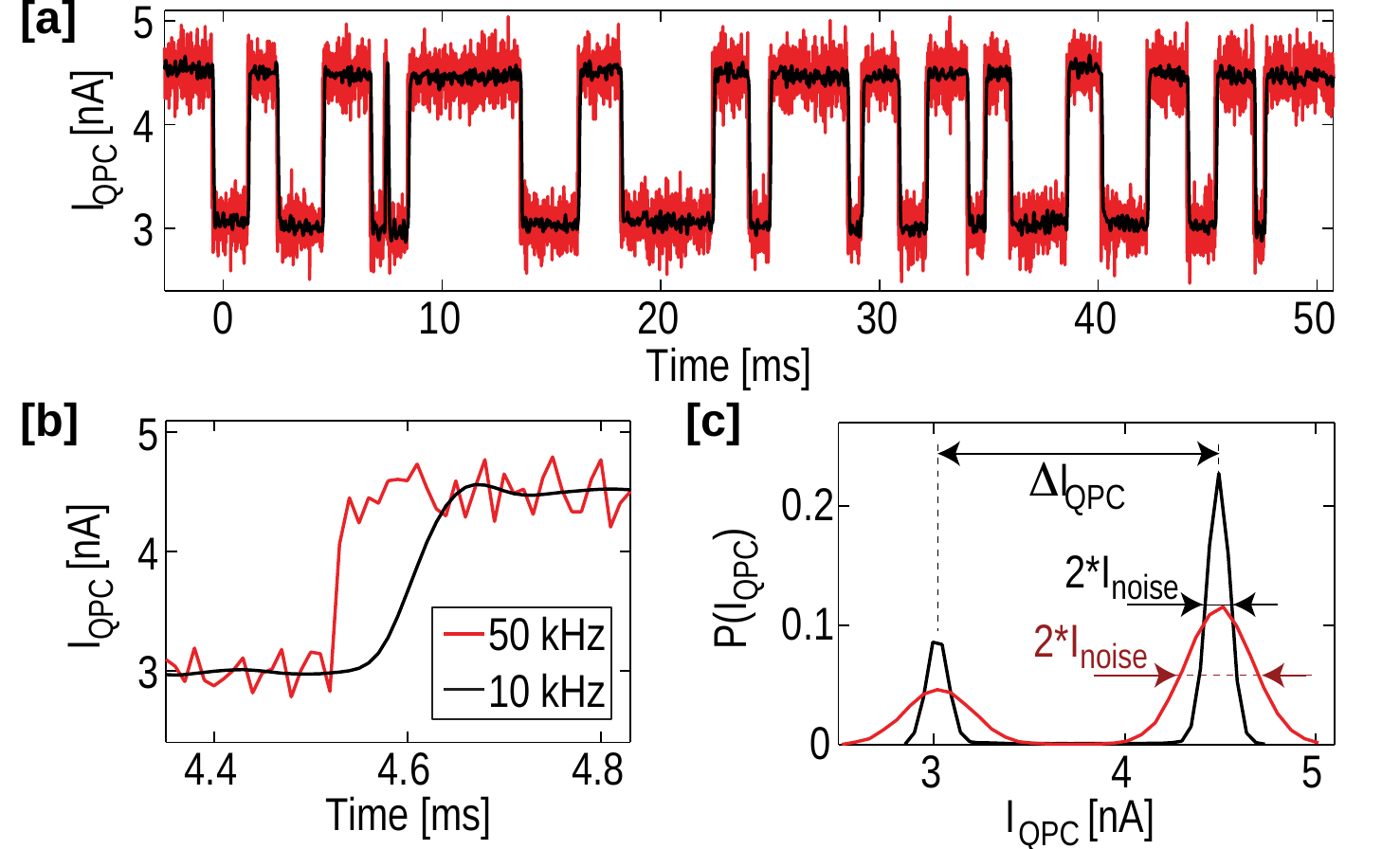}
\caption{(a) Time trace of the current through the quantum point
contact, showing a few transitions due to electrons tunneling into
and out of the QD. The two traces are taken with $10\kHz$ and
$50\kHz$ bandwidth. (b) Blow-up of one switching event from the trace in
(a). The rise time is clearly different for the data taken with
$10\kHz$ and $50\kHz$ bandwidth. (c) Histogram showing the distribution of the
current for the data in (a). The two levels are easily
distinguished.} \label{fig:QP_TraceHistRise}
\end{figure}

As described in the previous section, charge transitions in the QD may be detected in
real-time if tunnel couplings between the QD and its leads are tuned below
the QPC measurement bandwidth. This allows a wealth of experiments to be performed, like
investigating single-electron dynamics or probing interactions
between charge carriers in the system. We postpone the detailed investigation of single-electron
tunneling in quantum dots to section \ref{sec:TR_main}; here we focus on the experimental setup and how to optimize the QPC in order
to perform the best possible charge detection measurement.
Figure~\ref{fig:QP_TraceHistRise}(a) shows a time trace of the QPC
current, measured in a configuration where the coupling between the
QD and the source lead is below $1\kHz$, and the other lead is
completely pinched off. Again, the QPC current shows two levels,
corresponding to $(n)$ and $(n+1)$ electrons on the QD.

The time resolution available for detecting charge transitions as
seen in \FigRef{fig:QP_TraceHistRise}(a) is set directly by the
bandwidth of the QPC measurement circuit. On the other hand,
increasing the bandwidth also increases the noise in the
measurement, leading to a trade-off between noise and bandwidth. The
effect is visualized in \FigRef{fig:QP_TraceHistRise}, where the two
curves show the same set of data but filtered with different
bandwidths, $10\kHz$ (black) and $50\kHz$ (red). The filtering was
performed numerically with a 6th-order Bessel low-pass filter. In
\FigRef{fig:QP_TraceHistRise}(b), we zoom in on one of the
transitions of \FigRef{fig:QP_TraceHistRise}(a). The data taken with
lower bandwidth shows a considerably slower time response than the
trace taken with higher bandwidth. The lower-bandwidth filter also
introduces a time offset; this is not a major problem since we are
interested in determining the time intervals between transitions
rather than the absolute transition times.

In \FigRef{fig:QP_TraceHistRise}(c), we plot the distribution
functions for the two traces shown in
\FigRef{fig:QP_TraceHistRise}(a). The distributions contain two
peaks associated with the two QPC current levels. The distance
between peaks gives directly the change in QPC current ($\Delta
\Iqpc$) for one electron entering the QD, while the standard
deviation of the current distribution $p(I)$ around each peak
($I_\mathrm{noise}$) reflects the amount of noise in the measured
signal. As a consequence of the increased bandwidth for the red
trace, the data contains noise contributions from a broader
frequency spectrum and the peaks in the distribution function become
significantly broader.

\subsection{Signal-to-noise} \label{sec:QP_sn}
The ratio between the change in current $\Delta \Iqpc$ and the noise
$I_\mathrm{noise}$ is conveniently expressed as a signal-to-noise
(S/N) ratio. To maximize the useful information that can be
extracted from the measurement, we need to maximize the signal and
minimize the noise. In this subsection we consider the effects of
the noise, in the following sections we describe how to optimize the
signal by tuning the operation point of the QPC.

The noise of the QPC signal can be seperated into \emph{intrinsic}
and \emph{extrinsic} contributions. With intrinsic noise we refer to
noise generated by the QPC itself, while extrinsic noise is due to
amplifiers and other external noise sources.
It turns out that the main source of noise in the setup is given by
amplifier noise. Since the noise is extrinsic, it is essentially
independent of both operating point and biasing conditions of the
QPC. The only way to reduce this noise is to use an amplifier with
lower noise figures or to reduce cable capacitances. The amplifier
noise spectrum is not flat and depends on the details of the experimental setup
\cite{vandersypen:2004, gustavssonPhd:2008}; a rough estimate for the noise
contribution in the relevant frequency range is $\sim\!
400\fA/\sqrt{\Hz}$.

The fundamental intrinsic noise of the QPC current is the \emph{shot
noise}, which arises due to the fact that the current is carried by
discrete charged particles. The shot noise has a flat power spectrum
in the region of interest which scales linearly with the magnitude
of the current. For typical currents used here ($\sim\!10\nA$), the
white noise power is $\sim\!30~\fA/\sqrt{\Hz}$, which is
considerably lower than the amplifier noise.
The \emph{thermal} noise or \emph{Johnson-Nyquist} noise is
generated by the thermal agitation of the charge carriers in a
conductor and appears regardless of applied voltage. Since the sample is 
held at very low temperatures, its thermal noise becomes negligible compared to the
amplifier noise.
Another form of intrinsic noise arises because of fluctuations of
trapped charges close to the QPC. The charge traps sit at lattice
defects or at the heterostructure surface and may be activated by a
large current passing the QPC. Such noise is usually referred to as
\emph{burst} noise or \emph{popcorn} noise. In GaAs QDs, it is
believed that the \emph{1/f}-noise is generated by fluctuations in
an ensemble of charge traps distributed uniformly in the device
\cite{jung:2004}. The magnitude of the noise depends on the quality
of the heterostructure and on the abundance of traps close to the
QPC. As we will see later in this section, the charge detection technique provides a method for mapping out the charge traps near the QPC.

The noise is described by a power spectral density $S(\omega)$,
which depends on the physical process responsible for generating the
fluctuations. The amplitude of the current noise in a trace as shown
in \FigRef{fig:QP_TraceHistRise} is given by integrating the
spectral density over the measurement bandwidth
\begin{equation}
\label{eq:QP_integrateNoise}
 I_\mathrm{noise} \sim \sqrt{P_\mathrm{noise}} =
 \left( \int_0^{2 \pi f_\mathrm{BW}} S(\omega) \, d\omega \right)^{1/2}.
\end{equation}
Here, $P_\mathrm{noise}$ is the noise power and $f_\mathrm{BW}$ the
measurement bandwidth. If we assume for simplicity the spectrum to
be independent of frequency $(S(\omega) = \mathrm{const.})$, then
the current noise scales with the square root of the bandwidth,
\begin{equation}
\label{eq:QP_noiseBW}
 I_\mathrm{noise} \sim \sqrt{f_\mathrm{BW}}.
\end{equation}
Increasing the bandwidth thus increases the noise and lowers the
S/N, as visualized in \FigRef{fig:QP_TraceHistRise}. A
single-electron detector must be able to reliably detect transitions
between the two levels in the QPC current. How much can the
bandwidth and the noise be increased before the detection mechanism
becomes unreliable? A qualitative answer would be when
$I_\mathrm{noise}$ is comparable to the step height $\Delta \Iqpc$.
To investigate the issue quantitatively, we need to estimate the
probability of detecting false transitions due to the noise. The
problem is well understood in the language of information theory
\cite{shannon:1948}; here we make a simplified analysis to get a
quick estimate of the risk of detecting false counts.

For this purpose, we assume the distribution of the QPC current
$p(\Iqpc)$ to be Gaussian around each of its two levels and evaluate
the part of the distribution deviating by more than $\Delta \Iqpc/2$
from the peak value,
\begin{equation}
p_\mathrm{out} = \int_{\Delta \Iqpc/2}^{\infty} p(\Iqpc) \, d\Iqpc.
\end{equation}
This fraction is beyond the midline between the two peaks of the
distribution $p(\Iqpc)$ and gives rise to false counts. The number
of false counts $n_\mathrm{false}$ registered during a time interval
$\Delta t$ is equal to $p_\mathrm{out}$ multiplied with the number
of measurements performed in the interval, which according to
sampling theorem needs to be $n_\mathrm{meas} \sim 2\, \Delta t \,
f_\mathrm{BW}$. For the false counts, we get
\begin{equation}
\label{eq:QP_falseCounts} n_\mathrm{false} =p_\mathrm{out}  \,
n_\mathrm{meas}   \sim 2 \,p_\mathrm{out} \, \Delta t \,
f_\mathrm{BW}.
\end{equation}

\begin{figure}[tb]
\centering
\includegraphics[width=\linewidth]{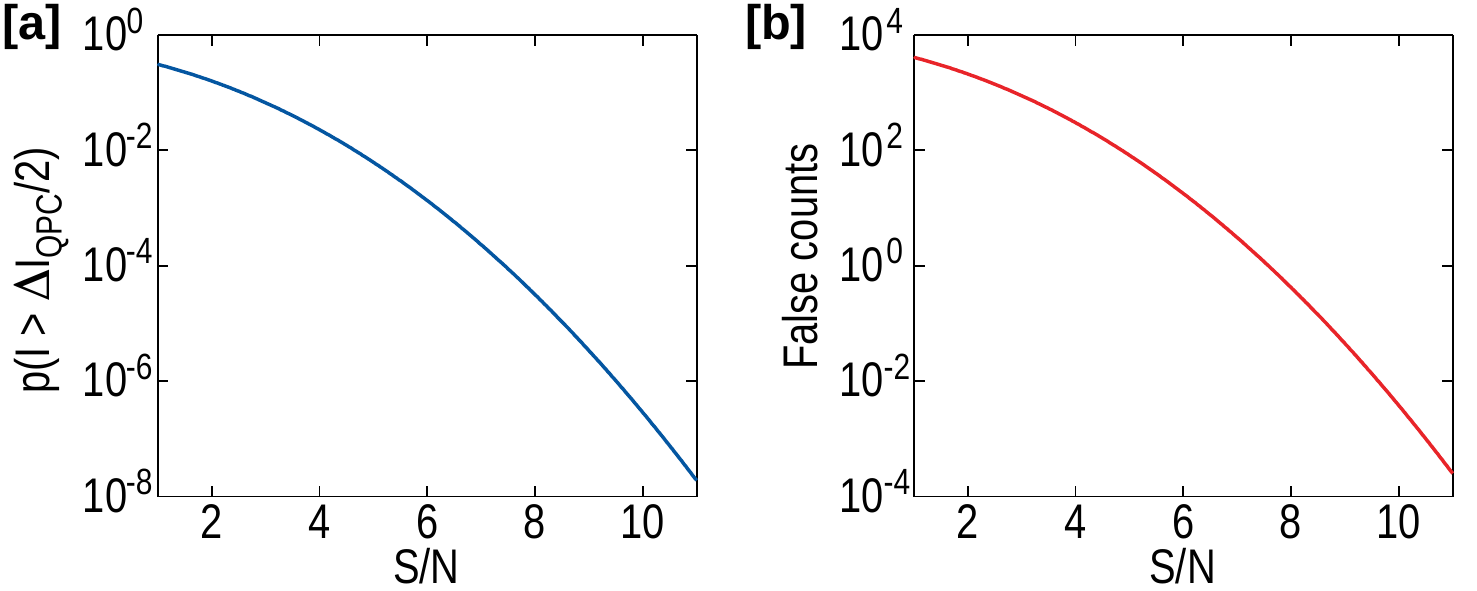}
\caption{(a) Part of the QPC current having $I > \Delta \Iqpc/2$,
assuming a Gaussian distribution. (b) Average number of false
counts, calculated from \EqRef{eq:QP_falseCounts} in the text with
$\Delta t = 1\s$ and $f_\mathrm{BW} = 10\kHz$. The risk off
detecting false events falls off rapidly with increased S/N.}
\label{fig:QP_FalseCounts}
\end{figure}

In \FigRef{fig:QP_FalseCounts}(a) we plot $p_\mathrm{out}$ as a
function of S/N. As a consequence of the Gaussian distribution, the
risk of detecting false counts falls off stronger than exponential
with increased S/N. Figure~\ref{fig:QP_FalseCounts}(b) shows the
risk of detecting a false count, calculated using
\EqRef{eq:QP_falseCounts} with $\Delta t = 1\s$ and $f_\mathrm{BW} =
10\kHz$.  For S/N=7, we find that the detector will register an
average of four false counts per second.

\subsection{Tuning the QPC operating point} \label{sec:QP_operating}
Next, we investigate the best regime for operating the QPC as a
charge detector.
The conductance of a QPC depends strongly on the confinement
potential $U_\mathrm{QPC}(\vec{r})$. When operating the QPC in the
region between pinch-off and the first plateau $(0 < G < 2e^2/h)$, a
small perturbation $\delta U_\mathrm{QPC}(\vec{r})$ leads to a large
change in conductance $\delta G$. If a QD is placed in close
vicinity to the QPC, we expect a fluctuation $\delta q$ in the QD
charge population to shift the QPC potential
$U_\mathrm{QPC}(\vec{r})$ and thus give rise to a measurable change
in QPC conductance. A figure of merit for using the QPC as a charge
detector is then
\begin{equation}\label{eq:QtoG}
\frac{\delta G}{\delta q}=  \frac{\delta
G\left[U_\mathrm{QPC}(\vec{r})\right]}{\delta
U_\mathrm{QPC}(\vec{r})} \, \frac{\delta
U_\mathrm{QPC}(\vec{r})}{\delta q}.
\end{equation}
The first factor describes how the conductance changes with
confinement potential, which depends strongly on the operating point
of the QPC. The second factor describes the electrostatic coupling
between the QD and the QPC and is essentially a geometric property
of the system.

\begin{figure}[tb]
\centering
\includegraphics[width=\linewidth]{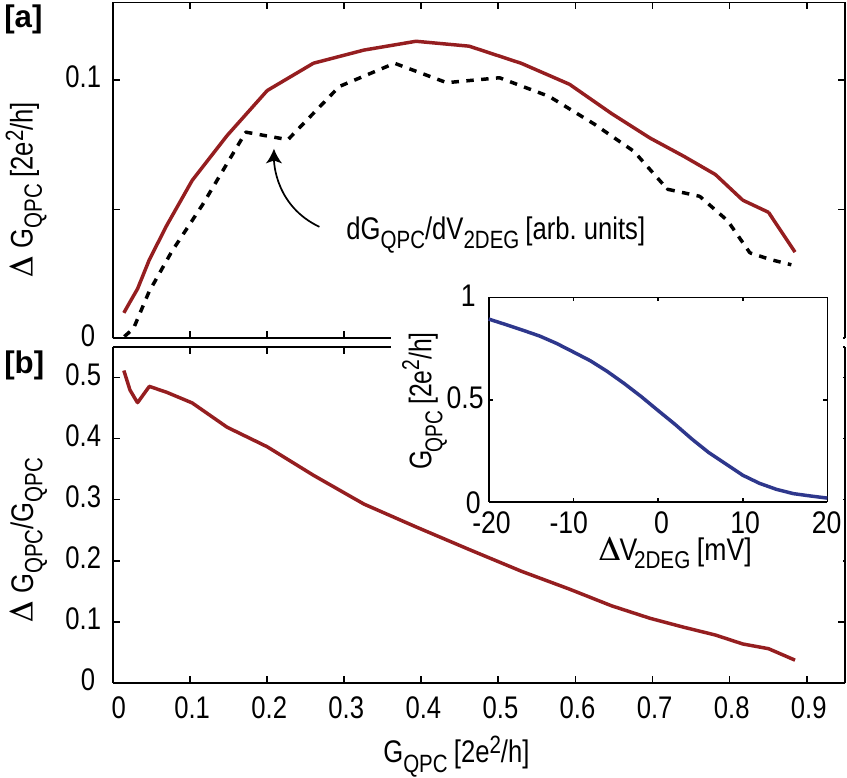}
\caption{ (a) Change of QPC conductance as one
electron enters the QD, measured for different values of average QPC
conductance.  The dashed line is the numerical derivative of the QPC
conductance with respect to gate voltage.  The change is maximal at
$G_\mathrm{QPC} = 0.4 \times 2 e^2/h$, which coincides with the
steepest part of the QPC conductance curve [see inset in (b)].
 (b) Relative change of QPC conductance for one electron entering the QD, defined as
$(G_\mathrm{high}-G_\mathrm{low})/G_\mathrm{high}$. The relative
change increases with decreased $G_\mathrm{QPC}$, reaching above
50\% at $G_\mathrm{QPC} = 0.02 \times 2 e^2/h$. The inset shows the
variation of $G_\mathrm{QPC}$ as a function of gate voltage.  Adapted from Ref.~\cite{gustavssonNWPRB:2008}.}
\label{fig:StepVsG}
\end{figure}

The performance of the charge detector depends strongly on the
operating point of the QPC. The best sensitivity for a device of
given geometry is expected when the QPC is tuned to the steepest
part of the conductance curve. This corresponds to maximizing the
factor $\delta G / \delta U_\mathrm{QPC}$ in \EqRef{eq:QtoG}. In
\FigRef{fig:StepVsG}(a) we plot the conductance change $\Delta G$
for one electron entering the QD versus QPC conductance, in the
range between pinch-off and the first conductance plateau ($0 <
G_\mathrm{QPC} <2e^2/h$). The change $\Delta G$ is maximal around $
G_\mathrm{QPC} \sim 0.4 \times 2 e^2/h$ but stays fairly constant
over a range from 0.3 to $0.6 \times 2 e^2/h$. The dashed line in
\FigRef{fig:StepVsG}(a) shows the numerical derivative of $\Gqpc$
with respect to gate voltage. The maximal value of $\Delta G$
coincides well with the steepest part of the QPC conductance curve.
The inset in the figure shows how
the conductance changes as a function of gate voltage. 

In \FigRef{fig:StepVsG}(b), we plot the relative change in
conductance $\Delta G/G_\mathrm{QPC}$ for the same set of data. The
relative change increases monotonically with decreasing conductance,
reaching above 50\% at $G_\mathrm{QPC} = 0.02 \times 2 e^2/h$.
The relative change in QPC conductance $\Delta \Gqpc/\Gqpc$ in this
particular device is extraordinarily large compared to top-gate
defined structures, where $\Delta \Gqpc/\Gqpc$ is typically around
one percent for the addition of one electron on the QD
\cite{vandersypen:2004, reilly:2007}. We attribute the large
sensitivity to the close distance between the QD and QPC ($\sim \!
50\nm$, due to the vertical arrangement of the QD and QPC) and to
the absence of metallic gates on the heterostructure surface, which
reduces screening.

The results of Fig.~\ref{fig:StepVsG} indicate that it may be
preferable to operate the charge detector close to pinch-off, where
the relative change in conductance is maximized.
The quantity relevant for optimal detector performance in the
experiment is the signal-to-noise (S/N) ratio between the change in
conductance $\Delta G$ and the noise level of the QPC conductance
measurement. We measure the conductance by applying a fixed bias
voltage $\Vsd$ across the QPC and monitoring the current. In the
linear response regime, both the average current $\Iqpc$ and the
change in current for one electron on the QD ($\Delta \Iqpc$) scale
linearly with applied voltage bias. The noise in the setup is dominated by
the voltage noise of the amplifier, which is essentially independent
of the QPC operating point and the applied bias in the region of
voltages discussed here. The S/N thus scales directly with $\Delta
\Iqpc$
\begin{equation}\label{eq:DeltaItoSN}
 S/N = \frac{\Delta \Iqpc^2}{\langle \Delta I^2_\mathrm{noise}\rangle}
 \propto \Vqpc^2  \, \Delta G^2  = \Iqpc^2  \left(\frac{\Delta
 G}{G}\right)^2.
\end{equation}
In practice the maximal usable QPC current is limited by effects
like heating or emission of radiation which can influence the
measured system. When considering heating effects, it becomes
important to minimize the power $P = \Vqpc \, \Iqpc$ dissipated in
the QPC. Putting the power dissipation as a constraint to
Eq.~(\ref{eq:DeltaItoSN}), the highest S/N is reached for the
maximal value of $(\Delta G)^2/G$. For the data shown in
Fig.~\ref{fig:StepVsG} this occurs at $\Gqpc = 0.2\times 2 e^2/h$.
However, this operation point requires a large voltage bias to be
applied to the QPC. If the QPC bias is larger than the
single-particle level spacing of the QD, the current in the QPC may
drive transitions in the QD and thus exert a back-action on the
measured device \cite{gustavssonPRL:2007} (see section \ref{sec:SP_main}).
Therefore, a better approach is to limit the QPC voltage. Here, the
best S/N is obtained when optimizing $\Delta G$ rather than $\Delta
G/G$ and operating the QPC close to $\Gqpc=0.5\times 2 e^2/h$. The
sensitivity of the QPC together with the bandwidth of the
measurement circuit allows a detection time of around $4~\mu s$
\cite{gustavssonAPL:2008}.

\subsection{Charge traps in the vicinity of the QPC} \label{sec:QP_ChargeTraps}
In the previous sections, we mentioned that charge fluctuations in traps in the vicinity of
the QPC may induce excess noise in the QPC current measurement \cite{jung:2004}.
If the trap is close enough and if the
fluctuations occur on a timescale slower than the measurement
bandwidth, the charge dynamics of the individual traps can be
investigated using the time-resolved charge detection methods. By
comparing the conductance change $\Delta G_\mathrm{trap}$ due to
charge fluctuations in a trap with the conductance change $\Delta
G_\mathrm{QD}$ due to an electron in the QD, we get an idea of the
position of the trap relative to the QD. The trap position may be
further pinned down by checking the influences of various gate
voltages \cite{gildemeister:2007}.

\begin{figure}[tb]
\centering
\includegraphics[width=\linewidth]{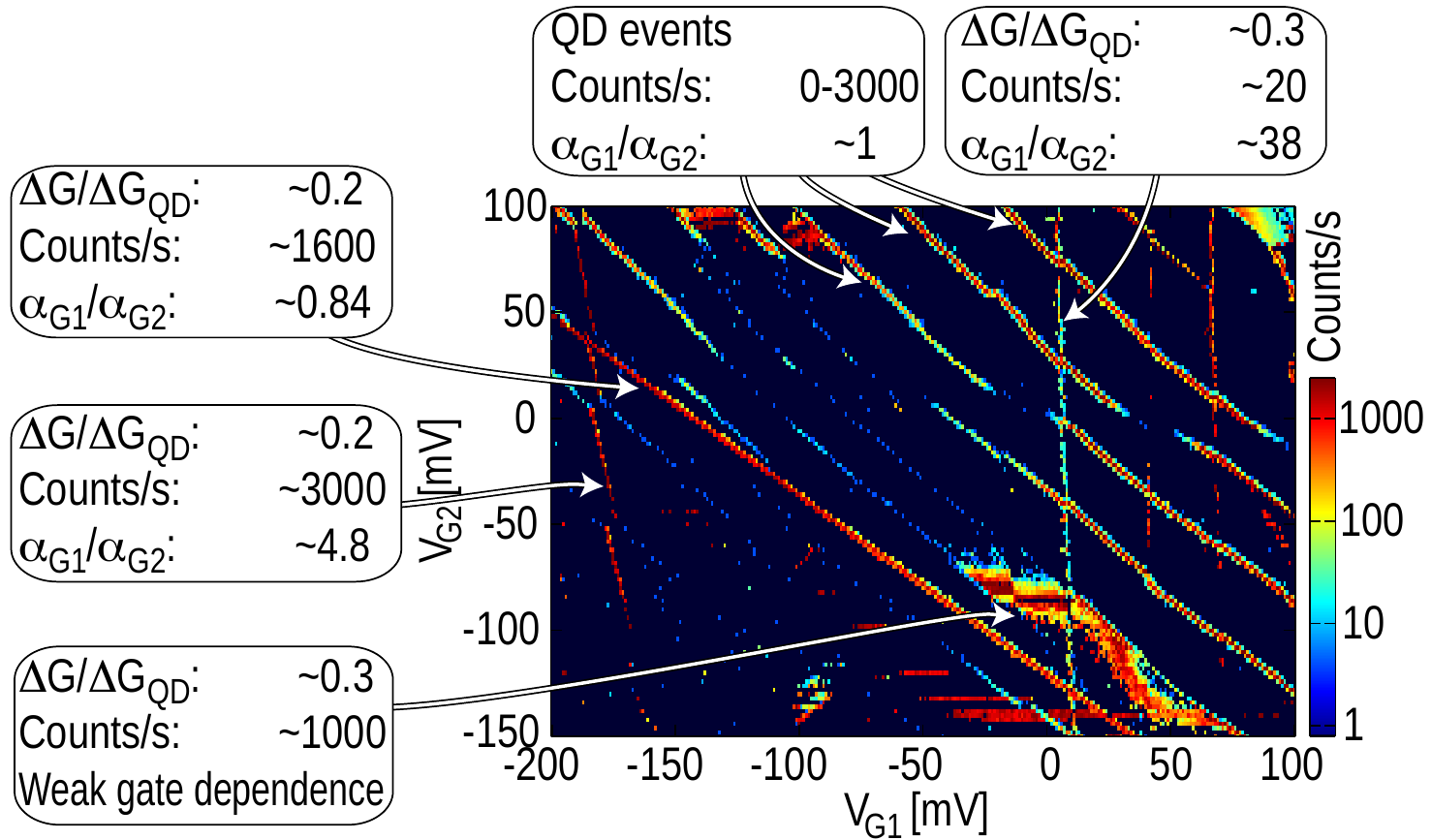}
\caption{Count rates for a single QD, measured vs voltage on the two
gates G1 and G2. Apart from the transitions due to electrons
tunneling into and out of the QD, there are several other lines
present in the figure. These originate from charge traps sitting in the
substrate close to the QPC. Such events can be distinguished from
tunneling in the QD by investigating the change of the QPC conductance
or looking at how the switches depend on gate voltages. The boxes
describe some of the transitions, where $\alpha_\mathrm{G1/G2}$ is
the capacitive lever arms of the gates relative to the trap. The electron temperature was 200 mK.
 } \label{fig:QP_TrapsQD}
\end{figure}

Figure~\ref{fig:QP_TrapsQD} shows electron counts registered by the
QPC charge detector for a QD-QPC structure defined by local
oxidation [see Fig.~\ref{fig:TR_sample}(a)]. The two
voltages $\Vgl$ and $\Vgr$ are applied to gates to the left and
right of the structure that have roughly the same capacitive lever
arms ($\aGl / \aGr \sim 1$) on the QD states. The lines with slope
$\Delta \Vgl/ \Delta \Vgr \sim -1$ in \FigRef{fig:QP_TrapsQD} all
give the same $\Delta \Gqpc$ and thus belong to tunneling in the QD.
In the lower-left region of the graph the tunneling in the QD
disappears due to pinch-off of the QD leads.

Various other lines are seen in the plot; their gate voltage
dependences and their influence on the QPC conductance are given in
the figure. Traps with $\aGl / \aGr > 1$ are situated closer to gate
G1, traps with $\aGl / \aGr < 1$ are closer to gate G2. The trap
with $\aGl / \aGr = 4.8$ seen to the left in the graph is probably
relatively close to the QD; the lines from the trap and the lines
from the QD anticross due to their mutual charging energy, similar
to a double quantum dot system. Almost all traps give a smaller
$\Delta \Gqpc$ compared to the QD, showing that the major influence
on QPC conductance still originates from the QD. We note that the
method only shows traps where the charge fluctuates on timescales
slower than the measurement bandwidth; traps with faster
fluctuations will give an overall increase in the noise floor.

\begin{figure}[tb]
\centering
\includegraphics[width=\linewidth]{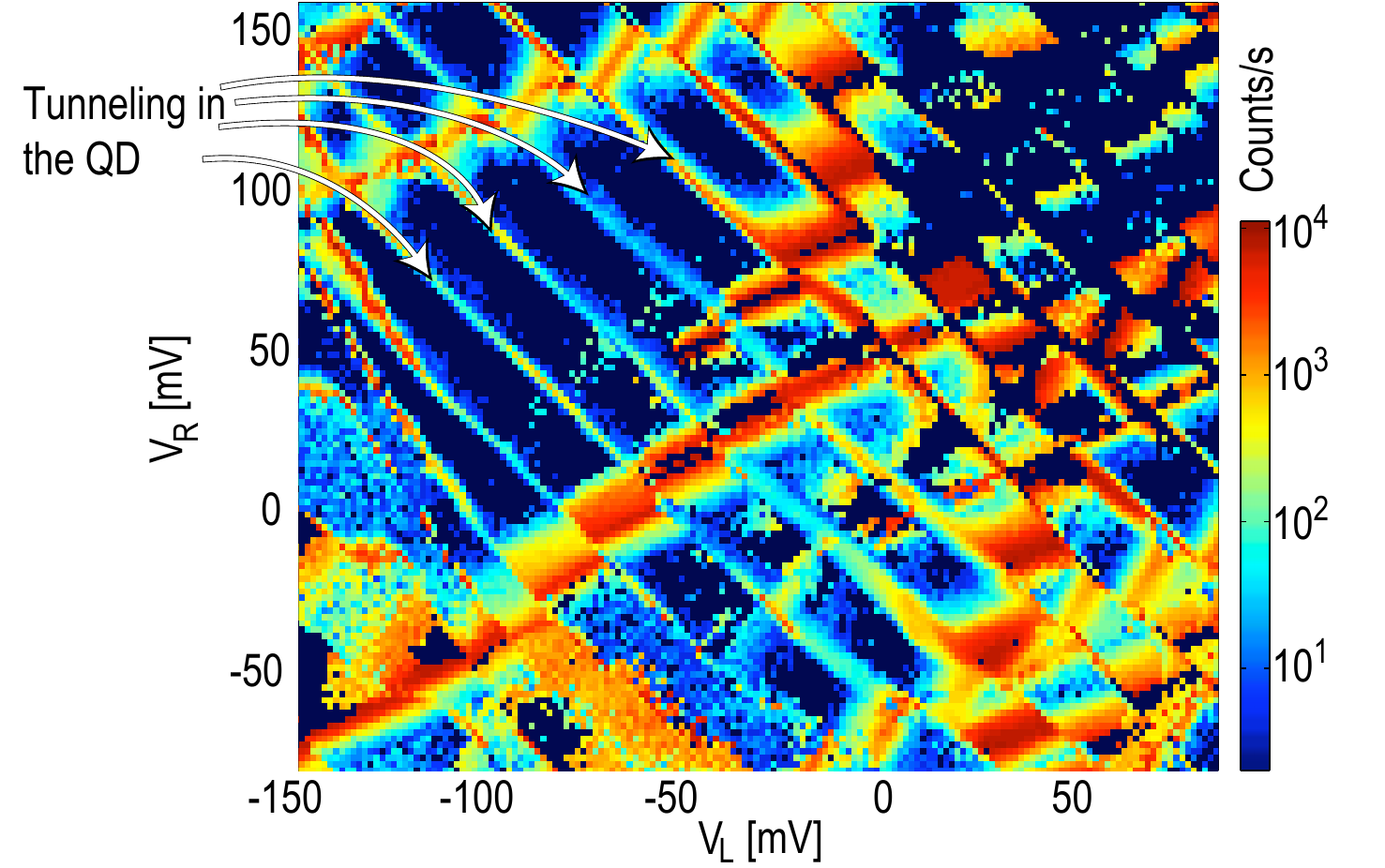}
\caption{Same as Fig.~\ref{fig:QP_TrapsQD}, this time measured for a
QPC defined by etching. The switches due to traps are more frequent
than in the AFM-defined sample, possibly because of surface states
formed in the etched trenches. The data was also taken at a higher electron temperature ($T=1.7\K$ instead of 200 mK)
 } \label{fig:QP_TrapsNW}
\end{figure}

It is not clear whether the charge traps are formed inside the
heterostructure or if they are sitting on the surface. In
\FigRef{fig:QP_TrapsNW}, we present a measurement similar to the one
shown in \FigRef{fig:QP_TrapsQD}, but this time for a QPC defined by
etching [see \FigRef{fig:QP_QuantizedC}(a)]. This sample shows a greater trap density
compared to the structure defined by local oxidation used in
\FigRef{fig:QP_TrapsQD}. The difference could be due to the
fabrication method; the etching procedure will bring surface states
closer to the QPC. For a structure defined by local oxidation, the
surface is kept further away. On the other hand, it is dangerous to
draw too far-going conclusions from the two sets of data; the
structures were fabricated on different (but similar) wafers, and
the measurement of \FigRef{fig:QP_TrapsNW} was performed at a higher
electron temperature ($T=1.7\K$ compared to $T=200\mK$). Further
experiments are necessary to clarify the issue.

\section{Time-resolved electron transport} \label{sec:TR_main}


In this section, we show how time-resolved charge-detection is
used to investigate properties of electron transport in a single
quantum dot. We start with describing the dynamics of electron
tunneling between one lead and a single QD state, before moving on
to more complex situations involving multiple leads, finite bias,
excited states and degenerate states. Finally, we show how the
potential landscape forming the tunnel barriers is influenced by
changing the gate voltages.

\subsection{Sample and experimental setup}
\begin{figure}[tb]
\centering
 \includegraphics[width=\linewidth]{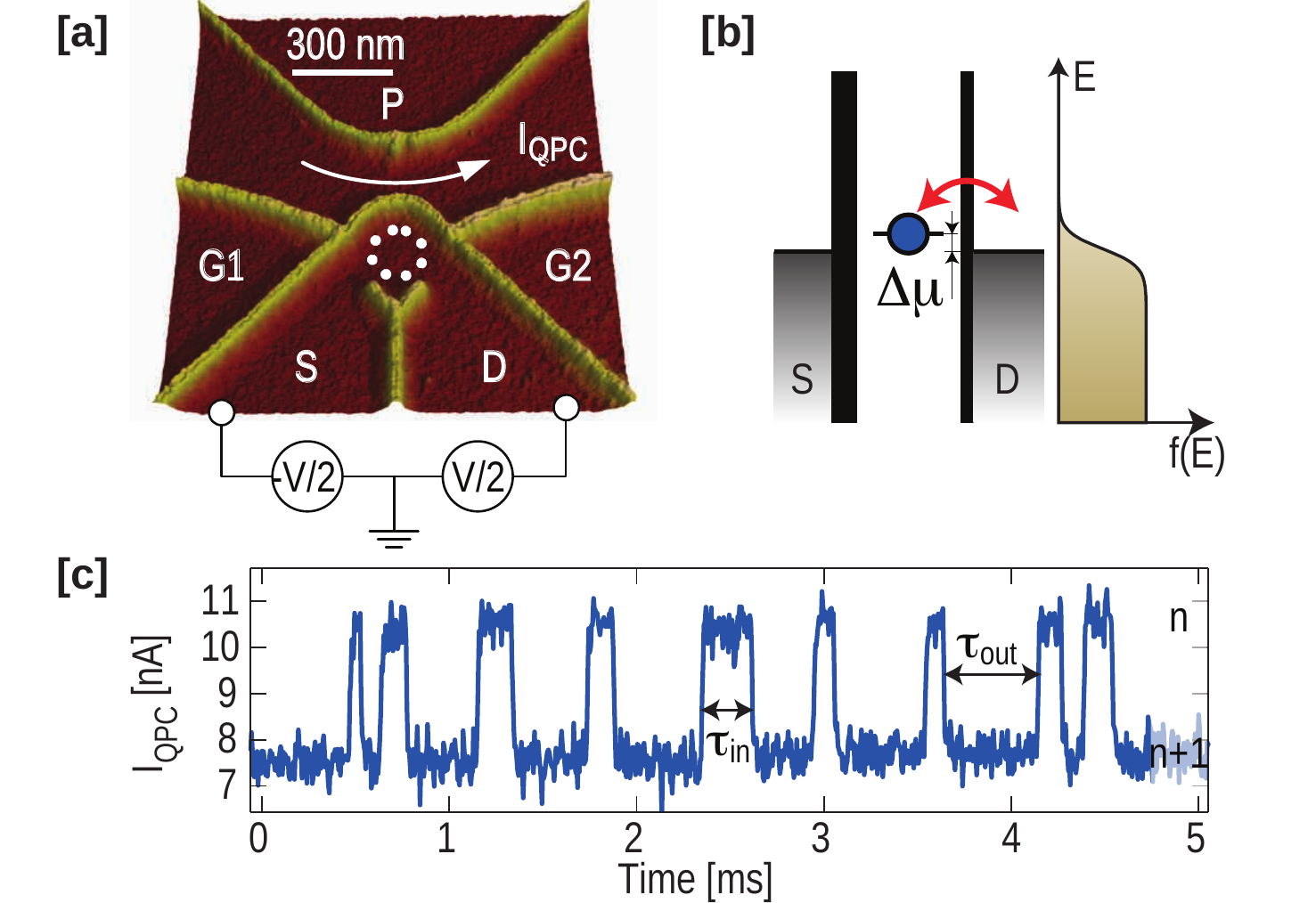}
 \caption{
 (a) Quantum dot with integrated charge read-out investigated in this section.
 (b) Schematic drawing depicting the electrochemical potentials of the system.
 By making the barrier between source and the QD very
 opaque, electron tunneling is only possible between the QD and the drain lead.
 (c) Current through the QPC as a function of time, showing a few
 electrons tunneling into and out of the QD. The lower current level corresponds to a situation
 where the QD holds one excess electron. Transitions between the
 two levels occur whenever an electron enters or leaves the QD. The quantities
 $\tau_{\mathrm{in}}$ and $\tau_{\mathrm{out}}$ specify the time
 it takes for an electron to tunnel into and out of the dot,
 respectively.
 }
\label{fig:TR_sample}
\end{figure}

The sample investigated in this section is shown in
Fig.~\ref{fig:TR_sample}(a). The structure was fabricated on a
GaAs-GaAlAs heterostructure containing a two-dimensional electron
gas 34 nm below the surface (density $4.5 \times 10^{15}$ m$^{-2}$,
mobility 25 m$^2$(Vs)$^{-1}$). An atomic force microscope (AFM) was
used to oxidize locally the surface, thereby defining depleted
regions below the oxide lines \cite{held:2002,fuhrer:2004}.

The sample consists of a QD [dotted circle in Fig.~\ref{fig:TR_sample}(a)]
and a nearby QPC. The charging energy of the QD is
$2.1~\mathrm{meV}$ and the mean level spacing is
$200-300~\mathrm{\mu eV}$. From the geometry and the characteristic
energy scales, we estimate that the QD contains about $30$
electrons. The QD is connected to source and drain leads through
tunnel barriers.
The transparency of the tunnel barriers is controlled by changing
the voltage on gates $G1$ and $G2$. In the experiment, we tune the
tunnel coupling rates between the QD and the leads to below 10 kHz.
This allows electron tunneling to be detected in real-time with the
low-bandwidth ($\sim\!30~\mathrm{kHz}$) detector.
The $P$ gate is used to tune the conductance of the QPC to a regime
where the sensitivity to changes in the QD charge is maximal. The voltage on gate $P$ is
adjusted to keep the QPC in the region of maximum sensitivity
whenever changing the voltage on another gate. The measurements were
performed in a dilution refrigerator with a base temperature of 60
mK.

\subsection{Electron tunneling with one lead connected to the quantum
dot} \label{sec:TR_oneLead}
First, we investigate the case of electron tunneling between a QD
and one lead. This is achieved by keeping the drain barrier open but
making the source barrier very opaque, allowing electron tunneling
only between the QD and the drain lead
[Fig.~\ref{fig:TR_sample}(b)].
Coulomb blockade prohibits the QD to hold more than one excess
electron.
When an electron enters the QD, the conductance through the QPC
is reduced due to the electrostatic coupling between the QD and the
QPC. As the electron leaves, the QPC conductance returns to the
original value.
This gives rise to a QPC current switching between two levels, as
shown in Fig.~\ref{fig:TR_sample}(c). The low level corresponds to a
situation where the dot holds an excess electron.
Transitions between the two levels occur whenever an electron enters
or leaves the QD. The duration between transitions gives directly
the time it takes for an electron to tunnel into or out of the QD.
In Fig.~\ref{fig:TR_sample}(c), these times are marked by
$\tau_{\mathrm{in}}$ and $\tau_{\mathrm{out}}$.


In the regime of single-level transport, the process of an electron
tunneling into or out of the dot is described by the rate equation
\begin{equation}\label{eq:TR_eqRate}
    \dot{p}_{\mathrm{in/out}}(t) = - \Gamma_{\mathrm{in/out}} \times
    p_{\mathrm{in/out}}(t).
\end{equation}
Here, $p_{\mathrm{in/out}}(t)$ is the probability density for an
electron to tunnel into or out of the dot at a time $t$ after a
complementary event. Solving the differential equation and
normalizing the resulting distribution gives
\begin{equation}\label{eq:TR_expDecay}
  p_{\mathrm{in/out}}(t) \mathrm{dt} = \mathrm{e}^{-\Gamma_{\mathrm{in/out}} t}
  \times \Gamma_{\mathrm{in/out}}\,\mathrm{dt}.
\end{equation}
The tunneling rates $\Gamma_{\mathrm{in/out}}$ in
Eqs.~(\ref{eq:TR_eqRate},~\ref{eq:TR_expDecay}) are effective rates
involving the dot-lead tunnel coupling $\Gamma$ and the thermal
population of the states in the lead, with
\begin{eqnarray}
 \Gamma_{\mathrm{in}} &=& \Gamma \times f(\Delta \mu/k_B T) \label{eq:TR_effGin} \\
 \Gamma_{\mathrm{out}} &=& \Gamma \times (1-f(\Delta \mu/k_B T)) \label{eq:TR_effGout}.
\end{eqnarray}
Here, $f(x) = 1/(1+\exp(x))$ is the Fermi distribution function, $T$
is the electron temperature in the lead and $\Delta \mu$ is the
energy difference between the electrochemical potential of the QD
and the Fermi level in the lead.
Equations~(\ref{eq:TR_effGin}-\ref{eq:TR_effGout}) are valid in a
small range around $\delta \mu=0$ where the tunnel coupling $\Gamma$
can be assumed to be independent of energy and gate voltages. The
gate-voltage influence on the tunnel coupling is investigated in
greater detail in section \ref{sec:TR_tuneCoupling}. Also,
Eqs.~(\ref{eq:TR_effGin}-\ref{eq:TR_effGout}) assume the QD state to
be non-degenerate. In the case of degenerate states, the rates
should be multiplied with the appropriate degeneracy factor. Here,
we assume non-degenerate states and postpone the discussion of
degenerate states to section \ref{sec:TR_degenerateStates}.

The method of time-resolved charge detection makes it possible to
test the validity of the model described in
Eqs.~(\ref{eq:TR_eqRate}-\ref{eq:TR_effGout}).
The tunneling rates $\Gamma_\mathrm{in}$, $\Gamma_\mathrm{out}$ are
determined directly from time traces such as the one shown in
Fig.~\ref{fig:TR_sample}(c). Using Eq. \ref{eq:TR_expDecay}, we find
\begin{equation}\label{eq:TR_tauToGamma}
\Gamma_{\mathrm{in}} = 1/\langle \tau_{\mathrm{in}} \rangle,
~\Gamma_{\mathrm{out}} = 1/\langle \tau_{\mathrm{out}} \rangle,
\end{equation}
where $\langle \tau_{\mathrm{in}} \rangle$ and $\langle
\tau_{\mathrm{out}} \rangle$ are the average tunneling times
extracted from the time trace.
It should be noted that the expression in
Eq.~(\ref{eq:TR_tauToGamma}) is valid only for an infinite-bandwidth
detector, and that the finite bandwidth of the detector leads to a
systematic under-estimation of the actual rates. However, knowing
the bandwidth makes it possible to correct for the deviations
\cite{naaman:2006}. The influence of the detector bandwidth is
discussed in greater detail in section~\ref{sec:ST_higherMoments}.

Combining Eqs.~(\ref{eq:TR_effGin}-\ref{eq:TR_tauToGamma}) gives an
expression for the Fermi function
\begin{equation}\label{eq:TR_fermiPop}
 f(\Delta E/k_B T) = \langle \tau_{\mathrm{out}}\rangle/(\langle
 \tau_{\mathrm{in}}\rangle + \langle \tau_{\mathrm{out}}\rangle) = \langle
 n_\mathrm{excess} \rangle,
\end{equation}
with $\langle n_\mathrm{excess}\rangle$ being the average excess
charge on the QD. The average dot population can be determined by
monitoring the average conductance of the QPC \cite{dicarlo:2004}.
By adding time resolution to the detector and counting electrons one
by one as they enter the QD, we can extract not only the Fermi
function of the lead but also the tunnel coupling $\Gamma$. Assuming
sequential tunneling and using
Eqs.~(\ref{eq:TR_effGin}-\ref{eq:TR_effGout}), we find that the rate
for electrons entering the dot $r_E$ is given by
\begin{equation}\label{eq:TR_countsPerSec}
    r_E = 1/(\langle \tau_{\mathrm{in}}\rangle + \langle \tau_{\mathrm{out}}\rangle)
    = \Gamma \times f (1-f).
\end{equation}
Measuring the count rate $r_E$ thus directly determines the tunnel
coupling $\Gamma$.

\begin{figure}[tb]
\centering
 \includegraphics[width=0.95\linewidth]{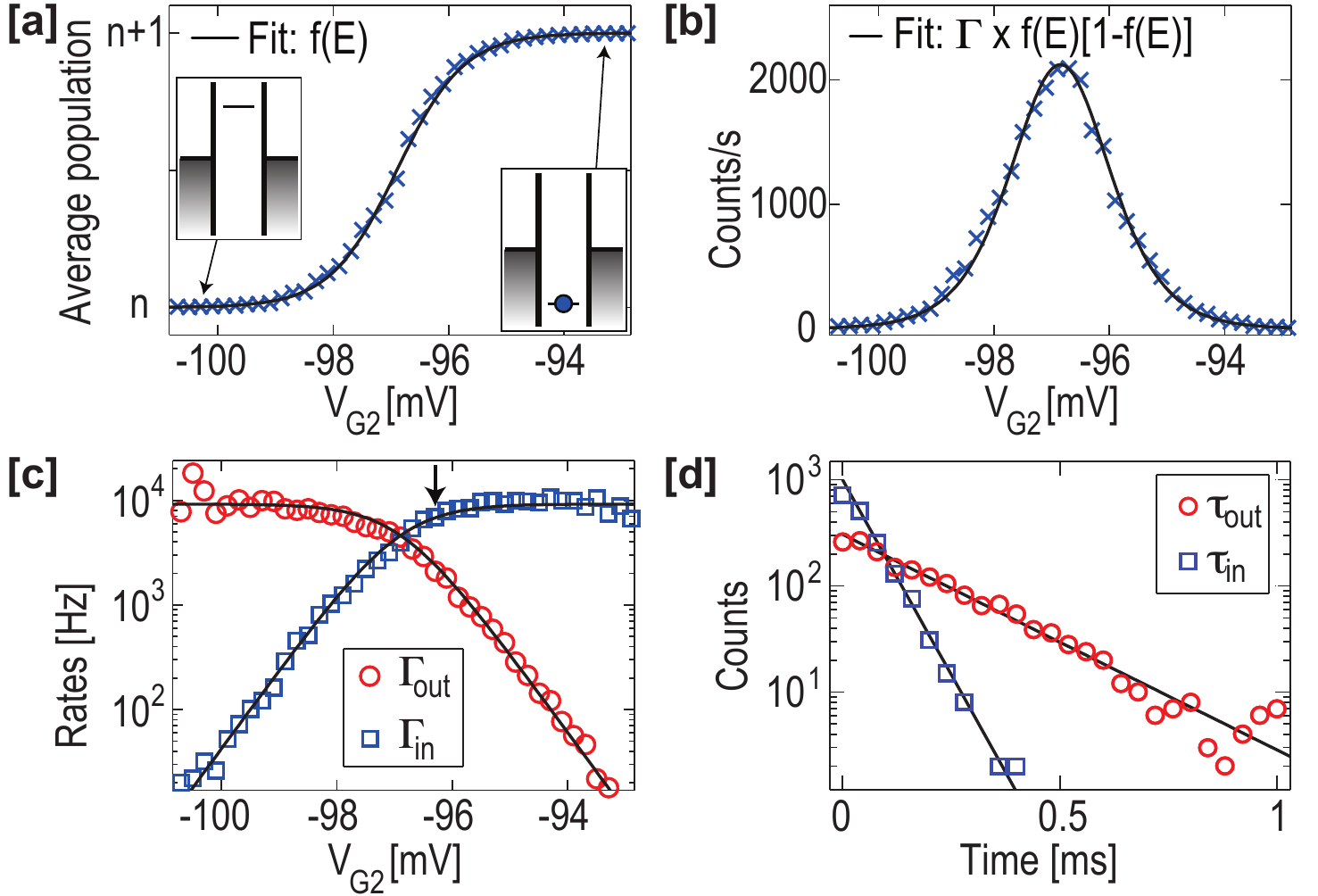}
 \caption{
 (a) Average dot population versus voltage on gate $G2$. The fit shows the Fermi distribution function
 with $T = 230~\mathrm{mK}$.
 (b) Counts per second for the same data as in (a). The data was fit to Eq.~(\ref{eq:TR_countsPerSec}), giving $\Gamma =
9.2~\mathrm{kHz}$ and $T = 230~\mathrm{mK}$.
 (c) Tunneling rates for electrons entering (squares) and leaving (circles) the
 QD, extracted from the same set of data as in (a,~b). The solid
 lines are the results of
 Eqs.~(\ref{eq:TR_effGin}-~\ref{eq:TR_effGout}) in the text,
 with $\Gamma = 9.2~\mathrm{kHz}$ and $T = 230~\mathrm{mK}$.
 (d) Distribution of tunneling times for electrons entering (squares) and
 leaving (circles) the QD, extracted at $V_\mathrm{G2} =
 -96.3~\mathrm{mV}$ [marked by arrow in (c)].
 The solid lines show the exponential behavior given by Eq. (\ref{eq:TR_expDecay}) in the text, with $\Gamma_\mathrm{in} = 1/\langle \tau_\mathrm{in}\rangle =
 7.2~\mathrm{kHz}$, $\Gamma_\mathrm{out} = 1 / \langle \tau_\mathrm{out}\rangle =
 2.0~\mathrm{kHz}$. The length of the time trace for the data shown in the figure is 0.5 s.
 }
\label{fig:TR_fermi}
\end{figure}

In Fig. \ref{fig:TR_fermi}(a,~b) we plot the average QD population
and the number of counts per second as gate $G2$ was used to change
the electrochemical potential of the QD. The solid lines are the
fits to Eq.~(\ref{eq:TR_fermiPop}) and
Eq.~(\ref{eq:TR_countsPerSec}), demonstrating the good agreement
between the data and the expected relations. By first determining
the lever arm between gate $G2$ and the dot from standard Coulomb
diamond measurements \cite{kouwenhoven:1997}, it was possible to
extract the electronic temperature ($T=230~\mathrm{mK}$) from the
width of the Fermi function. The same temperature was found by
checking the width of standard Coulomb blockade current peaks
\cite{kouwenhoven:1997}, measured with the QD in a more strongly
coupled regime.

The time-resolved detection method also allows the tunneling rates
$\Gamma_\mathrm{in}$ and $\Gamma_\mathrm{out}$ to be determined
separately. The rates are plotted in Fig.~\ref{fig:TR_fermi}(c),
extracted from the same set of data as shown in
Fig.~\ref{fig:TR_fermi}(a,~b). The solid lines are fits to
Eqs.~(\ref{eq:TR_effGin}-\ref{eq:TR_effGout}), with
$\Gamma=9.2~\mathrm{kHz}$ and $T=230~\mathrm{mK}$. The figure
clearly demonstrates an exponential falloff of the tunneling rates
as the QD electrochemical potential is shifted above or below the
Fermi level of the lead. This is a direct consequence of the Fermi
distribution for the electrons in the lead.
 The fact that both $\Gamma_\mathrm{in}$ and
$\Gamma_\mathrm{out}$ can be fitted with a single tunneling rate
$\Gamma$ shows that the QD state is non-degenerate. This is not
always the case, as will be seen in section
\ref{sec:TR_degenerateStates}.

The results presented so far rely on the assumption that
Eq.~(\ref{eq:TR_expDecay}) is correct. The validity of this
assumption can be tested by extracting the experimental distribution
function $p_{\mathrm{in/out}}(t)$ of tunneling times
$\tau_{\mathrm{in}}$, $\tau_{\mathrm{out}}$ from a time trace
containing a large number of events. Such distributions are shown in
Fig.~\ref{fig:TR_fermi}(d), taken at the position marked by the
arrow in Fig.~\ref{fig:TR_fermi}(c). The data exhibit the expected
exponential behavior of Eq.~(\ref{eq:TR_expDecay}), with dashed
lines being fits with $\Gamma_{\mathrm{in}} = 7.2~\mathrm{kHz}$ and
$\Gamma_{\mathrm{out}} = 2.0~\mathrm{kHz}$.

The measurements presented so far only involve tunneling between the
QD and one lead. These tunneling events are due to equilibrium
fluctuations and do not give rise to a net current through the QD.
Consequently, it is impossible to investigate such effects with
conventional current measurement techniques. This demonstrates the
power of time-resolved charge detection methods for probing
properties of mesoscopic structures.
The overall good agreement between
Eqs.~(\ref{eq:TR_expDecay}-\ref{eq:TR_effGout}) and the results of
Fig.~\ref{fig:TR_fermi} makes us confident that the model of
single-electron tunneling is well capable of describing the system.
Next, we move on to the case where the QD is connected to two leads.

\subsection{Electron tunneling with two leads connected to the quantum dot}
\begin{figure}[tb]
\centering
 \includegraphics[width=\linewidth]{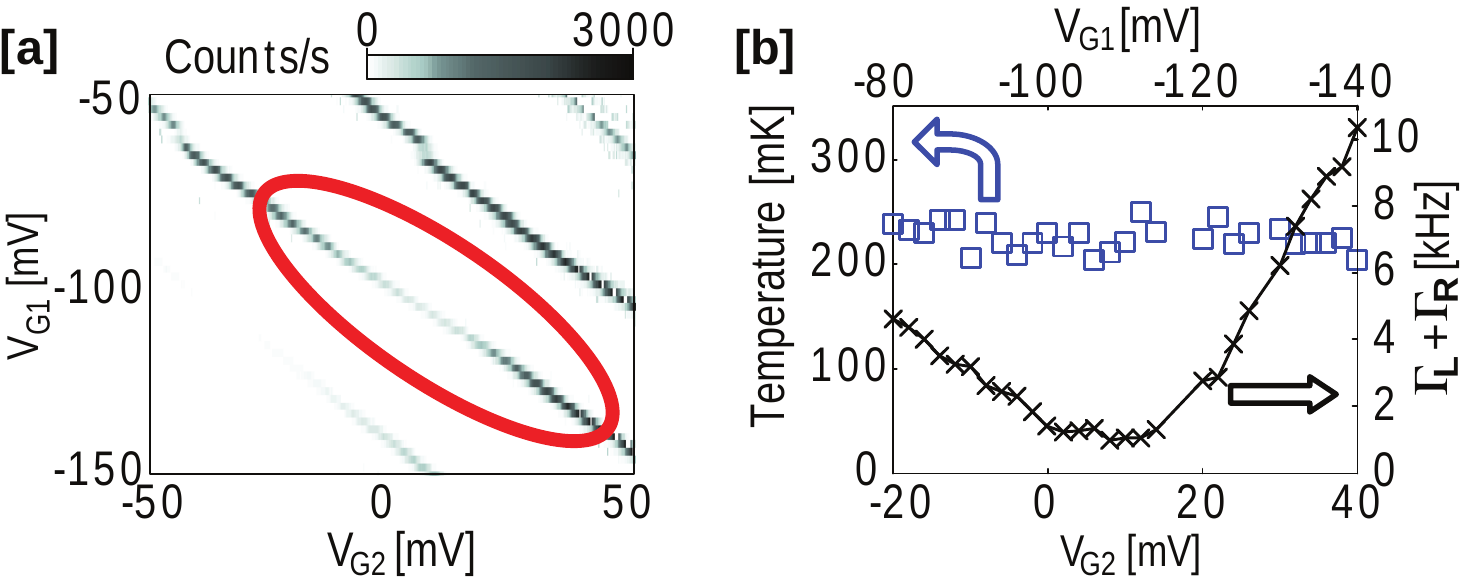}
 \caption{
 (a) Counts per second versus $V_\mathrm{G1}$ and $V_\mathrm{G2}$. For low values
 of $V_\mathrm{G1}$ and $V_\mathrm{G2}$, both the source lead and the drain lead are
 pinched off. For high voltages, the barriers open up so that tunneling occurs on a timescale
 faster than the measurement bandwidth.
 (b) Temperature (squares) and tunnel coupling (crosses), extracted from data shown within the ellipse in
 (a). As $V_{G2}$ is increased, $V_{G1}$ is decreased, in order
 to keep the dot at a constant potential. For low $V_{G2}$,
 tunneling occurs between the source lead and the dot, for high
 $V_{G2}$, the electrons tunnel between the drain and the dot. For
 intermediate gate values, both leads contribute to the tunneling.
 The electron temperature was found to be the same for both leads,
 within the accuracy of the experimental data.  Adapted from Ref.~\cite{gustavsson:2006}.
 }
\label{fig:TR_temp}
\end{figure}

In order to perform time-resolved measurements of electron transport
through the dot, the tunnel barriers have to be symmetrized so that
both give similar tunneling rates. The rates must be kept lower than
the bandwidth of the setup, but still high enough to give good
statistics. Figure~\ref{fig:TR_temp}(a) shows the number of events
per second as a function of the two gate voltages $V_{G1}$ and
$V_{G2}$. In the upper left corner of the figure, $V_{G1}$ is high
and $V_{G2}$ is low, corresponding to the case where the source lead
is open and the drain lead is closed. In the bottom right corner,
the configuration is inverted. For the region in between, marked by
the ellipse in Fig. \ref{fig:TR_temp}(a), the data indicate that
both leads are weakly coupled to the dot.

For zero voltage bias across the QD, the measurement method does not
enable us to distinguish whether an electron that tunnels into the
dot arrives from the left or from the right lead. Therefore, when
both leads are connected to the dot, the rates in Eqs.
(\ref{eq:TR_effGin}-\ref{eq:TR_effGout}) must be adjusted to contain
one part for the left lead and one part for the right lead,
\begin{eqnarray}\label{eq:TR_bothLeadGamma}
 \nonumber &\Gamma_{\mathrm{in}} = \Gamma^{\mathrm{in}}_L + \Gamma^{\mathrm{in}}_R =
 \Gamma_L f_L + \Gamma_R f_R, \\
 &\Gamma_{\mathrm{out}} = \Gamma^{\mathrm{out}}_L + \Gamma^{\mathrm{out}}_R =
 \Gamma_L (1-f_L) + \Gamma_R (1-f_R).
\end{eqnarray}
Here, $f_L$ and $f_R$ are the Fermi distribution functions of the
left and the right lead, respectively. Using Eq.
(\ref{eq:TR_bothLeadGamma}), we calculate the rate of events for the
case when both leads are coupled to the dot with rates accessible
for the detector,
\begin{equation}\label{eq:TR_eventPerSec2}
 r_E = \frac{[\Gamma_L f_L + \Gamma_R f_R] [\Gamma_L (1-f_L) + \Gamma_R (1-f_R)]}
 {\Gamma_L + \Gamma_R}.
\end{equation}
With no bias applied across the dot, the two distributions functions
$f_L$ and $f_R$ are identical except for a possible difference in
electronic temperature in the two leads. However, assuming $T_L =
T_R = T$, we have $f_L = f_R = f$, and Eq.
(\ref{eq:TR_eventPerSec2}) simplifies to $r_E = (\Gamma_L +
\Gamma_R) \times f(1-f)$. Fitting this expression to curves similar
to that shown in Fig. \ref{fig:TR_fermi}(b), we extract the
temperature and combined tunneling rate $\Gamma_L + \Gamma_R$ from
the data within the ellipse of Fig. \ref{fig:TR_temp}(a). The result
is presented in Fig. \ref{fig:TR_temp}(b). The rates and the
temperature shown in the graph are due to the combined tunneling to
and from both leads. Still, for low $V_\mathrm{G2}$/high
$V_\mathrm{G1}$, the drain lead is pinched off and tunneling occurs
mainly between the source lead and the dot. For high
$V_\mathrm{G2}$/low $V_\mathrm{G1}$, the source is pinched off and
the tunneling is dominated by electrons going between the drain and
the dot. The fact that the electronic temperatures extracted from
both regimes turn out to be the same ($T=230~\mathrm{mK}$) within the accuracy of the
analysis justifies the assumption that $T_L =
T_R$.

\subsection{Finite bias} \label{sec:TR_finiteBias}
\begin{figure}[tb]
\centering
 \includegraphics[width=\linewidth]{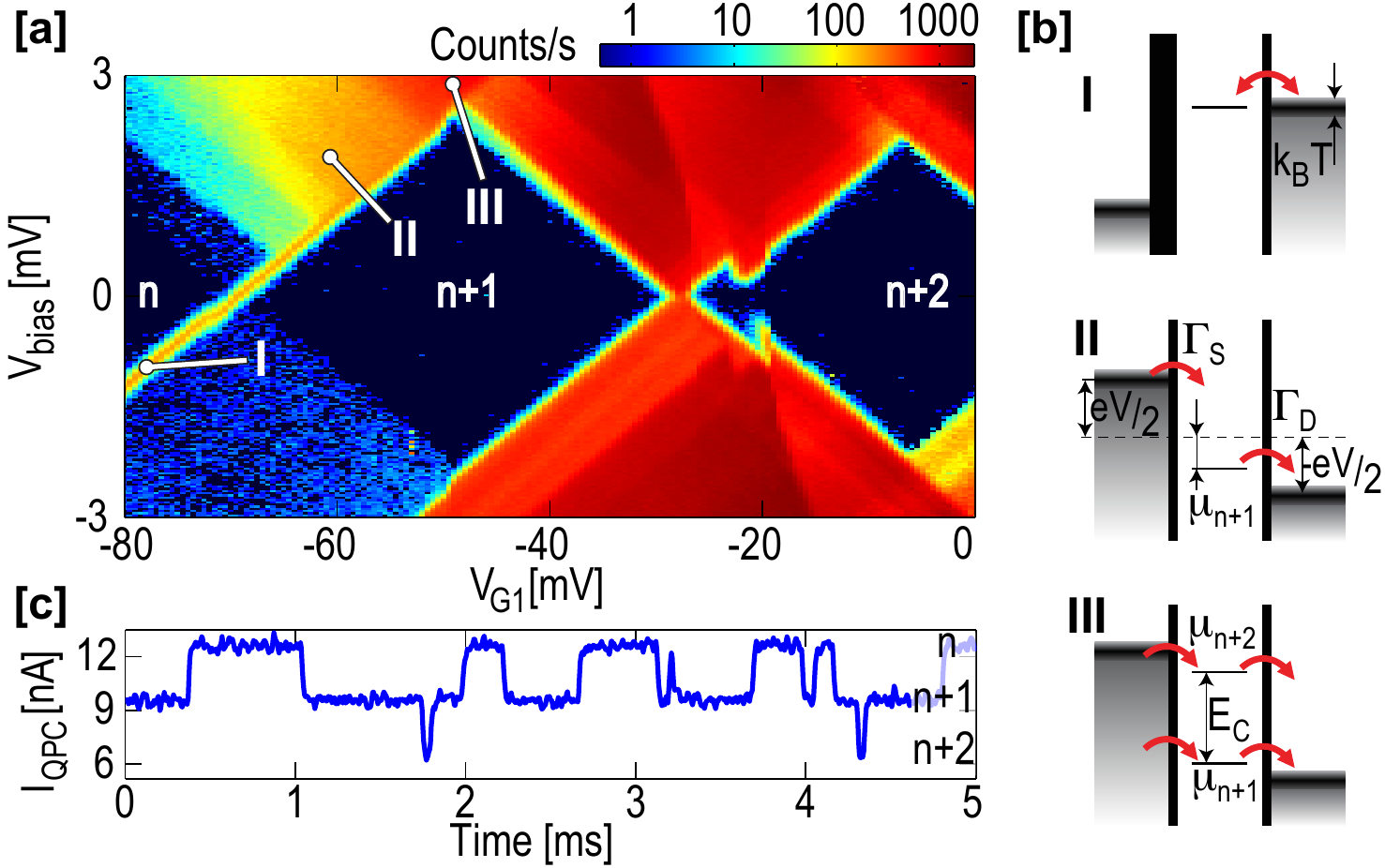}
 \caption{(a) Coulomb diamonds, measured by counting electrons entering the QD. For
 low values of $V_\mathrm{G1}$, the source lead is pinched off and
 tunneling can only occur between the dot and the
 drain lead. As $V_\mathrm{G1}$ increases, the source lead opens up and a current
 can flow through the dot.
 (b) Diagrams depicting the energy levels of the dot at points I, II and III. In case III,
 the bias is higher than the charging energy of the dot, meaning that
 the dot may hold 0, 1 or 2 excess electrons. (c) Time trace
 taken at point III. The three possible dot populations ($n$, $n+1$
 or $n+2$ electrons) are clearly resolvable. Adapted from Ref.~\cite{gustavsson:2006}.
 }
\label{fig:TR_CD}
\end{figure}

With the barriers properly symmetrized, we apply a finite bias
voltage between source and drain leads and measure electron
transport through the QD. Figure~\ref{fig:TR_CD}(a) shows Coulomb
blockade diamonds measured by counting electrons entering the QD.
The bias is applied symmetrically, with
\begin{equation}\label{eq:TR_symmBias}
    \mu_\mathrm{S} = |e|V_\mathrm{SD}/2,~~~~~~
    \mu_\mathrm{D} = -|e|V_\mathrm{SD}/2.
\end{equation}
The gate $G1$ is used as a plunger gate to control the dot
electrochemical potential. However, the gate also strongly affects
the source tunnel barrier. For low $G1$ voltages, the source lead is
closed, giving strong charge fluctuations only when the drain lead
is in resonance with the dot [see case I in
Fig.~\ref{fig:TR_CD}(a,~b)].

At higher gate voltages, the source lead opens up and a current can
flow through the dot. In point II of Fig. \ref{fig:TR_CD}(a), the QD
electrochemical potential $\mu_n$ lies within the bias window but
far away from the thermal broadening of the Fermi distribution in
the leads. The condition can be expressed as
\begin{equation}\label{eq:TR_condCurrent}
    |\!\pm \!eV/2-\mu_n| \gg k_B T,
\end{equation}
where the ''+'' case refers to the source contact and the ''-'' case
refers to the drain. Whenever Eq. (\ref{eq:TR_condCurrent}) is
fulfilled, electrons can only enter the dot from the source lead and
only leave through the drain. This makes it possible to determine
the individual tunnel couplings
$\Gamma_\mathrm{S}/\Gamma_\mathrm{D}$, with
\begin{equation}\label{eq:TR_gammaInSource}
 \Gamma_\mathrm{S} = \Gamma_{\mathrm{in}} = 1/\langle \tau_{\mathrm{in}} \rangle, ~~~
 \Gamma_\mathrm{D} = \Gamma_{\mathrm{out}} = 1/\langle \tau_{\mathrm{out}} \rangle.
\end{equation}
In this regime, we measure the current through the dot by counting
events. This opens up the possibility to use the QD as a very
precise current meter for measuring sub-fA currents
\cite{bylander:2005, fujisawa:2006}. Since the electrons are
detected one by one, the noise and higher order correlations of the
current can also be experimentally investigated. This is explained
in more detail in section \ref{sec:ST_subPoisson}

When the bias exceeds the dot charging energy, $E_\mathrm{C} \sim
2.1~\mathrm{meV}$, and the electrochemical potentials of the $(n+1)$
and the $(n+2)$ states are within the bias window [see case III of
Fig \ref{fig:TR_CD}(a,b)], transport processes are allowed where the
dot may contain 0, 1 or 2 excess electrons. A time trace measured at
point III of Fig. \ref{fig:TR_CD}(a) is shown in Fig.
\ref{fig:TR_CD}(c). The sensitivity of the QPC charge detector
allows to measure switching between three different levels,
corresponding to $(n)$, $(n+1)$ and $(n+2)$ electrons on the dot. It
is not possible to make this distinction in a standard current
measurement.

\subsection{Excited states} \label{sec:TR_excitedStates}
If there are excited states inside the bias window, tunneling may
occur into any of the available states. In this regime, the rates
$\Gamma_\mathrm{S}$ and $\Gamma_\mathrm{D}$ of
Eq.~(\ref{eq:TR_gammaInSource}) will not be the tunneling rates of
the single ground state but rather a sum of rates from all states
contributing to the tunneling process. A further complication with
excited states is that there may be equilibrium charge fluctuations
between the lead and the excited state, thereby removing the
unidirectionality of the electron motion. However, if the relaxation
rate of the excited state into the ground state is orders of
magnitude faster than the tunneling-out rate, the electron in the
excited state will have time to relax to the ground state before
equilibrium fluctuations can take place.

\begin{figure}[tb]
\centering
 \includegraphics[width=0.9\linewidth]{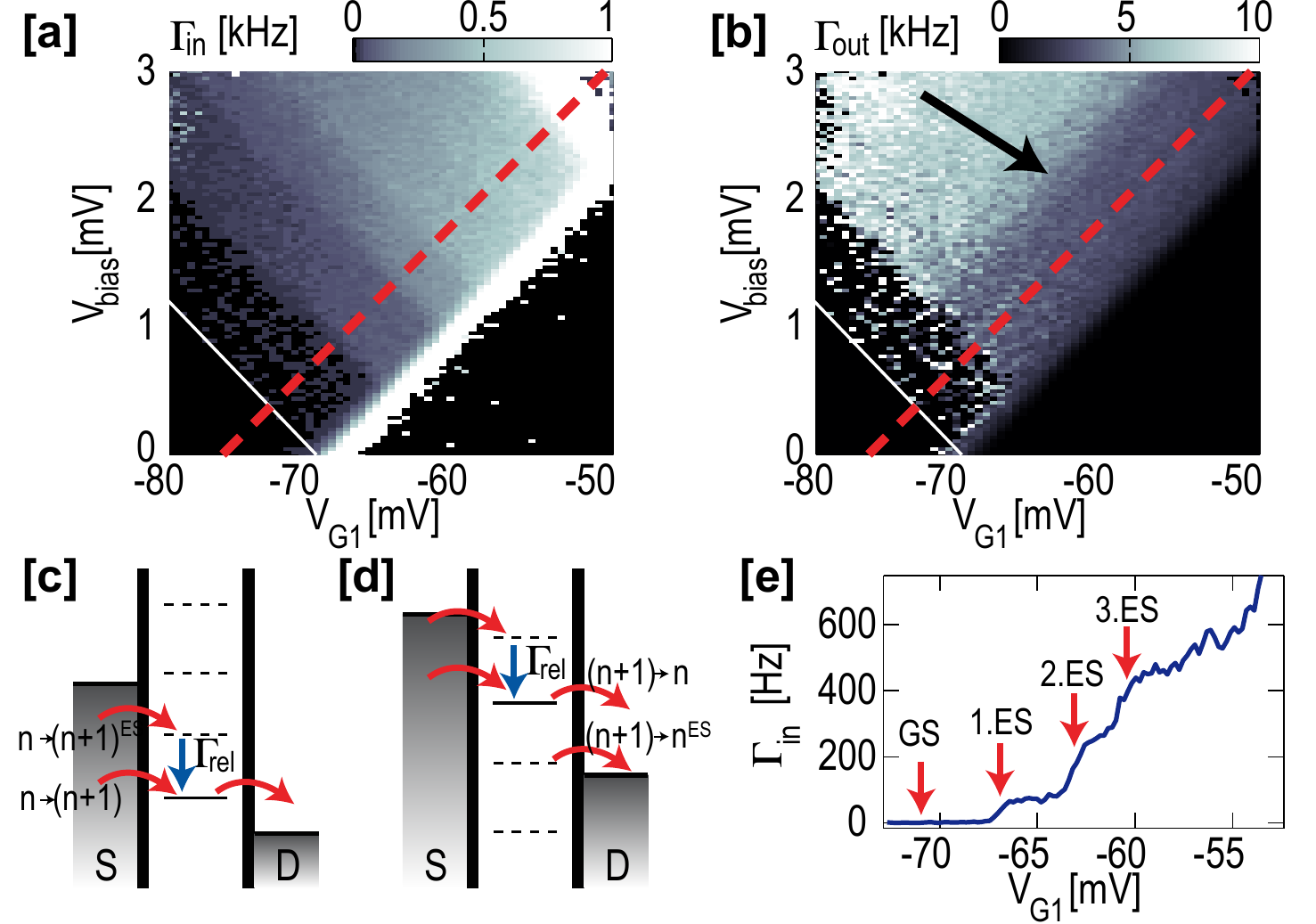}
 \caption{(a) and (b): Blow-up of the upper left region of Fig. \ref{fig:TR_CD}(a), showing the
 rates for electrons tunneling into (a) and out of (b) the QD,
 respectively. The white solid lines mark the positions where the
 source lead lines up with the electrochemical potential of the QD ground
 state. The dashed lines mark the lower edge of the region where condition of Eq. (\ref{eq:TR_condCurrent}) in the text is fulfilled. The color scales are different for the two figures,
 the rate for tunneling out is roughly 10 times faster than tunneling
 in.
 (c) Diagram depicting the energy levels along the dashed lines in
 (a) and (b). As the source lead is raised [corresponds to going upward
 along the dashed lines in (a)], excited states become
 available for tunneling.
 (d) Energy diagram for the configuration marked by the arrow in (b).
 Here, the excited states is visible in the rate for electrons tunneling out of the QD.
 (e) Tunneling rate for electrons entering the dot, measured along
 the dashed line in (a). Three excited states are clearly resolvable. Adapted from Ref.~\cite{gustavsson:2006}.
 }
\label{fig:TR_excitedStates}
\end{figure}

The separate rates $\Gamma_{\mathrm{in}}$ and
$\Gamma_{\mathrm{out}}$ for a close-up of the upper-left region of
Fig.~\ref{fig:TR_CD}(a) are plotted in
Fig.~\ref{fig:TR_excitedStates}(a,~b). It is important to note that
the requirement of Eq.~(\ref{eq:TR_condCurrent}) is met only for the
region along and above the dashed lines in the figures. At the lower
left end of the dashed lines, the energy levels of the dot are
aligned as shown in Fig.~\ref{fig:TR_excitedStates}(c). Going
diagonally upward along the lines corresponds to raising the Fermi
level of the source lead, while keeping the energy difference
between the dot and the drain lead fixed.

Starting at low bias and low voltage on the gate $V_\mathrm{G1}$,
the dot is in the Coulomb blockade regime, and no tunneling is
possible. Following the dashed line upwards, the QD ground state
becomes available for tunneling at
$V_{\mathrm{bias}}=0.3~\mathrm{mV}$. The transition is marked by the
white solid lines in Fig.~\ref{fig:TR_excitedStates}(a,~b). At these
low gate voltages, the source tunnel barrier is almost completely
pinched off, meaning that the rate for electrons entering the QD is
still low [Fig.~\ref{fig:TR_excitedStates}(a)]. Even so, some
electrons do enter the QD, as can be seen from the few points of
measurements of $\Gamma_\mathrm{out}$ within the corresponding
region of Fig.~\ref{fig:TR_excitedStates}(b).

We first concentrate on the tunneling-in rate in
Fig.~\ref{fig:TR_excitedStates}(a). As the source level is further
raised, excited states become available for transport. The first
excited state (at $V_{\mathrm{bias}}=0.85~\mathrm{mV}$ along the
dashed line) is more strongly coupled to the lead than the ground
state, giving a tunneling rate of $\sim\! 70~\mathrm{Hz}$ for
electrons entering the dot. The large difference in the tunneling-in
rate between the ground and the excited state can be understood if
the wavefunctions of the ground and excited state have different
spatial distributions. If the overlap with the lead wavefunction is
larger for the excited state, the tunneling rate will also be
larger. Similar differences in tunneling rates have been found
between the singlet and triplet states in a two-electron dot
\cite{hanson:2005, ciorga:2000}.

By further raising the source level, tunneling can also occur
through a second excited state. The measured tunneling-in rate will
now be the sum of the rates from both excited states; by subtracting
the contribution from the first state, the rate for the second state
can be determined. Using this method, we can resolve three excited
states, with excitations energies $\varepsilon_1 =
0.55~\mathrm{meV}$, $\varepsilon_2 = 1.0~\mathrm{meV} $,
$\varepsilon_3 = 1.3~\mathrm{meV}$ and with tunneling rates
$\Gamma_1 = 70~\mathrm{Hz}$, $\Gamma_2 = 190~\mathrm{Hz}$, $\Gamma_3
= 190~\mathrm{Hz}$. The excited states are clearly seen in
Fig.~\ref{fig:TR_excitedStates}(e), which is a cut along the dashed
diagonal line in Fig.~\ref{fig:TR_excitedStates}(a).


Focusing on the rates for electrons tunneling out of the QD
[Fig.~\ref{fig:TR_excitedStates}(b)], there is a noisy region where
the ground state but no excited states are within the bias window
($0.3<V_{\mathrm{bias}}<0.85~\mathrm{mV}$ along the dashed line). In
this regime, few electrons will enter the dot, meaning that the
statistics needed for measuring the rate of electrons leaving the
dot is not sufficient. However, for bias voltages higher than the
first excited state, the tunneling-out rate remains constant along
the dashed line. This is in contrast to the steps seen in the
tunneling-in rates, indicating that the rate for tunneling out of
the QD does not depend on the state used for tunneling into the QD.
Since the individual excited states are expected to have different
rates also for tunneling out of the dot, the data is consistent with
the interpretation that an electron entering the dot into an excited
state will always have time to relax to the ground state before it
tunnels out. The rate for tunneling out is $\sim\! 6~\mathrm{kHz}$,
giving an upper bound for the relaxation time of $\sim \!
170~\mathrm{\mu s}$.

The main relaxation mechanism in quantum dots is thought to be
electron-phonon scattering \cite{inoshia:1992}. Measurements on
few-electron vertical quantum dots have shown relaxation times of
$10~\mathrm{ns}$ \cite{fujisawa:2002}. Recent numerical
investigations have shown that the electron-electron interaction in
multi-electron dots can lead to reduced relaxation rates
\cite{bertoni:2005}. Still, the relaxation rate is expected to be
considerably faster than the upper limit we give here.

The rate for tunneling out is actually not constant for the whole
region of the Coulomb diamond, but shows a change at the position
marked by the arrow in Fig.~\ref{fig:TR_excitedStates}(b). This
transition occurs along a line perpendicular to the ones seen in
$\Gamma_\mathrm{in}$. This is expected assuming the transition seen
in Fig.~\ref{fig:TR_excitedStates}(b) involves changes in
$\Gamma_\mathrm{out}$ instead of in $\Gamma_\mathrm{in}$. Going
perpendicular to the dashed lines in
Figs.~\ref{fig:TR_excitedStates}(a,~b), we keep the QD and source
potential constant while lowering the drain lead. At some point, the
Fermi level of the drain is low enough so that an electron in the QD
$(n+1)$-electron ground state may tunnel out and leave the QD in an
$(n)$-electron excited state. The process is sketched in
Fig.~\ref{fig:TR_excitedStates}(d). Comparing
Figs.~\ref{fig:TR_excitedStates}(c-d), we see that the rate
$\Gamma_\mathrm{in}$ probes the excitation spectrum of the
$(n+1)$-electron QD, while $\Gamma_\mathrm{out}$ reflects the
spectrum of the $(n)$-electron QD.

\subsection{Tuning the tunnel couplings} \label{sec:TR_tuneCoupling}
\begin{figure}[tb]
\centering
 \includegraphics[width=\linewidth]{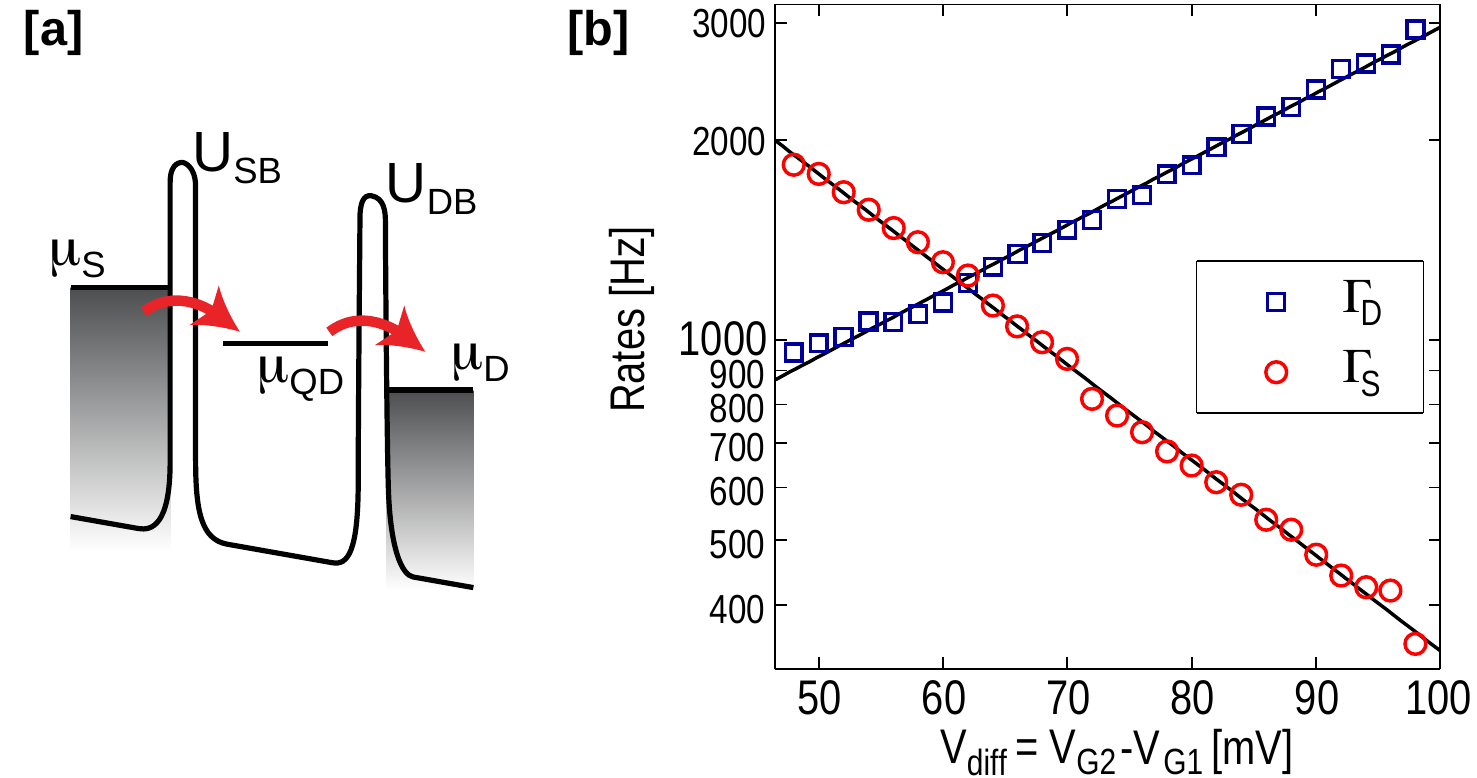}
 \caption{(a) Potential landscape of the QD when a fixed bias voltage
 is applied between the source and drain contacts.
 (b) Tunneling rates $\Gamma_\mathrm{S}/\Gamma_\mathrm{D}$ measured
 versus $V_\mathrm{diff} = V_\mathrm{G2} - V_\mathrm{G1}$. The sold
 lines are fits to Eq.~(\ref{eq_TR_GvsVdiff}) in the text. The
 measurements were performed by sweeping both gate voltages $V_\mathrm{G1}$, $V_\mathrm{G2}$,
 with $V_\mathrm{G1} = -0.142 - V_\mathrm{G2}$.
 }
\label{fig:TR_BarrierTuning}
\end{figure}

Changing a gate voltage does not only shift the electrochemical
potential of the QD, but also affects the height of the tunnel
barrier connecting the QD to the leads.
The effect was mentioned already in relation with the results of
Figs.~\ref{fig:TR_temp} and \ref{fig:TR_CD}. Here we investigate the
behavior more carefully.
Figure~\ref{fig:TR_BarrierTuning}(a) shows a sketch of the potential
landscape for a QD with a bias voltage applied between the source
and drain contacts. Electrons entering the QD from the source lead
need to tunnel through a potential barrier of height
$(U_\mathrm{SB}-\mu_\mathrm{QD})$, while the barrier height for
electrons tunneling from the QD to drain is
$(U_\mathrm{DB}-\mu_\mathrm{QD})$. By changing the voltages on gates
$G_1$ and $G_2$, we expect to be able to tune the potentials
$U_\mathrm{SB}$ and $U_\mathrm{DB}$ and thereby control the
tunneling rates.

The tunneling probability also strongly depends on the width of the
barrier as well as on the exact shape of the electrostatic potential
forming the QD and the barriers. These details are not known, but
for small perturbations to the barrier potential $\delta
U_\mathrm{SB/DB}$ and QD potential $\delta \mu_\mathrm{QD}$, the
tunneling rate is expected to depend exponentially on the energy
difference $(\delta U_\mathrm{SB/DB}-\delta \mu_\mathrm{QD})$
\cite{maclean:2007}
\begin{equation}\label{eq:TR_expTuneGamma}
 \Gamma \sim \Gamma_0 \exp[-\kappa(\delta U_\mathrm{SB/DB}-\delta
 \mu_\mathrm{QD})].
\end{equation}
Here, $\Gamma_0$ and $\kappa$ are constants given by the exact shape
of the potential.
To make quantitative comparisons with the experiments, we use a
capacitor model to estimate the influence that gate voltages have on
the different potentials in the system \cite{chen:2006}
\begin{eqnarray}\label{eq:TR_CMatrix}
     \left[
       \begin{array}{c}
         \delta \mu_\mathrm{QD} \\
         \delta U_\mathrm{SB} \\
         \delta U_\mathrm{DB} \\
       \end{array}
     \right]
     =  & \\ \nonumber
     \left[ \! \!
       \begin{array}{cccc}
         \alpha_\mathrm{S-QD} & \alpha_\mathrm{D-QD} & \alpha_\mathrm{G1-QD} & \alpha_\mathrm{G2-QD} \\
         \alpha_\mathrm{S-SB} & \alpha_\mathrm{D-SB} & \alpha_\mathrm{G1-SB} & \alpha_\mathrm{G2-SB} \\
         \alpha_\mathrm{S-DB} & \alpha_\mathrm{D-DB} & \alpha_\mathrm{G1-DB} & \alpha_\mathrm{G2-DB} \\
       \end{array}
     \! \! \right]  \! \! \! \! \! \! &
     \left[ \! \!
       \begin{array}{c}
         \delta \mu_\mathrm{S} \\
         \delta \mu_\mathrm{D} \\
         \delta (|e| V_\mathrm{G1}) \\
         \delta (|e| V_\mathrm{G2}) \\
       \end{array}
     \! \! \right]. 
     \end{eqnarray}
The coefficients $\alpha$ are the capacitive lever arms between the
gates and the various sample potentials. It should be noted that
both the gate voltages $V_\mathrm{G1}$, $V_\mathrm{G2}$ and the
source and drain potentials $\mu_\mathrm{S}$, $\mu_\mathrm{D}$ have
gating effects on the QD and on the barriers.
In the following, we focus on the influence of the gate voltages
$V_\mathrm{G1}$, $V_\mathrm{G2}$ and assume a fixed bias voltage
$V_\mathrm{SD}$ applied symmetrically across the QD, with
$\mu_\mathrm{S} = |e|V_\mathrm{SD}/2$, $\mu_\mathrm{D} =
-|e|V_\mathrm{SD}/2$.
Also, by operating the QD at fixed bias and ensuring that electron
transport is unidirectional [Eq.~(\ref{eq:TR_condCurrent})], we can
use the relations of Eq.~(\ref{eq:TR_gammaInSource}) to determine
the tunnel couplings $\Gamma_\mathrm{S}$ and $\Gamma_\mathrm{D}$
separately.

As seen in Eq.~(\ref{eq:TR_expTuneGamma}), the tunneling strength
depends on the difference $\delta U_\mathrm{SB/DB}-\delta
 \mu_\mathrm{QD}$. To simplify matters we want to fix the QD
potential $\mu_\mathrm{QD}$ and investigate only the influence that
the gate voltages have on the barrier potentials $U_\mathrm{SB}$ and
$U_\mathrm{DB}$.
This is done by sweeping the two gate voltages $V_\mathrm{G1}$ and
$V_\mathrm{G2}$ against each other in a way that $\mu_\mathrm{QD}$
remains constant. Setting $\delta \mu_\mathrm{QD}=0$ in
Eq.~(\ref{eq:TR_CMatrix}), assuming a fixed bias voltage ($\delta
\mu_\mathrm{S}=\delta \mu_\mathrm{D}=0$) and solving for $V_{G1}$
gives the prescription
\begin{equation}\label{eq:TR_G1G2relation}
    \delta V_\mathrm{G1} = -\frac{\alpha_\mathrm{G2-QD}}{\alpha_\mathrm{G1-QD}}
    \delta V_\mathrm{G2}.
\end{equation}
Due to the symmetry of the device, we have $\alpha_\mathrm{G1-QD}
\approx \alpha_\mathrm{G2-QD}$ so that the above expression reduces
to $\delta V_\mathrm{G1} \approx -\delta V_\mathrm{G2}$.
Introducing $V_\mathrm{diff} = V_\mathrm{G2}-V_\mathrm{G1}$ we find
from Eqs.~(\ref{eq:TR_expTuneGamma}-\ref{eq:TR_CMatrix})
\begin{eqnarray}
\nonumber \Gamma_\mathrm{S} & \sim & \exp[\kappa_S \frac{|e| \delta V_\mathrm{diff}}{2}(\alpha_\mathrm{G2-SB}
 - \alpha_\mathrm{G1-SB})] \equiv \\ & \equiv & \exp[\gamma_\mathrm{S} \, \delta V_\mathrm{diff}], \\
\nonumber \Gamma_\mathrm{D} & \sim & \exp[\kappa_D \frac{|e| \delta V_\mathrm{diff}}{2}(\alpha_\mathrm{G2-DB}
 - \alpha_\mathrm{G1-DB})] \equiv \\ & \equiv & \exp[\gamma_\mathrm{D} \, \delta V_\mathrm{diff}]. \label{eq_TR_GvsVdiff}
\end{eqnarray}
Thus we expect the tunneling rates to depend exponentially on the
voltage difference $V_\mathrm{diff}$.
The sign of the factors $\gamma_\mathrm{S}$ and $\gamma_\mathrm{D}$
determine if the rates increase or decrease with
$V_\mathrm{diff}$. From the geometry of the device we expect the
source barrier to be more strongly influenced by gate G1 than gate G2
$(\alpha_\mathrm{G1-SB} > \alpha_\mathrm{G2-SB})$, while the
opposite is true for the drain barrier.
This would make $\Gamma_\mathrm{S}$ decrease ($\gamma_\mathrm{S}<0$)
and $\Gamma_\mathrm{D}$ increase ($\gamma_\mathrm{D}>0$) with
$V_\mathrm{diff}$.

In Fig.~\ref{fig:TR_BarrierTuning}(b), we plot the tunneling rates
$\Gamma_\mathrm{S}$ and $\Gamma_\mathrm{D}$ measured while changing
$V_\mathrm{diff}$. The solid lines are fits to
Eq.~(\ref{eq_TR_GvsVdiff}), with fitting parameters
$\gamma_\mathrm{S} = -33.0~\mathrm{V^{-1}}$ and $\gamma_\mathrm{D} =
22.8~\mathrm{V^{-1}}$.
The results are consistent with Eq.~(\ref{eq_TR_GvsVdiff}), although
one would expect $\gamma_\mathrm{S} = -\gamma_\mathrm{D}$ from the
symmetry of the device.  However, the exact shapes of the confining
potential and the QD wavefunction are not known and it must be
considered unlikely that the potential barriers separating the QD
from source and drain contacts are geometrically exactly identical.

\subsection{Degenerate states}\label{sec:TR_degenerateStates}
\begin{figure}[tb]
\centering
 \includegraphics[width=\linewidth]{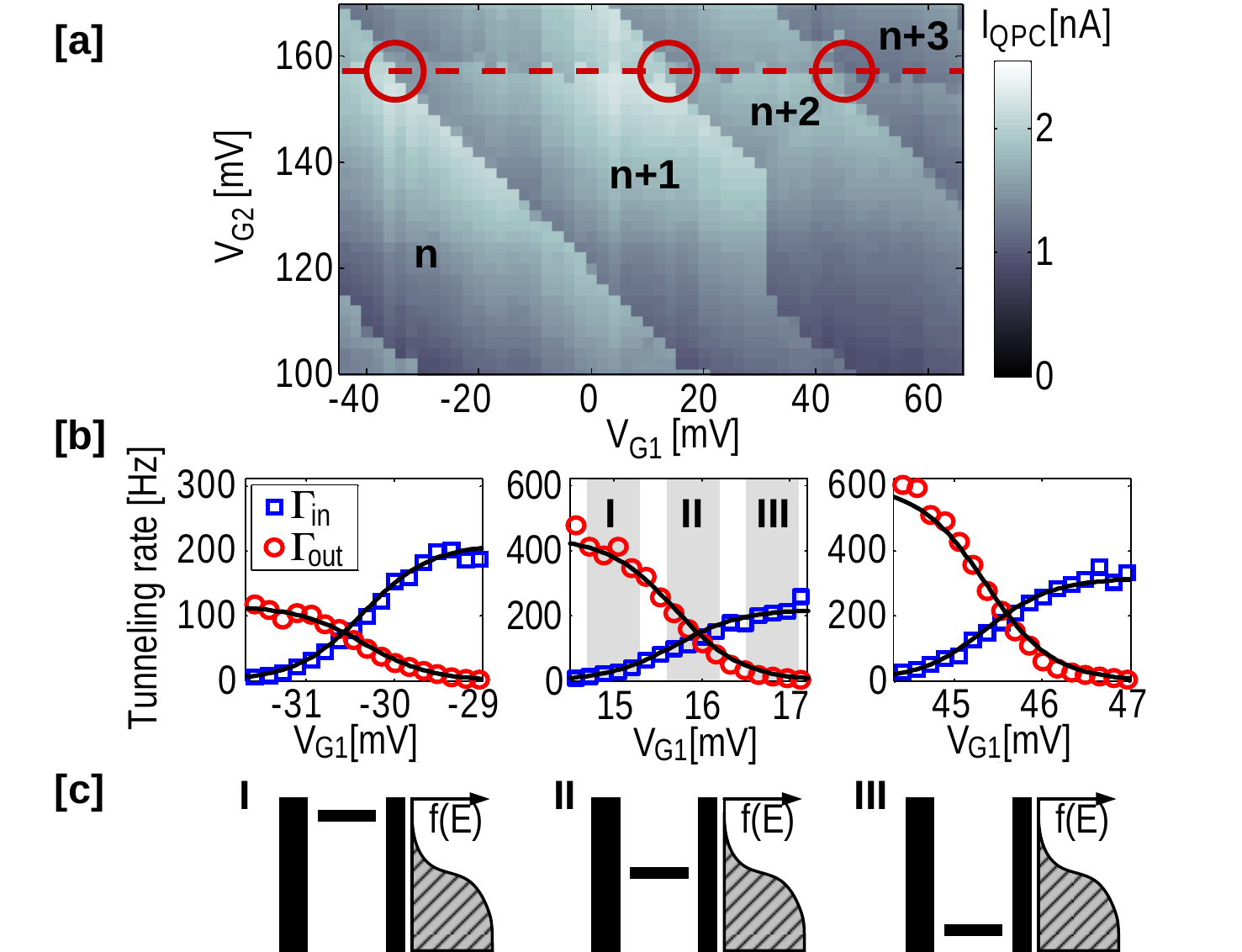}
 \caption{(a) Current through the QPC as a function of voltage on gates
 $G1$, $G2$. The diagonal lines show
 positions where the population of the QD changes by one electron. The
 numbers specify the dot occupation in the different regions.
 (b) Effective tunneling rates for electrons entering and leaving
 the dot, measured at the three charge degeneracy points marked by circles along the
 dashed line in (a). The solid lines are fits using Eqs. (\ref{eq:TR_effInDeg},
 \ref{eq:TR_effOutDeg}), with $T=230~\mathrm{mK}$ and the other fitting
 parameters given in Table \ref{tab:TR_gammas} in the text.
 (c) Alignment of the QD electrochemical potential relative to the
 Fermi level of the lead for the gate voltage configurations shown
 in the middle plot in (b). Adapted from Ref. \cite{gustavssonProcDresden:2007}.
 }
 \label{fig:TR_degeneracy}
\end{figure}

In this section, we discuss how degenerate states may influence the
measured statistics. For simplicity, we limit the discussion to the
case where the QD is connected only to one lead, with the other lead
being completely pinched off. In this configuration, the tunneling
is due to equilibrium fluctuations between the QD and the lead.
Fig.~\ref{fig:TR_degeneracy}(a) shows the average dc current through
the QPC when sweeping the two gates $G1$ and $G2$. The diagonal
lines correspond to electrons being loaded/unloaded from the QD.
Along these lines, the electrochemical potential of the QD is
aligned with the Fermi level of the right lead. From the slope of
the line we see that the voltages on the two gates $G1$ and $G2$
have roughly the same influence on the energy levels of the QD, as
expected from the device geometry. We now focus on determining the
tunneling rates for three electronic states along the dotted line in
Fig.~\ref{fig:TR_degeneracy}(a). Starting at low $V_{G1}$ voltages,
the dot gets successively populated as the voltage on $G1$ is
increased. At each charge degeneracy point, we use the time-resolved
measurement techniques to determine the rates for electrons entering
and leaving the dot. The results are shown in
Fig.~\ref{fig:TR_degeneracy}(b).

Taking the possibility of degenerate states into account, the
results of Eqs. (\ref{eq:TR_effGin}-\ref{eq:TR_effGout}) are
extended to
\begin{eqnarray}
  \Gamma_{\mathrm{in}} &=& g_{\mathrm{in}} \, \Gamma_R \times f_R(\Delta \mu/k_B T), \label{eq:TR_effInDeg}\\
  \Gamma_{\mathrm{out}} &=& g_{\mathrm{out}} \, \Gamma_R \times [1-f_R(\Delta \mu/k_B T)] \label{eq:TR_effOutDeg}.
\end{eqnarray}
Here, the factors $g_{\mathrm{in}}$ and $g_{\mathrm{out}}$ account
for possible degeneracies. For electrons entering the QD, the factor
$g_{\mathrm{in}}$ should include the number of degenerate
\emph{empty} states. For tunneling out, only the degeneracy of
\emph{occupied} states is relevant. The tunnel coupling  $\Gamma_R$
is assumed to be independent of energy and of the QD level within
the small gate voltage range considered here. The energy level for
three different gate voltages are drawn schematically in
Fig.~\ref{fig:TR_degeneracy}(c). The middle plot of
Fig.~\ref{fig:TR_degeneracy}(b) indicates the gate voltage ranges
corresponding to the drawings shown in
Fig.~\ref{fig:TR_degeneracy}(c).

The effective rates for electrons tunneling into and out of the QD
involve the density of states and the occupation probability in the
lead. This gives a strong dependence on the alignment between the
Fermi level in the right lead and the electrochemical potential of
the dot. Starting at low $V_{G1}$ voltages in
Fig.~\ref{fig:TR_degeneracy}(b) [case I in
Fig.~\ref{fig:TR_degeneracy}(c)], the QD potential is far above the
Fermi level of the lead. At this point, the density of occupied
states in the lead is low and the effective rate for tunneling into
the QD is low. If an electron eventually manages to tunnel in, the
effective rate for tunneling out again will be high, since there are
many empty states in the lead to tunnel into. As the gate voltage is
increased, the QD potential goes down to the Fermi level of the lead
[case II in Fig.~\ref{fig:TR_degeneracy}(b,~c)]. In this
configuration, the effective rates for tunneling into and out of the
QD are roughly equal. As the gate voltage is further increased, the
potential of the QD is pushed below the Fermi level. Here, the
density of occupied states in the lead is large, giving a high
effective rate for electrons entering the QD. Conversely, the
effective rate for leaving the dot is low [case III in
Fig.~\ref{fig:TR_degeneracy}(b,~c)].

Looking at the shape of the data in Fig.~\ref{fig:TR_degeneracy}(b),
we see that they indeed follow a Fermi function.  The solid lines in
the figure are fits using
Eqs.~(\ref{eq:TR_effInDeg}-\ref{eq:TR_effOutDeg}), with
$T=230~\mathrm{mK}$.
The parameters used in the fitting procedure are summarized in Table
\ref{tab:TR_gammas}.

\begin{table}[tbh]
 \centering
 \begin{tabular}{|c|c|c|c|}
  \hline
 $V_{G1}$  & \,\,\,\,\, $g_{\mathrm{in}} \, \Gamma_R$ \,\,\,\,\, & \,\,\,\,\, $g_{\mathrm{out}} \, \Gamma_R$\,\,\,\,\, & \,\,\,\,\, $g_{\mathrm{in}}/g_{\mathrm{out}}$\,\,\,\,\,\\
  \hline
  -30.35 mV& 210 Hz &  115 Hz & 1.8\\
  15.70 mV & 220 Hz & 440 Hz & 0.5\\
  45.35 mV & 315 Hz & 600 Hz & 0.5\\
  \hline
 \end{tabular}
 \caption{Fitting parameters for the solid lines in Fig.
 \ref{fig:TR_degeneracy}(b), fitted using Eqs.~(\ref{eq:TR_effInDeg}, \ref{eq:TR_effOutDeg}).
 }\label{tab:TR_gammas}
\end{table}

\begin{figure}[tb]
\centering
 \includegraphics[width=0.95\linewidth]{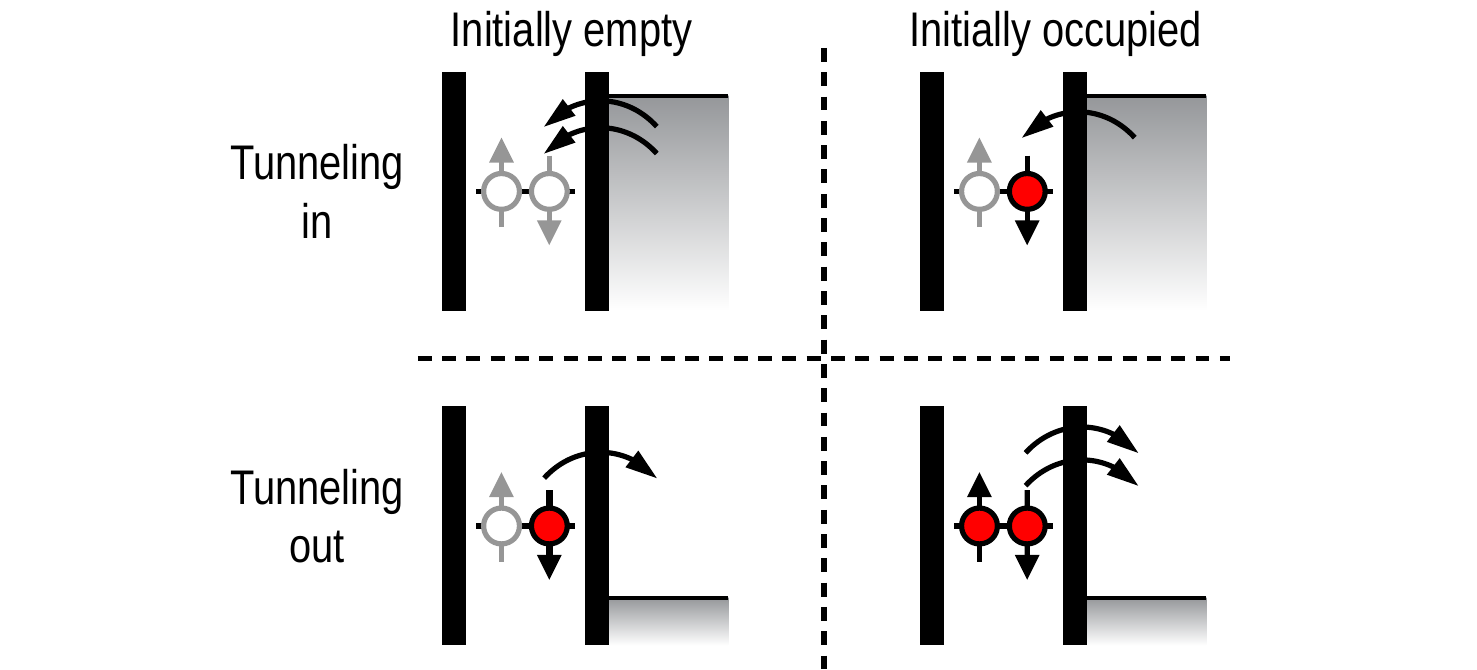}
 \caption{Effective tunneling rates for spin degenerate states
 in different configurations. The empty circles represents empty
 spin states, filled circles represent occupied ones. The
 arrows depict the number of possible tunnel processes.
 }
\label{fig:TR_GinOutSketch}
\end{figure}

Comparing the numbers of Table \ref{tab:TR_gammas}, we see that the
effective coupling $g_{\mathrm{in/out}} \, \Gamma_R$ differs
strongly depending on whether it was extracted from the tunneling in
or from the tunneling out data. One possible explanation for the
difference is degeneracy due to the electron spin. Assuming a
spin-degenerate state with both the spin-up and the spin-down state
initially empty, electrons from the lead could tunnel into either of
the two states. This makes $g_{\mathrm{in}}=2$. Once the electron
has tunneled into the QD, it sits in either the spin-up or the
spin-down state. Since only one of the spin-degenerate states is
occupied, the degeneracy for tunneling out will be
$g_{\mathrm{out}}=1$. The situation is different if we start with a
QD with one of the spin-degenerate states already occupied. For the
tunneling-in process, there is only one empty state available,
giving $g_{\mathrm{in}}=1$. For the tunneling-out process, any of
the two electrons sitting on the dot may tunnel out. This leads to
$g_{\mathrm{out}}=2$. The different situations are shown
schematically in Fig. \ref{fig:TR_GinOutSketch}. The model discussed
here assumes that the spin states are not influenced by Coulomb
interactions, which may be an oversimplification considering that we
are dealing with a many-electron system. Still, spin pairing has
been observed in chaotic QDs containing a large number of electrons
\cite{luscher:2001}.

The experimental method described here can only determine the ratio
$g_{\mathrm{in}}/g_{\mathrm{out}}$. In the following we assume the
degeneracies to be due to spin to be able to extract tunnel
couplings and absolute degeneracies from the data. The results of
this model are shown in Table \ref{tab:TR_gammasDeg}. For the first
resonance at $V_{G1}=-30.35~\mathrm{mV}$ we extract
$g_{\mathrm{in}}=2$ and $g_{\mathrm{out}}=1$, indicating a two-fold
degeneracy with both states initially empty. At the next resonance,
the degeneracy factors are exchanged, with $g_{\mathrm{in}}=1$ and
$g_{\mathrm{out}}=2$. For the third resonance, the degeneracy
factors are the same as for the second resonance, with
$g_{\mathrm{in}}=1$ and $g_{\mathrm{out}}=2$.

\begin{table}[tbh]
\centering
  \begin{tabular}{|c|c|c|c|}
  \hline
  $V_{G1}$ & $\Gamma_R$ & $g_{\mathrm{in}}$ & $g_{\mathrm{out}}$ \\
  \hline
  -30.35 mV& 110 Hz &  2 & 1  \\
  15.70 mV& 220 Hz  &  1 & 2  \\
  45.35 mV& 307 Hz & 1 & 2  \\
  \hline
  \end{tabular}
  \caption{Possible interpretation of the data shown in Table \ref{tab:TR_gammas},
  assuming spin-degenerate states.
   }\label{tab:TR_gammasDeg}
\end{table}

The first and second resonance could be attributed to consecutive
filling of spin states, meaning that the two first electrons would
form a so-called spin pair. The third electron does not follow the
rules expected from simple spin-filling. The reason could be due to
many-body effects between the electrons in the quantum dot or due to
a charge rearrangement taking place between the second and third
resonance (at $V\mathrm{G1} \sim 30\mV$). Also, we stress that there
are other possible explanations for the measurement results, like
energy-dependent tunneling rates or accidental degeneracies of
orbital states. To prove the spin degeneracy, one would need to
perform measurements at non-zero magnetic fields. This would lift
the spin degeneracy and make $g_{\mathrm{in}}=1$ and
$g_{\mathrm{out}}=1$.

As seen in section \ref{sec:TR_tuneCoupling}, changing a gate
voltage also affects the tunnel couplings in the system. Since the
tunneling rates $\Gamma_\mathrm{in}/\Gamma_\mathrm{out}$ are
measured at slightly different gate voltages, it could be that the
differences seen in Fig.~\ref{fig:TR_degeneracy}(b) are due to
tuning of the tunneling barrier. To avoid such influences, we used
gate $G1$ to tune the QD electrochemical potential, since it is
expected to have a smaller effect on the tunnel barrier between the
QD and drain than gate $G2$. From Eq.~(\ref{eq_TR_GvsVdiff}) and the
results of section \ref{sec:TR_tuneCoupling}, we estimate the change
of tunneling rates within the gate voltage range shown in
Fig.~\ref{fig:TR_degeneracy}(b) to be well below $10\%$.
Also, the gating effect of $G1$ on the tunnel barrier would make it
more likely for $\Gamma_R$ to increase with $V_\mathrm{G1}$. Since
$\Gamma_\mathrm{in}$ is determined at a slightly higher gate voltage
than $\Gamma_\mathrm{out}$, we would expect $\Gamma_\mathrm{in}$ to
be larger than $\Gamma_\mathrm{out}$. This is in contradiction with
the results of Table~\ref{tab:TR_gammas} and thus supports our interpretation of additional degeneracies of the QD states influencing the tunneling rates. 

\section{Statistics of electron transport} \label{sec:ST_main}
In this section, we investigate the statistical properties of
single-electron tunneling through a quantum dot. In the general
case, we find that current fluctuations due to shot noise
are suppressed because of Coulomb blockade. Electrons tend to avoid
each other, giving anti-bunching or \emph{sub-poissonian} noise. In
other regimes we find bunching of electrons, or
\emph{super-poissonian} noise. Finally, we investigate how the
finite bandwidth of the detector influences the measured statistics
and discuss the possibilities of using a quantum dot combined with a
charge-detector as a current meter.

\subsection{Electron transport and shot noise}
\label{sec:ST_generalNoise}
\begin{figure}[tb]
\centering
\includegraphics[width=\linewidth]{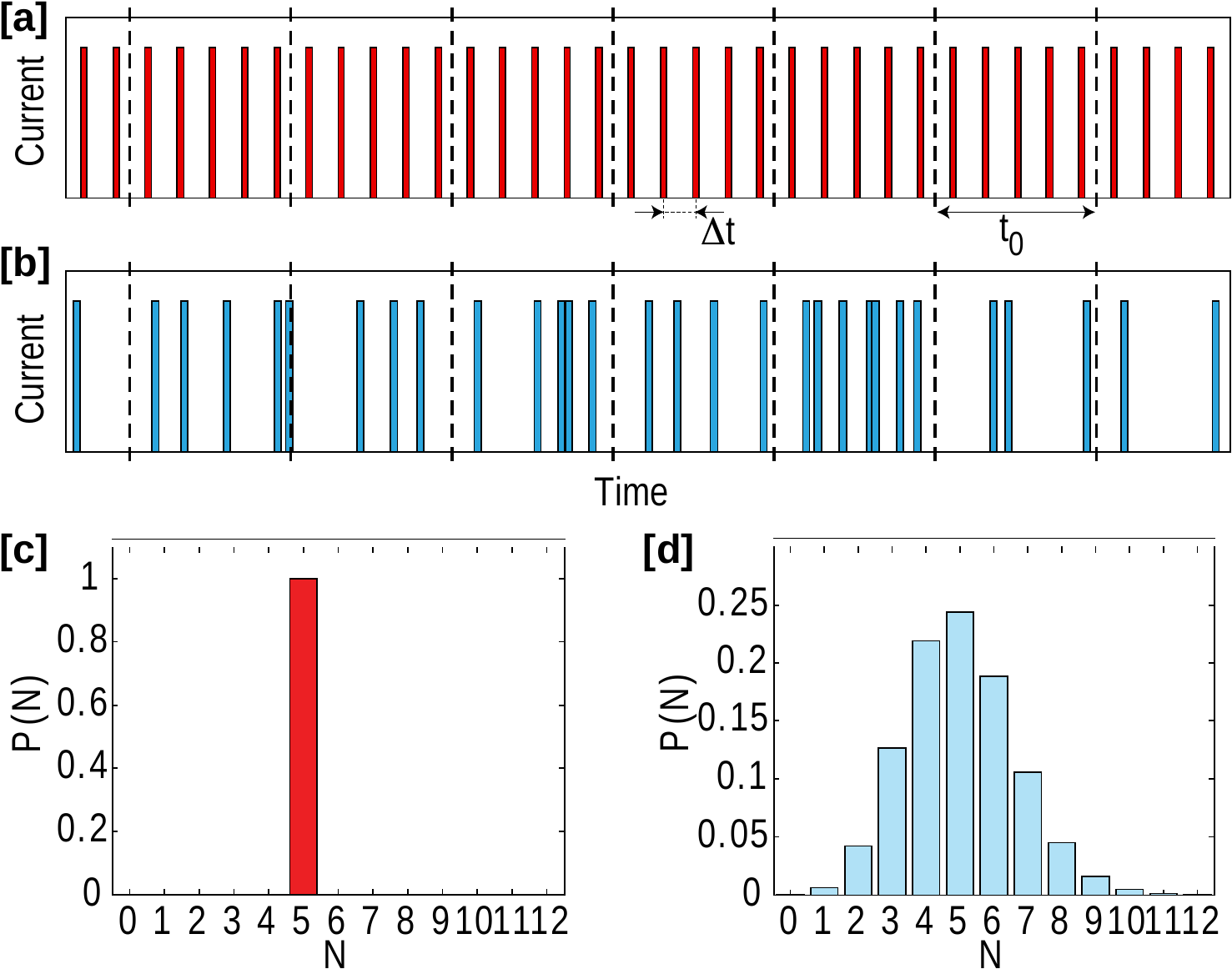}
\caption{ (a) Idealized current flow. Each spike corresponds to an
electron passing the conductor, with constant time intervals between
electrons. (b) Same as (a), but for a realistic
current. The electron flow shows random variations. (c-d)
Distribution function for the currents shown in (a-b). The
distribution function is formed by counting the number of electrons
passing the conductor within a time $t_0$. }
\label{fig:ST_idealizedTrace}
\end{figure}

Electrical current is carried by electrons passing through the
conductor. The current is given as
\begin{equation}\label{eq:ST_current}
   I = e/\langle t \rangle,
\end{equation}
where $e$ is the electron charge and $\langle t \rangle$ the average
time between electrons.
The discreteness of the charge carriers gives rise to temporal
fluctuations in the current. These statistical fluctuations are
called \emph{shot noise}.
The principle behind the shot noise is illustrated in
Fig.~\ref{fig:ST_idealizedTrace}. In
Fig.~\ref{fig:ST_idealizedTrace}(a), we show an idealized current
flow. Each spike corresponds to one electron passing the conductor.
The time interval between two electrons $\Delta t$ is constant, so
that the current is given as $I = e/\Delta t $.
Figure~\ref{fig:ST_idealizedTrace}(b) displays a more realistic
current, where the time intervals between electrons show random
fluctuations. If the average time between electrons $\langle t
\rangle = \Delta t$, then a measurement of the time-averaged current
in case (a) and (b) will give the same value.

Still, the currents in the two cases are obviously different. This
becomes clear in Fig.~\ref{fig:ST_idealizedTrace}(c-d), where we
plot the distribution function $P_{t_0}(N)$ of the number of
electrons $N$ that pass through the conductor within a fixed
time-interval $t_0$. The time $t_0$ is chosen so that the average
number of transmitted electron $\langle N \rangle = 5$ in both
cases. The distribution function for the idealized current in (a) is
simply a single peak with $P_{t_0}(5)=1$
[Fig.~\ref{fig:ST_idealizedTrace}(c)]. On the other hand, the
realistic current gives a broad distribution due to the statistical
fluctuations in the current.

In this way, the shape of the distribution function is a measure of
the statistical fluctuations of the current.
To be more quantitative, we calculate the central moments of the
distribution
\begin{equation}\label{eq:ST_centralMoments}
\mu_1 = \langle N \rangle, ~~~~~~ \mu_i= \langle (N- \langle N
\rangle)^i \rangle\mathrm{,~for~}i=2,3, \ldots
\end{equation}
Here, $\langle \ldots \rangle$ represents the mean over a large
number of periods of length $t_0$. The first moment (mean) gives
access to the average current, $I = e \mu_1 /t_0$. The second
central moment (variance) defines the shot noise power, with
\begin{equation}\label{eq:ST_shotNoise}
S_I = 2e^2 \mu_2 /t_0.
\end{equation}
Equation~(\ref{eq:ST_shotNoise}) is valid if $t_0$ is much larger
than correlation times in the system.
In the following section, we will also evaluate the third central
moment, $\mu_3$. It describes the asymmetry (skewness) of the
distribution function around its maximum.

The noise of a current is often expressed as the \emph{Fano} factor,
which is the width of the distribution divided by its mean,
\begin{equation}\label{eq:ST_fano}
F = S_I/2 e I = \mu_2 /\mu.
\end{equation}
For processes governed by Poisson statistics, like electron
tunneling through a single barrier, the Fano factor is equal to one.
If the Fano factor is smaller than one, we speak of sub-Poissonian
noise. This generally means lower noise power and electron
correlation in time. Conversely, if the Fano factor is greater than
one, the noise is super-Poissonian and the electron transport is
less regular than in the Poissonian case.

If the charge is transferred in units of $q$ instead of $e$, the
Fano factor will be modified by a factor $q/e$. By measuring both
the shot noise $S_I$ and current $I$, one can use this relation to
directly determine the fractionality of the charge of the carriers.
Such measurements have been performed to demonstrate the charge of
quasi-particles in the fractional quantum Hall effect
\cite{depicciotto:1997,saminadayar:1997} as well as the double
charge of Cooper pairs in superconductors \cite{jehl:2000}. These
are examples where noise measurements provide additional information
about the system that can not be extracted from a standard current
measurement \cite{blanter:2000}.

In electron transport through a semiconductor quantum dot (QD), the
noise is typically suppressed compared to the Poisson distribution.
This is due to Coulomb blockade, which enhances the temporal
correlation between successive electrons and thereby reduces the
noise
\cite{davies:1992,birk:1995,nauen:2002,nauen:2004,gustavsson:2005}.
The Pauli exclusion principle provides an additional noise suppression
mechanism \cite{lesovik:1989,buttiker:1990}. However, when several
channels with different coupling strengths contribute to electron
transport, interactions can lead to more complex processes and to an
enhancement of the noise \cite{sukhorukov:2001, cottet:2004,
belzig:2005, gustavsson:2006}. Furthermore, there are predictions
that entangled electrons may lead to super-Poissonian noise, thus
making noise measurements a possible way of detecting entanglement
in mesoscopic systems \cite{loss:2000, saraga:2003,hassler:2007}.

The above examples demonstrate that noise measurements are important
tools for characterizing properties of mesoscopic systems. However,
due to the very low current levels involved, it is difficult to
perform the experiments with conventional measurement techniques.
One has to carefully eliminate other noise sources like
Johnson-Nyquist thermal noise and the noise of the amplifiers.
Recent attempts include using a resonant circuit together with a
low-temperature amplifier \cite{zarchin:2007, mcclure:2007}, a
superconductor-insulator-superconductor junction \cite{onacNT:2006}
or a second QD acting as a high-frequency detector
\cite{onacQD:2006}.

A different approach is to use time-resolved charge detection
methods as described in Chapter \ref{sec:TR_main} to count the
electrons one-by-one as they pass through the conductor. From such a
measurement, one can directly determine the probability distribution
function $p_{t_0}(N)$. The distribution function is then used to
calculate both the shot noise as well as higher order moments. This
way of measuring is analogous to the theoretical concept of full
counting statistics (FCS), which was introduced as a new way of
examining current fluctuations \cite{levitov:1996}. In the following
sections, we investigate the experimental method in more detail.


\subsection{Sequential transport -- Sub-Poissonian noise}\label{sec:ST_subPoisson}
\begin{figure}[tb]
\centering
\includegraphics[width=\columnwidth]{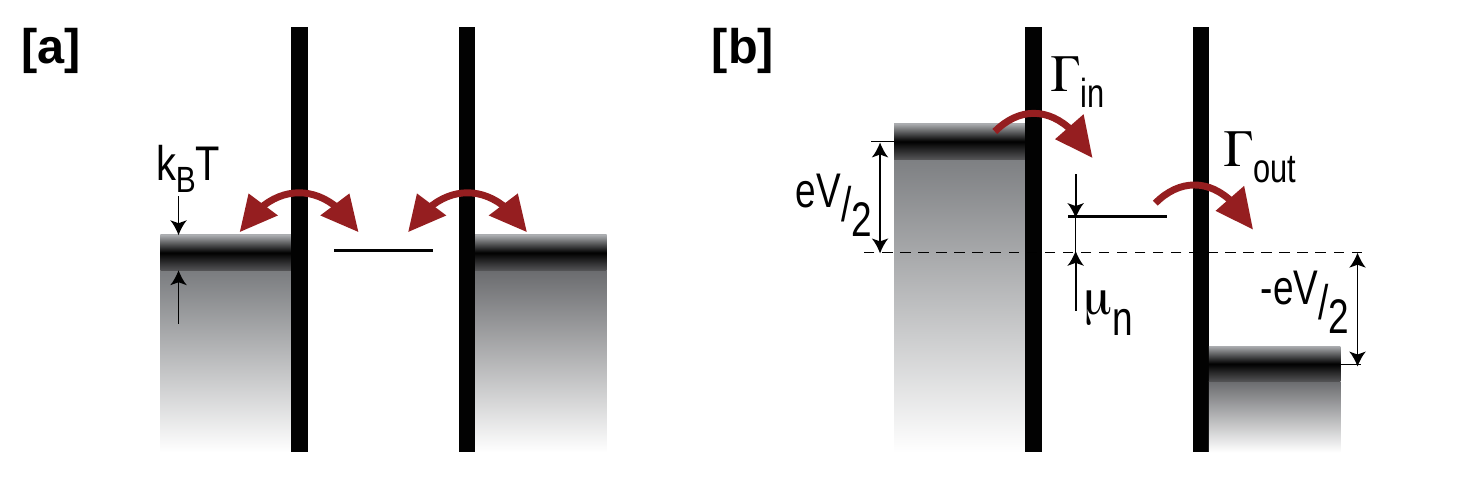}
\caption{(a) Energy level diagram for the quantum dot in the regime
of equilibrium fluctuations. Electrons may leave or enter the dot
from either of the two leads. (b) Energy level diagram for the
quantum dot in the high bias regime. With a large bias applied to
the QD, and with the Fermi levels of the leads far away from the
electrochemical potential of the dot, electrons can only enter the
QD from the source lead and only leave to the drain. }
\label{fig:ST_TunnelSketches}
\end{figure}

In order to use a charge detector for measuring current and current
noise, one has to avoid that electrons tunnel back and forth between
the dot and the source or drain lead due to thermal fluctuations
[Fig.~\ref{fig:ST_TunnelSketches}(a)]. This is achieved by applying
a finite bias voltage between source and drain, i.e.
\begin{equation}\label{eq:ST_highBias}
k_B T \ll |\!\: \pm eV/2 - \mu_n| \ll E_C.
\end{equation}
Here, $E_C$ is the charging energy, $\mu_{n}$ is the electrochemical
potential of the QD and $V$ is the bias voltage, symmetrically
applied to the QD [Fig. \ref{fig:ST_TunnelSketches}(b)]. With a
finite bias applied to the QD, and with the Fermi levels of the
leads far away from the electrochemical potential of the QD, the
probability for electrons to tunnel in the opposite direction is
exponentially suppressed. In this regime, we attribute each
transition $n \rightarrow n+1$ to an electron entering the QD from
the source contact, and each transition $n+1 \rightarrow n$ to an
electron leaving the QD to the drain contact. The charge
fluctuations in the QD then correspond to a non-equilibrium process,
and are directly related to the current through the dot.
The current is determined by the tunneling rates
$\Gamma_{\mathrm{in}}$ and $\Gamma_{\mathrm{out}}$, with
\begin{eqnarray}\label{eq:ST_currFromGamma}
 I = e \frac{\Gamma_{\mathrm{in}} \Gamma_{\mathrm{out}}}
 {\Gamma_{\mathrm{in}} + \Gamma_{\mathrm{out}}}.
\end{eqnarray}
From the tunneling rates, one could calculate all the higher moments
of the current distribution as well \cite{bagrets:2003}. However,
the results are only valid assuming that Eq.~(\ref{eq:TR_expDecay})
is correct. In order to measure the current and the current
distribution function for \emph{any} experimental configuration, we
instead focus on extracting the current distribution function
$p_{t_0}(N)$ from the experimental data.

\begin{figure}[tb]
\centering
\includegraphics[width=\columnwidth]{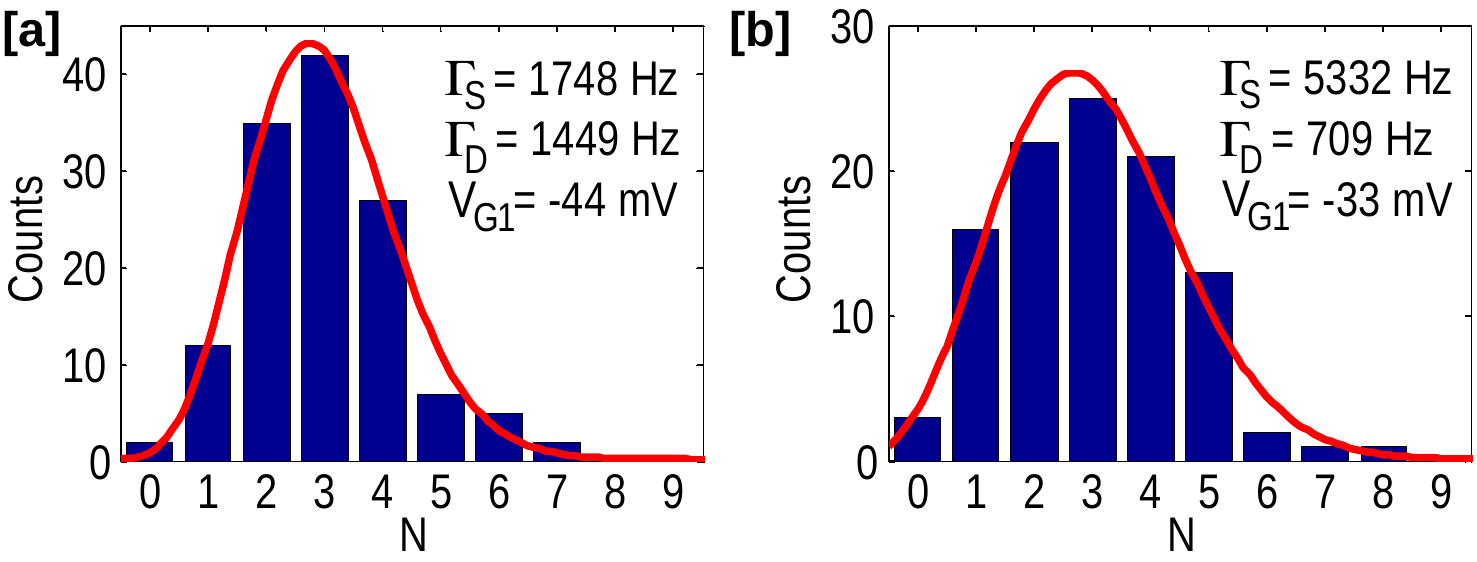}
\caption{Statistical distribution of the number $N$ of
electrons entering the QD during a given time $t_0$. The two panels
correspond to two different values of the tunneling rates, obtained
for different values of the gate voltage $V_{G1}$. The time $t_0$ is
chosen in order to have the same mean value of number of events,
$\langle N \rangle \approx 3$, for both graphs. The line shows the
theoretical distribution calculated from
Eqs.~(\ref{eq:ST_distribution}) and (\ref{eq:ST_generating}). The
tunneling rates are determined experimentally by the method
described in Chapter~\ref{sec:TR_main}, and no fitting parameters are
involved in the curves showing theoretical results. Adapted from Ref. \cite{gustavsson:2005}.
} \label{fig:ST_HistogramData}
\end{figure}

The distribution is found by splitting a time trace of length $T$
into $m=T/t_0$ intervals of length $t_0$ and counting the number of
electrons entering the QD within each interval. Examples of such
distributions are shown in Fig.~\ref{fig:ST_HistogramData}, taken at
two different gate configurations. The noise and the higher moments
are then extracted directly from the measured distribution using
Eqs.~(\ref{eq:ST_centralMoments}-\ref{eq:ST_fano}), giving for $I=
792~\mathrm{e/s}$, $F=\mu_1/\mu_2=0.52$ for case (a) and $I=
626~\mathrm{e/s}$, $F=0.89$ for case (b). The noise is relatively
close to Poissonian for case (b), but clearly sub-Poissonian in case
(a). This difference is easily seen by eye by comparing the width of
the two distributions. In order to understand why the shape is
different in the two situations in
Fig.~\ref{fig:ST_HistogramData}(a,~b), we need to calculated the
noise expected from the QD. This is the subject of the next section.

\subsection{Theory and model description} \label{sec:ST_subModel}
The noise properties of a QD in the sequential tunneling regime was
investigated in detail by Bagrets and Nazarov \cite{bagrets:2003},
using the framework of full counting statistics. Here, we summarize
their results, apply conditions appropriate for our experimental
configuration and compare the theoretical results with experimental
data.

\begin{figure}[b]
\centering
\includegraphics[width=\columnwidth]{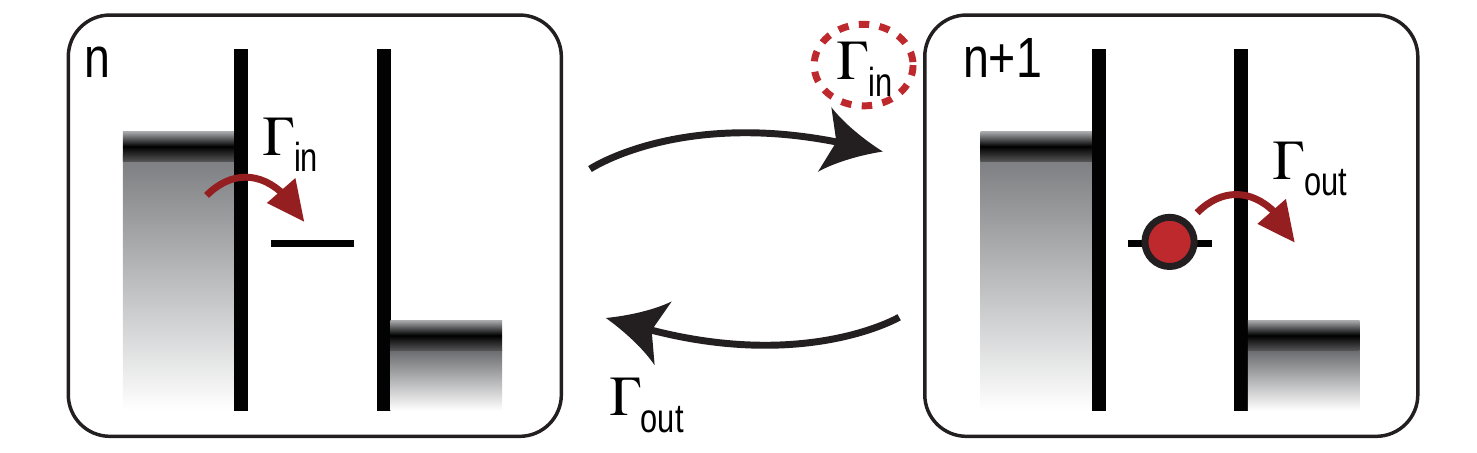}
\caption{State diagram of a two-state model describing electron
tunneling in a QD in the single-level regime. Transitions between
the states occur with rates $\Gamma_\mathrm{in}$ and
$\Gamma_\mathrm{out}$. The counting field $e^{i\chi}$ is introduced
for the transition involving an electron entering the QD, as marked
by the dashed circle. } \label{fig:ST_twoStateMaster}
\end{figure}

The QD occupancy in the low bias, single-level transport regime is
modeled by a two-state rate equation
\begin{equation}
\frac{d}{dt} \left( \begin{array}{c} p_n \\ p_{n+1} \end{array}
\right) =
\left( \begin{array}{cc} -\Gamma_{\mathrm{in}} & \Gamma_{\mathrm{out}} \\
\Gamma_{\mathrm{in}} & -\Gamma_{\mathrm{out}} \end{array} \right)
\left(
\begin{array}{c} p_n \\ p_{n+1}
\end{array} \right).
\label{eq:ST_master}
\end{equation}
Here, $p_n$ and $p_{n+1}$ give the
occupation probability for the the states with $n$ and $n+1$
electrons, respectively. The two states and the possible transitions are depicted in Fig.~\ref{fig:ST_twoStateMaster}.

To evaluate the counting statistics of the system, we need to
introduce a counting field $e^{i\chi}$ into the rate equation. We
choose to count electrons tunneling into the QD, which changes the
matrix in Eq. (\ref{eq:ST_master}) to:
\begin{equation}
M(\chi) = \left( \begin{array}{cc} -\Gamma_{\mathrm{in}} & \Gamma_{\mathrm{out}} \\
\Gamma_{\mathrm{in}} * e^{i\chi} & -\Gamma_{\mathrm{out}}
\end{array} \right).
\end{equation}
In the limit $t_0 \gg \Gamma^{-1}_{\mathrm{in}},
\Gamma^{-1}_{\mathrm{out}}$, the normalized distribution
$p_{t_0}(N/t_0)$ is independent of $t_0$. In the same limit, the
cumulant-generating function $S(\chi)$ is related to the lowest
eigenvalue of $M(\chi)$, $\lambda_0(\chi)$ as \cite{bagrets:2003}
\begin{eqnarray}
 S(\chi) &=& \lambda_0(\chi) t_0 = \frac{t_0}{2} \Big{[} \Gamma_{\mathrm{in}} +
\Gamma_{\mathrm{out}} - \nonumber \\
 &  &   \sqrt{(\Gamma_{\mathrm{in}}-
\Gamma_{\mathrm{out}})^2 + 4 \Gamma_{\mathrm{in}}
\Gamma_{\mathrm{out}} e^{-i \chi}} \Big{]}. \label{eq:ST_generating}
\end{eqnarray}
The distribution function for the number of electrons tunneling
through the quantum dot during a time $t_0$ is generated from the
cumulant-generating function $S(\chi)$ [see Appendix
\ref{sec:AP_cumulants}]:
\begin{equation}
p_{t_0}(N) = \int_{-\pi}^{\pi} \frac{d\chi}{2\pi} e^{-S(\chi)-i N
\chi}. \label{eq:ST_distribution}
\end{equation}
The solid lines in Fig.~\ref{fig:ST_HistogramData} are distributions
calculated from
Eqs.~(\ref{eq:ST_generating}-\ref{eq:ST_distribution}). The
tunneling rates $\Gamma_\mathrm{in}$ and $\Gamma_\mathrm{out}$ are
determined separately, as explained in Chapter \ref{sec:TR_main}.
The agreement with the experimental distribution is very good, in
particular, given that the curves involve no fitting parameters. As
mentioned earlier, the graphs show a clear qualitative difference:
Figure~\ref{fig:ST_HistogramData}(b) has a broader and more
asymmetric distribution than Fig.~\ref{fig:ST_HistogramData}(a). We
will see later that this difference comes from the different
asymmetries of the source and drain tunneling rates.

In order to perform a more quantitative analysis, we evaluate the
three first central moments $\mu_i$ of the current distribution,
which coincide with the first three cumulants $C_i$ [see Appendix
\ref{sec:AP_cumulants} for a discussion about the difference between
moments and cumulants]. The cumulants are generated directly from
the cumulant-generating function $S(\chi)$. The mean current is then
given by the first cumulant $C_1$ of the distribution:
\begin{equation}
I = \frac{e}{t_0} C_1 = \frac{e}{t_0} \left(-i \frac{dS}{d\chi}
\right)_{\chi=0} = e \frac{\Gamma_{\mathrm{in}}
\Gamma_{\mathrm{out}}}{\Gamma_{\mathrm{in}} + \Gamma_{\mathrm{out}}}.
\end{equation}

The symmetrized shot noise is calculated from the variance, or the
second cumulant $C_2$, of the distribution:
\begin{equation}
S_I = \frac{2 e^2}{t_0} C_2 = \frac{2 e^2}{t_0} \left(- \frac{d^2
S}{d\chi^2} \right)_{\chi=0} ~,
\end{equation}
from which we get the Fano factor:
\begin{equation}
F_2 = \frac{S_I}{2eI} = \frac{C_2}{
C_1}=\frac{\Gamma_{\mathrm{in}}^2+\Gamma_{\mathrm{out}}^2}{(\Gamma_{\mathrm{in}}
+ \Gamma_{\mathrm{out}})^2} = \frac{1}{2} \left( 1 +a^2 \right)
~,~ \label{eq:ST_mu2vsasym}
\end{equation}
where $a=(\Gamma_{\mathrm{in}} -
\Gamma_{\mathrm{out}})/(\Gamma_{\mathrm{in}} +
\Gamma_{\mathrm{out}})$ is the asymmetry of the coupling. This
result recovers the earlier calculations for the shot noise in a
quantum dot \cite{davies:1992}, and shows the reduction of the noise
by a factor $1/2$ for a QD symmetrically coupled to the leads, while
the Poissonian limit, $F_2 = 1$, is reached for an asymmetrically
coupled QD $(a=\pm 1)$.
The reduction of the noise is a direct consequence of Coulomb
blockade; when one electron occupies the QD, a second electron
cannot enter before the first one leaves. This leads to correlations
in the current fluctuations, and to a reduction of the noise. The
reduction is maximal when the tunnel barriers are symmetric. For an
asymmetrically coupled QD, the transport is essentially governed by
the weakly transparent barrier and the noise approaches the value
for a single tunneling barrier, $S_I=2 e I$. The results discussed
here assume tunneling with transmission coefficients much smaller
than one.

Finally, we want to calculate the third cumulant $C_3$, of the
fluctuations, which characterizes the asymmetry of the distribution
(skewness):
\begin{equation}
C_3 = i \left( \frac{d^3 S}{d\chi^3} \right)_{\chi=0}.
\end{equation}
The asymmetry can also be normalized to the mean of the
distribution:
\begin{eqnarray}
\!\!\!\!\!\!\!F_3 \!\!\!& \!\!\!\!\! = \!\! & \!\!\!\!\frac{C_3}{C_1}=\frac{\Gamma_{\mathrm{in}}^4 - 2
\Gamma_{\mathrm{in}}^3\Gamma_{\mathrm{out}} + 6
\Gamma_{\mathrm{in}}^2\Gamma_{\mathrm{out}}^2 - 2
\Gamma_{\mathrm{in}}\Gamma_{\mathrm{out}}^3 +
\Gamma_{\mathrm{out}}^4}{(\Gamma_{\mathrm{in}}
+ \Gamma_{\mathrm{out}})^4} \nonumber \\
&=& \frac{1}{4} \left( 1 + 3a^4 \right). \label{eq:ST_mu3vsasym}
\end{eqnarray}
The result shows that for a symmetrically coupled QD, the third
moment is reduced by a factor $1/4$ compared to the Poissonian
limit. For an asymmetrically coupled dot with $a \rightarrow \pm 1$,
we recover $F_3 \rightarrow 1$.

\subsection{Experimental results}
\begin{figure}[tb]
\includegraphics[width=1\linewidth]{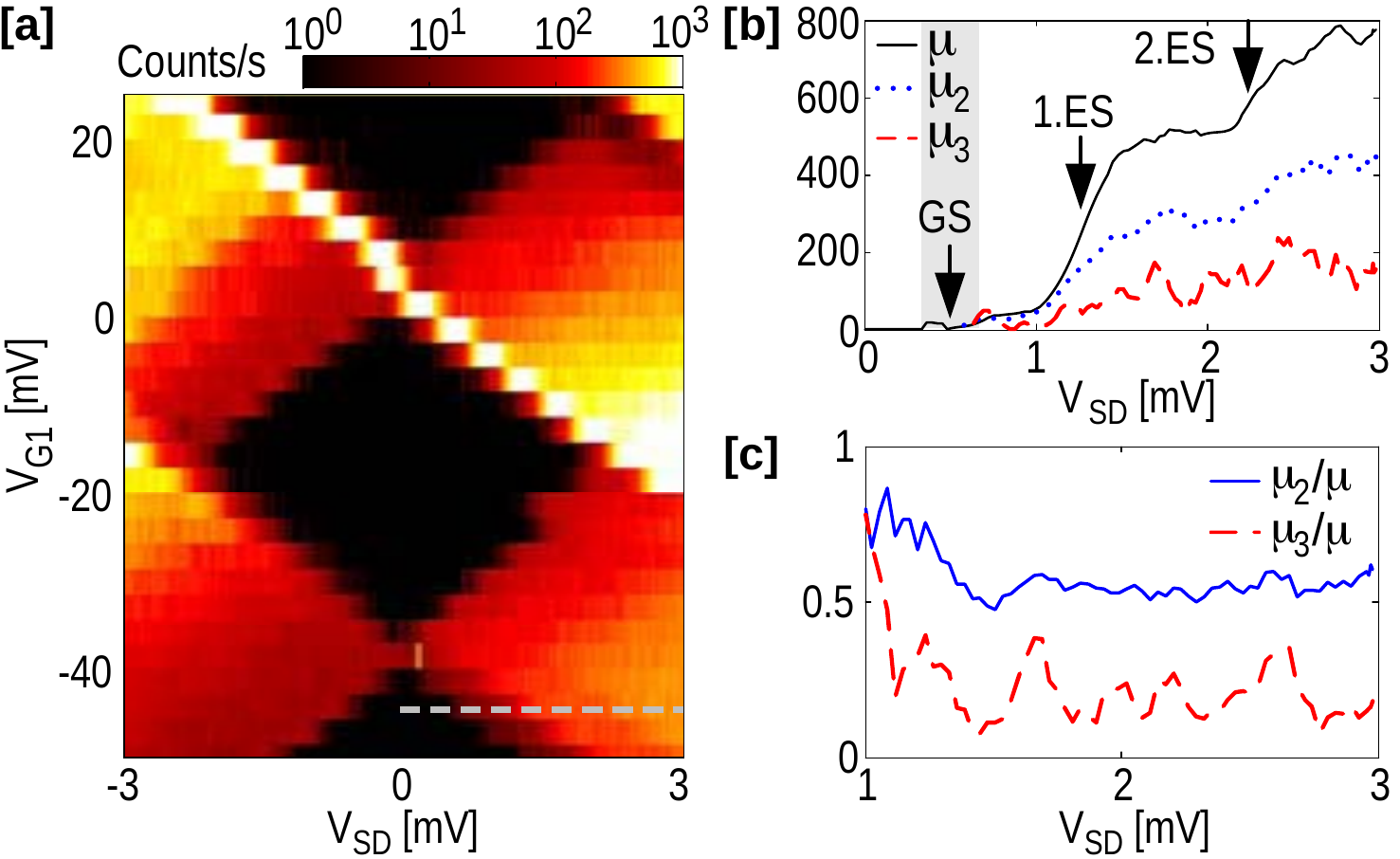}
\caption{\label{fig:ST_CoulombDiamonds} (a) Average number $\mu$ of
electrons entering the QD, measured as a function of the gate
voltage $V_{G1}$ and the bias voltage $V_\mathrm{SD}$. Far from the
edges of the Coulomb blockade region, i.e. for $|\pm
eV_\mathrm{SD}/2 - E_d| \gg k_B T$, the fluctuations of $n$ are
directly related to current fluctuations. The dashed line correspond
to the cross-section shown in panel (b). (b) Three first moments of
the fluctuations of $n$ as a function of the bias voltage
$V_\mathrm{SD}$ and at a given gate voltage $V_{G1}=-44$ mV. The
ground state (GS) as well as two excited states (ES) are clearly
visible. The moments are scaled so that $\mu$ corresponds to the
number of electrons entering the QD per second. In the gray region,
the condition $|\pm eV_\mathrm{SD}/2 - \mu_n| \gg k_B T$ is not
valid, and the number of electrons entering the QD cannot be taken
as the current flowing through the QD. The width of this region is
$9 \times k_B T/e \approx 300$ $\mu$V, determined from the width for
which the Fermi distribution is between 0.01 and 0.99. (c)
Normalized second and third moments as a function of the bias
voltage $V_\mathrm{SD}$ and at a given gate voltage $V_{G1}=-44$
mV. Adapted from Ref. \cite{gustavsson:2005}.}
\end{figure}

From experimental distributions as the ones shown in
Fig~\ref{fig:ST_HistogramData}, we can easily obtain moments of any
order using the relations in Eq.~(\ref{eq:ST_centralMoments}).
We first focus on the mean $\mu$ of the distribution. By measuring
$\mu$ as a function of the voltage applied on gate G1 and the bias
voltage $V$, we construct the Coulomb diamonds [see
Fig.~\ref{fig:ST_CoulombDiamonds}(a)]. We observe clear Coulomb
blockade regions as well as regions of finite current.
Figure~\ref{fig:ST_CoulombDiamonds}(b) shows a cross section taken
at $V_\mathrm{G1}=-44~\mathrm{mV}$, the position is indicated by the
dashed line in Fig.~\ref{fig:ST_CoulombDiamonds}(a)]. As the bias
voltage is increased, we see steps in the current. As explained in
Section~\ref{sec:TR_excitedStates}, the first step in
Fig.~\ref{fig:ST_CoulombDiamonds}(b) (see left arrow) corresponds to
the alignment of the chemical potential of the source contact with
the ground state in the QD, and the following steps with excited
states in the QD. From the resolution of the Coulomb diamonds, we
see that the sample is stable enough such that background charge
fluctuations do not play a significant role on the time scales
relevant for this experiment \cite{jung:2004}.

In addition to the mean, we evaluate the second and third central
moments from the measured counting statistics. These two moments are
plotted in Fig.~\ref{fig:ST_CoulombDiamonds}(b) as a function of the
bias voltage. The second moment (blue dotted line) reproduces the
steps seen in the current. These two moments can be represented by
their reduced quantities $F_2=\mu_2/\mu$ (Fano factor) and
$F_3=\mu_3/\mu$, as shown in Fig.~\ref{fig:ST_CoulombDiamonds}(c).
Both normalized moments are almost independent of the bias voltage,
and show a reduction compared to the values $\mu_2/\mu=\mu_3/\mu=1$
expected for classical fluctuations with Poissonian counting
statistics.

\begin{figure}[tb]
\includegraphics[width=1\linewidth]{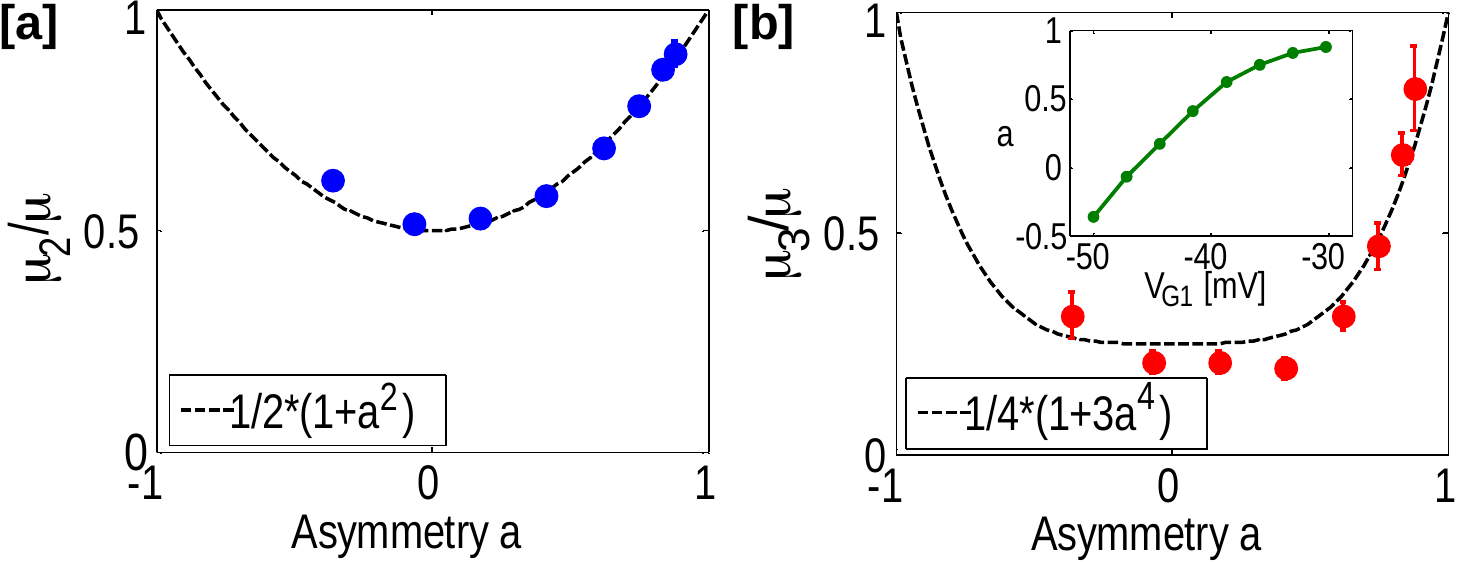}
\caption{\label{fig:ST_moments} (a) Second and (b) third normalized
central moments of the fluctuations of $n$ as a function of the
asymmetry of the tunneling rates, $a=(\Gamma_\mathrm{in} -
\Gamma_\mathrm{out})/(\Gamma_\mathrm{in} + \Gamma_\mathrm{out})$. To
increase the resolution, each point at a given asymmetry is obtained
by averaging over about $50$ points at a given voltage
$V_\mathrm{G1}$ and in a window of bias voltage $1.5 < V_\mathrm{SD}
< 3$ mV. Error bars correspond to the standard error of this
averaging process, and are of the size of the points if not shown.
The dashed lines are the theoretical predictions given by
Eqs.~(\ref{eq:ST_mu2vsasym},~\ref{eq:ST_mu3vsasym}). No fitting
parameters have been used, since the tunneling rates are fully
determined experimentally. Inset of (b): Variation of the asymmetry
of the tunneling rates, $a$, as a function of $V_\mathrm{G1}$. Adapted from Ref. \cite{gustavsson:2005}.}
\end{figure}

As described in section~\ref{sec:TR_tuneCoupling}, the tunnel
couplings can be tuned by adjusting the gate voltages
$V_\mathrm{G1}$ and $V_\mathrm{G2}$. In this way, we are able to
continuously change the symmetry of the barriers from symmetric to
very asymmetric coupling. In Fig.~\ref{fig:ST_moments}, we show the
normalized second and third central moments as a function of the
asymmetry $a$. The tunneling rates are directly measured as
described in section~\ref{sec:TR_main}, and the inset of Fig.~\ref{fig:ST_moments}(b) shows the variation of asymmetry
with gate voltage in the region of interest. As expected from the
discussion in the previous section, the noise is reduced for
symmetric barriers. The experimental data follow the theoretical
predictions given by
Eqs.~(\ref{eq:ST_mu2vsasym},~\ref{eq:ST_mu3vsasym}) very well. We
note in particular that no fitting parameters have been used since
the tunneling rates are determined separately.


\subsection{Time statistics}
A complementary way of investigating the correlations is to look at
the temporal statistics of electron transport. Instead of evaluating
the probability distribution for the number of electrons that are
transferred within a fixed time $t_0$, we examine the continuous
distribution $p_N(t)$ describing the time needed for a fixed number
of $N$ electrons to pass through the QD. With the rates for
tunneling into and out of the QD given by
Eq.~(\ref{eq:TR_expDecay}), we find for $N=1$
\begin{eqnarray}\label{eq:ST_antibunchingTime}
  p_\mathrm{\,N=1}(t) &=& \int_0^t p_\mathrm{in}(t')
    p_\mathrm{out}(t-t') dt' = \nonumber \\
   &=& \frac{\exp(-\Gamma_{\mathrm{in}}\,t) - \exp(-\Gamma_{\mathrm{out}}\,t)}
   {1/\Gamma_{\mathrm{in}} - 1/\Gamma_{\mathrm{out}}}.
\end{eqnarray}

\begin{figure}[b]
\includegraphics[width=0.95\linewidth]{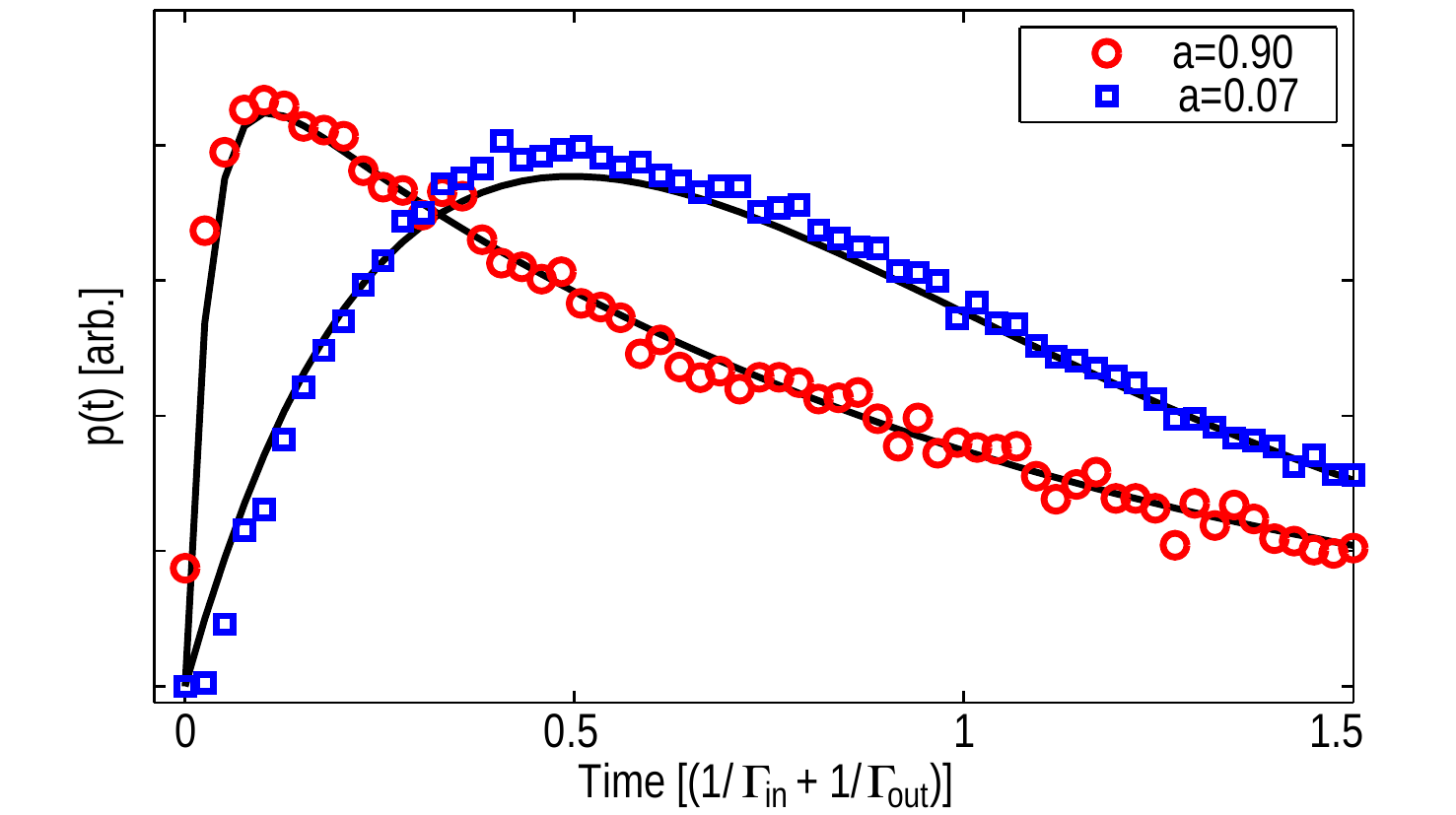}
\caption{\label{fig:ST_antibunching}
 Distribution of times needed for
 one electron to pass through
 the QD, measured with both symmetric and asymmetric tunnel couplings.
 To make qualitative comparisons of the two distributions easier,
 the two curves are plotted with different vertical scaling. The time scale
 is normalized to the average time needed for one electron to pass the dot.}
\end{figure}

In Fig.~\ref{fig:ST_antibunching}, we show the experimentally
determined distribution $p_{N=1}(t)$ for two different values of the
asymmetry together with the results of
Eq.~(\ref{eq:ST_antibunchingTime}). For the symmetric case
[$a=0.07$ in Fig.~\ref{fig:ST_antibunching}], there is a clear
suppression of transfer probability for short time scales. Again,
this is due to Coulomb blockade. We measure anti-bunching of
electrons and sub-Poisson noise levels. For the more asymmetric case
[$a=0.9$ in Fig.~\ref{fig:ST_antibunching}], anti-bunching is
less prominent and the probability distribution approaches the
exponential behavior expected for a single tunnel barrier.

The ability to measure the counting statistics of electron transport
relies on the high sensitivity of the QPC as a charge detector.
Given the bandwidth of our experimental setup, $\Delta f = 30$ kHz,
the method allows to measure currents up to $5$ fA, and we can
measure currents as low as a few electrons per second, i.e., less
than 1 aA. The low-current limitation is mainly given by the length
of the time trace and the stability of the QD, and is well below
what can be measured with conventional current meters. In addition,
as we directly count electrons one by one, this measurement is not
sensitive to the noise and drifts of the experimental setup. It is
also a very sensitive way of measuring low current noise levels. The
precision and limitations of the measurement method are described in
more details in sections \ref{sec:ST_higherMoments} and
~\ref{sec:ST_precision}.

\subsection{Bunching of electrons}
\begin{figure}[tb]
\centering
 \includegraphics[width=\columnwidth]{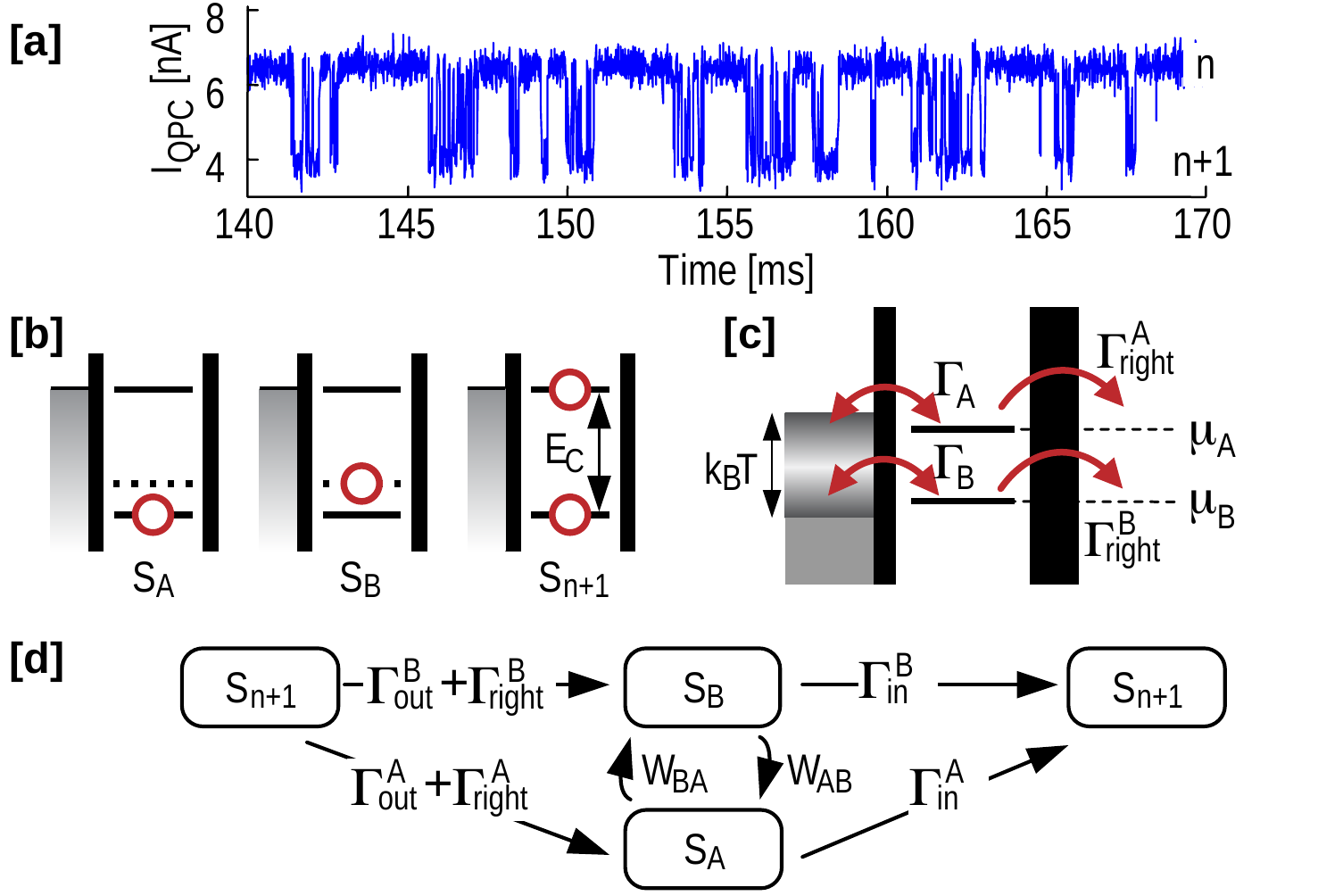}
 \caption{(a) Time trace of the QPC current showing bunching of
 electrons. (b) Dot states included in the model used to describe the bunching
 of electrons. The red circles correspond to electron occupation.
 State $S_A$ is the $n$-electron ground state, state $S_B$ is an excited
 $n$-electron state and state $S_{n+1}$ is the ground state when
 the dot contains $(n+1)$ electrons.
 (c) Energy diagram for the model. The two dot transitions are both
 within the thermal broadening of the lead.  Electrons enter the dot from the left
 lead and may leave through either the left or the right lead.
 (d) Possible transitions between the different
 states of the model.
 The rates $\Gamma^{\mathrm{A}}_{\mathrm{in}}$, $\Gamma^{\mathrm{B}}_{\mathrm{in}}$
 refer to
 electrons entering the QD from the left lead, thus taking the dot from state
 $S_{A/B}$ to state $S_{n+1}$. The rates
 $\Gamma^{\mathrm{A}}_{\mathrm{out}}$, $\Gamma^{\mathrm{B}}_{\mathrm{out}}$ describe electrons leaving
 the dot to the left lead, giving transitions from state $S_{n+1}$ to $S_{A/B}$. $W_{AB}$ and
 $W_{BA}$ are the direct transition rates between states $S_A$ and $S_B$.
 Finally, the rates $\Gamma^{\mathrm{A}}_{\mathrm{right}}$, $\Gamma^{\mathrm{B}}_{\mathrm{right}}$ refer to
 electrons leaving the dot through the right lead. Adapted from Ref. \cite{gustavsson:2006}.
 }\label{fig:ST_bunching}
\end{figure}

So far, we have analyzed data where the tunneling events can be well
explained by a rate equation approach with one rate for electrons
tunneling into and another rate for electrons leaving the dot. For
the trace shown in Fig. \ref{fig:ST_bunching}(a), the behavior is
distinctly different. The electrons come in bunches; there are
intervals where tunneling occurs on a fast time scale
($>\!10~\mathrm{kHz}$), in-between these intervals there are long
periods of time ($>\!1~\mathrm{ms}$) without any tunneling. The data
was taken with a bias applied so that the Fermi level of the source
lead lines up with the electrochemical potential of the dot, while
the drain lead is far below the electrochemical potential of the
dot, thus prohibiting electrons from entering the QD through the
drain lead. The voltage on gate $V_{G1}$ was set to
$34~\mathrm{mV}$, which is outside the range of the Coulomb diamonds
presented in Fig.~\ref{fig:ST_CoulombDiamonds}(a). Since the QPC
current is at the high level during the intervals without tunneling,
the dot contains one electron less when the fast tunneling is
blocked.

In order to explain the two different time scales, we assume a
mechanism where there are two almost energy-degenerate dot states
within the thermal broadening of the distribution in the source
lead. Because of Coulomb blockade, the dot may hold one or zero
excess electrons. Hence, the model includes three possible dot
states as shown in Fig. \ref{fig:ST_bunching}(b). State $S_A$ is the
$n$-electron ground state, state $S_B$ is an excited $n$-electron
state and state $S_{n+1}$ is the ground state when the dot contains
$(n+1)$ electrons. Transitions between the $S_A$/$S_B$ states and
the $S_{n+1}$ state occur whenever an electron tunnels into or out
of the dot.

The tunnel coupling strength between the dot and the lead is given
by the overlap of the dot and lead electronic wavefunctions. Since
the wavefunctions corresponding to the two states $S_A$ and $S_B$
may have different spatial distributions, the coupling strength
$\Gamma_A$ of the transition $S_A \Leftrightarrow S_{n+1}$ may
differ from the coupling $\Gamma_B$ of the $S_B \Leftrightarrow
S_{n+1}$ transition. The energy levels of the dot and the leads for
the configuration where we measure bunching of electrons are shown
in Fig.~\ref{fig:ST_bunching}(c), while the possible transitions of
the model are depicted in Fig.~\ref{fig:ST_bunching}(d).

Starting with one excess electron on the dot [state $S_{n+1}$ in
Fig.~\ref{fig:ST_bunching}(d)], at some point an electron will
tunnel out, leaving the dot in either state $S_A$ or state $S_B$.
Assuming $\Gamma_B \gg \Gamma_A$, it is most likely that the dot
will end up in the excited state $S_B$. If the tunneling rate
$\Gamma_B$ is faster than the relaxation process $S_B \Rightarrow
S_A$, an electron from the lead will have time to tunnel onto the
dot again and take the dot back to the initial $S_{n+1}$ state. The
whole process can then be repeated, leading to the fast tunneling in
Fig.~\ref{fig:ST_bunching}(a).

However, at some point the dot will end up in state $S_A$, either
through an electron leaving the dot via the $\Gamma_A$ transition,
or through relaxation of the $S_B$ state. To get out of state $S_A$,
there must be either a direct transition back to state $S_B$, or an
electron tunneling into the dot through the $S_A \Rightarrow
S_{n+1}$ transition. With $\Gamma_B \gg \Gamma_A$ and assuming
$\Gamma_B \gg ~W_{BA}$, both processes are slow compared to the
tunneling between the lead and state $S_B$.
This mechanism will block the fast tunneling and produce the
intervals without switching events seen in
Fig.~\ref{fig:ST_bunching}(a). Similar arguments can be used to show
that the blocking mechanism will be possible also if $\Gamma_B \ll
\Gamma_A$.

From the above reasoning, we see that the fast time scale is set by
the fast tunneling state, while the slow time scale is determined
either by the relaxation process $S_B \Rightarrow S_A$ or by the
slow tunneling rate, depending on which process is the fastest.
Either way, it is crucial that the relaxation rate is slower than
the fast tunneling rate (in our case $W_{AB} \ll \Gamma_B \sim
20~\mathrm{kHz}$). We speculate that the slow relaxation rate may be
due to different spin configurations of the two states. For a
few-electron QD, spin relaxation times of $T_1>1~\mathrm{ms}$ have
been reported \cite{hanson:2005, elzermanNature:2004, amasha:2008}.

To make quantitative comparisons between the model and the data, we
use the methods of full counting statistics to investigate how the
dot charge fluctuations change as the source lead is swept over a
Coulomb resonance. Theoretical investigations of multi-level quantum
dots have lead to predictions of electron bunching and
super-Poissonian noise \cite{belzig:2005}. Following the lines of
Refs. \cite{bagrets:2003, belzig:2005}, we first write the master
equation for the system,
\begin{equation}\label{eq:ST_masterBunching}
 \frac{d}{dt}
 \left[ \begin{array}{l}
 p_A \\
 p_B \\
 p_{n+1}
 \end{array}\right] = M
 \left[ \begin{array}{l}
 p_A \\
 p_B \\
 p_{n+1}
 \end{array}\right],
\end{equation}
with $M =$
\begin{equation}\label{eq:ST_defM}
 \left[\!\!\!\!
 \begin{array}{ccc}
 -\Gamma_{\mathrm{in}}^A - W_{BA} & W_{AB} & (\Gamma _{\mathrm{out}}^A+\Gamma
   _{\mathrm{right}}^A)  \, e^{i\chi} \\
 W_{BA} & -\Gamma _{\mathrm{in}}^B-W_{AB} & (\Gamma _{\mathrm{out}}^B+\Gamma
   _{\mathrm{right}}^B)  \, e^{i\chi} \\
 \Gamma _{\mathrm{in}}^A & \Gamma _{\mathrm{in}}^B & - \Gamma_{\mathrm{out}}
\end{array}
\!\!\! \right].
\end{equation}
Here $\Gamma_{\mathrm{out}} = (\Gamma
_{\mathrm{out}}^A+\Gamma_{\mathrm{out}}^B+\Gamma
_{\mathrm{right}}^A+\Gamma_{\mathrm{right}}^B)$ and $p_A$, $p_B$ and
$p_{n+1}$ are occupation probabilities for states $S_A$, $S_B$ and
$S_{n+1}$, respectively.
The effective tunneling rates are determined by multiplying the
tunnel coupling constants for each state with the Fermi distribution
of the electrons in the lead,
\begin{equation}\label{eq:ST_effGammaBunch}
 \Gamma _{\mathrm{in/out}}^{A/B} = f[\mp (eV - \mu_{A/B})] \,
 \Gamma_{A/B}.
\end{equation}
The tunneling rates $\Gamma _{\mathrm{right}}^A$ and $\Gamma
_{\mathrm{right}}^B$ are included to account for the possibility for
electrons to leave through the right barrier. The Fermi level of the
right lead is far below the electrochemical potential of the dot, so
that the states in the right lead can be assumed to be unoccupied.
Finally, $W_{AB}$ and $W_{BA}$ are the direct transition rates
between states $S_A$ and $S_B$. These rates obey detailed balance,
\begin{equation}\label{eq:ST_detBalBunch}
W_{AB}/W_{BA} = \exp\left[(\mu_A - \mu_B)/k_B T\right].
\end{equation}
The phenomenological relaxation rate between the two states is given
as $1/T_1 = W_{AB} + W_{BA}$.

In Eq.~(\ref{eq:ST_defM}), we introduce charge counting by
multiplying all entries of $M$ involving an electron leaving the dot
with the counting factor $\exp(i\chi)$ \cite{bagrets:2003}. We do
not distinguish whether the electron leaves the dot through the left
or the right lead. In this way we obtain the counting statistics
$p_{t_0}(N)$, which is the probability for counting $N$ events
within the time span $t_0$. The distribution describes fluctuations
of charge on the dot, which is exactly what is measured by the QPC
detector in the experiment. We stress that this distribution is
equal to the distribution of current fluctuations only if it can be
safely assumed that the electron motion is unidirectional. This is
the case if the condition in Eq.~(\ref{eq:ST_highBias}) is
fulfilled, i.e. if the tunneling due to thermal fluctuations is
suppressed. Here, we are in a regime where there is a mixture of
tunneling due to the applied bias and tunneling due to equilibrium
fluctuations. But since the model defined in Eq.~(\ref{eq:ST_defM})
is valid regardless of the direction of the electron motion, it can
still be used for analyzing the experimental data.

Using the method of Ref.~\cite{bagrets:2003}, we calculate the
lowest eigenvalue $\lambda_0(\chi)$ of $M$ and use it to obtain the
cumulant generating function (CGF) for $p_{t_0}(N)$,
\begin{equation}\label{eq:sOfChi}
  S(\chi) = -\lambda_0 (\chi) t_0.
\end{equation}
The CGF can then be used to obtain the cumulants of any order using
the relation $C_n = -(-i \partial_{\chi})^n S(\chi)|_{\chi=0}$. In
order to compare the theory with the experiment we extract the first
three cumulants of $p_{t_0}(N)$ from the experimental data.
The cumulants were found by taking a trace of length
$T=0.5~\mathrm{s}$ and splitting it into $m=T/t_0$ independent
traces. By counting the number of electrons $N$ leaving the dot in
each trace and repeating the procedure for all $m$ sub-traces, the
distribution function $p_{t_0}(N)$ could be experimentally
determined. The experimental cumulants were then calculated directly
from the measured distribution function. The time $t_0$ was chosen
such that $\langle N \rangle \approx 3$.

\begin{figure}[htb]
\centering
 \includegraphics[width=\columnwidth]{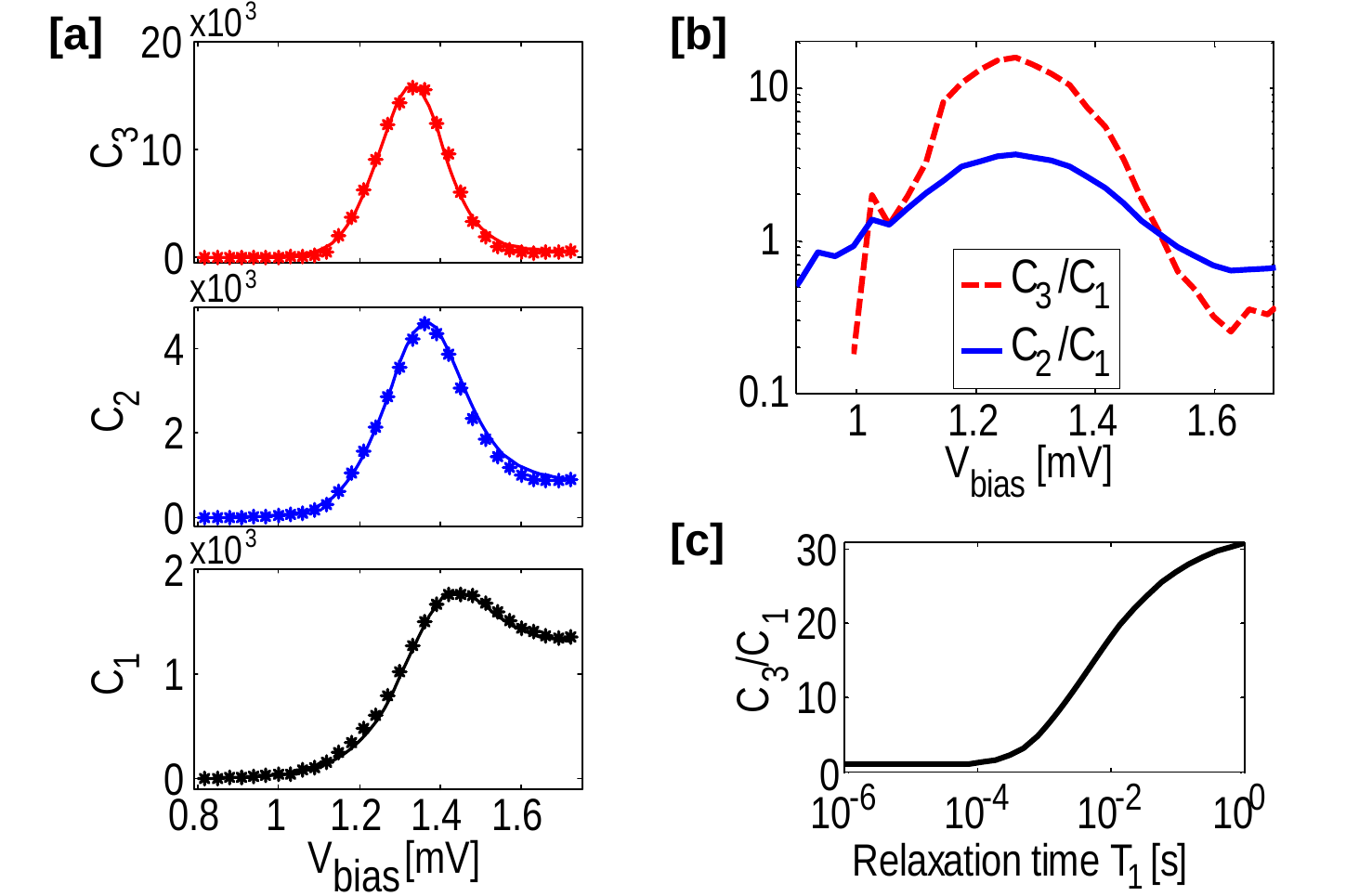}
 \caption{(a) First, second and third cumulants of the distribution
 of charge fluctuations. The symbols show values extracted from the experimental data, while
 the solid lines are calculated from the model given
 in the text. Fitting parameters are:
 $\Gamma_A = 1.6~\mathrm{kHz}$, $\Gamma_B = 20.5~\mathrm{kHz}$,
 $\Gamma_{\mathrm{right}}^A = 4.6~\mathrm{kHz}$,
 $\Gamma_{\mathrm{right}}^B = 310~\mathrm{Hz}$, $T_{1}=8~\mathrm{ms}$
 and $\mu_A-\mu_B =13~\mathrm{\mu eV}$. The electronic temperature was 400 mK.
 (b) Normalized cumulants $C_3/C_1$ and $C_2/C_1$
 versus bias voltage. The noise is clearly super-Poissonian in the
 central region of the graph. (c) Calculated maximal value of $C_3/C_1$ as a function
 of the relaxation time between the two states. The values are calculated
 by varying the relaxation time while keeping the other
 parameters to the values given by the fit shown in (a). The maximum value $C_3/C_1$
 extracted from the experimental data is $15.9$. Adapted from Ref. \cite{gustavsson:2006}.
 }\label{fig:ST_superPoisson}
\end{figure}

Figure \ref{fig:ST_superPoisson}(a) shows the first three cumulants
versus voltage applied to the source lead. The points correspond to
experimental data, while the solid lines show the cumulants
calculated from the CGF of our model, with fitting parameters
$\Gamma_A = 1.6~\mathrm{kHz}$, $\Gamma_B = 20.5~\mathrm{kHz}$,
$\Gamma_{\mathrm{right}}^A = 4.6~\mathrm{kHz}$,
$\Gamma_{\mathrm{right}}^B = 310~\mathrm{Hz}$, $T_{1}=8~\mathrm{ms}$
and $\mu_A-\mu_B =13~\mathrm{\mu eV}$. The electronic temperature in
this measurement was 400 mK. The figure shows good agreement between
the model and the experimental data.

Figure~\ref{fig:ST_superPoisson}(b) shows the normalized cumulants
$C_2/C_1$ and $C_3/C_1$ for the experimental data; we notice that
both the second and the third cumulants vastly exceed the first
cumulant when the Fermi level of the source lead is aligned with the
electrochemical potential of the dot ($V_{\mathrm{bias}} = 1.3~
\mathrm{mV}$). For a Poissonian process one expects $C_2/C_1 =
C_3/C_1 =1$; here, the noise is clearly of super-Poissonian nature,
as expected from the bunching behavior of the electrons.

When the bias voltage is further increased ($V_{\mathrm{bias}} >
1.5~ \mathrm{mV}$), the source lead is no longer in resonance with
the electrochemical potential of the dot and the equilibrium
fluctuations between the source and the dot are suppressed. In this
regime, the measured charge fluctuations are due to a current
flowing through the dot. Electrons enter the dot from the source
lead and leave the dot through the drain lead. The blocking
mechanism is no longer effective and the transport process will
predominantly take place through state $S_A$, since the tunnel
coupling to the drain lead is stronger for this state
($\Gamma^A_{\mathrm{right}} \gg \Gamma^B_{\mathrm{right}}$). The
transport through the dot can essentially be described by a rate
equation, with one rate for electrons entering and another rate for
electrons leaving the dot. For such systems, it has been shown in
section \ref{sec:ST_subPoisson} that the Coulomb blockade will lead
to an increase in correlation between the tunneling electrons
compared to a single-barrier structure, giving sub-Poissonian noise
\cite{davies:1992, gustavsson:2005}. The effect is seen for
$V_{\mathrm{bias}}>1.5~\mathrm{mV}$ in
Fig~\ref{fig:ST_superPoisson}(b); both the second and third
cumulants are reduced compared to the first cumulant.

The value of $T_{1}=8~\mathrm{ms}$ obtained from fitting the
experimental data is of the same order of magnitude as previously
reported values for the spin relaxation time $T_1$. We stress that
the bunching of electrons and the super-Poissonian noise can only
exist if the relaxation time is at least as long as the inverse
tunneling time. This is demonstrated in
Fig.~\ref{fig:ST_superPoisson}(c), which shows the maximum value
obtained for the ratio $C_3/C_1$ calculated for different $T_{1}$
while keeping the rest of the fitting parameters at the values given
in the caption of Fig.~\ref{fig:ST_superPoisson}.

\subsection{Higher order moments and limitations of the
detector}\label{sec:ST_higherMoments}
So far, we have presented measurements of the second and third
cumulants or central moments. As mentioned in
section~\ref{sec:ST_generalNoise}, the shot noise is a direct
consequence of the discreteness of the charge carriers in the
system. A measurement of the second moment (Fano factor) thus
provides a way to determine the charge of those discrete carriers.
The third moment of a tunneling current has been shown to be
independent of the thermal noise \cite{levitov:2004, fujisawa:2006},
thus making it a potential tool for investigating electron-electron
interactions even at elevated temperatures.

What about the higher order moments? In strongly interacting
systems, they are predicted to depend strongly on both the
conductance \cite{bagrets:2006} and on the internal level structure
\cite{belzig:2005} of the system. Determining higher order moments
may therefore give a more complete characterization of the electron
transport process. This can be of importance for realizing
measurements of electron correlation and entanglement effects in
quantum dots \cite{loss:2000, saraga:2003}. In quantum optics,
higher order moments are routinely measured in order to study
entanglement and coherence effects of the electromagnetic field
\cite{mandel:1995}.

In this section, we present measurements of the fourth and fifth
cumulant of the distribution function for charge transport through a
QD. As demonstrated in section \ref{sec:ST_subPoisson}, we determine
the cumulants by first generating the experimental probability
density function $p_{t_0}(N)$. This is done by splitting a time
trace of length $T$ into $m=T/t_0$ intervals and counting the number
of electrons entering the dot within each interval. The higher
cumulants describe more subtle features of the distribution
function. To extend the methods of section \ref{sec:ST_subPoisson}
to higher cumulants, it is therefore necessary to increase the
measurement time to collect more statistics. This requires a stable
sample without any fluctuating charge traps close to the QD.

In the experiment, we use a single QD with the same design as the one described in
section \ref{sec:TR_main} and section \ref{sec:ST_subPoisson}. The coupling between the QD and the QPC was weaker in
the sample used here, meaning that the bandwidth had to
be reduced below $10~\mathrm{kHz}$. On the other hand, the stability
of the structure allowed the measurement of time traces of length
$T=10~\mathrm{minutes}$. In the experiment, the QPC was voltage
biased with $V_{QPC} = 250~\mu V$. The current signal was sampled at
$100~\mathrm{kHz}$, software filtered at $4~\mathrm{kHz}$ using an
8th order Butterworth filter and finally resampled at
$20~\mathrm{kHz}$ in real-time to keep the amount of data
manageable.


\begin{figure}[htb]
\centering
 \includegraphics[width=.95\columnwidth]{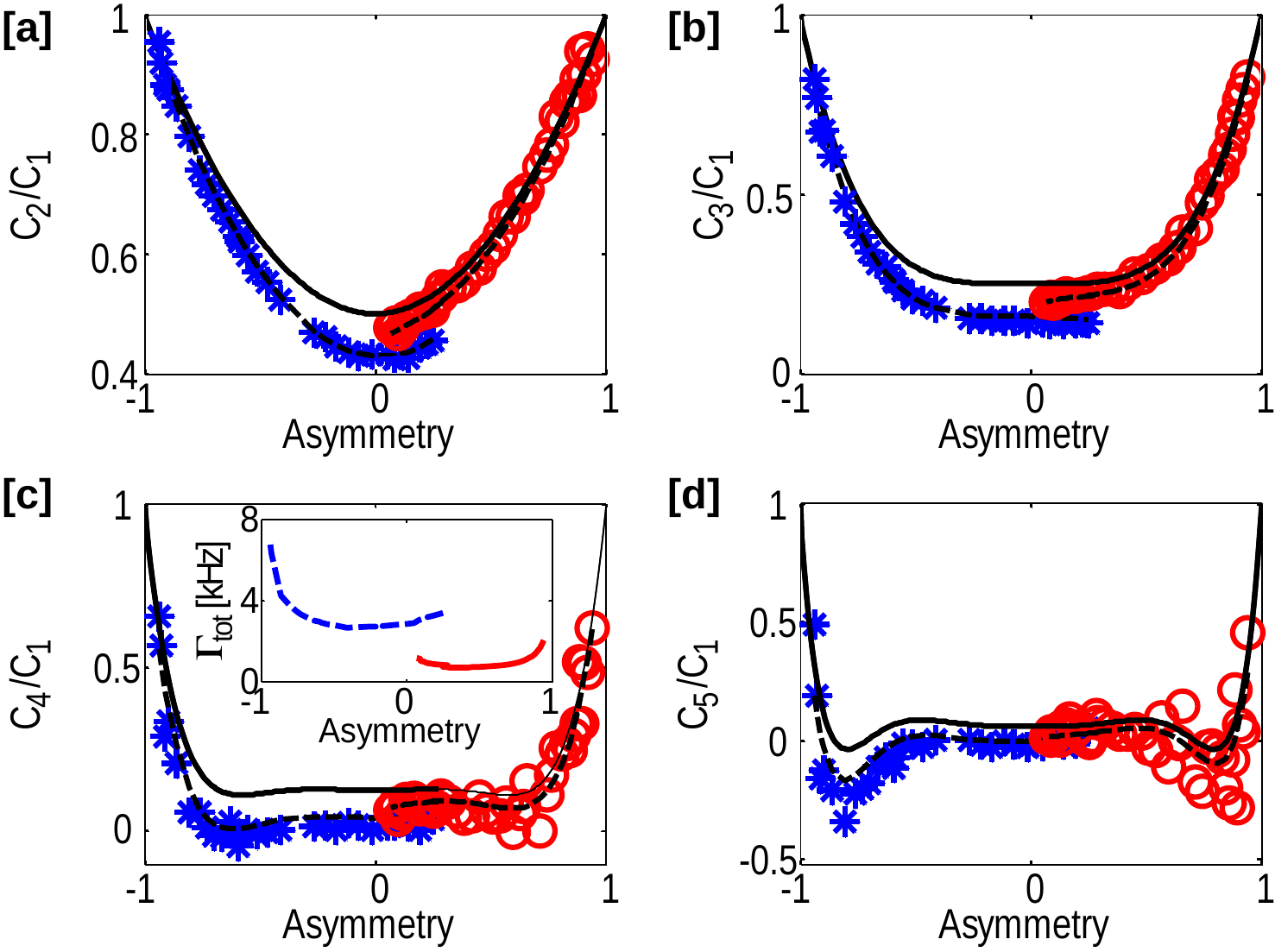}
 \caption{(a-d) Normalized cumulants $C_n/C_1$ versus dot asymmetry,
 $a = (\Gamma_{\mathrm{in}}-\Gamma_{\mathrm{out}})/(\Gamma_{\mathrm{in}}+\Gamma_{\mathrm{out}})$.
 The solid lines are theoretical predictions assuming a perfect
 detector, $C_2/C_1 = (1+a^2)/2$, $C_3/C_1 = (1+3a^4)/4$,
 $C_4/C_1 = (1+a^2-9 a^4+15 a^6)/8$ and $C_5/C_1 = (1+30 a^4-120 a^6+105 a^8)/16$.
 The dashed lines show the cumulants calculated from the
 model defined by Eq.~(\ref{eq:ST_FiniteBW:mainM}) in the text.
 The inset in (c) shows the variation of the total tunneling rate
 $\Gamma_{\mathrm{tot}} = \Gamma_{\mathrm{in}}
 +\Gamma_{\mathrm{out}}$ for the different measurement points. Adapted from Ref. \cite{gustavsson:2007}.
 } \label{fig:ST_FiniteBW:CvsA}
\end{figure}

The results are shown in Fig. \ref{fig:ST_FiniteBW:CvsA}, where we
plot the normalized cumulants for different values of the asymmetry
of the tunneling rates,
$a=(\Gamma_{\mathrm{in}}-\Gamma_{\mathrm{out}})/(\Gamma_{\mathrm{in}}+\Gamma_{\mathrm{out}})$.
The asymmetry is tuned by shifting the voltage on gate $G_1$ by an
amount $\Delta V$ and at the same time applying a compensating
voltage $-\Delta V$ on gate $G_2$. With the two gates having a
similar lever arm on the dot, the electrochemical potential of the
QD remains at the same level, but the height of the tunneling
barriers between the dot and the source and drain leads will change.
Doing so, we could tune the asymmetry from $a=-0.94$ to $a=+0.25$
while still keeping both tunneling rates within the measurement
bandwidth and avoiding charge rearrangements. To get data for the
full range of asymmetry, we did a second measurement at a different
gate voltage configuration. For the second set of data, the
asymmetry was tuned from $a=0.07$ to $a=0.93$. The stars and the
circles in Fig. \ref{fig:ST_FiniteBW:CvsA} represent data from the
two different sets of measurements. The measurements were performed
with a QD bias of $V_{\mathrm{bias}}=2.5~\mathrm{mV}$, with the
electrochemical potential of the dot far away from the Fermi levels
of the source and drain leads. This is to ensure that tunneling due
to thermal fluctuations is sufficiently suppressed.


The solid lines in Fig. \ref{fig:ST_FiniteBW:CvsA} depict the
theoretical predictions calculated from a two-state model
\cite{bagrets:2003}. The analytical expressions are given in the
figure caption. The higher cumulants show a complex behavior as a
function of the asymmetry, with local minima at $a = \pm 0.6$ for
$C_4/C_1$ and at $a = \pm 0.8$ for $C_5/C_1$. The fifth cumulant
even becomes negative for some configurations. The experimental data
qualitatively agrees with the theory, but for small values of the
asymmetry there are deviations from the expected behavior. The
deviations are stronger for the first set of data (stars). Since the
tunneling rates in the first measurement was about a factor of three
higher than in the second measurement [see inset of Fig.
\ref{fig:ST_FiniteBW:CvsA}(c)], we suspect the finite bandwidth of
the detector to be a possible reason for the discrepancies.

In general, experimental measurements of FCS for electrons are
difficult to achieve due to the need of a sensitive, high-bandwidth
detector capable of resolving individual electrons \cite{LuW:2003,
fujisawa:2004, bylander:2005}. However, a more fundamental
complication with the measurements is that most forms of the FCS
theory assume the existence of (1) a detector with infinite
bandwidth and (2) infinitely long data traces. Since no physical
detector or experiment can fulfill these requirements, every
experimental realization of the FCS will measure a distribution
which is influenced by the properties of the detector. In the
following, we investigate how the violation of the two assumptions
modifies the measured statistics.

%
%
%


Naaman et al.~\cite{naaman:2006} pointed out that measurements of
the transition rates of a Poisson two-state system using a finite
bandwidth detector always leads to an underestimate of the rates.
As a result, the measured probability distribution for the times
needed for an electron to tunnel into or out of the QD no longer
follow the expected exponential $p_{\mathrm{in/out}}(t) =
\Gamma_{\mathrm{in/out}} \exp(-\Gamma_{\mathrm{in/out}}\,t )$. Due
to the finite detection time, very fast tunneling events are less
likely to be detected, giving a cut-off for short time scales in the
measured distribution. Moreover, since the fast events are not
detected, the measurement will over-estimate the occurrence of slow
events. The long-time tail of the measured distribution will still
decay exponentially, but the tunneling rate extracted from the
distribution will be under-estimated. To determine the rates
correctly, the detection rate $\Gamma_{\mathrm{det}}$ of the
detector must be taken into account \cite{naaman:2006}.

\begin{figure}[tb]
\centering
 \includegraphics[width=\columnwidth]{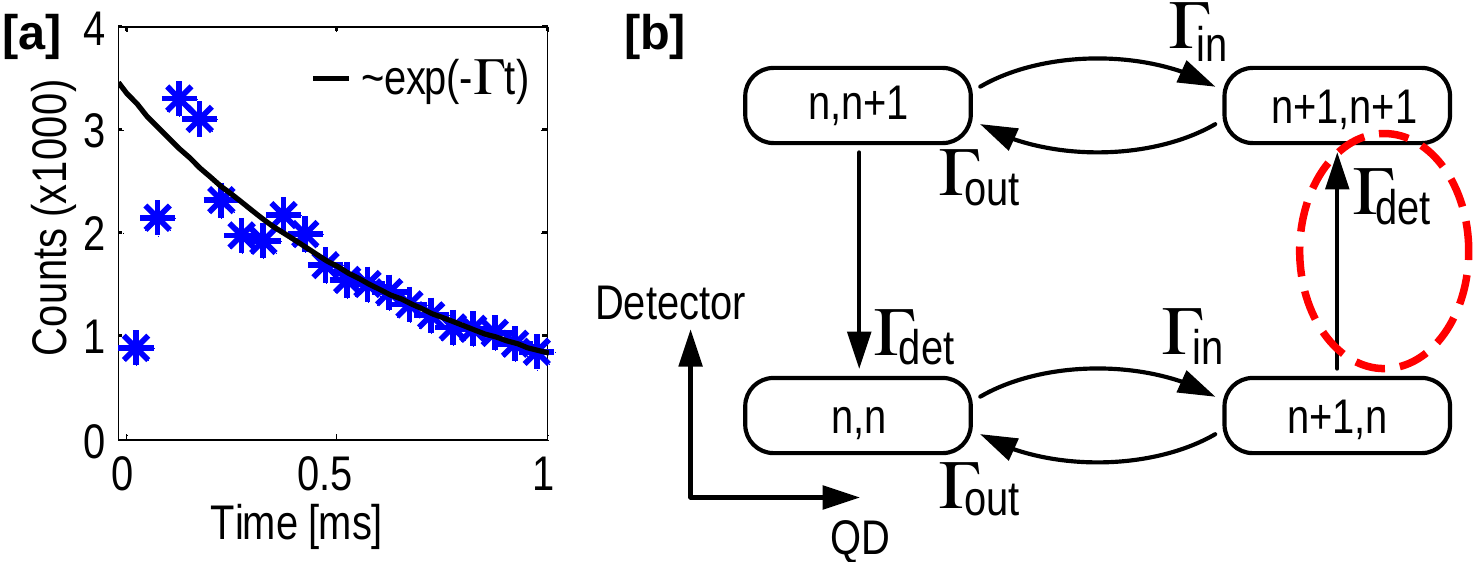}
 \caption{
 (a) Probability
 density of time needed for an electron to tunnel into the dot. Note
 the sharp decrease in counts for $t<100~\mathrm{\mu s}$ due to the
 finite bandwidth of the detector.
 The black curve is a long-time exponential fit with $\Gamma = 1.39~\mathrm{kHz}$.
 (b) Model for the dot-detector
 system. A state $(n,m)$ corresponds to $n$ electrons on the dot
 while the detector at the same time is measuring $m$ electrons. Adapted from Ref. \cite{gustavsson:2007}.
 } \label{fig:ST_FiniteBW:model}
\end{figure}

An example of a probability distribution taken from measured data is
shown in Fig. \ref{fig:ST_FiniteBW:model}(a). The long-time behavior
is exponential, but for times $t<100~\mathrm{\mu s}$ there is a
sharp decrease in the number of counts registered by the detector.
From the figure, we can estimate $\tau_{\mathrm{det}} =
1/\Gamma_\mathrm{det}$, which is the average time it takes for the
detector to register an event. We find $\tau_{\mathrm{det}} =
70~\mathrm{\mu s}$, giving a detection rate of
$\Gamma_{\mathrm{det}} = 1/\tau_{\mathrm{det}} = 14~\mathrm{kHz}$.
Note that the detection rate $\Gamma_{\mathrm{det}}$ does not only
depend on the measurement bandwidth but also on the signal-to-noise
ratio of the detector signal as well as the redundancy needed to
minimize the risk of detecting false events \cite{shannon:1949}. The
compensations for the tunneling rates are given as
\cite{naaman:2006}

\begin{eqnarray}\label{eq:ST_FiniteBW:compGamma1}
  \Gamma_{\mathrm{in}} &=&
 \Gamma^*_{\mathrm{in}} \frac{\Gamma_\mathrm{det}}
 {\Gamma_\mathrm{det}- \Gamma^*_\mathrm{in}-\Gamma^*_\mathrm{out}},\\ \label{eq:ST_FiniteBW:compGamma2}
  \Gamma_{\mathrm{out}} &=&
 \Gamma^*_{\mathrm{out}} \frac{\Gamma_\mathrm{det}}
 {\Gamma_\mathrm{det}- \Gamma^*_\mathrm{in}-\Gamma^*_\mathrm{out}}.
\end{eqnarray}
Here, $\Gamma_{\mathrm{in/out}}$ are the true tunneling couplings
and $\Gamma^*_{\mathrm{in/out}} = 1/\langle \tau_\mathrm{in/out}
\rangle$ are rates extracted from the measurement. All tunneling
rates presented in the following have been extracted using
Eqs.~(\ref{eq:ST_FiniteBW:compGamma1}-\ref{eq:ST_FiniteBW:compGamma2})
with $\Gamma_{\mathrm{det}}=14~\mathrm{kHz}$.



The finite bandwidth will also influence the FCS measured by the
detector. Following the ideas of Ref. \cite{naaman:2006}, we account
for the finite bandwidth by including the states of the detector
into the two-state model of section \ref{sec:ST_subModel}. Figure
\ref{fig:ST_FiniteBW:model}(b) shows the four possible states of the
combined dot-detector model. The state $(n+1,n)$ refers to a
situation where there are $n+1$ electrons on the dot, while the
detector at the same time reads $n$ electrons. The transition from
the state $(n+1,n)$ to the state $(n+1,n+1)$ occurs when the
detector registers the electron. This process occurs with the rate
of the detector, $\Gamma_{\mathrm{det}}$.

To calculate the FCS for the QD-detector system, we write the master
equation $\dot{P} = M\,P$, with $P =
[(n,n),(n+1,n),(n,n+1),(n+1,n+1)]$ and $M_{\chi}=$
\begin{equation}\label{eq:ST_FiniteBW:mainM}
\left[ \!\!\!
\begin{array}{cccc}
  -\Gamma_{\mathrm{in}} & \Gamma_{\mathrm{out}} & \Gamma_{\mathrm{det}} & 0 \\
  \Gamma_{\mathrm{in}} & -(\Gamma_{\mathrm{out}}+\Gamma_{\mathrm{det}}) & 0 & 0 \\
  0 & 0 & -(\Gamma_{\mathrm{in}} + \Gamma_{\mathrm{det}}) & \Gamma_{\mathrm{out}} \\
  0 & \Gamma_{\mathrm{det}}* e^{i\chi} & \Gamma_{\mathrm{in}} &
  -\Gamma_{\mathrm{out}}
\end{array}
\!\!\! \right].
\end{equation}
In the above matrix, we have included the counting factor
$e^{i\chi}$ at the element where the detector registers an electron
tunneling into the dot [see dashed circle in Fig.
\ref{fig:ST_FiniteBW:model}(b)]. The statistics obtained in this way
relates directly to what is measured in the experiment. Using the
methods of Ref. \cite{bagrets:2003}, we calculate the first few
cumulants for the above expression as a function of relative
bandwidth $k = \Gamma_\mathrm{det}/(\Gamma_{\mathrm{in}} +
\Gamma_{\mathrm{out}})$ and asymmetry $a = (\Gamma_{\mathrm{in}} -
\Gamma_{\mathrm{out}})/(\Gamma_{\mathrm{in}} +
\Gamma_{\mathrm{out}})$. The normalized second and third cumulants
take the form
\begin{eqnarray}
 \label{eq:ST_FiniteBW:C2BW}
 C_2/C_1 &=&
 \frac{1+a^2}{2} - \frac{k (1-a^2)}{2(1+k)^2},\\
 \label{eq:ST_FiniteBW:C3BW}
 C_3/C_1 &=&
 \frac{1+ 3a^4}{4} -
 \frac{3 k (1 + k + k^2 )}{4(1+k)^4} - \nonumber\\
 & & \frac{6 \, a^2 k^2}{4(1+k)^4} + \frac{3\, a^4 k (1 + 3 k + k^2
 )}{4(1+k)^4}.
\end{eqnarray}

\begin{figure}[tb]
\centering
 \includegraphics[width=.95\columnwidth]{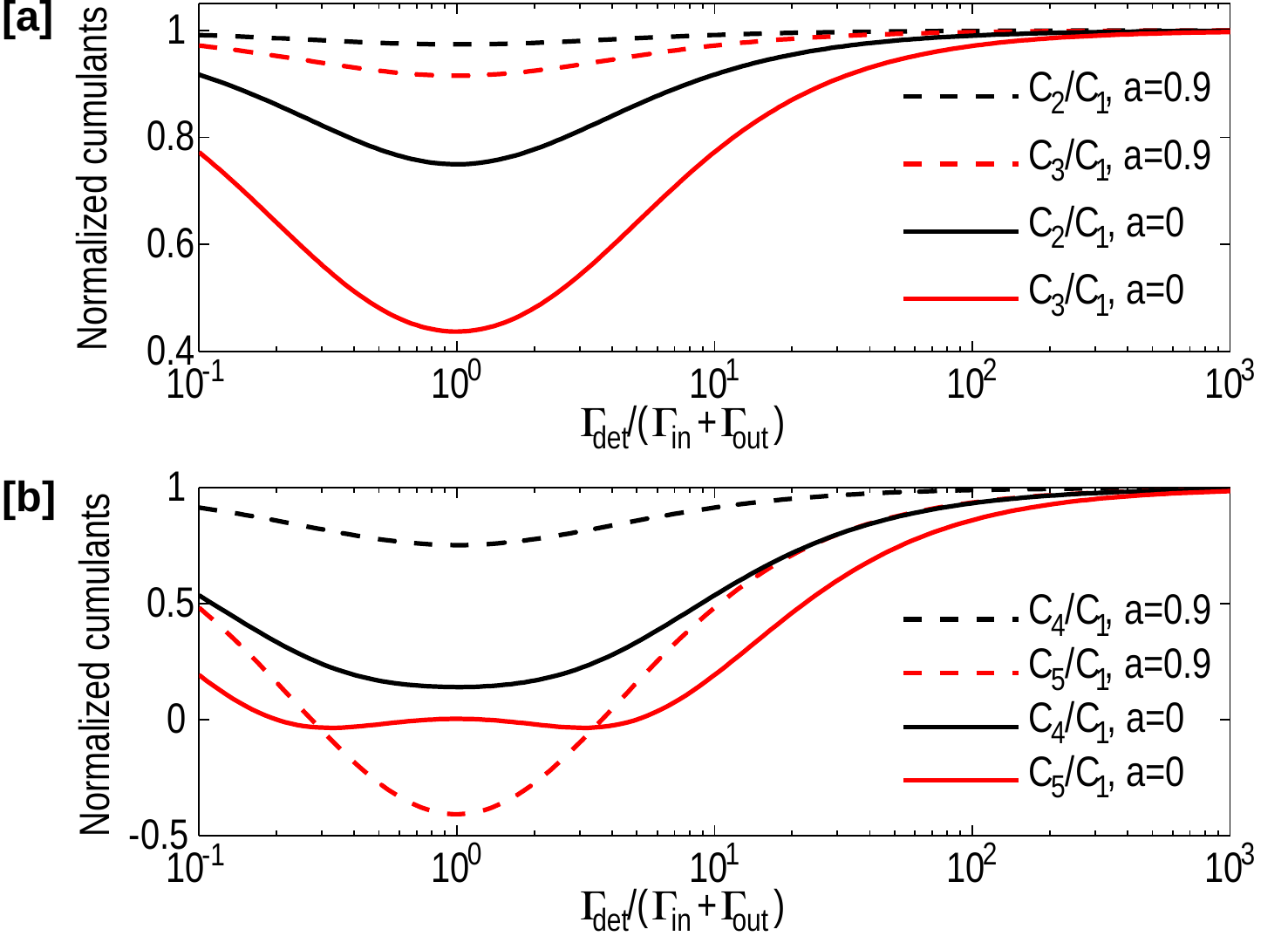}
 \caption{
 Higher cumulants versus relative detection bandwidth
 $\Gamma_{\mathrm{det}}/(\Gamma_{\mathrm{in}}+\Gamma_{\mathrm{out}})$,
 calculated from the model in Fig.~\ref{fig:ST_FiniteBW:model}(b).
 The cumulants are normalized with respect to the results for the infinite-bandwidth case. The
 influence of the finite bandwidth is maximal when the asymmetry
 $a=(\Gamma_{\mathrm{in}}-\Gamma_{\mathrm{out}})/(\Gamma_{\mathrm{in}}+\Gamma_{\mathrm{out}})$
 is zero. Adapted from Ref. \cite{gustavsson:2007}.
 } \label{fig:ST_FiniteBW:corr}
\end{figure}

In Fig.~\ref{fig:ST_FiniteBW:corr}(a) we plot the second and third
cumulants from Eq. (\ref{eq:ST_FiniteBW:C2BW}) and Eq.
(\ref{eq:ST_FiniteBW:C3BW}) for different values of asymmetry $a$
and relative bandwidth $k$. The cumulants have been normalized to
the values for the infinite bandwidth detector. Fig.
\ref{fig:ST_FiniteBW:corr}(b) shows the corresponding results for
the forth and fifth cumulants. With $\Gamma_\mathrm{det} \gg
\Gamma_{\mathrm{in}} + \Gamma_{\mathrm{out}}$, the cumulants
approach the infinite bandwidth result, as expected. However, even
with $\Gamma_\mathrm{det} = 10(\Gamma_{\mathrm{in}} +
\Gamma_{\mathrm{out}})$ and perfect symmetry ($a=0$), the second
cumulant deviates by almost 10\% and the third cumulant by more than
20\% from the perfect detector values. As the bandwidth is further
decreased, the deviations grow stronger and reach a maximum as
$\Gamma_\mathrm{det} = \Gamma_{\mathrm{in}} +
\Gamma_{\mathrm{out}}$. With $\Gamma_\mathrm{det} \ll
\Gamma_{\mathrm{in}} + \Gamma_{\mathrm{out}}$, the cumulants once
again approach the perfect detector values. When the detector is
much slower than the underlying tunneling process, it will only
sample the average population of the two states. In this limit, the
dynamics of the system does not interfere with the dynamics of the
detector and we recover the correct relative noise levels. It should
be noted that this is true only for the noise relative to the
detected mean current. Since the detector will miss most of the
tunneling events, the absolute values of both the current and the
noise will be underestimated.

Over the full range of bandwidth and asymmetry, we find that the
noise detected with the finite bandwidth system is always lower than
for the ideal detector case. The reduction can be qualitatively
understood by considering the probability distribution $p_{t_0}(N)$.
The finite bandwidth makes it less probable to detect fast events,
meaning that the probability of detecting a large number of
electrons within the interval $t_0$ will decrease more than the
probability of detecting few electrons. This will cut the high-count
tail of the distribution and thereby reduce its width ($C_2$) and
its skewness ($C_3$). An interesting feature is that the cumulants
calculated for a less symmetric configurations [$a=0.9$ in Fig.
\ref{fig:ST_FiniteBW:model}(c)] show less influence of the
finite bandwidth. 

\begin{figure}[tb]
\centering
 \includegraphics[width=\columnwidth]{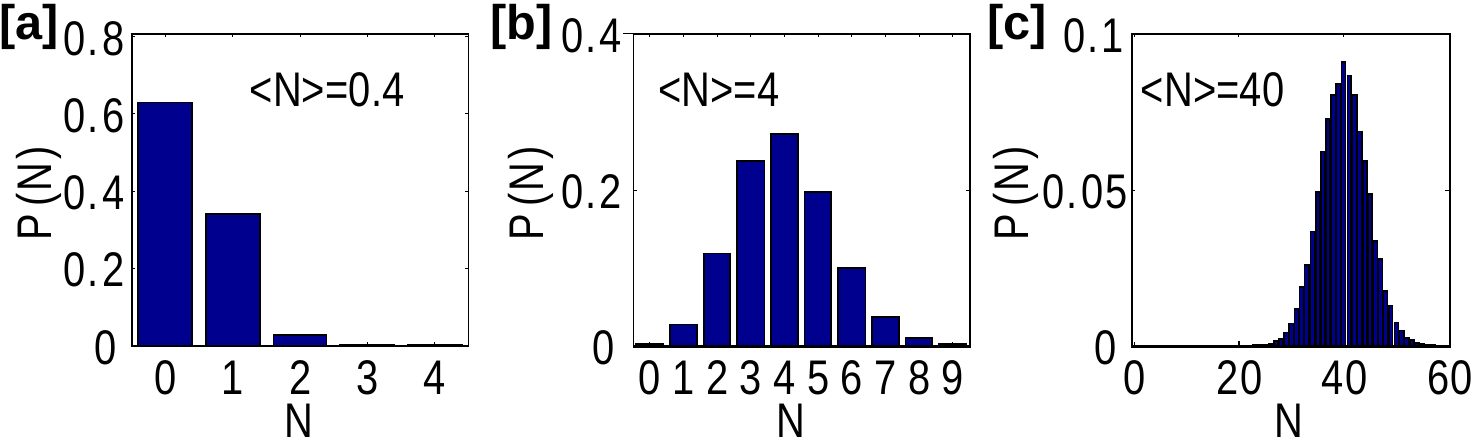}
 \caption{ Current distribution functions, extracted
 for different values of $t_0$. The underlying data is the same in
 all three figures. For very short $t_0$ [case (a)], the distribution clearly
 has different properties compared to case (c).
 } \label{fig:ST_FiniteBW:histDiffT0}
\end{figure}

A second limitation of a general FCS measurement is the finite
length of each time trace. In order to generate the experimental
probability density function $p_{t_0}(N)$, the total trace of length
$T$ must be split into $m=T/t_0$ intervals, each of length $t_0$.
Most FCS theories only predict results for the case $t_0 \gg
1/\Gamma$, where $\Gamma$ is a typical transition rate of the
system.
In the experiment, it is favorable to make $t_0$ as short as
possible in order to increase the number of samples $m=T/t_0$. This
will improve the quality of the distribution and help to minimize
statistical errors. However, if $t_0$ is made too short, this will
influence the extracted distribution. This is visualized in
Fig.~\ref{fig:ST_FiniteBW:histDiffT0}, where distribution functions
for different $t_0$ are extracted from the same set of experimental
data. The distributions give the same current $I=e \langle N \rangle
/ t_0$, but their properties are clearly different. In the extreme
case of $t_0 \ll 1/\Gamma~(\langle N \rangle \ll 1)$, the
distribution approaches the Bernoulli distribution, for which only
$p_{t_0}(0)$ and $p_{t_0}(1)$ are non-zero.

The condition $t_0 \gg 1/\Gamma$ is imposed by the approximation
that the cumulant generating function (CGF) $S(\chi)$ for
$p_{t_0}(N)$ only depends on the lowest eigenvalue
$\Lambda_{\mathrm{min}}$ of the master equation matrix $M_{\chi}$,
with $S(\chi) = -t_0 \Lambda_{\mathrm{min}}$. A FCS valid for finite
$t_0$ must include all eigenvalues and eigenvectors of $M_{\chi}$
\cite{bagrets:2003}. The corresponding expression is
\begin{equation}\label{eq:ST_FiniteBW:SAll}
    \exp[S(\chi)] = \langle q_0 | p^{(n)}\rangle
    \exp(-t_0 \Lambda_n)
    \langle q^{(n)} | p_0\rangle,
\end{equation}
where $\langle q^{(n)}|$ and $|p^{(n)}\rangle$ are the left and
right eigenvectors of the matrix $M_{\chi}$, $\Lambda_{n}$ are the
eigenvalues of $M_{\chi}$ and $\langle q_0|,~|p_0\rangle$ are the
eigenvectors corresponding to the lowest eigenvalue
$\Lambda_{\mathrm{min}}$. The cumulants generated from the CGF in
Eq.~(\ref{eq:ST_FiniteBW:SAll}) will in general be a function of
$t_0$.

\begin{figure}[tb]
\centering
 \includegraphics[width=\columnwidth]{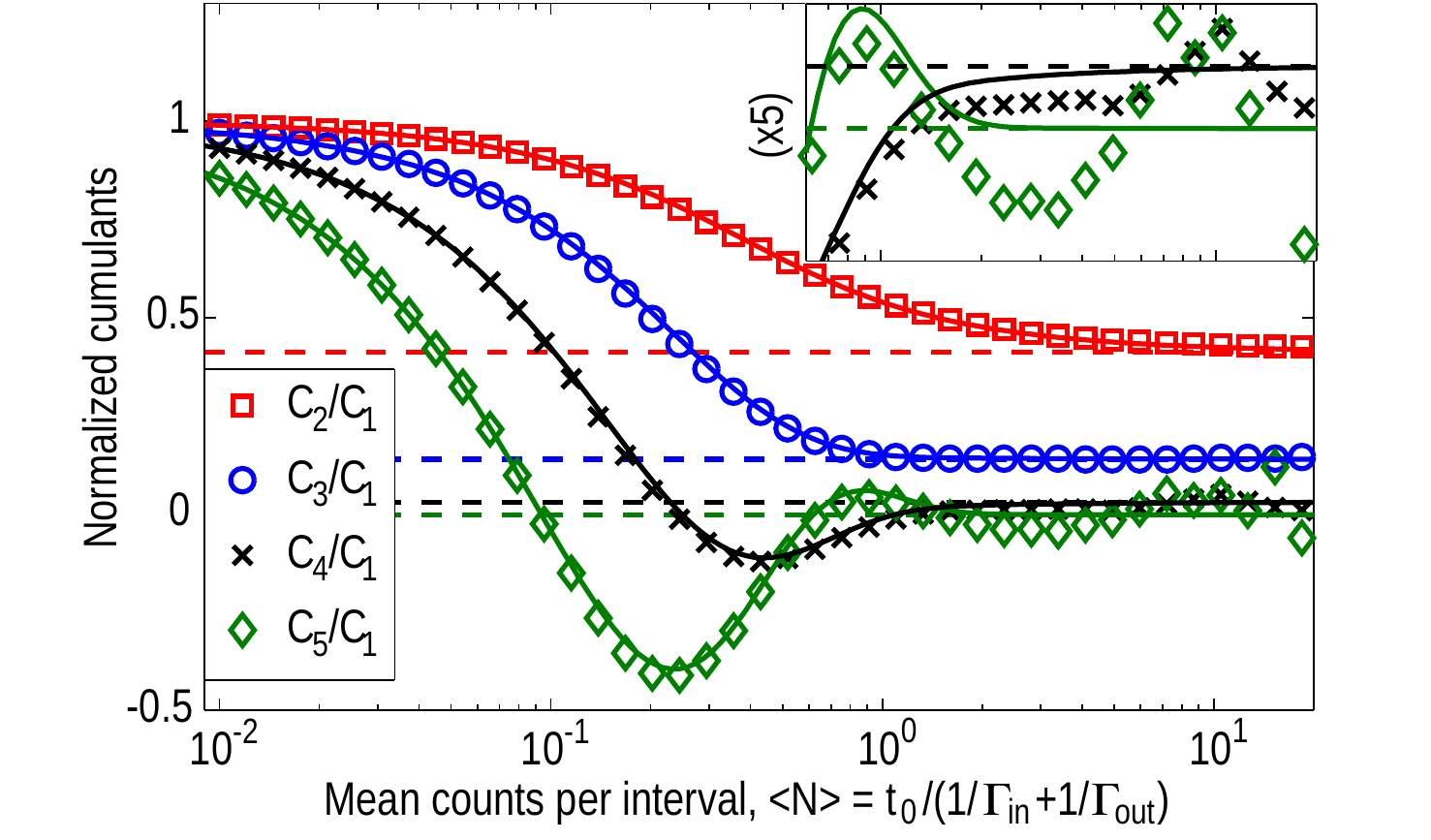}
 \caption{Normalized cumulants evaluated for different lengths of the time
 interval $t_0$. The symbols show the experimental
 data, extracted from a time trace of length $T=10~\mathrm{minutes}$,
 containing 350595 events, with $a= 0.053$, and $\Gamma_{\mathrm{tot}} =
 3062~\mathrm{Hz}$. The solid lines are calculations from the
 FCS given by Eq. (\ref{eq:ST_FiniteBW:SAll}) in the text, while the
 dashed lines are the asymptotes for $t_0 \rightarrow \infty$. The
 inset shows a magnification of the vertical axis (horizontal
 axis unchanged) for $C_4/C_1$ and $C_5/C_1$ for $\langle N \rangle >
 0.6$. Adapted from Ref. \cite{gustavsson:2007}.
 }
\label{fig:ST_FiniteBW:CvsN}
\end{figure}

To investigate how small $t_0$ can be before systematic errors
become relevant, we calculate the cumulants from the CGF of Eq.
(\ref{eq:ST_FiniteBW:SAll}) with the master equation matrix
$M_{\chi}$ of Eq. (\ref{eq:ST_FiniteBW:mainM}). The results are
shown in Fig. \ref{fig:ST_FiniteBW:CvsN}, where we plot the
normalized cumulants as a function of the mean number of counts per
interval, $\langle N \rangle = t_0/(1/\Gamma_{\mathrm{in}} +
1/\Gamma_{\mathrm{out}})$. The symbols show cumulants extracted from
measured data ($T = 10~\mathrm{minutes}$, $a=0.053$,
$\Gamma_{\mathrm{in}} + \Gamma_{\mathrm{out}} = 3062~\mathrm{Hz}$
and $\Gamma_{\mathrm{det}}=14~\mathrm{kHz}$), while the solid lines
are results from the CGF for the same set of parameters. The dashed
lines are the asymptotes for the limiting case $t_0 \rightarrow
\infty$.

In general, data and theory are in good agreement. There are
some deviations in the fourth and fifth cumulants for large $t_0$
($\langle N \rangle>6$ in Fig. \ref{fig:ST_FiniteBW:CvsN}), but
these are statistical errors in the experiment due to the finite length
of the total time trace. For short $t_0$, all cumulants converge to
$C_n/C_1 \rightarrow 1$. This is because as $\langle N \rangle \ll
1$, the probability distribution $p_{t_0}(N)$ will be non-zero only
for $N=0$ and $N=1$, with $p_{t_0}(0)=1-q$, $p_{t_0}(1)=q$ and
$q=\langle N \rangle$. This is the definition of a Bernoulli
distribution, for which the normalized cumulants $C_n/C_1
\rightarrow 1$ as $q\rightarrow 0$ \cite{mathWorldBernoulli}.

Focusing on the other regime, $\langle N \rangle > 1$, we see that
cumulants of different orders converge to their asymptotic limits
for different values of $t_0$. The second cumulant needs a longer
interval $t_0$ to reach a specified tolerance compared to the higher
cumulants. This is of interest for the experimental determination of
higher cumulants.
By choosing a shorter value of $t_0$ when calculating higher
cumulants, the amount of samples $m = T/t_0$ can be increased. For
the data in Fig.~\ref{fig:ST_FiniteBW:CvsA}, the cumulants were
calculated with intervals $t_0$ giving $\langle N \rangle = 15$ for
$C_2$, $\langle N \rangle = 6$ for $C_3$, $\langle N \rangle = 3$
for $C_4$ and $\langle N \rangle = 2$ for $C_5$.
The maximal deviations between the correct cumulants and the ones
determined with a finite length $t_0$ can be estimated by checking
the convergence for all values of the asymmetry. For the data shown
in Fig.~\ref{fig:ST_FiniteBW:CvsA}, we find $\Delta C_2/C_1 =
0.007$, $\Delta C_3/C_1 = 0.009$, $\Delta C_4/C_1 = 0.01$ and
$\Delta C_5/C_1 = 0.03$.

Coming back to the results of Fig. \ref{fig:ST_FiniteBW:CvsA}, we
are now able to explain why the measured cumulants show lower values
compared to the perfect-detector theory.
The dashed lines in Fig. \ref{fig:ST_FiniteBW:CvsA} are the
cumulants calculated from the combined QD-detector model of Eq.
(\ref{eq:ST_FiniteBW:mainM}), with $\Gamma_{\mathrm{det}} =
14~\mathrm{kHz}$. The overall agreement is good, especially since no
fitting parameters are involved. Higher cumulants end up to be
slightly lower than theory predicts. We speculate that the
deviations could be due to low-frequency fluctuations of the
tunneling rates over the time of measurement.

\subsection{Measurement precision} \label{sec:ST_precision}
In this section we investigate the precision possible to achieve
with a current meter based on single-electron counting. For this
purpose, we assume a QD in the high-bias regime with a single state
available for transport, i.e., the model defined by
Eq.~(\ref{eq:ST_master}) in section~\ref{sec:ST_subModel}. As
derived in section~\ref{sec:ST_subModel}, the current $I$ and the
shot noise are
\begin{equation} \label{eq:ST_precision:curr}
I = e \frac{\Gamma_{\mathrm{in}}
\Gamma_{\mathrm{out}}}{\Gamma_{\mathrm{in}} +
\Gamma_{\mathrm{out}}},
\end{equation}
\begin{equation} \label{eq:ST_precision:noise}
S_I = 2e^2
\frac{\Gamma_\mathrm{in}\,\Gamma_\mathrm{out}\,(\Gamma_\mathrm{in}^2+\Gamma_\mathrm{out}^2)}{(\Gamma_\mathrm{in}
+ \Gamma_\mathrm{out})^3}.
\end{equation}
When counting electrons passing through the QD, we use the tunneling
electrons to probe the tunnel couplings
$\Gamma_\mathrm{in}/\Gamma_\mathrm{out}$. Since tunneling is a
statistical process, it involves a certain degree of randomness and
we need to detect an ensemble of electrons in order to be able to
form the average $\Gamma_\mathrm{in/out} = 1/\langle
\tau_\mathrm{in/out} \rangle$. The statistical variations of the
tunneling times imply that there is relation between the duration
and the precision of the measurement. More precisely, assuming that
the tunneling rates $\Gamma_\mathrm{in}/\Gamma_\mathrm{out}$ in
Eqs.~(\ref{eq:ST_precision:curr}-\ref{eq:ST_precision:noise}) are
constant, for how long is it necessary to measure in order to reach
a certain precision in the current or the noise level? This is
investigated in the following section. The theoretical findings are
then compared with experimental results.

\subsection{Theoretical precision}
In the single-level regime, the process of an electron tunneling
into or out of the dot is described by the rate equation
\begin{equation}\label{app:eqRate}
    \dot{p}_{\mathrm{in/out}}(t) = - \Gamma_{\mathrm{in/out}} \times
    p_{\mathrm{in/out}}(t).
\end{equation}
Here, $p_{\mathrm{in/out}}(t)$ is the probability density for an
electron to tunnel into or out of the dot at a time $t$ after a
complementary event. Since the expressions for electrons entering
and leaving the dot are the same, we drop the subscripts ($in/out$)
and use the notations $p(t)$ and $\Gamma$ to describe either one of
the two processes. Solving the differential equation and normalizing
the resulting distribution gives
\begin{equation}\label{app:expDecay}
p(t) \mathrm{dt} = \Gamma \mathrm{e}^{-\Gamma t} \mathrm{dt}.
\end{equation}

In the experiment, we measure a time trace containing a sequence of
tunneling times $\tau_k$, $k=1,2,3,\ldots$ To estimate $\Gamma$ and
its relative accuracy from such a sequence, we need to calculate the
probability distribution for extracting a certain value $\Gamma$,
given a fixed sequence of tunneling times. We start by dividing the
time axis into bins of width $\Delta \tau$ and number them with
$i=0,1,2,\ldots$ A tunneling event $\tau_{k}$ will be counted in bin
$i$ if $i\Delta \tau \leq \tau_{k} < (i+1)\Delta \tau$. Using Eq.
(\ref{app:expDecay}) and assuming $\Delta \tau \ll 1/\Gamma$, we
find that the probability to get a count in bin $i$ for a given
value of $\Gamma$ is equal to
\begin{equation}\label{app:eqPinOut}
 p(i|\Gamma) = \Gamma \Delta \tau
 \mathrm{e}^{-\Gamma \Delta \tau\,i}.
\end{equation}
A certain sequence $\{i_n\}$ is realized with probability
\begin{eqnarray}\label{app:eqSeq}
  p(\{i_n\}|\Gamma) &=& \prod_{n=1}^{N} \Gamma \Delta \tau
 \mathrm{e}^{-\Gamma \Delta \tau\,i_n} = \nonumber \\
 &=& (\Gamma \Delta \tau)^N \mathrm{e}^{-\Gamma \Delta \tau \sum_{n=1}^N
 i_n} = \nonumber \\
 &=& (\Gamma \Delta \tau)^N \mathrm{e}^{-\Gamma \Delta \tau \sum_{i=0}^{\infty} n_i i} = \nonumber \\
 &=& (\Gamma \Delta \tau)^N \mathrm{e}^{-\Gamma \Delta \tau N \langle i
 \rangle}.
\end{eqnarray}
Here, $n_i$ is the number of times an event falls into bin $i$,
$\sum_{i=0}^{\infty} n_i = N$ is the total number of events in the
trace and $\langle i \rangle = \frac{1}{N}\sum_{i=0}^{\infty} n_i i$
is the average of $i$. A certain set of bin occupations $\{n_i\}$
can be achieved with many different $\{i_n\}$-series, namely
$N!/\prod_{i=0}^{\infty} n_i !$. Assuming that they all occur with
the same probability $p(\{i_n\}|\Gamma)$, we find
\begin{equation}\label{app:eqSeq2}
  p(\{n_i\}|\Gamma) = \frac{N!}{\prod_{i=0}^{\infty} n_i !}
  (\Gamma \Delta \tau)^N \mathrm{e}^{-\Gamma \Delta \tau N \langle i
 \rangle}.
\end{equation}
This is our sampling distribution. For an estimate of $\Gamma$ we
use Bayes theorem
\begin{equation}\label{app:pToP}
  p(\Gamma|\{n_i\}) = p(\Gamma) \frac{p(\{n_i\}|\Gamma)}
  {p(\{n_i\})}.
\end{equation}
Because we have no information on the prior probabilities
$p(\Gamma)$ and $p(\{n_i\})$, the principle of indifference requires
them to be constants, giving
\begin{equation}\label{app:pGammaN1}
  p(\Gamma|\{n_i\}) = C \, (\Gamma \Delta \tau)^N \mathrm{e}^
  {-\Gamma \Delta \tau N \langle i \rangle},
\end{equation}
where $C$ is constant. Normalization $\int_0^{\infty}
p(\Gamma|\{n_i\}) \mathrm{d}\Gamma = 1$ leads to
\begin{eqnarray}\label{app:pGammaN2}
  p(\Gamma|\{n_i\}) &=& \frac{N^N \langle i \rangle^{N+1} \Delta \tau}{N!}
  (\Gamma \Delta \tau)^N \mathrm{e}^
  {-\Gamma \Delta \tau N \langle i \rangle} \nonumber \\
  &=& \frac{N^N}{N!} \langle \tau \rangle
  (\Gamma \langle \tau \rangle)^N \mathrm{e}^
  {-N \Gamma \langle \tau \rangle}.
\end{eqnarray}
The most likely value of $\Gamma$ is therefore $\Gamma^* = 1/\langle
\tau \rangle$. The relative accuracy of this estimate is given by
the width of the distribution. Setting $x=\Gamma \langle \tau
\rangle$ and evaluating the width at half maximum gives
\begin{eqnarray}\label{app:pGammaN3}
 x^N \mathrm{e}^{-x N} &=& \frac{1}{2} \mathrm{e}^{-N} \nonumber \\
 \Rightarrow \ln(x) &=& x -1 -\frac{1}{N} \ln(2).
\end{eqnarray}
For large $N$ we can expand $\ln(x)$ in a Taylor series around
$x=1$. Keeping only the first two terms, it follows
\begin{eqnarray}\label{app:pGammaN4}
 \frac{1}{2}(x-1)^2 &=& \frac{1}{N}\ln(2)  \nonumber \\
 \Rightarrow x &=& 1 \pm \sqrt{\frac{2 \ln(2)}{N}}.
\end{eqnarray}
Thus the relative accuracy is
\begin{equation}\label{app:relAccuracy}
    \Delta \Gamma / \Gamma = \sqrt{2 \ln(2)/N}.
\end{equation}

\subsection{Experimental precision}
In order to compare the results of Eq.~(\ref{app:relAccuracy}) with
the measurement, we take a data set
$\{t^\mathrm{in}_i,~t^\mathrm{out}_i\}$ containing 120000 events,
extracted from a trace such as the one shown in
Fig.~\ref{fig:TR_sample}(c). The tunneling rates are
$\Gamma_\mathrm{in} = 1/\langle \tau_\mathrm{in} \rangle =
594~\mathrm{Hz}$ and $\Gamma_\mathrm{out} = 1/\langle
\tau_\mathrm{out} \rangle = 494~\mathrm{Hz}$. In the following, we
choose to investigate the precision of $\Gamma_\mathrm{in}$ and drop
the subscript. We have also performed the analysis for
$\Gamma_\mathrm{out}$, with similar results.

\begin{figure}[tb]
\centering
 \includegraphics[width=1\columnwidth]{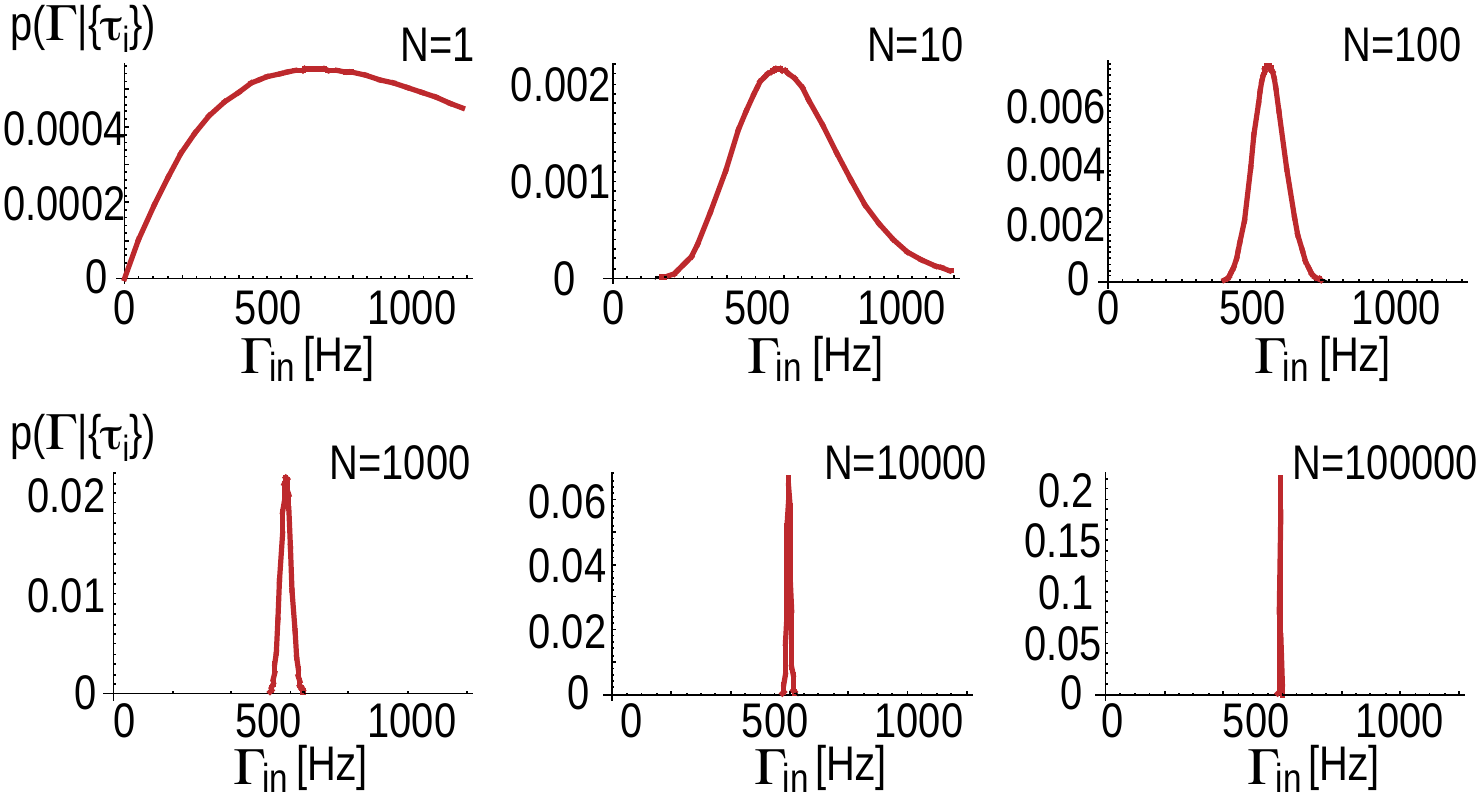}
 \caption{Probability distributions for a sequence of tunneling
 times $\{\tau_i\}$ to belong to a physical process being characterized by the tunnel coupling
$\Gamma$. The different graphs correspond to different lengths of
the data set.
 } \label{app:precSubset}
\end{figure}

To proceed, we use Eq.~(\ref{app:expDecay}) to calculate the
probability that a certain set of tunneling times $\{\tau_i\}$
belongs to a physical process characterized by the tunnel coupling
$\Gamma$
\begin{equation}\label{app:probGforSet}
    p(\Gamma|\{\tau_i\}) = \prod_{i=1}^{N} \Gamma \tau_{i}\,
 \mathrm{e}^{-\Gamma \tau_{i}}.
\end{equation}
In Fig.~\ref{app:precSubset} we plot the probability distributions
of Eq.~(\ref{app:probGforSet}) for subsets of $\{\tau_i\}$ with
different lengths $N$. As the size of the subset is increased, the
probability distribution gets focused around $\Gamma =
\Gamma_\mathrm{in} = 594~\mathrm{Hz}$. This simply reflects the fact
that the larger the amount of experimental evidence available, the
less likely it becomes that the data is generated by a tunneling
process with $\Gamma \neq \Gamma_\mathrm{in}$.

\begin{figure}[tb]
\centering
 \includegraphics[width=.95\columnwidth]{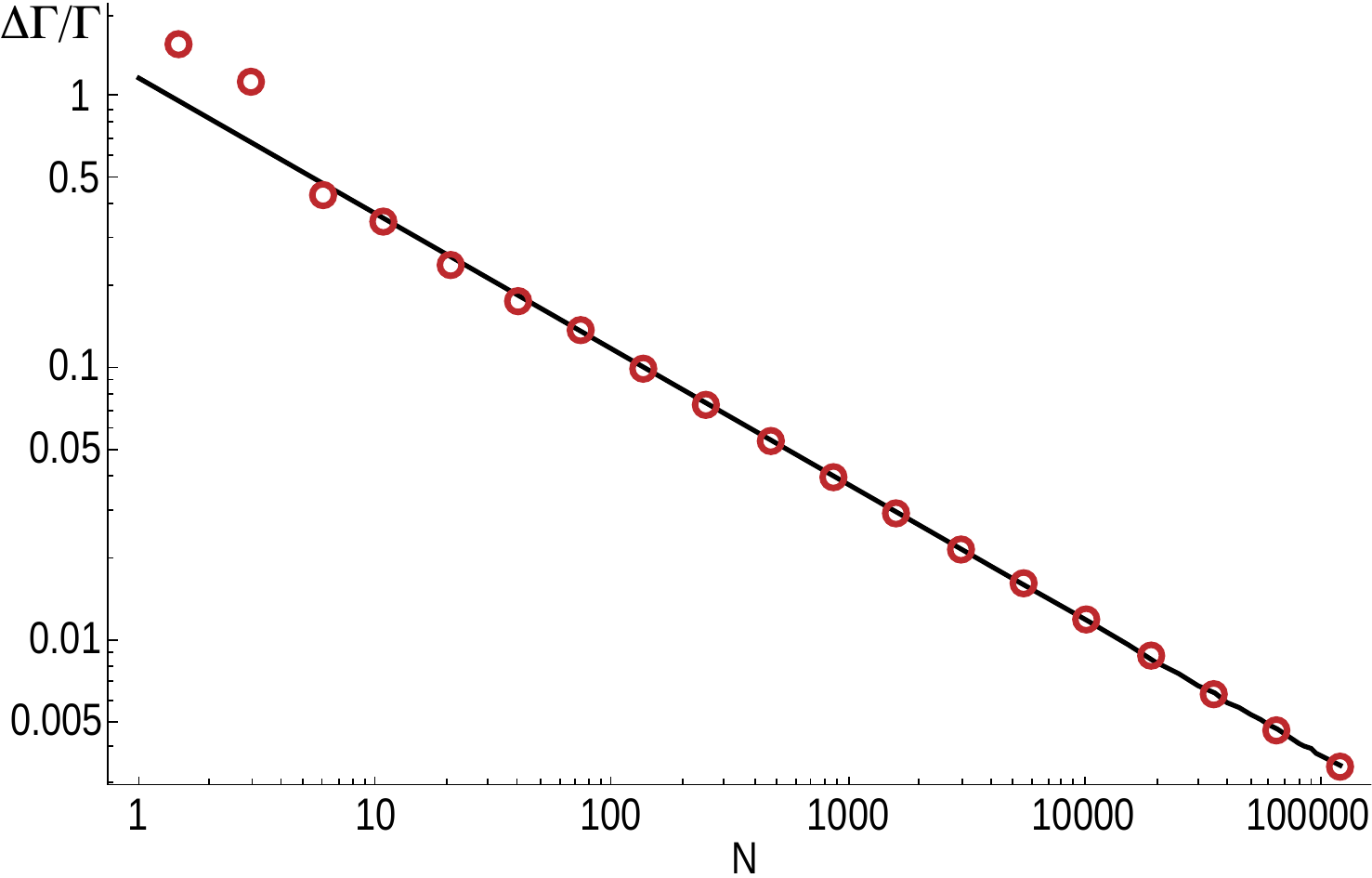}
 \caption{Accuracy achieved when determining the tunnel coupling
 $\Gamma$ versus the size of the data set. The accuracy was measured by taking
 the width of the distributions as shown in Fig.~\ref{app:precSubset}.
 The solid line is the result of Eq.~(\ref{app:relAccuracy})} \label{app:precVsN}
\end{figure}

The experimental uncertainty in $\Gamma$ is given by the width of
the distributions in Fig.~\ref{app:precSubset}.
Figure~\ref{app:precVsN} shows the normalized uncertainty $\Delta
\Gamma / \Gamma$ versus subset size $N$. The solid line is the
result of Eq.~(\ref{app:relAccuracy}), showing very good agreement
with the experimental data. The results validate
Eq.~(\ref{app:expDecay}) and demonstrate the stability of the
sample; a sudden change in the tunnel coupling $\Gamma$ during the
relatively long measurement time of 10 minutes would introduce
deviations between Eq.~(\ref{app:relAccuracy}) and the measured
precision. For the full data set $N=120000$, we find $\Gin = 593.8
\pm 1.7~\mathrm{Hz}$ and $\Gout = 494.2 \pm 1.4~\mathrm{Hz}$.

For simplicity, we have assumed a perfect detector with infinite
bandwidth. We have also performed the analysis for a model
incorporating the detector bandwidth as explained in
section~\ref{sec:ST_higherMoments}, and we obtain very similar
results. The analysis is slightly more involved since a tunnel
coupling $\Gamma_\mathrm{in}$ will depend not only on the set
$\{\tau^\mathrm{in}_i\}$, but also on $\{\tau^\mathrm{out}_i\}$.

\subsection{Current meter precision}
Knowing the precision of the tunneling rates $\Gin / \Gout$, we use
the relations in
Eqs.~(\ref{eq:ST_precision:curr}-\ref{eq:ST_precision:noise}) to
determine the precision of the current and the noise. For the data
set with $N=120000$ discussed in the previous section, we find
\begin{equation}\label{app:precCurr}
    I = (292.87 \pm 0.64)~\mathrm{e/s} = (46.917 \pm 0.10)~\mathrm{aA}.
\end{equation}
The shot noise of the current is equal to
\begin{equation}\label{app:precNoise}
    S_I =  (7.5772 \pm 0.017) \times 10^{-36}~\mathrm{A^2/Hz}.
\end{equation}
Conventional measurement techniques are usually limited by the
current noise of the amplifiers (typically $10^{-29}$ A$^2$/Hz)
\cite{depicciotto:1997,saminadayar:1997,birk:1995,nauen:2004}: here
we demonstrate a measurement of the noise power with a sensitivity
better than $10^{-37}$ A$^2$/Hz. The limits in precision
investigated here are not due to a measurement apparatus but appears
because of the discreteness of charge; the precision of the shot
noise measurement is limited by the shot noise itself. In the
experiment more uncertainty occurs if (1) the correction for the
finite bandwidth in \EqRef{eq:ST_FiniteBW:compGamma1} is incorrect
or (2) because of the detection of false events due to an
insufficient signal-to-noise ratio in the measurement of the QPC
conductance (see section~\ref{sec:QP_sn}).

\section{Double quantum dots} \label{sec:DQ_main}

The double quantum dot is the mesoscopic analogue of a diatomic
molecule. In weakly coupled dots, the electrons are well localized
within the individual dots, their wavefunctions are spatially
separated and electron transport is described by sequential
tunneling between discrete single-dot states. With increased
interdot coupling, the single-dot wavefunctions hybridize and form
molecular states extending over both dots. The ability to tune both
the interdot coupling and the energy levels of the individual QDs
make the double quantum dot an interesting model system for studying
interactions in coupled quantum systems. In this section we show how
to use time-resolved charge detection techniques to probe various
properties of double quantum dots.

\begin{figure}[tb]
\centering
\includegraphics[width=\linewidth]{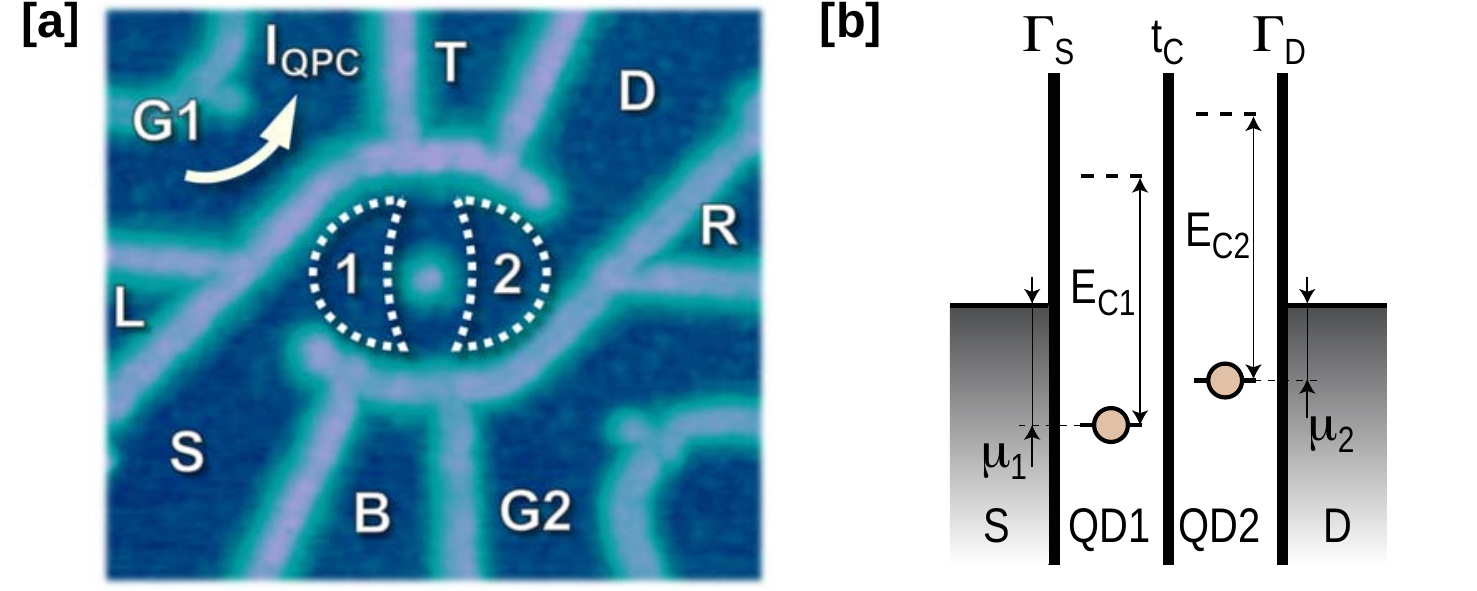}
\caption{(a) AFM-image of the sample investigated in this section.
The structure consists of a double quantum dot (DQD) (marked by 1
and 2) with a near-by quantum point contact. (b) Energy level
diagram of the DQD. The QDs are tunnel coupled to source and drain
leads with tunneling rates $\Gs$ and $\Gd$, while the interdot
transitions are characterized by the coupling energy
$t_\mathrm{C}$.} \label{fig:DQ_Sample}
\end{figure}

The measurements presented in this section were performed on the
sample shown in \FigRef{fig:DQ_Sample}(a). The structure is
fabricated with local oxidation techniques and consists of two QDs
(marked by 1 and 2 in the figure) connected by two separate tunnel
barriers. For the results presented here only the upper tunnel
barrier was kept open; the lower was pinched-off by applying
appropriate voltages to the surrounding gates. For the purpose of
this section, the system may be described as a standard serial
double quantum dot (DQD); the ring-shape properties of the sample
are investigated and utilized in section \ref{sec:AB_main}.

The DQD is coupled to source and drain leads via tunnel barriers.
Several in-plane gates [marked by T, B, L and R in
\FigRef{fig:DQ_Sample}(a)] are used to tune the various tunnel
couplings. Two quantum point contacts are located in the upper-left
and lower-right parts of \FigRef{fig:DQ_Sample}(a). In the
measurement, it was only possible to operate the upper-left QPC as a
charge detector; the one in the lower-right corner was always
pinched off. The conductance of the upper-left QPC was measured by
applying a bias voltage of $200-400\uV$ and monitoring the current
($\Iqpc$ in the figure). The QPCs were also used as in-plane gates
to control the electron population in the DQD. This was achieved by
applying fixed voltages $\Vgl$, $\Vgr$ to both sides of the QPCs in
addition to the bias voltage.

In \FigRef{fig:DQ_Sample}(b) we sketch the energy levels in the
system. The QD states are coupled to source and drain leads with
tunneling rates $\Gs$ and $\Gd$, while the interdot coupling is
described by a coupling energy $t_\mathrm{C}$. The electrochemical
potentials of the two QDs are denoted by $\mul$ and $\mur$, measured
relative to the Fermi levels of the source and drain leads. The next
unoccupied QD states are separated by the charging energies $\Ecl$
and $\Ecr$.

\begin{figure}[tb]
\centering
\includegraphics[width=\linewidth]{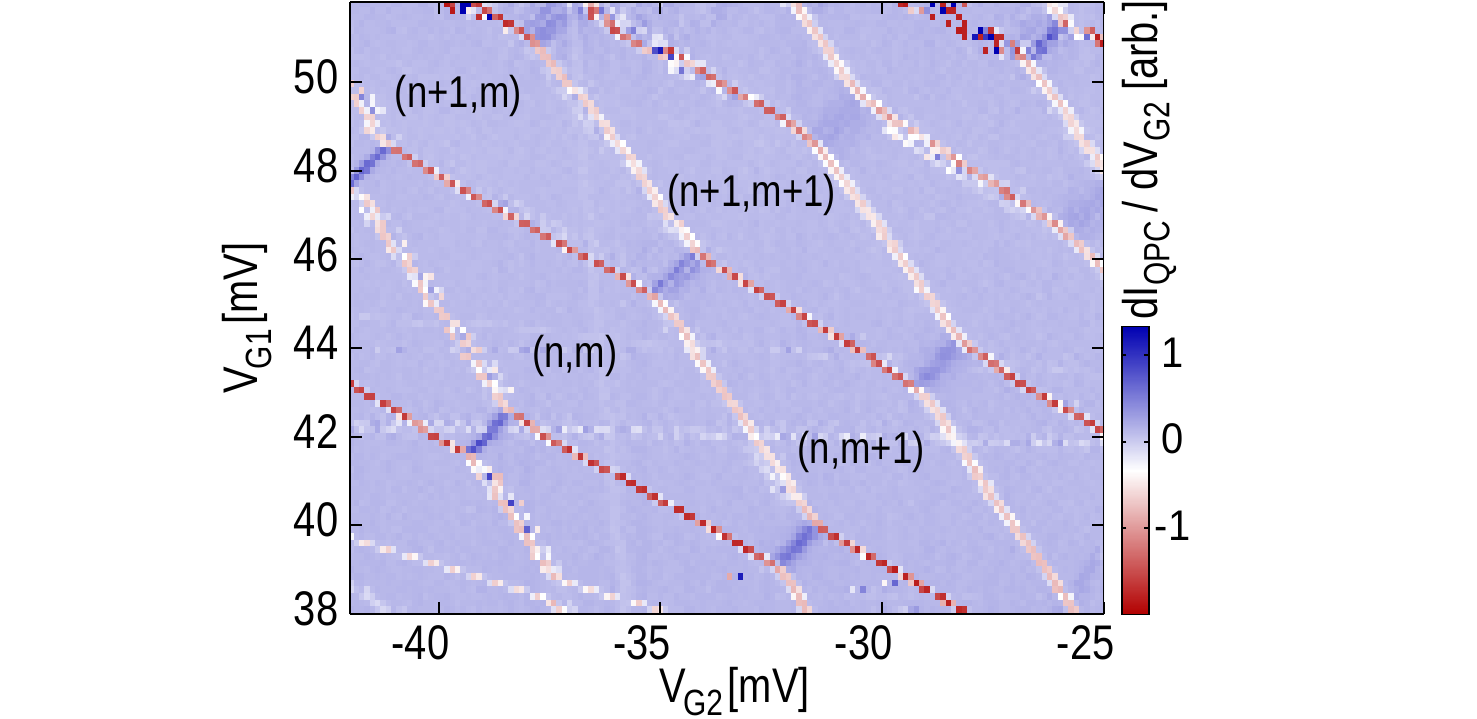}
\caption{Numerical derivative of the QPC current with respect to the
voltage on gate G2. A positive derivative reflects an increase in
QPC conductance, which means that an electron is moving away from
the QPC. For a negative derivative, an electron is coming closer to
the QPC. The horizontal white line most likely originates from
electron fluctuations of a charge trap. The numbers in the figure
refer to the number of electrons in the two QDs. The data was taken
with QPC bias $V_\mathrm{QPC-SD} = 400\uV$ and zero bias across the
DQD.} \label{fig:DQ_IQPCClosed}
\end{figure}

To reach a well-defined DQD configuration we apply negative
gate voltages in order to close the constrictions between QD1 and
QD2. The gate voltages also influence the tunneling coupling to
source and drain; as a consequence the DQD current $\Idqd$ drops
below the measurable limit and we need to operate the charge
detector to measure charge transitions in the QDs.
Figure~\ref{fig:DQ_IQPCClosed} shows the numerical derivative of the
QPC current with respect to the gate voltage $\Vgr$. A compensation
voltage was applied to the QPC gate [upper-leftmost part of
\FigRef{fig:DQ_Sample}(a)] to keep the QPC conductance relatively
constant within the gate voltage of interest. This gives the uniform
light-bluish background of \FigRef{fig:DQ_IQPCClosed}. On top of
that there is a clear hexagon pattern emerging, with all features
expected from a DQD \cite{vanderwiel:2002}.

The numbers in brackets denote the electron population of the two
QDs. The charge transitions occurring at the borders between
different regions of fixed charge give rise to different changes of
$d\Iqpc/d\Vgr$. To understand these features we first note that the
QPC is asymmetrically positioned with respect to the DQD, with QD1
being much closer than QD2. Charge fluctuations in QD1 are therefore
expected to give a stronger influence on the QPC conductance than
fluctuations in QD2. Now, starting within the hexagon marked by
(n,m) and increasing $\Vgr$ will lower the DQD potentials $\mul$ and
$\mur$ and eventually allow an additional electron to enter the DQD.
As the transition takes place, the QPC conductance decreases, giving
a sharp peak with negative $d\Iqpc/d\Vgr$ in
\FigRef{fig:DQ_IQPCClosed}. Depending on the energy level
configuration of the two QDs, the electron may enter into either QD1
or QD2. The dip in $d\Iqpc/d\Vgr$ is stronger for the transition
$\mathrm{(n,m)} \rightarrow \mathrm{(n+1,m)}$ than for
$\mathrm{(n,m)} \rightarrow \mathrm{(n,m+1)}$, reflecting the
stronger coupling between the QPC and QD1.

Since the gate G2 is located closer to QD2 the gate voltage $\Vgr$
has a larger influence on $\mur$ than on $\mul$. Increasing $\Vgr$
may thus lead to a situation where $\mur+\Ecr$ is shifted below
$\mul$. At the transition an electron will tunnel from QD1 over to
QD2. The process takes an electron further away from the QPC,
leading to an increase in QPC conductance and a positive peak in
$d\Iqpc/d\Vgr$. The effect is clearly seen at the transition
$\mathrm{(n+1,m)} \rightarrow \mathrm{(n,m+1)}$ in
\FigRef{fig:DQ_IQPCClosed}.

An interesting feature of \FigRef{fig:DQ_IQPCClosed} is that the
blue lines corresponding to interdot transitions grow broader and
fainter at higher gate voltages. This is a consequence of increased
interdot coupling $\tc$; if the coupling is strong enough the
interdot transition is smeared out over the gate voltage region
where the electron is delocalized over both QDs. Measuring the width
of these transitions thus provides a convenient way to determine the
tunnel coupling between the two QDs that works even if the electron
tunneling occurs on timescales much faster than the detector
bandwidth. The method is investigated in more detail in
section~\ref{sec:DQ_tunnelCoupling}.

\subsection{Time-resolved detection}
The electron population of the DQD is monitored by operating the QPC
in the lower-right corner of \FigRef{fig:DQ_Sample}(a) as a charge
detector \cite{field:1993}. By tuning the tunneling rates of the DQD
below the detector bandwidth, charge transitions can be detected in
real-time \cite{vandersypen:2004, schleser:2004, fujisawa:2004}.
In the experiment, the tunneling rates $\Gs$ and $\Gd$ to source and
drain leads are kept around 1 kHz, while the interdot coupling $t$
is set much larger ($t \sim 20\ueV \sim 5~\mathrm{GHz}$). Interdot
transitions thus occur on timescales much faster than it is possible
to register with the detector ($\tau_\mathrm{det} \sim 50 \us$)
\cite{naaman:2006}, but the coupling energy may still be determined
from charge localization measurements \cite{dicarlo:2004}.
The conductance of the QPC was measured by applying a bias voltage
of $200-400\uV$ and monitoring the current [$\Iqpc$ in
\FigRef{fig:DQ_Sample}(a)]. We ensured that the QPC bias voltage was
kept low enough to avoid charge transitions driven by current
fluctuations in the QPC \cite{gustavssonPRL:2007}. The sample is
realized without metallic gates so that the coupling between dots
and QPC detectors is not screened by metallic structures.

\begin{figure}[tb]
\centering
\includegraphics[width=\linewidth]{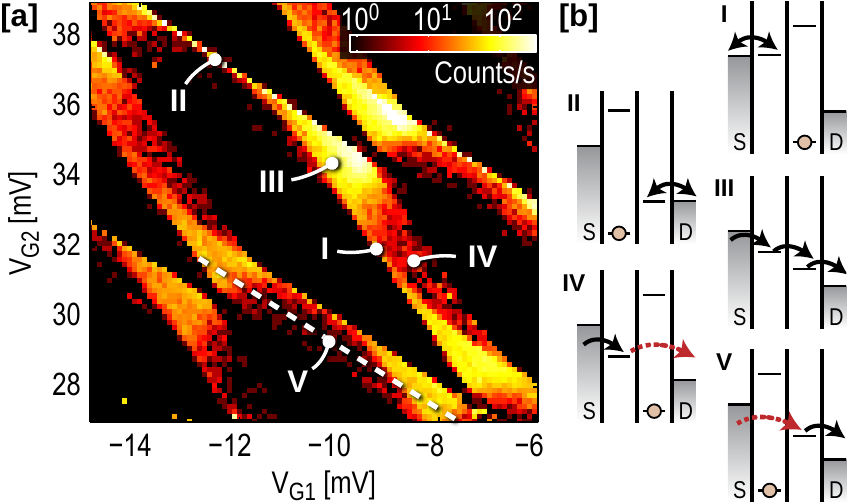}
\caption{(a) Charge stability diagram of the DQD, measured by
counting electrons entering and leaving the DQD. The data was taken
with a voltage bias of $V_\mathrm{DQD-SD} = 600\uV$ applied over the
DQD. The QPC conductance was measured with $V_\mathrm{QPC-SD} =
300\uV$. The count rates were extracted from traces of length
$T=0.5\s$. (b) Energy level diagrams for different configurations in
(a). Adapted from Ref. \cite{gustavssonCotPRB:2008}.
} \label{fig:DQ_HexCount}
\end{figure}

Figure~\ref{fig:DQ_HexCount}(a) shows a charge stability diagram for the
DQD, measured by counting electrons tunneling into and out of the
DQD. The data was taken with a bias voltage of $600\uV$ applied
across the DQD, giving rise to finite-bias triangles of sequential
transport \cite{vanderwiel:2002}. The diagrams in
Fig.~\ref{fig:DQ_HexCount}(b) show schematics of the DQD energy levels
for different positions in the charge stability diagram. Depending
on energy level alignment, different kinds of electron tunneling are
possible.

At the position marked by I in \FigRef{fig:DQ_HexCount}(a), the
electrochemical potential $\mul$ of QD1 is aligned with the Fermi
level of the source lead. The tunneling is due to equilibrium
fluctuations between source and QD1. A measurement of the count rate
as a function of $\mul$ provides a way to determine both the
tunneling rate $\Gs$ and the electron temperature in the source lead
\cite{gustavsson:2006}. The situation is reversed at point II in
\FigRef{fig:DQ_HexCount}(a). Here, electron tunneling occurs between QD2
and the drain, thus giving an independent measurement of $\Gd$ and
the electron temperature of the drain lead.
At point III within the triangle of \FigRef{fig:DQ_HexCount}(a), the
levels of both QD1 and QD2 are within the bias window and the
tunneling is due to sequential transport of electrons from the
source lead into QD1, over to QD2 and finally out to the drain. The
electron flow is unidirectional and the count rate relates directly
to the current flowing through the system \cite{fujisawa:2006}.
Between the triangles, there are broad, band-shaped regions with low
but non-zero count rates where sequential transport is expected to
be suppressed due to Coulomb blockade [cases IV and V in
\FigRef{fig:DQ_HexCount}(a,b)]. The finite count rate in this region is
attributed to electron tunneling involving virtual processes. These
features will be investigated in more detail in the forthcoming
sections.

\begin{figure}[tb]
\centering
\includegraphics[width=\linewidth]{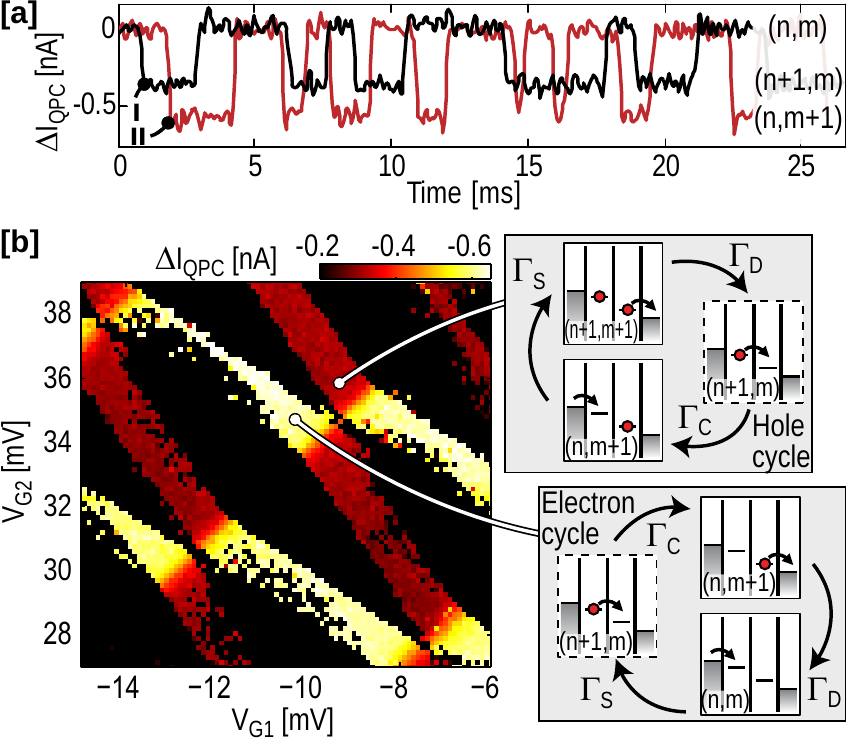}
\caption{(a) Typical time traces of the QPC current from
configurations I and II in \FigRef{fig:DQ_HexCount}. (b) Change of QPC
current $\Delta \Iqpc$ as one electron enters the DQD, extracted
from the same set of data as shown in \FigRef{fig:DQ_HexCount}. The two
levels correspond to the QPC detector registering electron tunneling
in QD1 and QD2, respectively. The energy level diagrams describe the
hole and the electron cycle of sequential transport within the
finite bias triangles.  Adapted from Ref. \cite{gustavssonCotPRB:2008}.} \label{fig:HexagonHeight}
\end{figure}

To begin with, we use the time-resolved charge detection methods to
characterize the system. Typical time traces of the QPC current for
DQD configurations marked by I and II in \FigRef{fig:DQ_HexCount}(a) are
shown in \FigRef{fig:HexagonHeight}(a). The QPC current switches
between two levels, corresponding to electrons entering or leaving
QD1 (case I) or QD2 (case II). The change $\Delta \Iqpc$ as one
electron enters the DQD is larger for charge fluctuations in QD2
than in QD1. This reflects the stronger coupling between the QPC and
QD2 due to the geometry of the device. A measurement of $\Delta
\Iqpc$ thus gives information about the charge localization in the
DQD.


In \FigRef{fig:HexagonHeight}(b) we investigate the charge
localization in more detail by plotting the absolute change in QPC
current $\Delta \Iqpc$ for the same set of data as in
\FigRef{fig:DQ_HexCount}(a).
The detector essentially only measures two different values of
$\Delta \Iqpc$; either $\Delta \Iqpc \sim -0.3\nA$ or $\Delta \Iqpc
\sim -0.6\nA$. Comparing the results of
\FigRef{fig:HexagonHeight}(b) with the sketches in
\FigRef{fig:DQ_HexCount}(b), we see that regions with high $\Delta \Iqpc$
match with the regions where we expect the counts to be due to
electron tunneling in QD2, while the regions with low $\Delta \Iqpc$
come from electron tunneling in QD1.

The regions inside the bias triangles are described in detail in the
energy level diagrams of \FigRef{fig:HexagonHeight}(b). We assume
each QD to hold n and m electrons, respectively. In the lower
triangle, the current is carried by a sequential \emph{electron
cycle}. Starting from the (n,m)-configuration, an electron will
tunnel in from the source lead at a rate $\Gs$ making the transition
$(\mathrm{n,m})\rightarrow(\mathrm{n+1,m})$. The electron then
passes on to QD2 at a rate $\Gc \sim t/h$
$[(\mathrm{n+1,m})\rightarrow(\mathrm{n,m+1})]$ before leaving to
drain at the rate $\Gd$
$[(\mathrm{n,m+1})\rightarrow(\mathrm{n,m})]$. Since the rate $\Gc$
is much faster than the detector bandwidth (and $\Gc\gg \Gs,~\Gc\gg
\Gd$), the detector will only register transitions between the two
states $(\mathrm{n,m})$ and $(\mathrm{n,m+1})$. Therefore, we expect
the step height $\Delta \Iqpc$ within the lower triangle to be equal
to $\Delta \Iqpc$ measured for electron fluctuations in QD2, in
agreement with the results of \FigRef{fig:HexagonHeight}.

For the upper triangle, the DQD holds an additional electron and the
current is carried by a \emph{hole cycle}. Starting with both QDs
occupied $[(\mathrm{n+1,m+1})]$, an electron in QD2 may leave to the
drain $[(\mathrm{n+1,m+1})\rightarrow(\mathrm{n+1,m})]$, followed by
a fast interdot transition from QD1 to QD2
$[(\mathrm{n+1,m})\rightarrow(\mathrm{n,m+1})]$. Finally, an
electron can tunnel into QD1 from the source lead
$[(\mathrm{n,m+1})\rightarrow(\mathrm{n+1,m+1})]$. In the hole
cycle, the detector is not able to resolve the time the system stays
in the $(\mathrm{n+1,m})$ state; the measurement will only register
transitions between $(\mathrm{n+1,m+1})$ and $(\mathrm{n,m+1})$.
This corresponds to fluctuations of charge in QD1, giving the low
value of $\Delta \Iqpc$ in \FigRef{fig:HexagonHeight}(b).
Finally, we note that at the transition between regions of low and
high $\Delta \Iqpc$ the electron wavefunction delocalizes onto both
QDs. This provides a method for determining the interdot coupling energy
$\tc$, which is the subject of the next section.


\subsection{Determining the coupling energy}
\label{sec:DQ_tunnelCoupling}

\begin{figure}[tb]
\centering
\includegraphics[width=\linewidth]{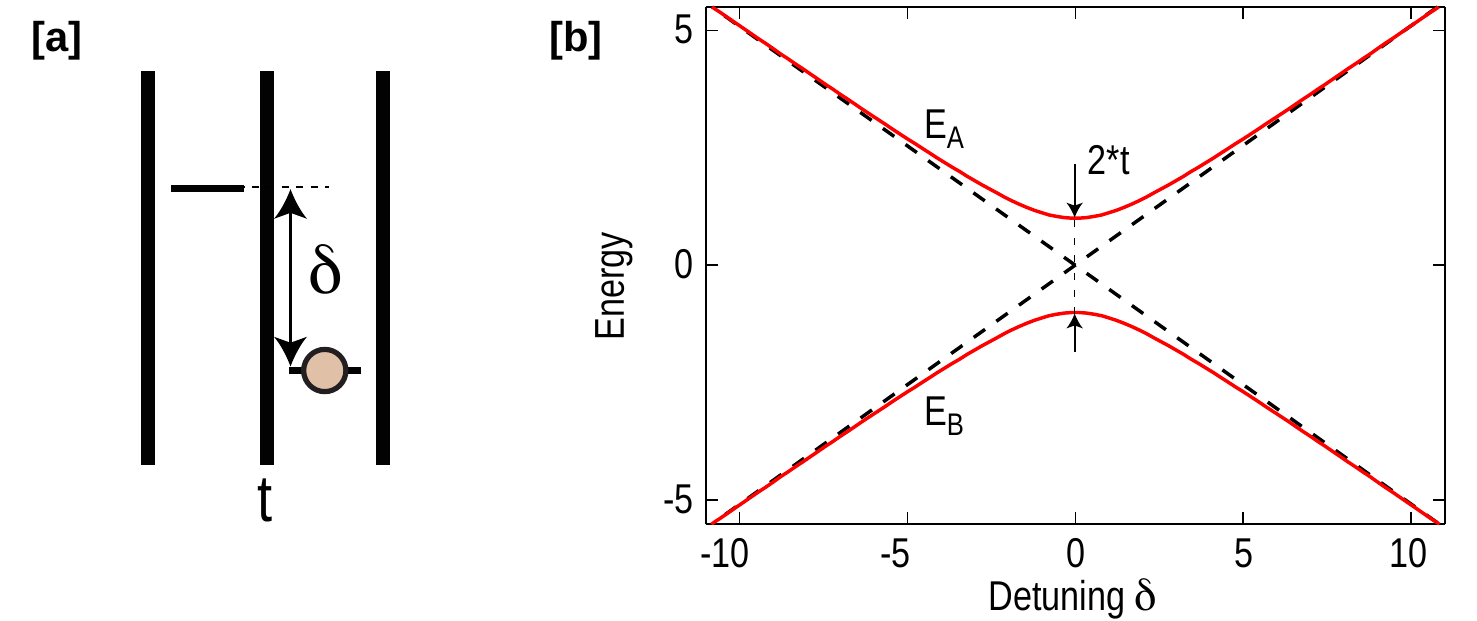}
\caption{(a) Schematics of a tunnel coupled two-level system. (b)
The energy levels of the two-level system, calculated from
\EqRef{eq:DQ_TLSEnergy} with $t=1$. The dashed lines show the energy
levels for isolated QDs ($t=0$).} \label{fig:DQ_TwoLevelSystem}
\end{figure}

In the previous section we have shown that interdot transitions
occur much faster than the detector bandwidth, but so far we did not
try to quantify the tunnel coupling. As already mentioned, the
coupling can be determined by looking at the delocalization of
charge as a function of energy separation of the QD states
\cite{dicarlo:2004}. To simplify the problem, we consider the DQD as
a tunnel-coupled two-level system containing one electron, isolated
from the environment [see \FigRef{fig:DQ_TwoLevelSystem}(a)]. We
introduce the basis states $\{\Psi_\mathrm{1},\Psi_\mathrm{2}\}$
describing the electron sitting on the left or the right QD,
respectively. The two states are tunnel coupled with coupling $t$
and separated in energy by the detuning $\delta = \mul - \mur$. The
Hamiltonian of the system is
\begin{equation}\label{eq:DQ_TLSHamiltonian}
    H = \left[    \begin{array}{cc}
                   -\delta/2 & t \\
                   t & \delta/2 \\
                 \end{array}
               \right].
\end{equation}
The eigenvectors of the Hamiltonian in \EqRef{eq:DQ_TLSHamiltonian}
form the bonding $\Psi_\mathrm{B}$ and antibonding states
$\Psi_\mathrm{A}$ of the system. The eigenvalues give the energies
$E_\mathrm{B}$, $E_\mathrm{A}$ of the two states, with
\begin{equation}\label{eq:DQ_TLSEnergy}
 E_\mathrm{B} = -\frac{1}{2} \sqrt{4 t^2 + \delta^2}, ~~~~
 E_\mathrm{A} = \frac{1}{2} \sqrt{4 t^2 + \delta^2}.
\end{equation}
The energies are plotted in \FigRef{fig:DQ_TwoLevelSystem}(b); at
zero detuning, the states anticross due to the coupling energy. For
a finite temperature $T$, the system will be in a statistical
mixture of the bonding and antibonding states. The occupation
probabilities $p_\mathrm{B}$ and $p_\mathrm{A}$ of the two states
are determined by detailed balance,
\begin{eqnarray}\label{eq:DQ_TLSTemp}
 p_\mathrm{B} &=& 1 - \frac{1}{1+e^\frac{E_\mathrm{A}-E_\mathrm{B}}{k_B\, T}} =
 1 - \frac{1}{1+e^\frac{\sqrt{4 t^2 + \delta^2}}{k_B\, T}}, \nonumber \\
 p_\mathrm{A} &=& \frac{1}{1+e^\frac{E_\mathrm{A}-E_\mathrm{B}}{k_B\, T}} =
 \frac{1}{1+e^\frac{\sqrt{4 t^2 + \delta^2}}{k_B\, T}}.
\end{eqnarray}
In the measurement, we use the change of the QPC current ($\Delta
\Iqpc$) when one electron enters the DQD to determine the amount of
charge localized in the individual QDs. To evaluate this quantity
from Eqs.~(\ref{eq:DQ_TLSHamiltonian}-\ref{eq:DQ_TLSTemp}), we take
the thermal population of the bonding and antibonding states and
project them onto the states $\Psi_1$ and $\Psi_2$
\begin{eqnarray}\label{eq:DQ_TLSpL}
 p_1 &=& \left( p_\mathrm{B} \Psi_\mathrm{B} + p_\mathrm{A} \Psi_\mathrm{A}
 \right) \cdot \Psi_1 = \nonumber \\
 &=& \frac{1}{2}
 \left(1 - \frac{\delta \tanh\left( \frac{\sqrt{4 t^2 + \delta^2}} {2k_B \, T}\right)}{\sqrt{4 t^2 +
 \delta^2}} \right),  \nonumber \\
 p_2 &=& \left( p_\mathrm{B} \Psi_\mathrm{B} + p_\mathrm{A} \Psi_\mathrm{A}
 \right) \cdot \Psi_2 = \nonumber \\
 &=& \frac{1}{2}
 \left(1 + \frac{\delta \tanh\left( \frac{\sqrt{4 t^2 + \delta^2}} {2k_B \, T}\right)}{\sqrt{4 t^2 +
 \delta^2}} \right).
\end{eqnarray}

\begin{figure}[tb]
\centering
\includegraphics[width=\linewidth]{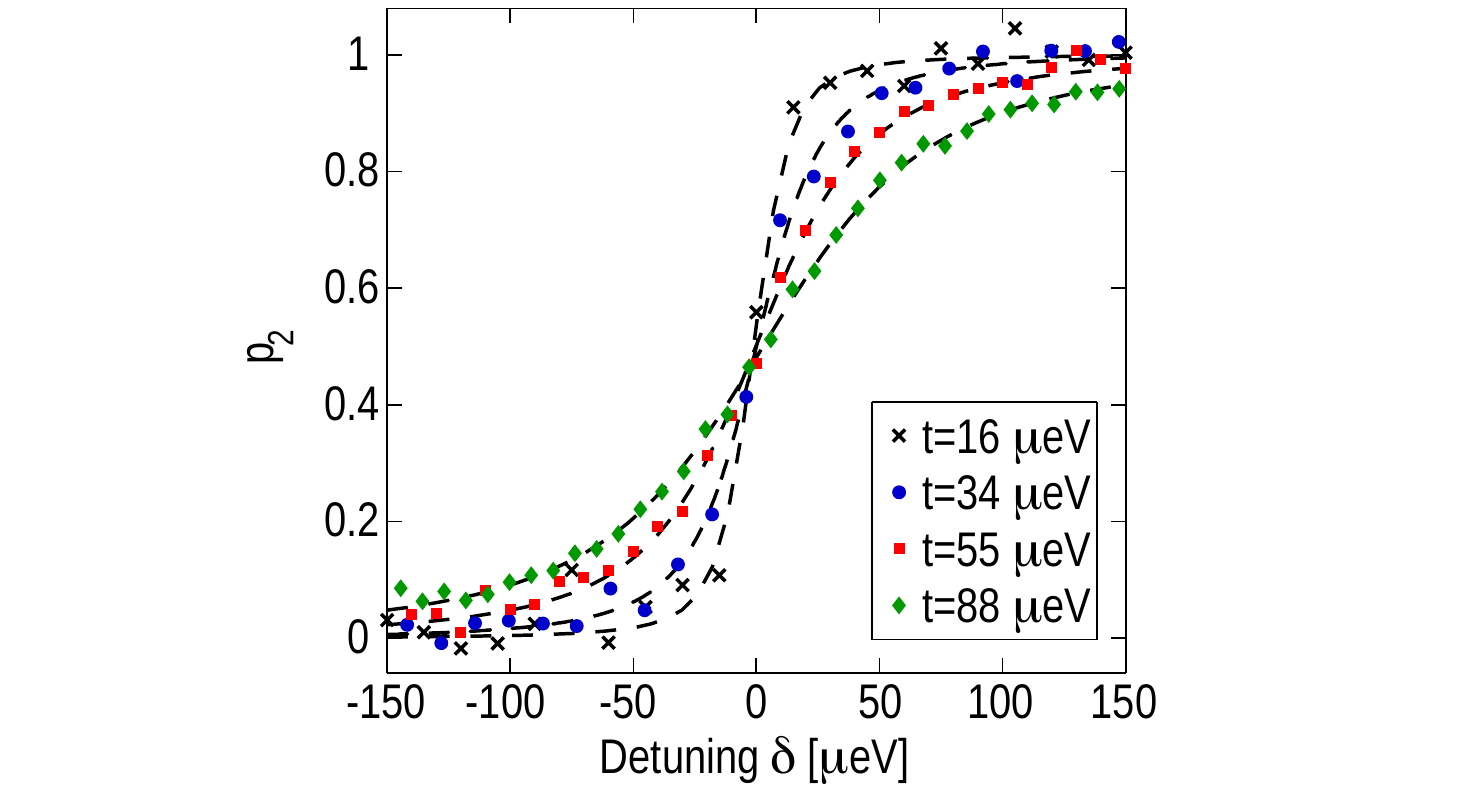}
\caption{Charge population on QD2, evaluated from the change in QPC
current $\Delta \Iqpc$ for one electron entering the DQD. The
different traces were taken at different gate voltages. The dashed
lines are fits to \EqRef{eq:DQ_TLSpL}, with $T=100~\mK$.}
\label{fig:DQ_HeightDetuning}
\end{figure}

Next, we compare the results of \EqRef{eq:DQ_TLSpL} with
experimental data. Figure~\ref{fig:DQ_HeightDetuning} shows the
measured electron population on QD2 versus detuning, extracted from
the change in QPC current $\Delta \Iqpc$. The signal has been
normalized to the levels measured for complete localization in QD1
and QD2. The different data sets are taken for different gate
voltages, demonstrating the possibility to tune the tunnel coupling.
The dashed lines are fits to \EqRef{eq:DQ_TLSpL}, showing good
agreement with the data. It should be noted that this method for
determining the tunneling coupling can only be used as long as the
coupling is larger than the thermal broadening. For a temperature of
$T=100\mK$, the limit corresponds to $t \gtrsim 10\ueV$.

\subsection{Cotunneling} \label{sec:DQ_cotunneling}
We now focus on the regions of weak tunneling occuring in regions
outside the boundaries expected from sequential transport. In case
IV, the electrochemical potential of QD1 is within the bias window,
but the potential of QD2 is shifted below the Fermi level of the
source and not available for transport. We attribute the non-zero
count rate for this configuration to be due to electrons
\emph{cotunneling} from QD1 to the drain lead. The time-energy
uncertainty principle still allows electrons to tunnel from QD1 to
drain by means of a higher order process. In case V, the situation
is analogous but the roles of the two QDs are reversed; electrons
cotunnel from the source into QD2 and leave sequentially to the
drain lead.

\begin{figure}[tb]
\centering
\includegraphics[width=\linewidth]{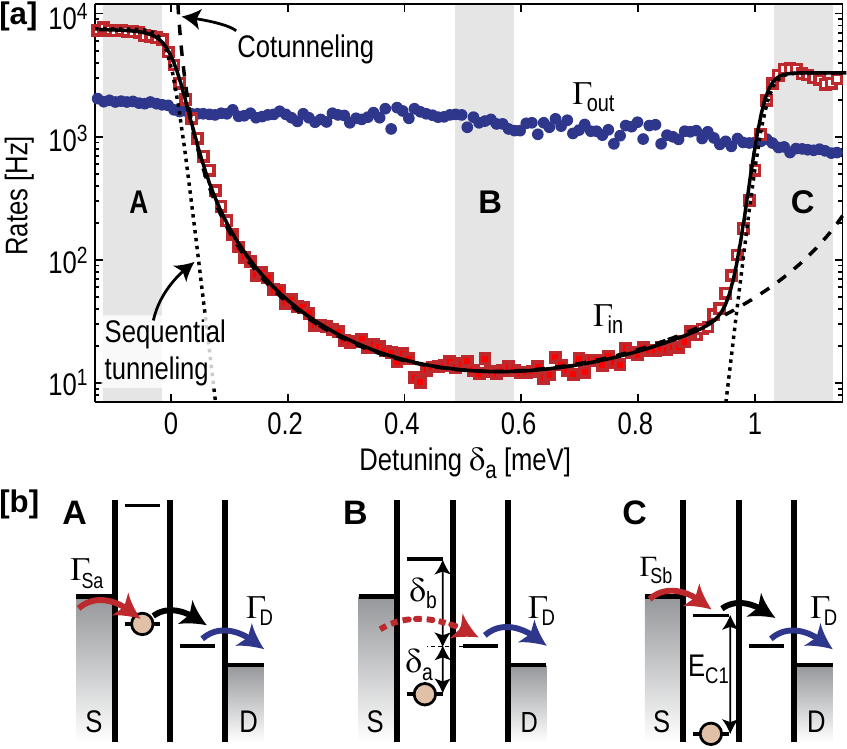}
\caption{Tunneling rates for electrons entering and leaving the DQD,
measured while keeping the potential of QD2 fixed and sweeping the
electrochemical potential of QD1. The data is measured in a
configuration similar to going along the dashed line in
\FigRef{fig:DQ_HexCount}(a).
The dotted lines are tunneling rates expected from sequential
tunneling, while the dashed line is a fit to the cotunneling model
of \EqRef{eq:cotunneling}. The solid line corresponds to the model
involving molecular states [\EqRef{eq:cotMolecular}]. Parameters are
given in the text. (b) Schematic drawings of the DQD energy levels
for three different configurations in (a). At point A, electrons
tunnel sequentially through the structure. Moving to point B, the
energy levels of QD1 are shifted and the electron in QD1 is trapped
due to Coulomb blockade. Electron transport from source to QD2 is
still possible through virtual processes, but the rate for electrons
entering the DQD drops substantially due to the low probability of
the virtual processes. At point C, the next level of QD1 is brought
inside the bias window and sequential transport is again possible. Adapted from Ref. \cite{gustavssonCotPRB:2008}.}
\label{fig:CotunnelingTrace}
\end{figure}

To investigate the phenomenon more carefully, we measure the rates
for electrons tunneling into and out of the DQD in a configuration
similar to the configuration along the dashed line in
\FigRef{fig:DQ_HexCount}(a). The line corresponds to keeping the
electrochemical potential of QD2 fixed within the bias window and
sweeping $\mu_1$. The data is presented in
\FigRef{fig:CotunnelingTrace}.
In the region marked by A in \FigRef{fig:CotunnelingTrace},
electrons tunnel sequentially from source into QD1, relax from QD1
down to QD2 and finally tunnel out from QD2 to the drain lead.
Proceeding from region A to region B, the electrochemical potential
$\mu_1$ is lowered so that an electron eventually gets trapped in
QD1. At point B, the electrons lack an energy $\delta_a = \mu_2 -
\mu_1$ to leave to QD2. Still, electron tunneling is possible by
means of a virtual process \cite{averin:1990}. Due to the
energy-time uncertainty principle, there is a time-window of length
$\sim\!\hbar/\delta_a$ within which tunneling from QD1 to QD2
followed by tunneling from the source into QD1 is possible without
violating energy conservation. An analogous process is possible
involving the next unoccupied state of QD1, occuring on timescales
$\sim\! \hbar/\delta_b$, where $\delta_b = E_\mathrm{C1}-\delta_a$
and $E_{C1}$ is the charging energy of QD1. The two processes
correspond to electron cotunneling from the source lead to QD2.
Continuing from point B to point C, the unoccupied state of QD2 is
shifted into the bias window and electron transport is again
sequential.

In the sequential regime (regions A and C), we fit the rate for
electrons entering the DQD to a model involving only sequential
tunneling [dotted lines in \FigRef{fig:CotunnelingTrace}(a)]
\cite{kouwenhoven:1997}. The fit allows us to determine the tunnel
couplings between source and the occupied
($\Gamma_\mathrm{Sa}$)/unoccupied ($\Gamma_\mathrm{Sb}$) states of
QD2, giving $\Gamma_\mathrm{Sa} = 7.5\kHz$, $\Gamma_\mathrm{Sb} =
3.3\kHz$ and $T = 100\mK$. Going towards region B, the rates due to
sequential tunneling are expected to drop exponentially as the
energy difference between the levels in QD1 and QD2 is increased. In
the measurement, the rate $\Gin$ initially decreases with detuning,
but the decrease is slower than exponential and flattens out as the
detuning gets larger. This is in strong disagreement with the
behavior expected for sequential tunneling. Instead, in a region
around point B we attribute the measured rate $\Gamma_\mathrm{in}$
to be due to electrons cotunneling from source to QD2.

The rate for cotunneling from source to QD2 is given as
\cite{singleCharge:1992}:
\begin{equation}\label{eq:cotunneling}
  \Gamma_\mathrm{cot} = \Gamma_\mathrm{Sa}\, \frac{t_a^2}{\delta_a^2}
  + \Gamma_\mathrm{Sb}\, \frac{t_b^2}{\delta_b^2}
  + \cos \phi \,\sqrt{\Gamma_\mathrm{Sa}\,\Gamma_\mathrm{Sb}} \, \frac{t_a\, t_b}{\delta_a\,\delta_b}.
\end{equation}
Here, $t_a$, $t_b$ are the tunnel couplings between the
occupied/unoccupied states in QD1 and the state in QD2. The first
term describes cotunneling involving the occupied state of QD1, the
second term describes the cotunneling over the unoccupied state and
the third term accounts for possible interference between the two.
The phase $\phi$ defines the phase difference between the two
processes. To determine $\phi$ one needs to be able to tune the
phases experimentally, which is not possible from the measurement
shown in \FigRef{fig:CotunnelingTrace}(a). In the following we
therefore assume the two processes to be independent ($\phi=\pi/2$).
Interference effects between cotunneling processes have been studied
in detail in Ref.~\cite{gustavssonNL:2008}.

The dashed line in \FigRef{fig:CotunnelingTrace}(a) shows the
results of \EqRef{eq:cotunneling}, with fitting parameters
$t_a=15~\mathrm{\mu eV}$ and $t_b=33~\mathrm{\mu eV}$. These values
are in good agreement with values obtained from charge localization
measurements. The values for $\Gamma_\mathrm{Sa}$ and
$\Gamma_\mathrm{Sb}$ are taken from measurements in the sequential
regimes. We emphasize that \EqRef{eq:cotunneling} is valid only if
$\delta_a, \delta_b \gg t_a, t_b$ and if sequential transport is
sufficiently suppressed. The data points used in the fitting
procedure are marked by filled squares in the figure. It should be
noted that the sequential tunneling in region C prevents
investigation of the cotunneling rate at small $\delta_b$. This can
easily be overcome by inverting the DQD bias.
The rate for electrons tunneling out of the DQD
[$\Gamma_\mathrm{out}$ in \FigRef{fig:CotunnelingTrace}(a)] shows
only slight variations over the region of interest. This is expected
since $\mu_2$ stays constant over the sweep. The slight decay of
$\Gamma_\mathrm{out}$ with increased detuning comes from tuning of
the tunnel barrier between QD2 and the drain \cite{maclean:2007}.

The cotunneling may be modified by the existence of a near-by QPC.
If the QPC were able to detect the presence electron in QD2 during
the cotunneling we would expect this to influence the cotunneling
process. For the measurements in \FigRef{fig:CotunnelingTrace}(a)
the QPC current was kept below 10 nA. This gives an average time
delay between two electrons passing the QPC of $e/I_\mathrm{QPC}
\sim \! 16~\mathrm{ps}$. Since this is larger than the typical
cotunneling time, it is unlikely that the electrons in the QPC are
capable of detecting the cotunneling process. The influence of the
QPC may become important for larger QPC currents. However, when the
QPC bias voltage is larger than the detuning ($e
V_\mathrm{QPC}>\delta$), the fluctuations in the QPC current may
start to drive inelastic charge transitions between the QDs
\cite{gustavssonPRL:2007, gustavssonNL:2008}. Such transitions will
compete with the cotunneling. For this reason it was not possible to
extract what effect the presence of the QPC may have on the
cotunneling process.

\subsection{Molecular states}
The overall good agreement between \EqRef{eq:cotunneling} and the
measured data demonstrates that time-resolved charge detection
techniques provide a direct way of quantitatively using the
time-energy uncertainty principle. However, a difficulty arises as
$\delta \rightarrow 0$; the cotunneling rate in
\EqRef{eq:cotunneling} diverges, as visualized for the dashed line
in \FigRef{fig:CotunnelingTrace}(a).
The problem with \EqRef{eq:cotunneling} is that it only takes
second-order tunneling processes into account. For small detuning
$\delta$ the cotunneling described in \EqRef{eq:cotunneling} must be
extended to include higher order processes \cite{pohjola:1997}.
%

\begin{figure}[b!] \centering
\includegraphics[width=\linewidth]{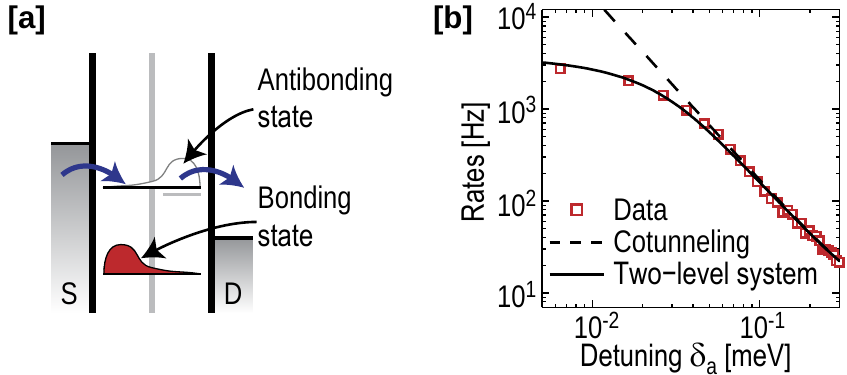}
\caption{(a) Cotunneling described using molecular states. Due to
the large detuning the empty antibonding state is mainly localized
on QD2, but a small part of the wavefunction is still present in QD1
which allows an electron to enter from the source. (b) The rate for
electrons tunneling into the DQD ($\Gin$) as a function of DQD
detuning $\delta_a$. The figure shows the same data as in
\FigRef{fig:CotunnelingTrace}, but plotted on a log-log scale to
enhance the features at small detuning. The dashed line is the
results of the cotunneling model in \EqRef{eq:cotunneling}, the
solid line shows the result of the molecular-state model
[\EqRef{eq:cotMolecular}]. Adapted from Ref. \cite{gustavssonCotPRB:2008}.} \label{fig:CotunnelingTLS}
\end{figure}

A different approach is to assume the coupling between the QDs to be
fully coherent and describe the DQD in terms of the bonding and
antibonding molecular states \cite{graeber:2006,pedersen:2007}. Both
the sequential tunneling and the cotunneling can then be treated as
first-order tunneling processes into the molecular states; what we
in \FigRef{fig:CotunnelingTrace} referred to as cotunneling would be
tunneling into an antibonding state. The model is sketched in
\FigRef{fig:CotunnelingTLS}(a). The bonding state is occupied and in
Coulomb blockade. Still, an electron may tunnel from drain into the
antibonding state. Due to the large detuning, the antibonding state
is mainly located on QD2, the overlap with the electrons in the
source lead is small and the tunneling is weak. Changing the
detuning will have the effect of changing the shape of the molecular
states and shift their weights between the two QDs.

To calculate the rate for electrons tunneling from source into the
antibonding molecular state of the DQD as visualized in
\FigRef{fig:CotunnelingTLS}(a), we use the formalism from
section \ref{sec:DQ_tunnelCoupling} and project the thermal population
$p_\mathrm{B}$, $p_\mathrm{A}$ of the molecular states
$\Psi_\mathrm{B}$ and $\Psi_\mathrm{A}$ onto the unperturbed state
of QD1, $\Psi_1$. This gives the probability $p_1$ of finding an
electron in QD1 if making a projective measurement in the
$\Psi_1$-basis.
The measured rate $\Gin$ is equal to the probability of finding QD1
being empty $(1-p_1)$ multiplied with $\Gs$, the tunneling rate
between the source and the unperturbed state in QD1.
\begin{eqnarray}\label{eq:cotMolecular}
    \Gin &=& \Gs \, (1-p_1) = \Gs \,
    \left( 1 - (p_\mathrm{B} \Psi_\mathrm{B} + p_\mathrm{A}
    \Psi_\mathrm{A}) \cdot \Psi_1 \right) \nonumber \\
    & = & \Gs \, \frac{1}{2}
 \left(1 - \frac{\delta \tanh\left( \frac{\sqrt{4 t^2 + \delta^2}} {2k_\mathrm{B} \, T}\right)}{\sqrt{4 t^2 +
 \delta^2}} \right)
\end{eqnarray}
For large detuning, the bonding and antibonding states are well
localized in QD1 and QD2, respectively. Here, we should recover the
results for the cotunneling rate obtained for the second-order
process [\EqRef{eq:cotunneling}]. First, we assume low temperature
$k_\mathrm{B} T \ll \delta$, so that the electron only populates the
bonding ground state ($p_\mathrm{B}=1$ and $p_\mathrm{A}=0$):
\begin{equation}\label{eq:cotMolLowT}
    \Gin = \Gs \,
    \frac{1}{2} \left( 1+ \frac{\delta}{\sqrt{4t^2 +
    \delta^2}}\right).
\end{equation}
In the limit $\delta \gg t$ the relation reduces to $\Gin \approx
\Gs \, t^2/\delta^2$ and the rate approaches the result of the
second-order cotunneling processes in \EqRef{eq:cotunneling}. The
advantage of the molecular-state model is that it is valid for any
detuning, both in the sequential and in the cotunneling regime.

The solid line in \FigRef{fig:CotunnelingTrace}(a) shows the results
of \EqRef{eq:cotMolecular}. The equation has been evaluated twice,
once for the occupied [(n,m)] and once for the unoccupied state in
QD2 [(n,m+1)]; the curve in \FigRef{fig:CotunnelingTrace}(a) is the
sum of the two rates. The same parameters were used as for the
cotunneling fit of \EqRef{eq:cotunneling}. The model shows very good
agreement with data over the full range of the measurement.
To compare the results of the molecular-state and the cotunneling
model in the regime of small detuning, we plot the data in
\FigRef{fig:CotunnelingTrace}(a) on a log-log scale
[\FigRef{fig:CotunnelingTLS}(b)]. For large detuning, the tunneling
rate follows the $1/\delta^2$ predicted by both the
mole\-cular-state and the cotunneling model. For small detuning, the
deviations become apparent as the cotunneling model diverges whereas
the molecular-state model still reproduces the data well.

\subsection{Excited states}
\label{sec:excitedStates}
So far, we have only considered cotunneling involving the ground
states of the two QDs. The situation is more complex if we include
excited states in the model; the measured rate may come from a
combination of cotunneling processes involving different QD states.
To investigate the influence of excited states experimentally, we
start by extracting the DQD excitation spectrum using finite bias
spectroscopy \cite{vanderwiel:2002}.
If the coupling between the QDs is weak ($\tc \ll \Delta E_1,~\Delta
E_2$, with $\Delta E_{1,2}$ being the mean level spacing in each
QD), the DQD spectrum essentially consists of the combined
excitation spectrum of the individual QDs. For a more strongly
coupled DQD the QD states residing in different dots will hybridize
and delocalize over both QDs. In this section we consider a
relatively weakly coupled configuration ($t \sim 25\ueV$) and assume
the excited states to be predominantly located within the individual
QDs.

\begin{figure}[tb]
\centering
\includegraphics[width=\linewidth]{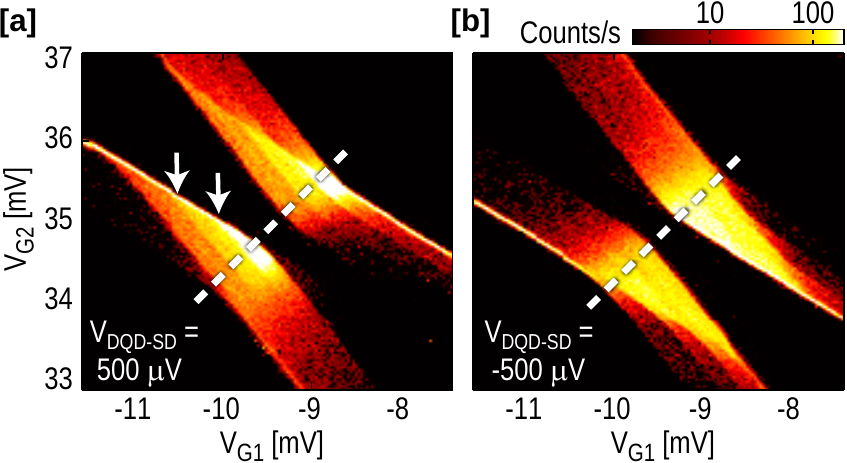}
\caption{Finite-bias spectroscopy of the DQD, taken with positive
(a) and negative (b) bias. The figures are constructed by counting
electrons entering and leaving the DQD. Excited states are visible,
especially for the positive bias data [marked with arrows in (a)].
The data was taken with $V_\mathrm{DQD-SD} = \pm 500 \uV$,
$V_\mathrm{QPC-SD} = 250 \uV$.  Adapted from Ref. \cite{gustavssonCotPRB:2008}.} \label{fig:TrianglesPosNeg}
\end{figure}

\begin{figure}[tb]
\centering
\includegraphics[width=\linewidth]{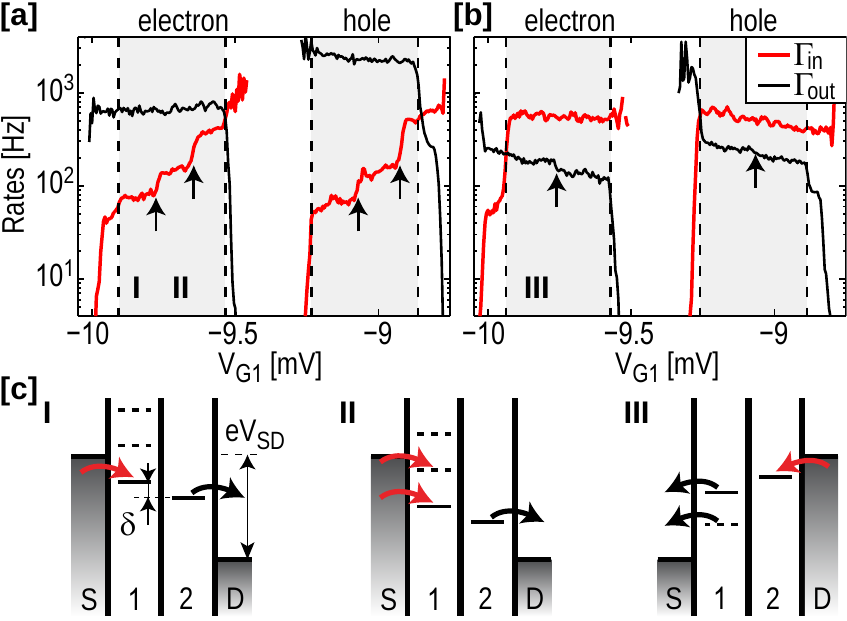}
\caption{(a,~b) Tunneling rates for electrons entering and leaving
the DQD, measured along the dashed lines in
\FigRef{fig:TrianglesPosNeg}(a,~b). In (a), we show the results for
positive bias across the DQD, in (b) the results for negative bias.
The shaded areas mark the regions where electron transport is
sequential, either in the electron or the hole transport cycle. The
arrows indicate the positions of excited states. The data was
extracted from QPC conductance traces of length $T=5\s$, taken with
$\V_\mathrm{QPC-SD} = 250\mV$. (c) Schematics of the DQD energy
configuration at three different positions in (a,~b).  Adapted from Ref. \cite{gustavssonCotPRB:2008}.}
\label{fig:ExcitedStates}
\end{figure}

Figure~\ref{fig:TrianglesPosNeg} shows a magnification of two
triangles from \FigRef{fig:DQ_HexCount}(a), measured with both negative
and positive bias applied across the DQD.
Excited states are visible within the triangles, especially for the
case of positive bias [marked with arrows in
\FigRef{fig:TrianglesPosNeg}(a)]. Transitions between excited states
occur along parallel lines at which the potential of QD1 is held
constant; this indicates that the excited states are located in QD1.
To investigate the states more carefully, we measure the separate
tunneling rates $\Gin$ and $\Gout$ along the dashed lines in
\FigRef{fig:TrianglesPosNeg}. The results are presented in
\FigRef{fig:ExcitedStates}, together with a few sketches depicting
the energy level configuration of the system.

We begin with the results for the positive bias case, which are
plotted in \FigRef{fig:ExcitedStates}(a). Going along the dashed
line in \FigRef{fig:TrianglesPosNeg}(a) corresponds to keeping the
detuning $\delta$ between the QDs fixed and shifting the total DQD
energy. The measurements were performed with a small detuning
($\delta \approx 100 \ueV$) to ensure that electron transport is
unidirectional. Because of this, the outermost parts of the traces
in \FigRef{fig:ExcitedStates}(a) correspond to regions where
transport is due to cotunneling [compare the dashed line with the
position of the triangle in \FigRef{fig:TrianglesPosNeg}(a)]; the
regions where transport is sequential are shaded gray in
\FigRef{fig:ExcitedStates}(a).

Starting in the regime marked by I in
\FigRef{fig:ExcitedStates}(a,c), electrons may tunnel from source
into the ground state of QD1, relax down to QD2 and tunnel out to
the drain lead. Assuming the relaxation process to be much faster
than the other processes, the measured rates $\Gin$ and $\Gout$ are
related to the tunnel couplings of the source and drain $\Gin
\approx \Gs$ and $\Gout = \Gd$. Going to higher gate voltages lowers
the overall energy of both QDs. At the position marked by an arrow
in \FigRef{fig:ExcitedStates}(a), there is a sharp increase in the
rate for tunneling into the DQD. We attribute this to the existence
of an excited state in QD1; as shown in case II in
\FigRef{fig:ExcitedStates}(c), the electron tunneling from source
into QD1 may enter either into the ground $\mathrm{(n+1,m)}$ or the
excited state $\mathrm{(n+1^*,m)}$, giving an increase in $\Gin$.
When further lowering the DQD energy another excited state comes
into the bias window and $\Gin$ increases even more [second arrow in
\FigRef{fig:ExcitedStates}(a)]. The rate for tunneling out of the
DQD shows only minor variations within the region of interest. This
supports the assumption that the excited states quickly relax and
that the electron tunnels out of the DQD from the ground state of
QD2

Finally, continuing to the edge of the shaded region ($\Vgl \sim
-9.55\mV$), the potential of QD2 goes below the Fermi level of the
drain. Here, electrons get trapped in QD2 and the tunneling-out rate
drops drastically. At the same time, $\Gin$ increases; when the
electron in QD2 eventually tunnels out, the DQD may be refilled from
either the source or the drain lead.
The picture described above is repeated in the triangle with hole
transport ($-9.25\mV < \Vgr < -8.9\mV$). This is expected, since the
hole transport cycle involves the same QD states as in the electron
case. An interesting feature is that $\Gin$ shows essentially the
same values in both the electron and the hole cycle, while $\Gout$
increases by a factor of three. The presence of the additional
electron in QD1 apparently affects the tunnel barrier between drain
and QD2 more than an additional electron in QD2 affects the barrier
between QD1 and source.

Next, we move over to the case of negative bias
[\FigRef{fig:ExcitedStates}(b)]. Here, the roles of QD1 and QD2 are
inverted, meaning that electrons enter the DQD into QD2 and leave
from QD1. Following the data and the arguments presented for the
case of positive bias, one would expect this configuration to be
suitable for detecting excited states in QD2. However, looking at
the tunneling rates within the sequential region of
\FigRef{fig:ExcitedStates}(b), the rate for entering QD2 ($\Gin$)
stays fairly constant, while the rate for tunneling out decreases at
the point marked by the arrow. Again, we attribute the behavior to
the existence of an excited state in QD1.

The situation is described in sketch III of
\FigRef{fig:ExcitedStates}(c). The electrochemical potential of QD1
is high enough to allow the electron in the $\mathrm{(n+1,m)}$-state
to tunnel out to the source and leave the DQD in an excited state
$\mathrm{(n^*,m)}$. Since the energy difference
$E[\mathrm{(n^*,m)}]-E[\mathrm{(n+1,m)}]$ is smaller than
$E[\mathrm{(n,m)}]-E[\mathrm{(n+1,m)}]$, the transition involving
the excited state appears \emph{below} the ground state transition.
As the overall DQD potential is lowered, the transition energy
involving the excited state goes below the Fermi level of the drain,
resulting in a drop of $\Gout$ as only the ground state transition
is left available. Similar to the single QD case
\cite{gustavsson:2006}, the tunneling-in rate samples the excitation
spectrum for the {(n+1,m)}-configuration, while the tunneling-out
rate reflects the excitation spectrum of the $\mathrm{(n,m)}$-DQD.

To conclude the results of \FigRef{fig:ExcitedStates}, we find two
excited states in QD1 in the $\mathrm{(n+1,m)}$ configuration with
$\Delta E_1^{\alpha}=180\ueV$ and $\Delta E_1^{\beta}=340\ueV$, and
one excited state in QD1 in the $\mathrm{(n,m)}$ configuration, with
$\Delta E_1=220\ueV$. No clear excited state is visible in QD2. This
does not necessarily mean that such states do not exist; if they are
weakly coupled to the lead they will only have a minor influence on
the measured tunneling rates. Excited states in both QDs have been
measured in other configurations; there, we find similar spectra of
excited states for both QDs.

\subsection{Inelastic cotunneling} \label{sec:inelastic} Next, we
investigate the cotunneling process in the presence of excited
states. Looking carefully at the lower-right regions of the
negative-bias triangles in \FigRef{fig:TrianglesPosNeg}(b), we see
that the count rates in the cotunneling regions outside the
triangles are not constant along lines of fixed detuning
(corresponds to going in a direction parallel to the dashed line).
Instead, the cotunneling regions seem to split into three parallel
bands.

\begin{figure}[tb]
\centering
\includegraphics[width=\linewidth]{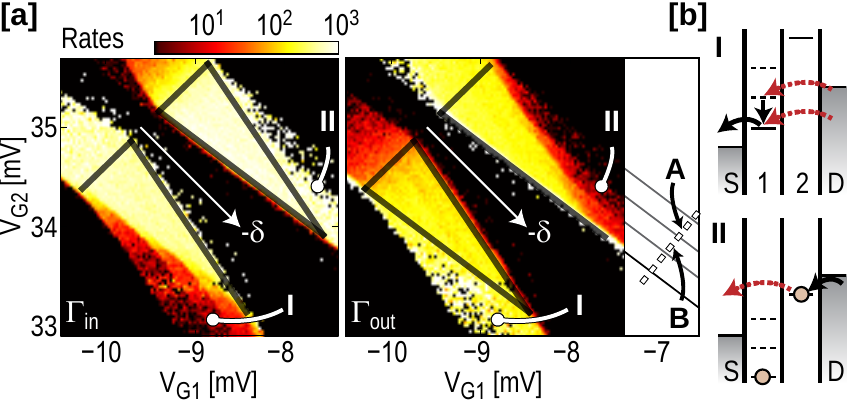}
\caption{(a) Tunneling rates for electrons entering and leaving the
DQD, extracted from the same set of data as used in
\FigRef{fig:TrianglesPosNeg}(b). The data was measured with
$V_\mathrm{DQD-SD}=-500\uV$. The solid lines mark the position of
the finite-bias triangles. The plot region in the right-hand panel
has been extended to include the regime investigated in
Fig.~\ref{fig:InelasticTrace}. (b) Energy-level diagrams for the two
positions marked in (a). In case I, the cotunneling itself is
elastic, with energy relaxation occurring after the cotunneling has
taken place. In case II, inelastic cotunneling processes are
possible.  Adapted from Ref. \cite{gustavssonCotPRB:2008}.} \label{fig:InelasticMap}
\end{figure}

In \FigRef{fig:InelasticMap}(a), we plot the tunneling rates $\Gin$
and $\Gout$ for electrons entering and leaving the DQD, extracted
from the same set of data as used in
\FigRef{fig:TrianglesPosNeg}(b). The thick solid lines mark the
edges of the finite-bias triangles. Again, the cotunneling rates
outside the triangles are not uniform; parallel bands appear in
$\Gin$ for the position marked by I and in $\Gout$ for the position
marked by II in the figures.

To understand the data we draw energy level diagrams for the two
configurations [see \FigRef{fig:InelasticMap}(b)]. Focusing first on
case I, we see that the electrochemical potential of QD1 is within
the bias window, whereas QD2 is detuned and in Coulomb blockade. The
cotunneling occurs via QD2 states; electrons cotunnel from drain
into QD1, followed by sequential tunneling from QD1 to the source
lead. The picture is in agreement with what is measured in
\FigRef{fig:InelasticMap}(a); the cotunneling rate ($\Gin$) is low
and strongly depends on detuning $\delta$, while the sequential rate
$\Gout$ is high and essentially independent of detuning. The three
bands seen in $\Gin$ occur because of the excited states in QD1;
depending on the average DQD energy, electrons may cotunnel from
drain into one of the excited states, relax to the ground state and
then leave to the source lead. The state of QD2 remains unaffected
by the cotunneling process. For this configuration, we speak of
\emph{elastic} cotunneling.

The situation is different in case II. Here, cotunneling occurs in
QD1 as electrons tunnel directly from QD2 into the source lead. This
means that $\Gin$ is sequential while $\Gout$ describe the
cotunneling process. As in case I, the cotunneling rate $\Gout$
splits up into three bands; we attribute this to cotunneling where
the state of QD1 is changed during the process. QD1 ends up in one
of its excited states. The energy of the electron arriving in the
source lead is correspondingly decreased compared to the
electrochemical potential of QD2. Here, the cotunneling is
\emph{inelastic}.

\begin{figure}[b!]
\centering
\includegraphics[width=\linewidth]{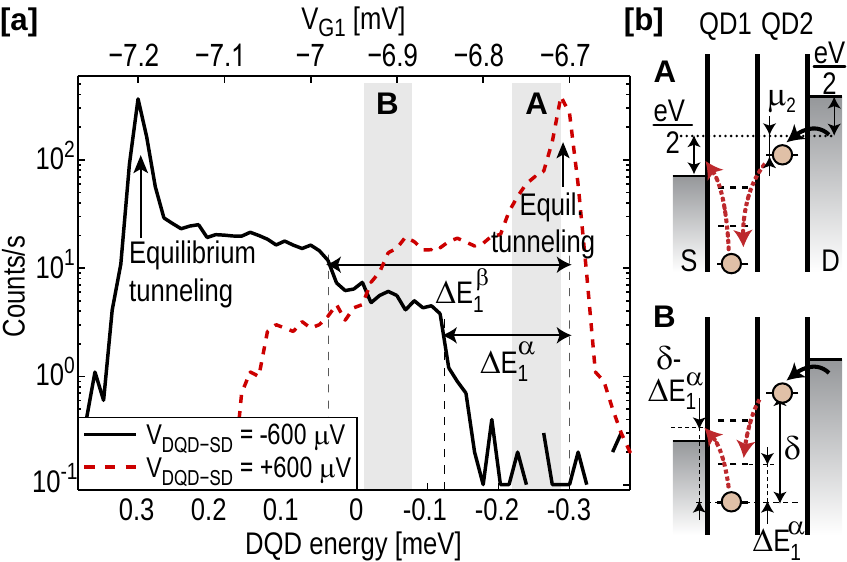}
\caption{(a) Electron count rate along the dashed line in
\FigRef{fig:InelasticMap}(a), measured for both positive and
negative DQD bias. In the trace, the detuning $\delta$ stays
constant and we sweep the average DQD energy. The DQD energy is
defined from the position where the electrochemical potential of QD2
is right in the middle between the Fermi levels of the source and
drain leads [see the dotted line in the energy level diagram in
(b)]. The steps in the count rate are due to the onset of inelastic
cotunneling processes in QD1. The data was extracted from traces of
length $T=10\s$, measured with $V_\mathrm{QPC-SD} = 200\uV$. (b)
Energy level diagrams for the two configurations marked in (a).  Adapted from Ref. \cite{gustavssonCotPRB:2008}.}
\label{fig:InelasticTrace}
\end{figure}

The inelastic cotunneling is described in greater detail in
\FigRef{fig:InelasticTrace}. In \FigRef{fig:InelasticTrace}(a) we
plot the count rate for positive and negative DQD bias, measured
along the dashed line at the right edge of
\FigRef{fig:InelasticMap}(a). Figure~\ref{fig:InelasticTrace}(b)
shows energy level diagrams for negative bias at two positions along
this line. The bias voltage is applied symmetrically to the DQD,
meaning that the Fermi levels in source and drain leads are shifted
by $\pm eV/2$ relative to the Fermi energy at zero bias [dotted line
in Fig.~\ref{fig:InelasticTrace}(b)]. In the measurement of
\FigRef{fig:InelasticTrace}(a) we sweep the average DQD energy while
keeping the detuning $\delta$ constant. The average DQD energy is
defined to be zero when $\mu_2$ aligns with the zero-bias Fermi
energy in the leads [i.e. when $\mu_2 =
(\mu_\mathrm{S}+\mu_\mathrm{D})/2$].

Starting in the configuration marked by A, cotunneling is only
possible involving the QD2 ground state. Cotunneling is weak, with
count rates being well below 1 count/s. Continuing to case B, we
raise the average DQD energy. When the electrochemical potential of
QD2 is sufficiently increased compared to the Fermi level of the
source, inelastic cotunneling becomes possible leading to a sharp
increase in count rate. The process is sketched in
\FigRef{fig:InelasticTrace}(b); it involves the simultaneous
tunneling of an electron from QD2 to the first excited state of QD1
with an electron in the QD1 ground state leaving to the source. The
process is only possible if
\begin{equation}\label{eq:CondInel}
    \delta-\Delta E^\alpha_1 = \mu_1 - \mu_2 - \Delta E^\alpha_1 > \mu_\mathrm{S} - \mu_1.
\end{equation}
Here, $\Delta E^\alpha_1$ is the energy of the first excited state
in QD1. The position of the step in \FigRef{fig:InelasticTrace}(a)
directly gives the energy of the first excited state, and we find
$\Delta E^\alpha_1=180\ueV$.

Further increasing the average DQD energy makes an inelastic process
involving the second excited state in QD2 possible, giving $\Delta
E^\alpha_2=340\ueV$. Finally, as the DQD energy is raised to become
equal to half the applied bias, the electrochemical potential of QD2
aligns with Fermi level of the drain lead. Here electron tunneling
mainly occurs due to equilibrium fluctuations between drain and QD2,
giving a sharp peak in the count rate. The excited states energies
extracted from the inelastic cotunneling give the same values as
obtained from finite-bias spectroscopy within the triangles, as
described in the previous section. The good agreement between the
two measurements demonstrates the consistency of the model.

The dashed line in \FigRef{fig:InelasticTrace}(a) shows data taken
with reversed DQD bias; for this configuration the Fermi levels of
the source and drain leads are inverted, the electrons cotunnel from
source to QD2 and the peak due to equilibrium tunneling occurs at
$\mu_2= \mu_\mathrm{D} = -300\ueV$.

\subsection{Noise in the cotunneling regime}
Using time-resolved charge detection methods, we can extract the
noise of electron transport in the cotunneling regime. For a
weakly-coupled single QD in the regime of sequential tunneling,
transport in most configurations is well-described by independent
tunneling events for electrons entering and leaving the QD
\cite{gustavsson:2005}. The Fano factor becomes a function of the
tunneling rates \cite{davies:1992}:
\begin{equation}
F_2 = \frac{S_I}{2eI}
 = \frac{\Gamma_{\mathrm{in}}^2+\Gamma_{\mathrm{out}}^2}{(\Gamma_{\mathrm{in}}
+ \Gamma_{\mathrm{out}})^2} = \frac{1}{2} \left( 1 +a^2 \right),
\label{eq:mu2vsasym}
\end{equation}
with $a=(\Gamma_{\mathrm{in}} -
\Gamma_{\mathrm{out}})/(\Gamma_{\mathrm{in}} +
\Gamma_{\mathrm{out}})$. For symmetric barriers ($a=0$), the Fano
factor is reduced to 0.5 because of an increase in electron
correlation due to Coulomb blockade. In the case of cotunneling, the
situation is more complex. As described in the previous section,
cotunneling may involve processes leaving either QD in an excited
state. The excited state has a finite lifetime $\tau_\mathrm{rel}$;
during this time, the tunneling rates may be different compared to
the ground-state configuration \cite{schleserPRL:2005}. We therefore
expect that the existence of an electron in an excited state may
induce temporal correlations on time scales on the order of
$\tau_\mathrm{rel}$ between subsequent cotunneling events. In this
way, the noise of the cotunneling current has been proposed as a
tool to probe excited states and relaxation processes in QDs
\cite{aghassi:2006, aghassi:2008}.

\begin{figure}[t]
\centering
\includegraphics[width=\linewidth]{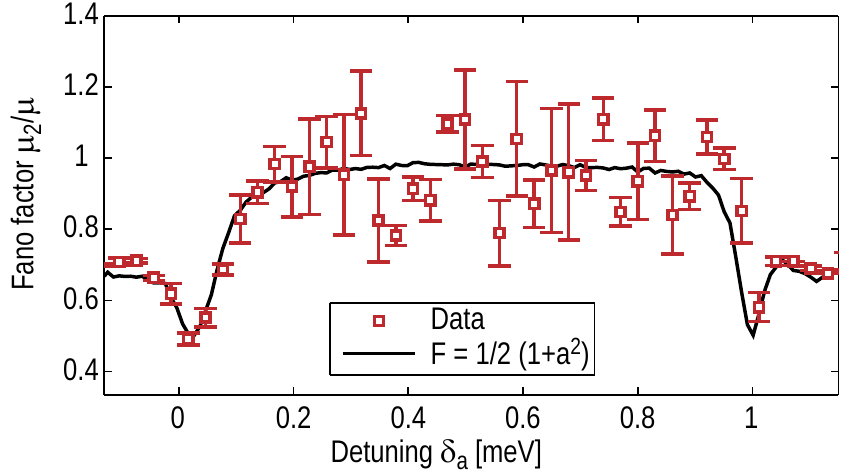}
\caption{Fano factor for electron transport in the cotunneling
regime of \FigRef{fig:CotunnelingTrace}. The data was extracted from
traces of length $T=30\s$. The solid line is the result of
\EqRef{eq:mu2vsasym}, which assumes independent tunneling events.
The minima in Fano factor occur at positions where the tunneling
rates $\Gin$ and $\Gout$ are equal (see
\FigRef{fig:CotunnelingTrace}). The error bars show standard error,
extracted by splitting the data into six subsets of length $T=5\s$
and evaluating the noise for each subset.  Adapted from Ref. \cite{gustavssonCotPRB:2008}.}
\label{fig:CotunnelingNoise}
\end{figure}

In \FigRef{fig:CotunnelingNoise}, we plot the Fano factor measured
from the same region as that of \FigRef{fig:CotunnelingTrace}. The
Fano factor was extracted by measuring the distribution function for
transmitted charge through the system \cite{gustavsson:2005}. The
solid line shows the result of \EqRef{eq:mu2vsasym}, with tunneling
rates extracted from the measured traces. In the outermost regions
of the graph, the electrons tunnel sequentially through the DQD.
Here, the Fano factor is reduced due to Coulomb blockade, similar to
the single QD case. At the edges of the cotunneling regions, the
Fano factor drops further down to $F=0.5$. This is because the
injection rate $\Gin$ drops drastically as sequential transport
becomes unavailable, while $\Gout$ stays approximately constant. At
some point we get $\Gin = \Gout$, which means that the asymmetry $a$
is zero and the Fano factor of \EqRef{eq:mu2vsasym} shows a minimum.
Further into the cotunneling region, the Fano factor approaches one
as transport essentially becomes limited by a single rate; the
cotunneling rate ($\Gin$) is two orders of magnitude smaller than
the sequential rate $\Gout$.

We do not see any major deviation from the results of
\EqRef{eq:mu2vsasym}, which is only valid assuming independent
tunneling events. We have performed similar measurements in several
inelastic and elastic cotunneling regimes, without detecting any
clear deviations from \EqRef{eq:mu2vsasym}. As it turns out, there
are two effects that make it hard to detect correlations due to the
internal QD relaxations. For the first, the correlation time is
essentially set by the relaxation time $\tau_\mathrm{rel}$, which
typically occurs on a $\sim\!10~\mathrm{ns}$ timescale. This is
several orders of magnitude smaller than a typical tunneling time of
$\sim 1/\Gin \sim 100 \ms$ \cite{fujisawa:2002}. Secondly, the slow
cotunneling rate limits the amount of experimental data available
within a reasonable measurement time. This explains the large spread
between the data points in \FigRef{fig:CotunnelingNoise} in the
cotunneling regime. We conclude that the measurement bandwidth
currently limits the possibility of examining correlations in the
cotunneling regime using time-resolved detection techniques. A
higher-bandwidth detector would solve both the above mentioned
problems. It would allow a general increase in the tunneling rates
in the system, which would both decrease the difference between
$\tau_\mathrm{cot}$ and $\tau_\mathrm{rel}$ as well as provide
faster acquisition of sufficient statistics.

\subsection{Spin effects in many-electron dots}
So far, we have neglected the spin properties of the electrons by
considering them to be spin-less particles. In few-electron double
quantum dots, spin effects have been shown to lead to \emph{Pauli}
spin blockade \cite{ono:2002,koppens:2005}; the current is strongly
suppressed in configurations where a spin flip is required for
electrons to traverse the DQD. The Pauli blockade configuration has
been utilized for performing electron spin resonance experiments
\cite{koppens:2006, nowack:2007} as well as for studying
interactions between the electron and nuclei spin systems
\cite{ono:2004,koppens:2005,pfund:2007,reillyNucl:2007}.

In the system investigated here, the DQD contains a relatively large
number of electrons; from the energy scales and from the geometry of
the device we estimate each QD to hold $\sim \! 30$ electrons. This
makes the observation of spin blockade more difficult, since neither
the excitation spectrum nor the exact QD spin configuration is well
known. For few-electron QDs, the first two electrons fill up
spin-degenerate single-particle states and form a spin-pair
\cite{tarucha:1996}. Spin pairing has also been reported in
many-electron chaotic dots \cite{luscher:2001} and quantum rings
\cite{fuhrer:2001}. If spin-pairing occurs, it is possible to get a
spin-zero many-electron ground state and we may neglect the
spin-less core of electrons and only consider the spin of the
outermost electrons \cite{pfund:2007}.

\begin{figure}[tb]
\centering
\includegraphics[width=\linewidth]{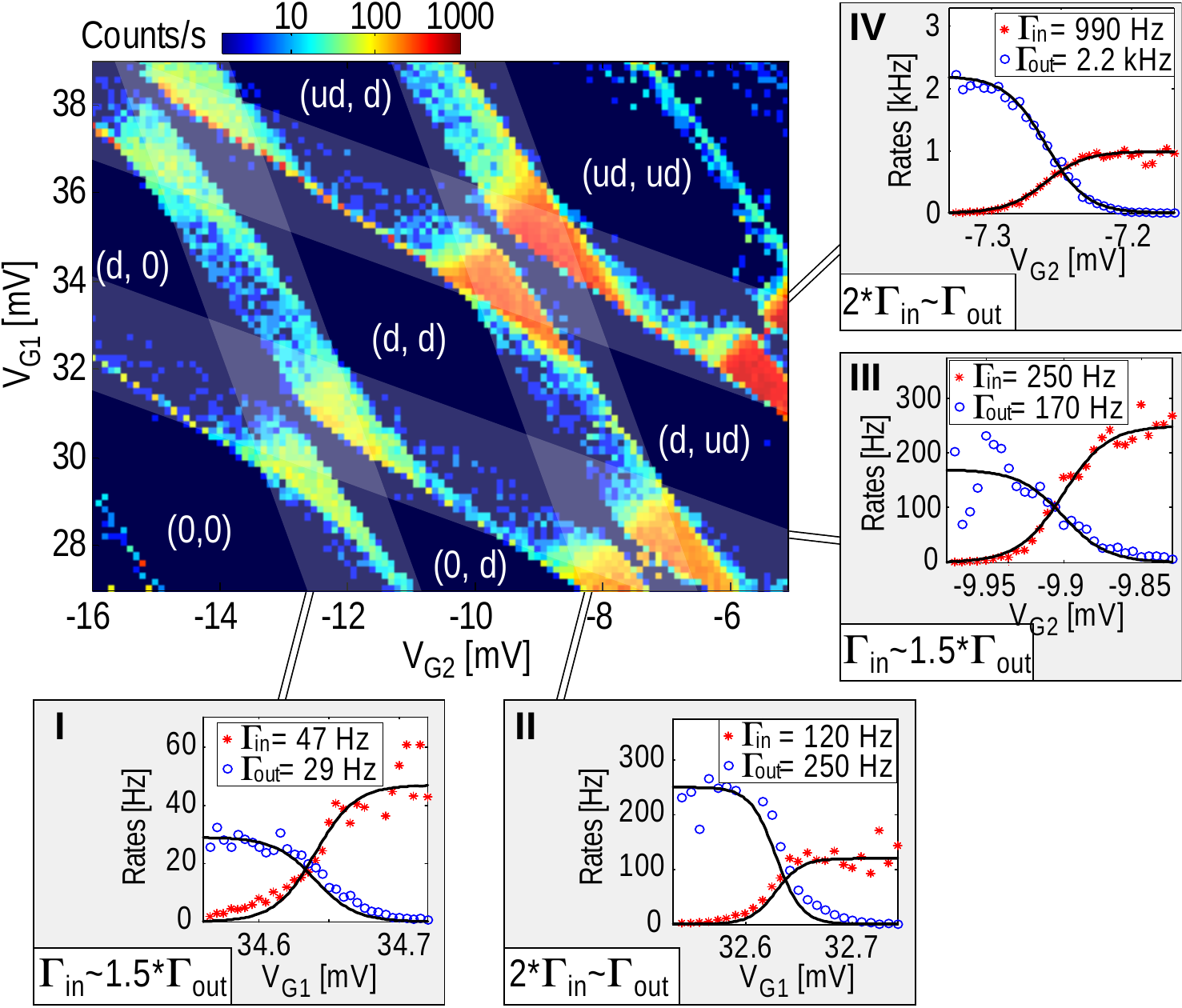}
\caption{Charge stability diagram of the DQD, together with a possible spin
configuration extracted from analyzing rates for electrons tunneling
into and out of the QDs. The numbers do not refer to the absolute
DQD electron population but to the excess electron population
relative to the configuration marked by (0,0). Cases I-II shows
degeneracies measured when filling QD2, cases III-IV refer to
filling of QD1. The measurements shown in cases I-IV were taken with
zero bias over the DQD in order to minimize the influence of
cotunneling.} \label{fig:DQ_HexagonDegen}
\end{figure}

To investigate the occurrence of spin pairing and spin blockade in
our device, we use the methods of
section~\ref{sec:TR_degenerateStates} to determine the degeneracy of
the QD ground states. Depending on the occupancy of a
spin-degenerate state, the rates for electrons tunneling into and
out of a QD should differ by a factor of two. By performing such
measurements for consecutive electron filling in the DQDs, we can
extract a possible spin configuration of the DQD. The method is
visualized in Fig.~\ref{fig:DQ_HexagonDegen}, together with a
possible spin configuration for the hexagons from
\FigRef{fig:DQ_HexCount}(a). The numbers in the figure do not
refer to the absolute DQD electron population but to the number of
excess electrons relative to the configuration indicated by (0,0).

Starting in the Coulomb-blockaded region marked by (0,0), we
increase the gate voltage $\Vgr$ to add an electron into QD2. At the
transition to (0,d) (case I in \FigRef{sec:TR_degenerateStates}), we
find that the tunneling rate for electrons entering QD2 is larger
than the rate for electrons tunneling out. Increasing the gate
voltage further to add another electron to QD2 (case II), the
relation between $\Gin$ and $\Gout$ is reversed. This is in
agreement with successive filling of electrons into a degenerate
state; if both degenerate states are initially unoccupied (case I),
an incoming electron may tunnel into either state with an effective
tunneling rate $\Gin = g \times \Gamma$. Here, $g$ is the degeneracy
factor and $\Gamma$ is the tunnel coupling to the lead. On the other
hand, the rate for electrons leaving the QD is determined by the
number of occupied degenerate states. With only one electron in the
QD we get $\Gout = \Gamma$ and thus expect $\Gin/\Gout = g$. The
situation is reversed if the QD is initially occupied (case II);
here we expect $\Gin/\Gout = 1/g$.

If we assume the degeneracy to be due to spin states, the data
indicates that QD2 is successively filled up with one spin-down and
one spin-up electron. The pattern is repeated if we perform similar
measurements for QD1 (cases III-IV). From these measurements we
extract the spin configurations shown in
\FigRef{fig:DQ_HexagonDegen}. It should be noted that the degeneracy
factors in cases I and III are lower than the factor $g=2$ expected
from spin-degenerate states. This might be due to changes in the
tunneling coupling $\Gamma$ within the gate voltage region of
interest, although the coupling normally only changes slightly
within the small voltage range used here (see
section~\ref{sec:TR_tuneCoupling}). Therefore, the spin
configurations marked in \FigRef{fig:DQ_HexagonDegen} should not be
considered as definite; we can not rule out other explanations for
the data.

\begin{figure}[tb]
\centering
\includegraphics[width=\linewidth]{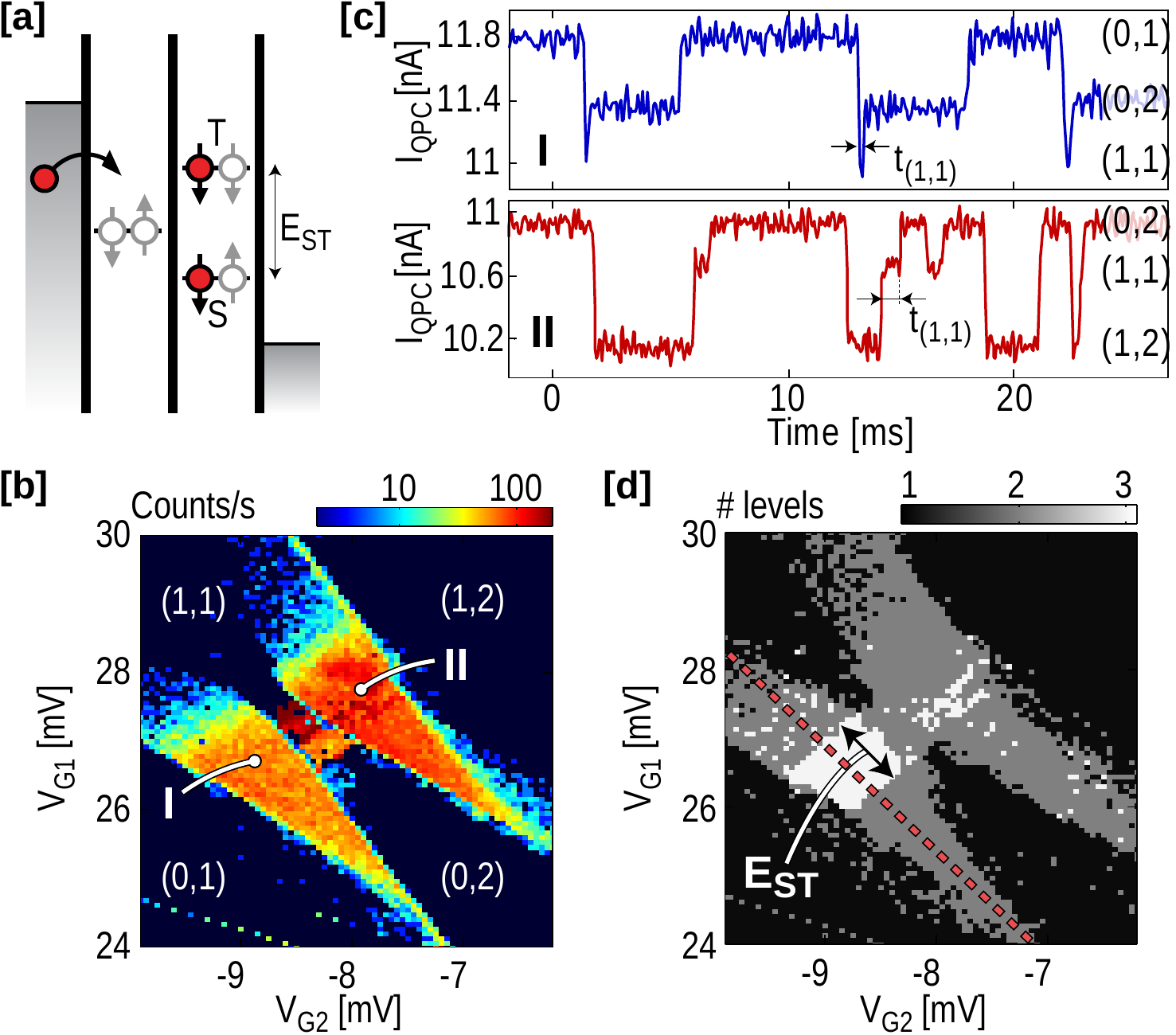}
\caption{(a) Sketch of a DQD in the regime of Pauli spin blockade.
If the electron entering QD1 has the same spin orientation as the
electron sitting in QD2, transport is blocked until either electron
flips its spin to allow a singlet state to form in QD2. (b) Charge
stability diagram measured by counting electrons entering the DQD.
The numbers refer to the assumed excess charge population relative
to a state where both QDs have zero spin. The data was extracted
from QPC conductance traces of length $T=0.5\s$, taken with
$V_\mathrm{DQD-SD} = 600\uV$ and $V_\mathrm{QPC-SD} = 400\uV$. (c)
QPC conductance traces, taken at the points marked in (b). For both
positions, a third level appears which we attribute to the
transition $(1,1)\rightarrow(0,2)$. (d) Regions of the charge
stability diagram of (b) where the charge detector finds more then
two levels in the QPC conductance traces. In the spin blockade model
of (a), the width of regions with three levels corresponds to the
singlet-triplet spacing in QD2.}
 \label{fig:DQ_SpinBlockade}
\end{figure}

Keeping this reservation in mind but still assuming the spin
configuration of \FigRef{fig:DQ_HexagonDegen} to be correct, we
expect spin blockade to occur in the transport triangle involving
the configurations (0,d), (d,d) and (0,ud). The principle of the
blockade is explained in \FigRef{fig:DQ_SpinBlockade}(a). We start
in the configuration (0,d), where QD1 is empty and QD2 contains one
excess electron. An electron may tunnel from source into QD1 and
since the QD is initially empty, the incoming electron may be either
spin-up or spin-down. If the spin is opposite to the spin of the
electron in QD2, the electron in QD1 can continue to QD2 to form the
spin singlet ground state (0,ud). Finally, an electron may leave to
the drain which takes the system back to the state (0,d) and the
cycle can be repeated.

However, if the electron tunneling from source into QD1 has the same
spin orientation as the electron in QD2, it cannot continue from QD1
to QD2. This is because of the singlet-triplet splitting in QD2; due
to the exchange energy the system favors the formation of a spin
singlet and the energy of the spin triplet is raised by the
single-triplet splitting $E_\mathrm{ST}$. The electron in QD1 is
thus blocked until a spin-flip occurs in either QD1 and QD2. Since
spin relaxation is slow, the effect leads to a sharp decrease in the
current through the DQD \cite{ono:2002}. In our case, we do not
measure the average current but rather count the electrons as they
pass through the structure. As mentioned in
section~\ref{sec:DQ_tunnelCoupling}, the tunnel coupling between the
QDs is too strong to allow interdot charge transitions to be
resolved in time. However, in the spin-blockade regime interdot
charge transition from QD1 to QD2 should be limited by the spin
relaxation rate, which for GaAs QDs has been reported to be several
milliseconds or even seconds for magnetic fields of $1\T$
\cite{elzermanNature:2004, amasha:2008}. This is within the
bandwidth of the charge detector and we thus expect spin blockade to
help make the interdot charge transitions resolvable.

Figure~\ref{fig:DQ_SpinBlockade}(b) shows the finite-bias charge
stability diagram measured by counting electrons in the regime
located between the (d,d) and (0,ud)-region of
\FigRef{fig:DQ_HexagonDegen}. The data shows two triangle-shaped
regions of electron and hole transport expected from the applied
voltage bias. Figure~\ref{fig:DQ_SpinBlockade}(c) shows examples of
QPC current traces taken at the two positions marked in
\FigRef{fig:DQ_SpinBlockade}(b). Taking a closer look at the data
from position I, we see that the time trace actually contains three
levels; starting at the QPC current level labeled (0,1), the QPC
current drops to level (1,1) as an electron tunnels into QD1. The
electron relatively quickly continues to QD2 [level (0,2) in
\FigRef{fig:DQ_SpinBlockade}(c)], before tunneling out to the drain
and taking the QPC conductance back to level (0,1). The ability of
the QPC to determine if the electron is sitting in QD1 or QD2 comes
directly from the geometry of the device; the QPC is located closer
to QD1.

For case II of \FigRef{fig:DQ_SpinBlockade}(b,c), transport is
governed by the hole process
$(0,2)\rightarrow(1,2)\rightarrow(1,1)\rightarrow(0,2)$. As for the
electron process in case I, the transition that possibly involves
spin relaxation is the one where the electron hops from QD1 to QD2
[$(1,1)\rightarrow(0,2)$]. The timescale of the interdot transitions
is marked by $t_{(1,1)}$ in the traces of
\FigRef{fig:DQ_SpinBlockade}(c). From the above discussion, we
expect $t_{(1,1)}$ to be long enough to be measurable as long as the
DQD detuning is smaller than the singlet-triplet spacing, $\delta <
E_\mathrm{ST}$. If $\delta > E_\mathrm{ST}$, the electron in QD1 may
tunnel to QD2 regardless of the spin direction and we expect to
resolve only two levels in the QPC conductance traces. This is
visualized in \FigRef{fig:DQ_SpinBlockade}(d), where we plot the
number of current levels detected by an automatic level detection
algorithm. Focusing first on the
electron transport cycle, there is a region of three-level traces
situated at the base of the triangle. In the model of spin-blockade,
the width of the region in direction of detuning is equal to the
singlet-triplet splitting in QD2, giving $E_\mathrm{ST} \sim 200
\ueV$. For the hole cycle, the region showing three levels is less
regular. However, this is actually an artifact due to imperfections
of the level detection algorithm; for the electron cycle (case I),
the third, fast level occurs below the other levels, which makes it
relatively easy to detect. It is much harder to reliably detect a
fast third level if it occurs in-between the two main levels as in
the hole region (case II). Manual inspection of the traces confirms
that the region of three-level traces indeed has the same extension
for the hole as for the electron cycle.

\begin{figure}[tb]
\centering
\includegraphics[width=\linewidth]{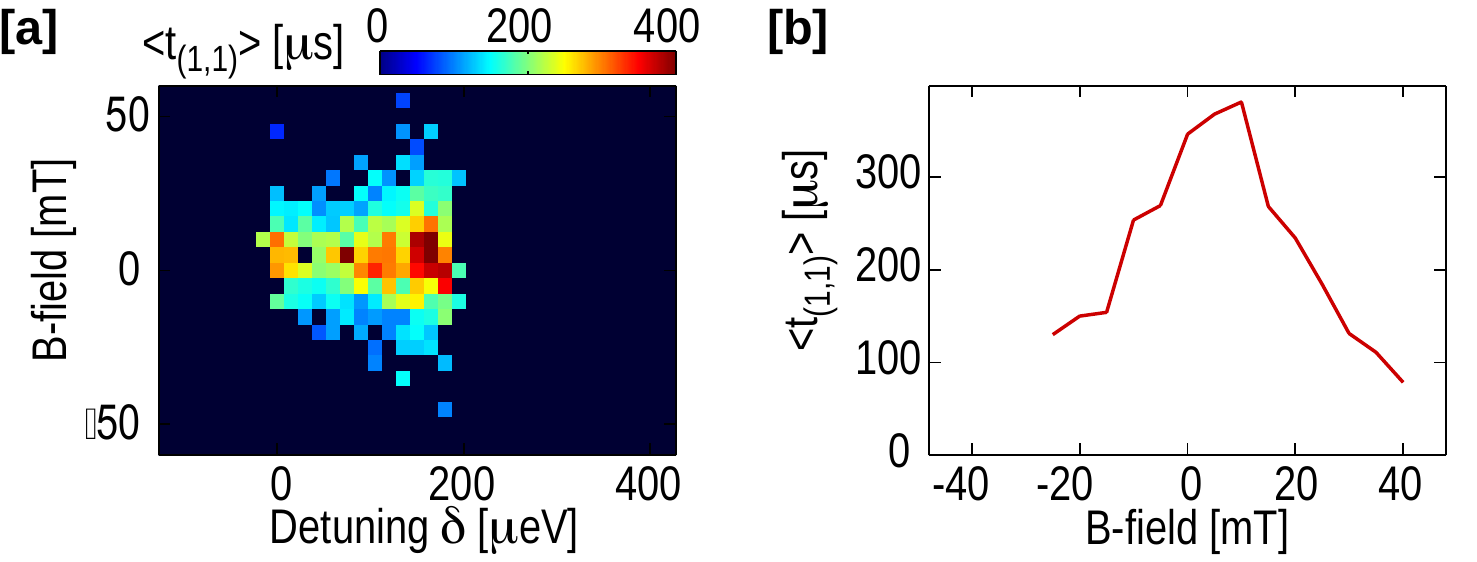}
\caption{(a) Average time spent in the (1,1) state, measured vs DQD
detuning and magnetic field. The third level in the QPC conductance
traces corresponding to the (1,1) state is only visible in the
region $0 < \delta < E_\mathrm{ST} \sim 200\ueV$. The data was taken
along the dashed line in Fig.~\ref{fig:DQ_SpinBlockade}(d). (b)
Cross-section of the graph in (a), taken at $\delta = 170\ueV$. The
time $t_{(1,1)}$ falls of quickly with increased B-field and drops
below the time resolution of the detector as $|B|>30\mT$.}
\label{fig:DQ_SpinLifetime}
\end{figure}

To get more quantitative concerning the time scale of the
$(1,1)\rightarrow(0,2)$ transition 
and to investigate if it is really related to spin relaxation, we
measure the average of $t_{(1,1)}$ as a function of detuning [dashed
line in \FigRef{fig:DQ_SpinBlockade}(d)] and magnetic field. The
result is presented in \FigRef{fig:DQ_SpinLifetime}(a). As mentioned
in the previous paragraph, the third level is only visible in the
region of $0 < \delta < E_\mathrm{ST} \sim 200\ueV$.
Figure~\ref{fig:DQ_SpinLifetime}(b) shows a cross section taken at
$\delta = 170\ueV$. The time spent in the (1,1)-state is around
$400\us$ at zero magnetic field, but decays rapidly with increased
B-field and disappears below the time resolution of the detector as
$|B|>30\mT$.

The spin blockade can be conveniently expressed using a model
involving two-electron spin singlet (S) and triplet (T) states
distributed over both QDs. In this language, the electron tunneling
into the DQD from the source lead can enter either the singlet
S(1,1) or the triplet T(1,1) state. The singlet S(1,1) quickly
relaxes to S(0,2) followed by an electron leaving the DQD to the
drain. On the other hand, if the electron enters into the triplet
T(1,1), it can not proceed to T(0,2) since this state is raised by
the singlet-triplet splitting in QD2. The triplet T(1,1) first needs
to relax to S(1,1) before proceeding to S(0,2), leading to spin
blockade.

For GaAs quantum dots, the spin blockade has been observed to be
lifted at zero magnetic field because of mixing of the T(1,1) and
S(1,1) states due to hyperfine interactions with the nuclear spin
bath. The mixing energy is given by the magnitude of the random
magnetic field $\vec{B}_\mathrm{N}$ generated by the fluctuating
nuclear spins, with $E_\mathrm{N} = g \mu_B |\vec{B}_\mathrm{N}|
\sim 0.1\ueV$ for a typical quantum dot containing $n \sim 10^6$
nuclei. The mixing can be removed by applying an external magnetic
field so that the electron Zeeman splitting becomes larger than the
mixing energy $E_\mathrm{N}$.
This typically occurs on a magnetic field scale of a few mT
\cite{koppens:2005}.
In our case, we observe the opposite behavior; the relaxation rate
$\Gamma_\mathrm{rel} = 1/\langle t_{(1,1)} \rangle$ is minimal at
zero magnetic field and increases with external magnetic field. In
contrast to the setup of Ref.~\cite{koppens:2005}, we are in the
strong coupling regime, with $t \sim 30 \ueV \gg E_\mathrm{N}$. As
discussed in section~\ref{sec:DQ_tunnelCoupling}, the tunnel
coupling will hybridize the S(1,1) and S(0,2) singlet configurations
and thereby keep the energy separation to the T(1,1) triplet larger
than $E_\mathrm{N}$ over the full range of detuning in
\FigRef{fig:DQ_SpinLifetime}(a). This suppresses the relaxation due
to hyperfine mixing, even at zero external magnetic field
\cite{pfundPRB:2007}.

A strong increase in the relaxation rate for small magnetic field
has been seen in InAs DQD \cite{pfund:2007}. The behavior was
attributed to the strong spin-orbit interactions of that material
system. The main spin relaxation mechanism in few-electron single
QDs in GaAs is also due to spin-orbit coupling, with relaxation
rates increasing with external magnetic field \cite{meunier:2007,
amasha:2008}. However, the relaxation times seen in
\FigRef{fig:DQ_SpinLifetime}(b) are much shorter and the B-field
dependence much stronger than reported for few-electron single
quantum dots. It is unclear how the existence of additional
electrons in our DQD influences the relaxation process and it is
uncertain if it is reasonable to assume electron-electron
interactions to be weak enough to allow the QDs to be modeled using
independent single-particle states. From the measurements presented
here, one can not make a clear statement whether the observed
features are due to spin relaxation or not. It would certainly be
interesting to repeat the time-resolved measurements on a DQD
containing only two electrons.

\subsection{Weak interdot coupling}
In the last part of this section, we treat the case where the
three barriers of the DQD are tuned so that all tunneling processes
occur on timescales slower than the bandwidth of the charge
detector. In this regime it is possible to detect electrons
tunneling back and forth between the QDs and thus determine the
direction of the tunneling electrons \cite{fujisawa:2006}.

\begin{figure}[tb]
\centering
\includegraphics[width=\linewidth]{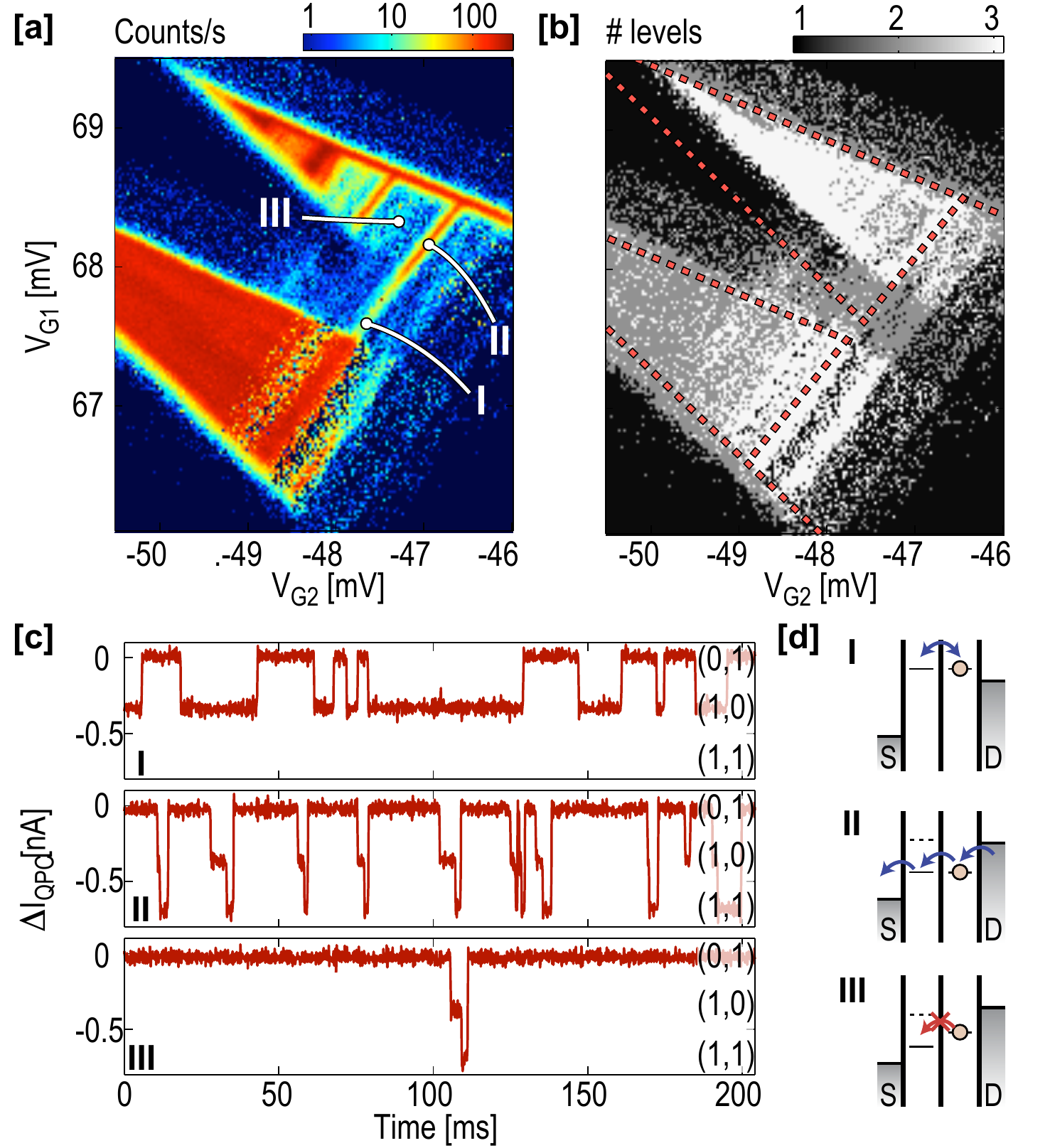}
\caption{(a) Charge stability diagram of the DQD measured by
counting electrons entering the DQD. The data was taken with
$V_\mathrm{DQD-SD}= - 700\uV$ and $V_\mathrm{QPC-SD}=300\uV$. (b)
Number of levels in the QPC conductance traces, extracted from the
same data as in (a). The dashed lines show the extension of the
triangle expected from the applied bias and the capacitive lever
arms of the gates. (c) QPC conductance traces, taken at three
positions marked in (a). In case I, the tunneling is due to
equilibrium fluctuations between QD1 and QD2. In cases II-III, a
current flows through the DQD. (d) Energy level diagrams
depicting the DQD configuration for the three position in (a,c).}
\label{fig:DQ_ThreeLevelMap}
\end{figure}

It turned out to be difficult to reach this regime for the
ring-shaped DQD of \FigRef{fig:DQ_Sample}(a). The constrictions
between the QDs were generally much more open than the constrictions
to source and drain leads, which made it hard to pinch off the
middle constriction while at the same time keeping source and drain
open and forming well-defined dots. A measurement from one of the
few cases where we were partly successful is shown in
\FigRef{fig:DQ_ThreeLevelMap}(a). The plot shows the charge
stability diagram measured with $-700\uV$ bias applied across the
DQD. The transport triangles due to electron and hole transport are
well visible. There are a few striking things in this measurement
compared to charge stability diagrams shown previously in this
section. First, the size of the triangle due to hole transport is
considerably smaller than the electron triangle. Although this is
not quantitatively understood, we speculate that it is due to a
weakly coupled state in QD2 that blocks transport in parts of the
triangles. Second, there are bands of weak tunneling occurring
outside the triangles. We attribute this to photon absorption
processes driven by the current flowing in the QPC; this is the
subject of section~\ref{sec:SP_main}. Finally, there are stripes
occurring parallel to the base line of the triangles; these are
excited states in the QDs probed by interdot transitions.

Figure~\ref{fig:DQ_ThreeLevelMap}(b) shows the number of QPC current
levels found with the automatic level detection algorithm. Three levels are found in most of
the hole transport triangle as well as in large parts of the
electron transport triangle, showing that tunneling between the QDs
is indeed slow enough to be detected by the detector.
Figure~\ref{fig:DQ_ThreeLevelMap}(c) shows three QPC conductance
traces taken at the positions marked in
\FigRef{fig:DQ_ThreeLevelMap}(a). Energy level diagrams for the
corresponding configurations are shown in
\FigRef{fig:DQ_ThreeLevelMap}(d).

Starting at the position marked by I, the two QD levels are aligned
but shifted outside the bias window. Here, equilibrium fluctuations
occur between the QDs. The QPC conductance trace shows transitions
between two levels corresponding to an electron sitting on QD1 and
QD2, respectively. The transitions occur on a relatively slow
timescale of $\sim \! 10\ms$.

Continuing to case II, we keep the alignment of the levels in the
two QDs but shift them inside the bias window of the hole transport
cycle. Looking at the trace in \FigRef{fig:DQ_ThreeLevelMap}(c), we
see that the transition of electrons from QD2 to QD1
[$(0,1)\rightarrow(1,0)$] still occurs on a timescale comparable to
case I. However, before the electron in QD1 has time to tunnel back
to QD2, an electron is quickly refilled into QD2 from the drain lead
and takes the QPC conductance to the (1,1) level. Afterwards, an
electron may leave from QD1 to source and the system is back in the
(0,1) state. Each cycle corresponds to one electron being
transferred through the DQD.

The timescale for interdot transition is clearly slower than the
tunneling involving the source or drain lead. The DQD current is
thus limited by the central barrier. This is clearly visualized if
we continue to case III, which corresponds to a slightly lowered
electrochemical potential of QD1 relative to QD2. Here, the interdot
transition can not occur resonantly; the tunneling electron needs to
lose parts of it energy to the environment. This makes the tunneling
process less probable and reduces the count rate in the region of
case III in \FigRef{fig:DQ_ThreeLevelMap}(a). The QPC conductance
trace \FigRef{fig:DQ_ThreeLevelMap}(c) shows that the electron
indeed spends most of the time in QD2; once a transition to QD1
occurs it is immediately followed by tunneling from drain to QD2 and
from QD1 to source, as discussed for case II. Finally, by further
lowering the electrochemical potential of QD1 an excited state of
QD1 lines up with QD2 and tunneling may again occur resonantly. This
is the reason for the stripes parallel to the triangle baseline
occurring inside the triangles.

\begin{figure}[tb]
\centering
\includegraphics[width=\linewidth]{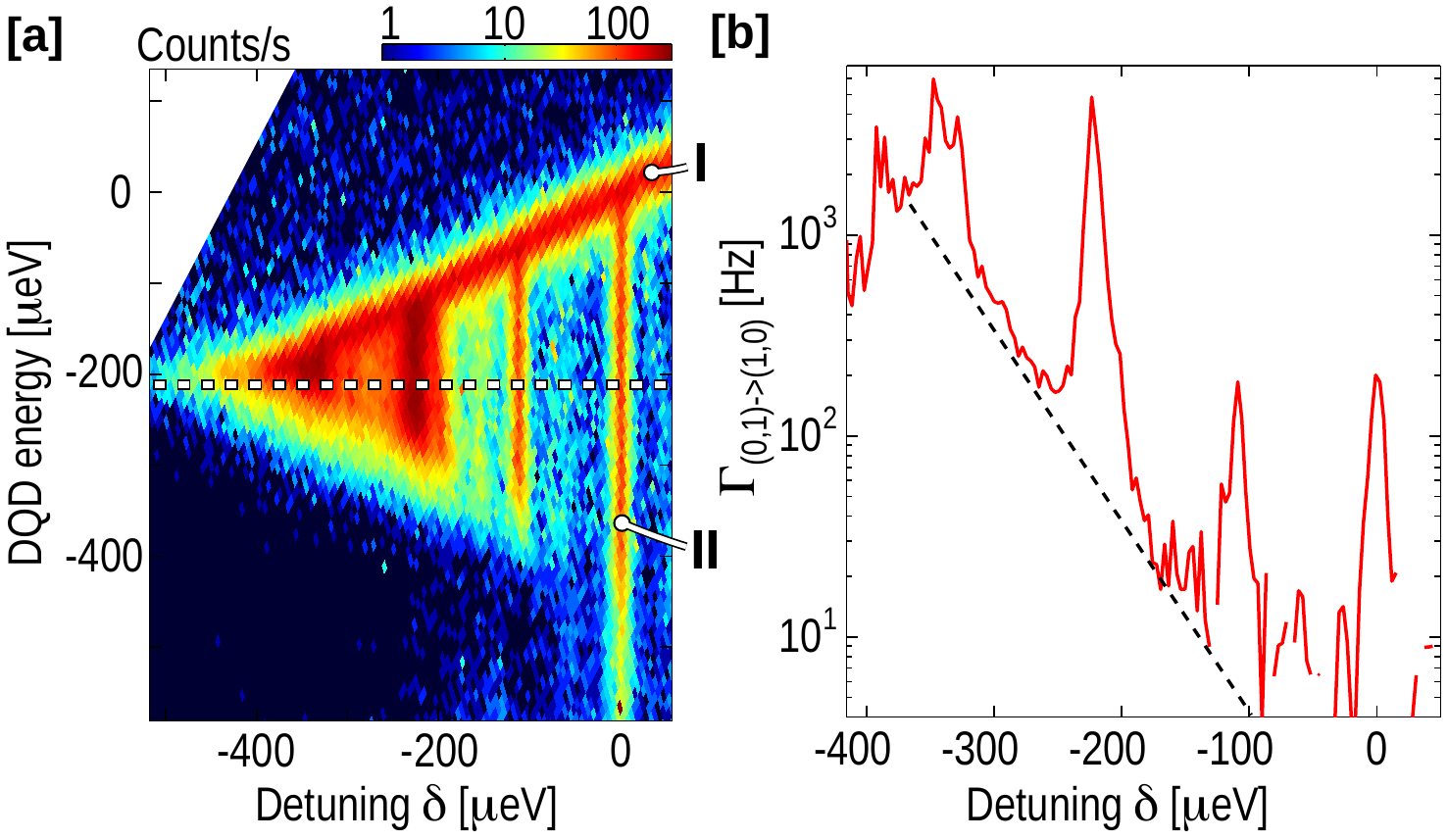}
\caption{(a) Same as \FigRef{fig:DQ_ThreeLevelMap}(a), but with the
gate voltages converted to average energy and detuning of the double
quantum dot. The line coming from tunneling between QD1 and the lead
(case I) is considerably broader than the lines due to interdot
transitions (case II). (b) Rate for the interdot transition
$(0,1)\rightarrow(1,0)$, measured along the dashed line in (a). The
peaks come from resonant tunneling through excited states in either
QD. The inelastic tunneling between resonant peaks increases
strongly with increased detuning. The dashed line is a guide to the
eye depicting exponential increase. } \label{fig:DQ_ThreeLevelTrace}
\end{figure}

The above discussion raises a few interesting questions concerning
the interdot tunneling. First, what sets the width of the regime
with resonant interdot tunneling? In case I, the electrons in the
DQD are isolated from the leads and it seems unlikely that the DQD
transitions should be influenced by the thermal distribution of the
electrons in the leads. Second, what are the relaxation processes
leading to the slow but non-zero tunneling rates in the non-resonant
regime? To answer the first question, we take the data from
\FigRef{fig:DQ_ThreeLevelMap}(a) and use the known capacitive lever
arms to convert the gate voltages into energy of the DQD. The
result is presented in \FigRef{fig:DQ_ThreeLevelTrace}(a), where we
plot the count rate for the hole transport triangle vs average DQD
energy and detuning energy $\delta$. The two axes have the same
scaling, which makes it easier to compare energy scales of different
processes.

We first focus on the tilted line with slope $1/2$ marked by I in
\FigRef{fig:DQ_ThreeLevelTrace}(a). The line is due to equilibrium
fluctuations between QD1 and the source lead; the broadening of the
line is a direct measure of the electron temperature in the source
lead (see section \ref{sec:TR_oneLead}). By converting the energy to
temperature we find that the electron temperature in the lead is
around $T=100\mK$. However, the width of the thermally-broadened
line stands in sharp contrast to the narrow vertical lines coming
from interdot transitions (case II). The width of these lines is
only around a quarter of the thermal-broadened line, which would
correspond to a temperature of $25\mK$.
This energy scale matches relatively well to the base temperature of
the cryostat. A possible explanation could therefore be that the
broadening occurs because of scattering with thermally excited
acoustic phonons. A straightforward experimental check of this
hypothesis would be to investigate how the broadening changes when
raising the base temperature of the cryostat. Unfortunately, shortly
after measuring the data in \FigRef{fig:DQ_ThreeLevelMap} we had to
warm up the cryostat, and we were not able to reach the same regime
in subsequent cool-downs.


A different energy scale is given by the tunnel coupling between the
two QDs. If we assume the transport between the QDs to be coherent
and convert the measured tunneling rate of $\Gamma \sim 100\Hz$ to a
coupling energy, we find
\begin{equation}\label{eq:DQ_interdotEnergy}
    t \sim h f \sim 0.4~\mathrm{peV}.
\end{equation}
This is obviously several orders of magnitudes smaller than the
width measured in the experiment. Still, the discussion raises some
interesting questions concerning coherence and projective
measurements. For a fully coherent system, the electron
wavefunctions in the two QDs hybridize and form bonding and
antibonding states that delocalize over both dots. At zero detuning
both the bonding and antibonding wavefunctions have the same spatial
extent, which means that a charge detector would not be able to
resolve transitions between the two states independently of how
slowly the transitions occur. The very fact that we detect electrons
tunneling back and forth between the QDs even at zero detuning is an
obvious indication that the system is not coherent. The decoherence
rate is faster than the tunnel coupling, meaning that the coherent
evolution of an electron between the two QDs is interrupted by a
projective measurement taking the electron back into the states of
the individual QDs. The rate at which we observe transitions between
the two QDs thus depends not only on the tunnel coupling but also on
the decoherence in the system. It would certainly be interesting to
perform measurements in a regime where the tunnel coupling and the
decoherence rate are comparable, and to investigate how the measured
transition rates are affected by the presence of the QPC and its
ability to perform projective measurements. One would expect an
increased QPC bias to introduce additional effects compared to the
intrinsic decoherence.

Finally, we come back to the question of the relaxation mechanism
leading to the finite count rate between the lines of resonant
tunneling in \FigRef{fig:DQ_ThreeLevelMap}(a) and
\FigRef{fig:DQ_ThreeLevelTrace}(a).
Figure~\ref{fig:DQ_ThreeLevelTrace}(b) shows the interdot transition
rate $\Gamma_{(0,1)\rightarrow(1,0)} = 1/\langle t_{(0,1)}\rangle$
measured along the dashed line in
\FigRef{fig:DQ_ThreeLevelTrace}(b), extracted from traces similar to
the ones shown in \FigRef{fig:DQ_ThreeLevelMap}(c). The ground state
transition as well as transitions due to three exited states give
rise to clear peaks in the figure. In between the peaks, the rate of
the non-resonant transition increases strongly as the detuning gets
larger.

Spontaneous energy relaxation in a DQD has been investigated
previously using conventional current measurement techniques
\cite{fujisawa:1998}. In that work, the authors find that the
emission rate \emph{decreases} with increased detuning and attribute
the mechanism behind the relaxation to phonon emission. This is in
disagreement with the results of \FigRef{fig:DQ_ThreeLevelTrace}(b),
where the emission rate clearly increases with detuning. It would
therefore be interesting to perform further experiments in this
regime and investigate the inelastic tunneling of
\FigRef{fig:DQ_ThreeLevelTrace}(b) in more detail. In addition to
checking the obvious influence of the temperature of the phonon bath
there could be other explanations for the relaxation such as photon
emission to the nearby quantum point contact \cite{aguado:2000} or
to anywhere else in the environment.

\section{Detector back-action} \label{sec:SP_main}

In the previous sections, we used quantum point contacts to
measure charge transitions in various mesoscopic structures. While
doing so we assumed the point contact to be an idealized detector
that does not exert any back-action on the measured object. In
reality, this is not true. The scattering of electrons in the
quantum point contact leads to emission of microwave radiation. In
this section, we show that the radiation may drive transitions in a
double quantum dot. Turning the perspectives around, the double
quantum dot can be seen as a frequency-selective microwave detector.
The frequency of the absorbed radiation is set by the energy
separation between the levels in the dots, which is easily tuned
with gate voltages. By combining this with time-resolved charge
detection techniques, we can directly relate the detection of a
tunneling electron to the absorption of a single photon. 

\subsection{Using the double quantum dot as a frequency-selective detector}

The interplay between quantum optics and mesoscopic physics opens up
new horizons for investigating radiation produced in nanoscale
conductors \cite{beenaaker:2001, gabelli:2004}. Microwave photons
emitted from quantum conductors are predicted to show non-classical
behavior such as anti-bunching \cite{beenakker:2004} and
entanglement \cite{emary:2005}. Experimental investigations of such
systems require sensitive, high-bandwidth detectors operating at
microwave-frequency \cite{zakka:2007}. On-chip detection schemes,
with the device and detector being strongly capacitively coupled,
offer advantages in terms of sensitivity and large bandwidths. In
previous work, the detection mechanism was implemented utilizing
photon-assisted tunneling in a
superconductor-insulator-superconductor junction \cite{deblock:2003,
onacNT:2006} or in a single quantum dot (QD) \cite{onacQD:2006}.

\begin{figure}[tb]
\centering
 \includegraphics[width=\linewidth]{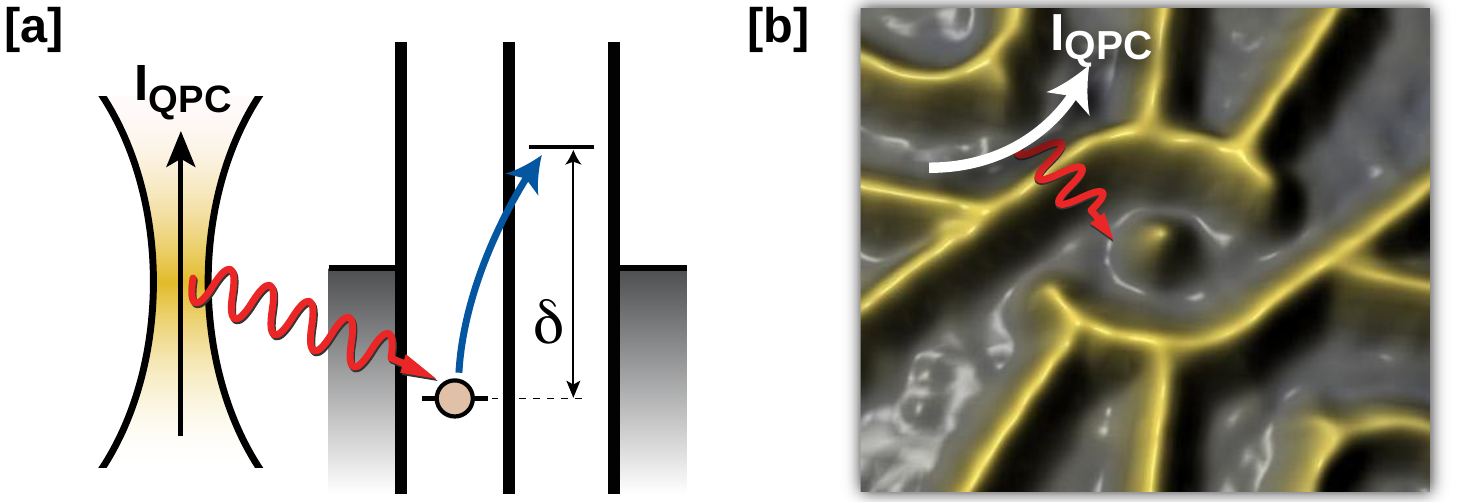}
 \caption{Schematic for operating a double quantum dot (DQD) as a high-frequency noise detector. The tunable level separation
 $\delta$ of the DQD allows frequency-selective detection.
 (b) The double quantum dot used in the experiment.
 }
\label{fig:SP_model}
\end{figure}

Aguado and Kouwenhoven proposed to use a double quantum dot (DQD) as
a frequency-tunable quantum noise detector \cite{aguado:2000}. The
idea is sketched in Fig.~\ref{fig:SP_model}(a), showing the energy
levels of the DQD together with a quantum point contact acting as a
noise source. The DQD is operated with a fixed detuning $\delta$
between the electrochemical potentials of the left and right QDs.
For an isolated system, the DQD is in the Coulomb blockade regime
and there will be no current flowing. However, if the system absorbs
an energy $E = \delta$ from the environment, the electron in QD1 is
excited to QD2. This electron may leave to the drain lead, a new
electron enters from the source contact and the cycle can be
repeated. The process induces a current flow through the system.
Since the detuning $\delta$ may be varied continuously by applying
appropriate gate voltages, the absorption
energy is tunable. 

The scheme is experimentally challenging, due to low current levels
and fast relaxation processes between the QDs \cite{khrapai:2006}.
Here, we show that these problems can be overcome by using
time-resolved charge-detection techniques to detect single electrons
tunneling into and out of the DQD.
%
%
Apart from giving higher sensitivity than conventional current
measurement techniques, the method also allows us to directly relate
a single-electron tunneling event to the absorption of a single
photon. The system can thus be viewed as a frequency-selective
single-photon detector for microwave energies.
This, together with the fact that the charge-detection methods allow
precise determination of the device parameters, provide major
advantages compared to other setups \cite{gabelli:2004, zakka:2007,
deblock:2003, onacNT:2006, onacQD:2006}.

\begin{figure}[tb]
\centering
 \includegraphics[width=0.97\linewidth]{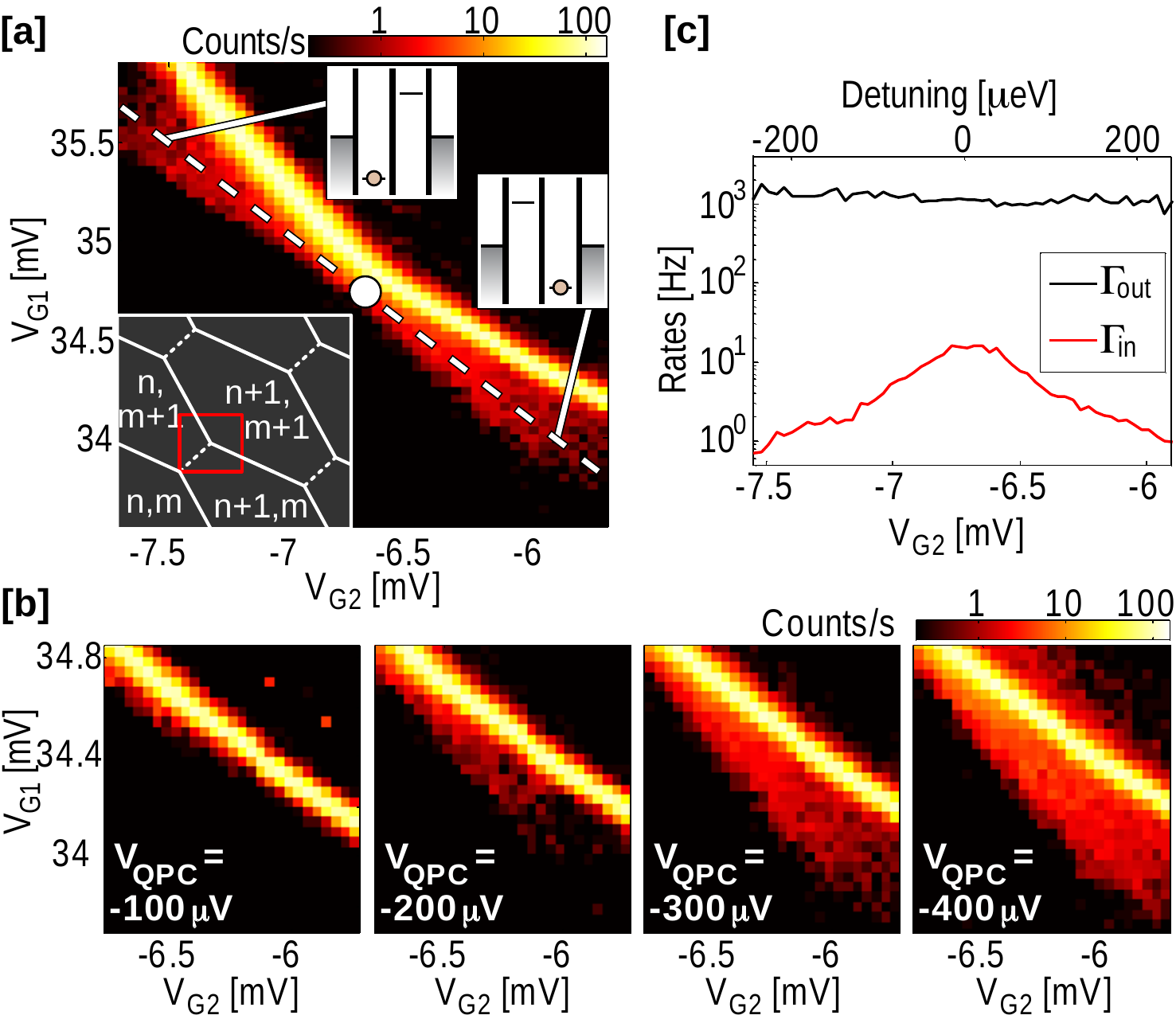}
 \caption{(a) Count rate for electrons leaving the DQD, measured for a small region close to
 a triple point (marked by a white point). The inset shows a sketch of the surrounding hexagon pattern.
 The dashed line denotes the detuning axis, with zero detuning occurring at the triple
 point. The data was taken with $V_\mathrm{QPC}=-300\uV$.
 (b) Blow-up of the lower-right region of (a), measured for
 different QPC bias voltages.
 (c) Rates for electron tunneling into and out of the DQD, measured
 along the dashed line in (a). $\Gamma_{\mathrm{in}}$ falls of rapidly with detuning, while $\Gamma_{\mathrm{out}}$
 shows only minor variations.  Adapted from Ref. \cite{gustavssonPRL:2007}.
   }
\label{fig:SP_TriNoBias}
\end{figure}

The measurements were performed on the structure shown in
Fig.~\ref{fig:SP_model}(b), which consists of two quantum dots
embedded in a ring, together with a nearby QPC. As described in section \ref{sec:DQ_main}, we
tune the surrounding gates so that only the upper tunnel barrier
connecting the two QDs is kept open. The tunnel coupling between the
QDs was set to $t = 32~\mathrm{\mu eV}$, as determined using charge
localization measurements explained in section
\ref{sec:DQ_tunnelCoupling}. The tunneling barriers between the DQD
and the source and drain contacts were tuned to a few kHz to enable
electron counting in real-time.
In the following, we present measurements taken with zero bias
across the DQD. Fig.~\ref{fig:SP_TriNoBias}(a) shows count rates
close to the triple point where the $(n+1,m)$, $(n,m+1)$ and
$(n+1,m+1)$ states are degenerate [see inset of
Fig.~\ref{fig:SP_TriNoBias}(a)]. The arguments presented below are
applicable also for the triple point between the $(n,m)$, $(n+1,m)$,
$(n,m+1)$ states, but for simplicity we consider only the first
case. At the triple point [marked by a white dot in
Fig.~\ref{fig:SP_TriNoBias}(a)], the detuning $\delta$ is zero and
both dots are aligned with the Fermi level of the leads.
The two strong, bright lines emerging from this point come from
resonant tunneling between QD1 and the source lead (lower-right
line) or between QD2 and the drain lead (upper-left line). The
amplitude of the count rate at the lines gives directly the strength
of the tunnel couplings to source and drain leads
\cite{schleser:2004, naaman:2006}, and we find the rates to be
$\Gamma_\mathrm{S} = 1.1~\mathrm{kHz}$ and $\Gamma_\mathrm{D} =
1.2~\mathrm{kHz}$.

Along the white dashed line in Fig.~\ref{fig:SP_TriNoBias}(a), there
are triangle-shaped regions with low but non-zero count rates where
tunneling is expected to be strongly suppressed due to Coulomb
blockade. The DQD level arrangements inside the triangles are shown
in the insets. Comparing with the sketch in
Fig.~\ref{fig:SP_model}(a), we see that both regions have DQD
configurations favorable for noise detection. The dashed line
connecting the triangles is the detuning axis, with zero detuning
occuring at the triple point. We define the detuning as $\delta =
\mul-\mur$, so that the detuning is negative in the upper-left part
of the figure.

In Fig.~\ref{fig:SP_TriNoBias}(b), the lower-right part of
Fig.~\ref{fig:SP_TriNoBias}(a) was measured for four different QPC
bias voltages. The resonant line stays the same in all four
measurements, but the triangle becomes both larger and more
prominent as the QPC bias is increased. This is a strong indication
that the tunneling is due to absorption of energy from the QPC. The
counts observed above the resonance line for $V_\mathrm{QPC} =
-400~\mathrm{\mu V}$ are due to electrons being excited from the
ground state to the first excited state of the DQD.

\begin{figure}[tb]
\centering
 \includegraphics[width=\linewidth]{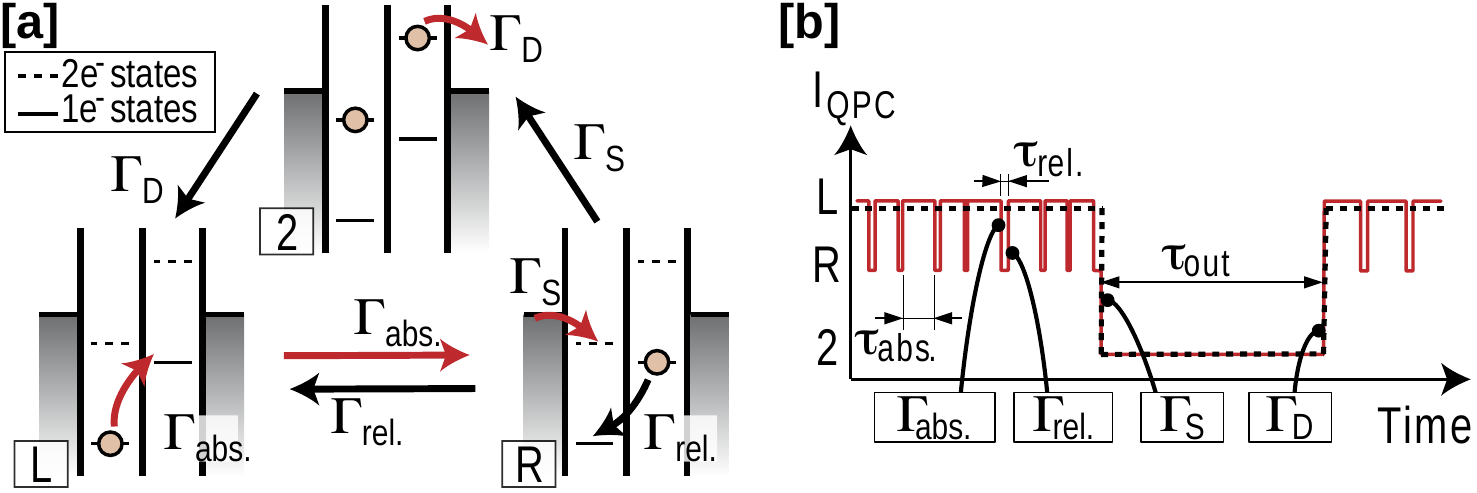}
 \caption{(a) Energy level diagrams for the three states of the DQD. The labels $L$, $R$ and $2$
 denote the excess charge population. The levels are raised by the intradot charging energy
 $E_{Ci}$ when the DQD holds two excess electrons.
 (b) Schematic changes of the detector signal as electrons tunnel into, between and out of the DQD.
  Adapted from Ref. \cite{gustavssonPRL:2007}.}
\label{fig:SP_modelTrace}
\end{figure}

The time-resolved measurement technique allows the rates for
electron tunneling into and out of the DQD to be determined
separately \cite{gustavsson:2005}. Figure \ref{fig:SP_TriNoBias}(c)
shows the rates $\Gamma_{\mathrm{in}}$ and $\Gamma_{\mathrm{out}}$
measured along the dashed line of Fig.~\ref{fig:SP_TriNoBias}(a).
The rate for tunneling out stays almost constant along the line, but
$\Gamma_{\mathrm{in}}$ is maximum close to the triple point and
falls of rapidly with increased detuning. This suggests that only
the rate for electrons tunneling into the DQD is related to the
absorption process. To explain the experimental findings we model
the system using a rate-equation approach. For a configuration
around the triple point, the DQD may hold $(n+1,m)$, $(n,m+1)$ or
$(n+1,m+1)$ electrons. We label the states $L$, $R$ and $2$ and draw
the energy diagrams together with possible transitions in
Fig.~\ref{fig:SP_modelTrace}(a). The figure shows the case for
negative detuning, with $\delta \gg k_B T$. Note that when the DQD
holds two excess electrons, the energy levels are raised by the
mutual charging energy, $E_\mathrm{Cm}=800~\mathrm{\mu eV}$.

In Fig \ref{fig:SP_modelTrace}(b) we sketch the time evolution of
the system. The red curve shows the expected charge detector signal
assuming a detector bandwidth much larger than the transitions
rates. Starting in state $L$, the electron is trapped until it
absorbs a photon and is excited to state $R$ (with rate
$\Gamma_{\mathrm{abs.}}$). From here, the electron may either relax
back to state $L$ (rate $\Gamma_{\mathrm{rel.}}$) or a new electron
may enter QD1 from the source lead and put the system into state $2$
(rate $\Gamma_{\mathrm{S}}$). Finally, if the DQD ends up in state
$2$, the only possible transition is for the electron in the right
dot to leave to the drain lead.

The relaxation rate for a similar DQD system has been measured to be
$1/\Gamma_{\mathrm{rel.}} = 16~\mathrm{ns}$ \cite{petta:2004}, which
is much faster than the available measurement bandwidth. Therefore,
the detector will not be able to register the transitions where the
electron is repeatedly excited and relaxed between the dots. Only
when a second electron enters from the source lead [transition
marked by $\Gamma_{\mathrm{S}}$ in Fig.~\ref{fig:SP_modelTrace}(a,
b)], the DQD will be trapped in state $2$ for a sufficiently long
time ($\sim\! 1/\Gamma_D \sim\! 1~\mathrm{ms}$) to allow detection.
The measured time trace will only show two levels, as indicated by
the dashed line in Fig.~\ref{fig:SP_modelTrace}(b). Such a trace
still allows extraction of the effective rates for electrons
entering and leaving the DQD, $\Gamma_{\mathrm{in}} = 1/\langle
\tau_{\mathrm{in}} \rangle$ and $\Gamma_{\mathrm{out}} = 1/\langle
\tau_{\mathrm{out}} \rangle$. To relate $\Gamma_{\mathrm{in}}$,
$\Gamma_{\mathrm{out}}$ to the internal DQD transitions, we write
down the master equation for the occupation probabilities of the
states:
\begin{equation}\label{eq:SP_master} \frac{d}{dt }\left[ \!\!
\begin{array}{c}
             p_L \\
             p_R \\
              p_2 \\
              \end{array} \!\! \right] =
                        \left[\!\!
          \begin{array}{ccc}
            -\Gamma_{\mathrm{abs.}} & \Gamma_{\mathrm{rel.}} & \Gamma_{\mathrm{D}} \\
            \Gamma_{\mathrm{abs.}} & -(\Gamma_{\mathrm{S}}+\Gamma_{\mathrm{rel.}}) & 0 \\
            0 & \Gamma_{\mathrm{S}} & -\Gamma_{\mathrm{D}} \\
          \end{array}
        \!\!\right] \!\!\! \left[\!\!
                        \begin{array}{c}
                          p_L \\
                          p_R \\
                          p_2 \\
                        \end{array}
                      \!\!\right].
\end{equation}
Again, we assume negative detuning, with $|\delta| \gg k_B T$. The
measured rates $\Gamma_{\mathrm{in}}$, $\Gamma_{\mathrm{out}}$ are
calculated from the steady-state solution of Eq.~\ref{eq:SP_master}:
\begin{eqnarray}
  \Gamma_{\mathrm{in}} &=& \Gamma_{\mathrm{S}} \, \frac{p_R}{p_L+p_R} =
  \frac{\Gamma_{\mathrm{S}} \Gamma_{\mathrm{abs.}}}
  {\Gamma_{\mathrm{S}} + \Gamma_{\mathrm{abs.}} + \Gamma_{\mathrm{rel.}}}, \\
  \Gamma_{\mathrm{out}} &=& \Gamma_{\mathrm{D}}. \label{eq:SP_GoutEqGsGd}
\end{eqnarray}
In the limit $\Gamma_{\mathrm{rel.}} \gg \Gamma_{\mathrm{S}},\,
\Gamma_{\mathrm{abs.}}$, the first expression simplifies to
\begin{equation}\label{eq:SP_GinGabs}
 \Gamma_{\mathrm{in}}=\Gamma_{\mathrm{S}} \,
 \Gamma_{\mathrm{abs.}}/\Gamma_{\mathrm{rel.}}.
\end{equation}
The corresponding expressions for positive detuning are found by
interchanging $\Gamma_{\mathrm{S}}$ and $\Gamma_{\mathrm{D}}$ in
Eqs.~(\ref{eq:SP_master}-\ref{eq:SP_GinGabs}). Coming back to the
experimental findings of Fig.~\ref{fig:SP_TriNoBias}(c), we note
that $\Gamma_\mathrm{out}$ only shows small variations within the
region of interest. This together with the result of
Eq.~(\ref{eq:SP_GoutEqGsGd}) suggest that we can take
$\Gamma_\mathrm{S}$, $\Gamma_\mathrm{D}$ to be independent of
detuning.
The rate $\Gamma_\mathrm{in}$ in Eq.~(\ref{eq:SP_GinGabs}) thus
reflects the dependence of
$\Gamma_\mathrm{abs.}/\Gamma_\mathrm{rel.}$ on detuning. Assuming
also $\Gamma_\mathrm{rel.}$ to be constant, a measurement of
$\Gamma_\mathrm{in}$ gives directly the absorption spectrum of the
DQD. The measurements cannot exclude that $\Gamma_\mathrm{rel.}$
also varies with $\delta$, but as we show below the model assuming
$\Gamma_\mathrm{rel.}$ independent of detuning fits the data well.

Equation~(\ref{eq:SP_GinGabs}) shows that the low-bandwidth detector
can be used to measure the absorption spectrum, even in the presence
of fast relaxation. Moreover, the detection of an electron entering
the DQD implies that a quantum of energy was absorbed immediately
before the electron was detected. The charge detector signal thus
relates directly to the detection of a single photon.
The efficiency of the detector is currently limited by the bandwidth
of the charge detector. However, it should be possible to increase
the bandwidth significantly by operating the QPC in a mode analogous
to the radio-frequency single-electron transistor
\cite{schoelkopf:1998, muller:2007, reilly:2007, cassidy:2007}.

\begin{figure}[tb]
\centering
 \includegraphics[width=\linewidth]{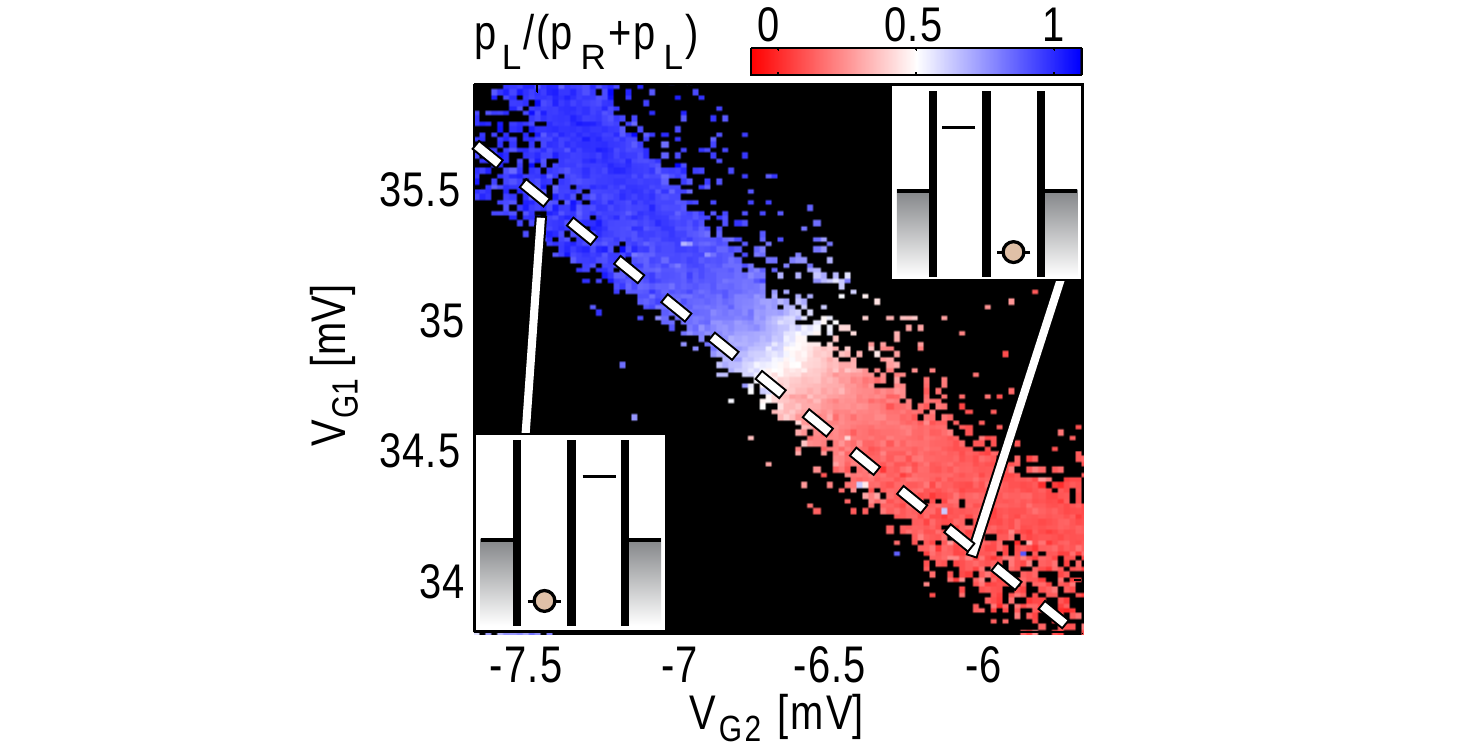}
 \caption{Average population of state $L$, measured for the
same region as shown in \FigRef{fig:SP_TriNoBias}(a). The insets
show the DQD level configurations for positive and negative
detuning, the dashed line defines the detuning axis.
 }
\label{fig:SP_stepHeight}
\end{figure}

To justify the assumption $\Gamma_{\mathrm{rel.}} \gg
\Gamma_{\mathrm{abs.}}$, we note that even when the detector is too
slow to detect individual transitions between the states $L$ and
$R$, its dc-response still gives the average population of the two
states. In Fig.~\ref{fig:SP_stepHeight}, we plot the relative
population of state $L$, $p_L/(p_L+p_R)$, for the same gate voltage
configuration as in \FigRef{fig:SP_TriNoBias}(a). The data was
extracted by analyzing the absolute change in the QPC conductance
for one electron tunneling into DQD (see
section~\ref{sec:DQ_tunnelCoupling}).
Looking at the region of negative detuning (upper-left part of
\FigRef{fig:SP_stepHeight}), the average DQD population within the
regions of photon-assisted tunneling is very close to the pure
$L$-state. The electron spends most of the time in QD1, which
validates the assumption $\Gamma_{\mathrm{rel.}} \gg
\Gamma_{\mathrm{abs.}}$. Similar arguments can be applied for the
region of positive detuning.

For fixed DQD detuning, the processes described above only pump
electrons in one direction. The system may therefore thought of as a
ratchet, giving unidirectional electron flow even at zero bias
\cite{khrapai:2006}.

\subsection{Measuring the QPC emission spectrum}
\begin{figure}[tb]
\centering
 \includegraphics[width=\linewidth]{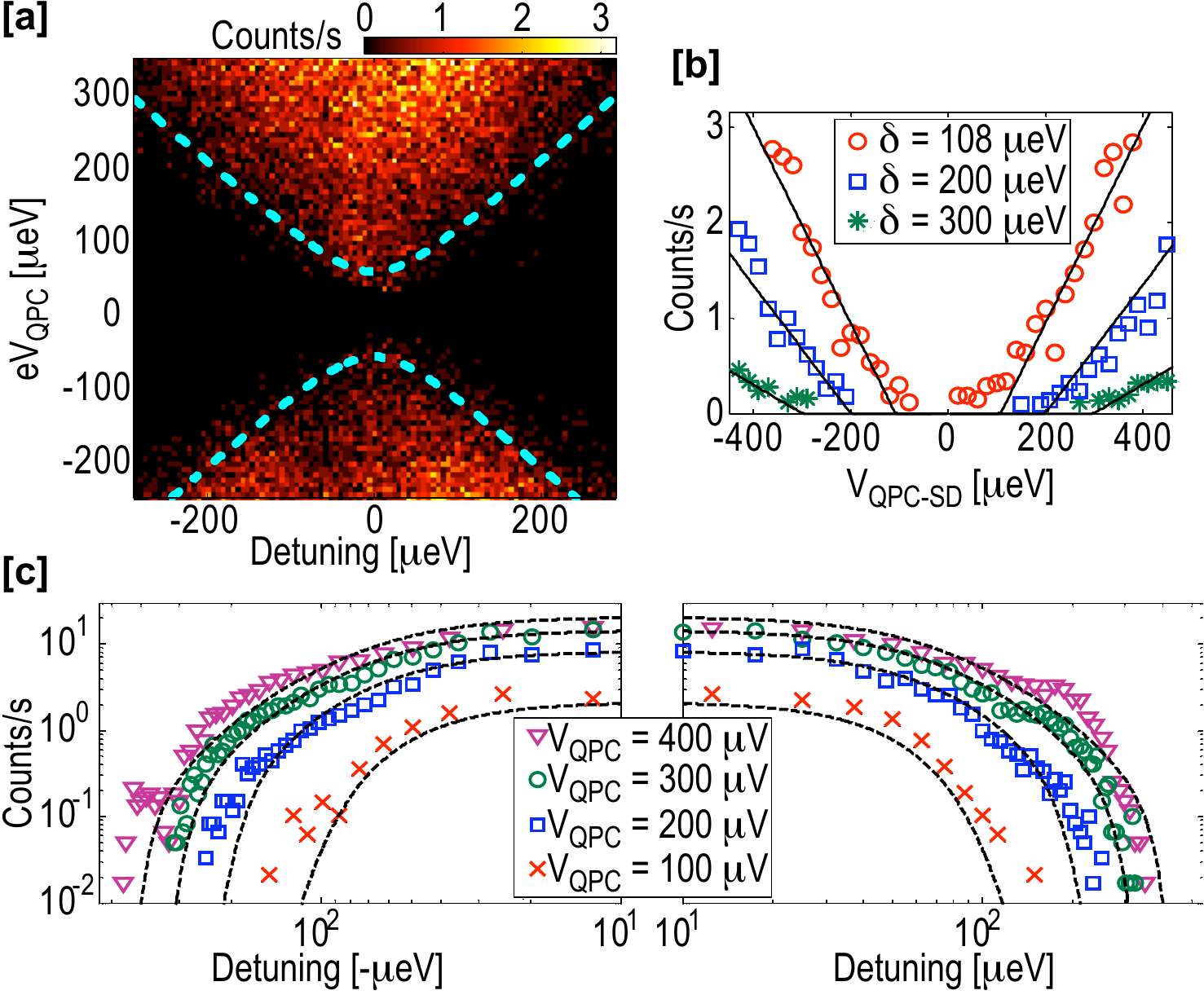}
 \caption{Count rate measured versus detuning and QPC bias voltage.
 The dashed line shows the level separation for a two-level
 system, with $\Delta_{12}= \sqrt{4\,t^2 + \delta^2}$.
 There are only counts in the region where $|e V_\mathrm{QPC}| > \Delta_{12}$.
 (b) Count rate versus QPC bias for different values of
 detuning. The solid lines are guides to the eye.
 (c) DQD absorption spectrum, measured for different QPC bias. The
  dashed lines are the results of Eq.~(\ref{eq:SP_absRate}), with
 parameters given in the text.  Adapted from Ref. \cite{gustavssonPRL:2007}.
 }
\label{fig:SP_CntsVsDet}
\end{figure}

In the following, we use the DQD to quantitatively investigate the
microwave radiation emitted from the nearby QPC. Figure
\ref{fig:SP_CntsVsDet}(a) shows the measured count rate for
electrons leaving the DQD versus detuning and QPC bias. The data was
taken along the dashed line of Fig.~\ref{fig:SP_TriNoBias}(a), with
gate voltages converted into energy using lever arms extracted from
finite bias measurements. Due to the tunnel coupling $t$ between the
QDs, the energy level separation $\Delta_{12}$ of the DQD is given
by $\Delta_{12} =\sqrt{4\,t^2 + \delta^2}$. The dashed lines in
\ref{fig:SP_CntsVsDet}(a) show $\Delta_{12}$, with $t =
32~\mathrm{\mu eV}$. A striking feature is that there are no counts
in regions with $|eV_\mathrm{QPC}| < \Delta_{12}$. This originates
from the fact that the voltage-biased QPC can only emit photons with
energy $\hbar \omega \le eV_\mathrm{QPC}$ \cite{aguado:2000,
onacQD:2006, zakka:2007}. The result presents another strong
evidence that the absorbed photons originate from the QPC.


To describe the results quantitatively, we consider the emission
spectrum of a voltage biased QPC with one conducting channel. In the
low-temperature limit $k_B T \ll \hbar \omega  $, the spectral noise
density $S_I(\omega)$ for the emission side ($\omega>0$) takes the
form (see \cite{aguado:2000} for the full expression)
\begin{equation}\label{eq:SP_SI}
 S_I(\omega) =  \frac{4 e^2}{h} D (1-D) \frac{e V_\mathrm{QPC} - \hbar \omega}{1-e^{-(e V_\mathrm{QPC} -  \hbar \omega)/k_B
 T}},
\end{equation}
where $D$ is the transmission coefficient of the channel.
Using the model of Ref. \cite{aguado:2000}, we find the absorption
rate of the DQD in the presence of the QPC:
\begin{equation}\label{eq:SP_absRate}
 \Gamma_\mathrm{abs.} = \frac{4 \pi e^2 k^2 t^2 Z_l^2}{h^2}
 \frac{S_I(\Delta_{12}/\hbar)}{\Delta_{12}^2}.
\end{equation}
The constant $k$ is the capacitive lever arm of the QPC on the DQD
and $Z_l$ is the zero-frequency impedance of the leads connecting
the QPC to the voltage source.
Equation (\ref{eq:SP_absRate}) states how well fluctuations in the
QPC couple to the DQD system.

Figure \ref{fig:SP_CntsVsDet}(b) shows the measured absorption rates
versus $V_\mathrm{QPC}$, taken for three different values of
$\delta$. As expected from Eqs.~(\ref{eq:SP_SI},
\ref{eq:SP_absRate}), the absorption rates increase linearly with
bias voltage as soon as $|eV_\mathrm{QPC}| > \delta$. The different
slopes for the three data sets are due to the
$1/\Delta_{12}^2$-dependence in the relation between the emission
spectrum and the absorption rate of Eq.~(\ref{eq:SP_absRate}). In
Fig.~\ref{fig:SP_CntsVsDet}(c), we present measurements of the
absorption spectrum for fixed $V_\mathrm{QPC}$. The rates decrease
with increased detuning, with sharp cut-offs as $|\delta| > e
V_\mathrm{QPC}$. In the region of small detuning, the absorption
rates saturate as the DQD level separation $\Delta_{12}$ approaches
the limit set by the tunnel coupling. The dashed lines show the
combined results of Eqs.~(\ref{eq:SP_GinGabs}-\ref{eq:SP_absRate}),
with parameters $T=0.1~\mathrm{K}$, $Z_l = 0.7~\mathrm{k\Omega}$,
$D=0.5$, $t=32~\mathrm{\mu eV}$, $k = 0.15$, $\Gamma_{\mathrm{S}} =
1.1~\mathrm{kHz}$ and $\Gamma_{\mathrm{D}} = 1.2~\mathrm{kHz}$.
Using $\Gamma_{\mathrm{rel.}}$ as a fitting parameter, we find
$1/\Gamma_{\mathrm{rel.}} = 5~\mathrm{ns}$. This should be seen as a
rough estimate of $\Gamma_\mathrm{rel.}$ due to uncertainties in
$Z_l$, but it shows reasonable agreement with previously reported
measurements \cite{petta:2004}. The overall good agreement between
the data and the electrostatic model of Eq.~(\ref{eq:SP_absRate})
supports the assumption that the interchange of energy between the
QPC and the DQD is predominantly mediated by photons instead of
phonons or plasmons.

\begin{figure}[tb]
\centering
 \includegraphics[width=\linewidth]{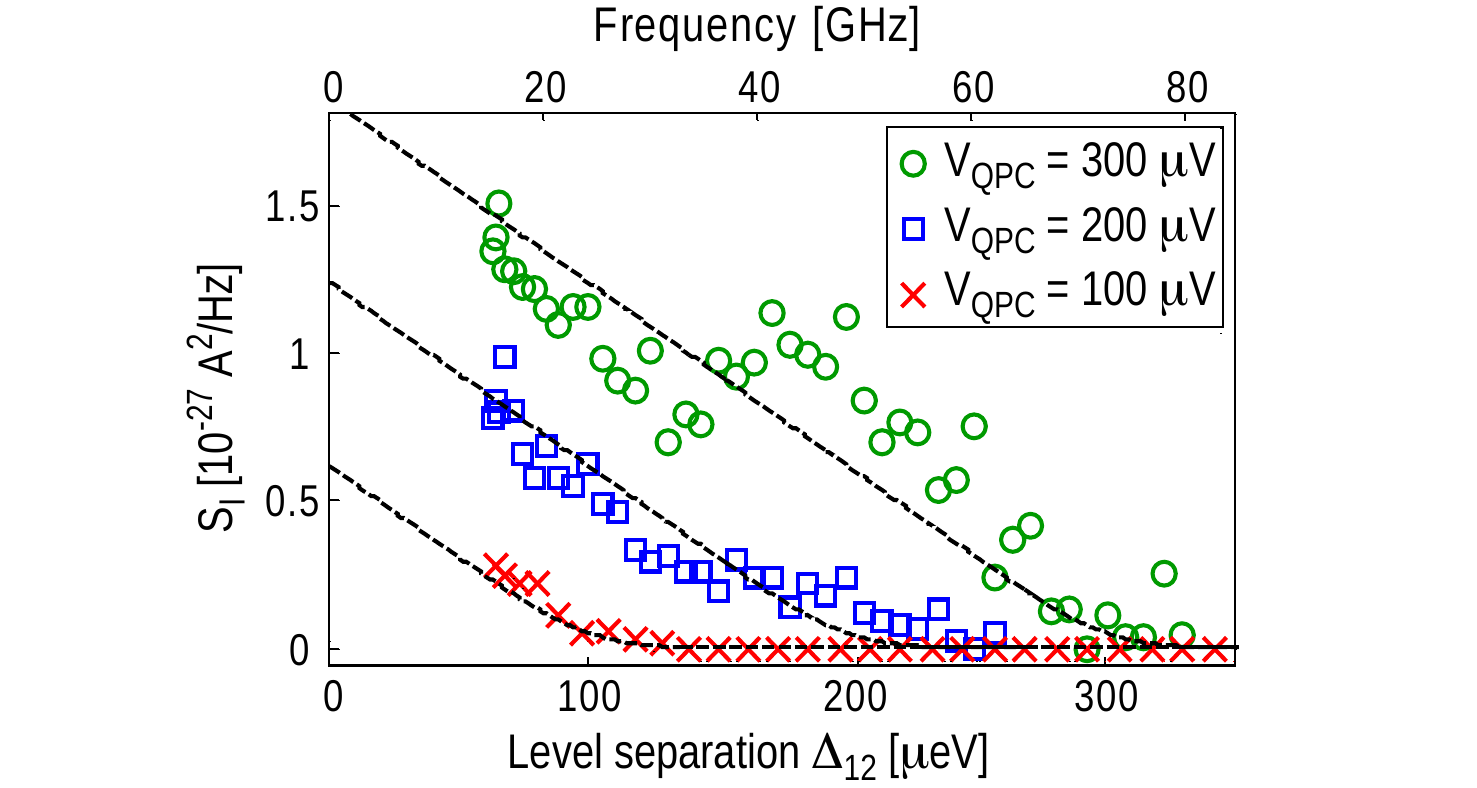}
 \caption{Noise spectrum of the QPC, extracted from the data in \FigRef{fig:SP_CntsVsDet}(c). The
 dashed lines show spectra expected from Eq.~(\ref{eq:SP_SI}).  Adapted from Ref. \cite{gustavssonPRL:2007}.}
\label{fig:SP_spectrum}
\end{figure}

The data for $V_\mathrm{QPC}=400~\uV$ shows some irregularities
compared to theory, especially at large positive detuning. We
speculate that the deviations are due to excited states of the
individual QDs, with excitation energies smaller than the detuning. In Fig.~\ref{fig:SP_spectrum}, we convert
the detuning $\delta$ to level separation $\Delta_{12}$ and use
Eq.~(\ref{eq:SP_absRate}) to extract the noise spectrum $S_I$ of the
QPC. The linear dependence of the noise with respect to frequency
corresponds well to the behavior expected from Eq.~(\ref{eq:SP_SI}).
Again, the deviations at $\Delta_{12}=190~\mathrm{\mu eV}$ are
probably due to an excited state in one of the QDs. The excited
states are also visible in finite-bias spectroscopy, giving a
single-level spacing of $\Delta E \approx 200~\mathrm{\mu eV}$. This
sets an upper bound on frequencies that can be detected with the
detector. The frequency-range can be extended by using DQD in carbon
nanotubes \cite{mason:2004} or InAs nanowires \cite{fasth:2005,
pfund:2006}, where the single-level spacing is significantly larger \cite{gustavssonNWPRB:2008}.

\subsection{Finite DQD bias regime}
\begin{figure}[tb]
\centering
 \includegraphics[width=\linewidth]{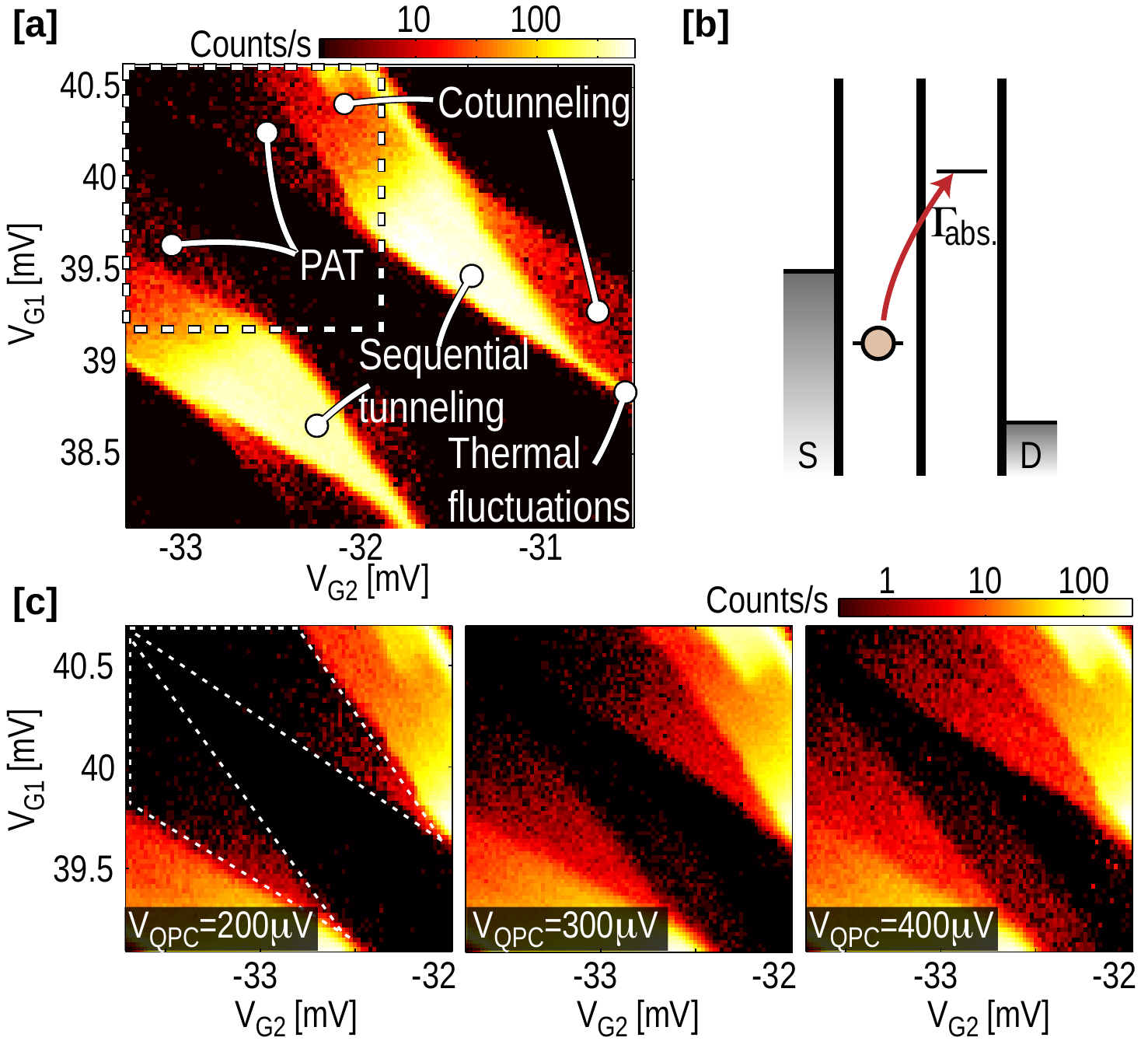}
 \caption{(a) Charge stability diagram for the DQD, measured with a bias voltage
 $V_\mathrm{DQD-SD}=300\uV$ applied over the DQD. Tunneling due
 thermal fluctuations, sequential transport, cotunneling and photon-assisted
 processes (PAT) are visible.
 The data was taken with $V_\mathrm{QPC-SD}=300\uV$. (b) DQD energy
 level diagram for the upper region of photon-assisted tunneling in (a). The detuning is
 opposite to the bias direction.
 (c) Magnifications of the region marked by the dashed rectangle in
 (a), measured for three different QPC bias voltages. The dashed lines
 in the leftmost figure show the regions where we expect photon-assisted tunneling.
 As the QPC bias voltage is increased, the count rate goes up inside the
 PAT regions.
 }
\label{fig:SP_trianglesBias}
\end{figure}

Finally, we apply a voltage bias over the DQD in order to compare
the tunneling originating from sequential transport with the
tunneling due to photon absorption processes.
Figure~\ref{fig:SP_trianglesBias}(a) shows a charge stability
diagram measured with DQD bias $V_\mathrm{DQD-SD}=300\uV$. The two
triangles associated with electron and hole transport cycles are
clearly visible. Besides that, we have regions of cotunneling (see
section~\ref{sec:DQ_cotunneling}) as well as sharp lines with
tunneling due to equilibrium fluctuations whenever the
electrochemical potential of QD1 or QD2 lines up with the Fermi
levels in the source or drain, respectively. In addition, there are
faint triangles appearing in the detuning direction \emph{opposite}
to the transport triangles; we attribute these features to
photon-assisted tunneling (PAT).

The DQD energy level configuration in the upper region with faint
tunneling (next to the hole transport triangle) is depicted in
\FigRef{fig:SP_trianglesBias}(a). In this regime the DQD may hold
one or two excess electrons. For this energy level alignment neither
sequential tunneling nor cotunneling is possible. The DQD can only
change its state if an electron in QD1 absorbs a photon and is
excited to QD2. From this configuration, an electron may enter QD1
from the source lead followed by the electron in QD2 leaving to the
drain. In \FigRef{fig:SP_trianglesBias}(c), we present blow-ups of
the region marked by the dashed rectangle in
\FigRef{fig:SP_trianglesBias}(a), measured for different QPC bias
voltages. The dashed lines in the leftmost panel in
\FigRef{fig:SP_trianglesBias}(c) show the regions where we expect
photon-assisted tunneling. As the QPC bias is increased, we see that
the count rate inside these regions indeed goes up significantly.
For the highest QPC bias voltage, there are extra features appearing
outside the anticipated PAT-region. Again, we attribute this to an
excited state in QD2.

\section{Single-electron interference} \label{sec:AB_main}
A central concept of quantum mechanics is the wave-particle
duality; matter exhibits both wave- and particle-like properties and
can not be described by either formalism alone. Up to this point, we
have treated the electrons as particles tunneling back and forth
between quantum dots. In this chapter, we investigate their wave
properties by studying interference of individual electrons taking
two different paths in an Aharonov-Bohm interferometer. The
time-resolved charge detection technique enables us to count
electrons one-by-one as they pass the interferometer. In this way we
make a direct measurement of the self-interference of a single
electron. With increased bias voltage across the quantum point
contact a back-action is exerted on the interferometer leading to
dephasing. We attribute this to emission of radiation from the
quantum point contact, which drives non-coherent electronic
transitions in the quantum dots.

\subsection{The Aharonov-Bohm effect}
\begin{figure}[tb] \centering
 \includegraphics[width=\linewidth]{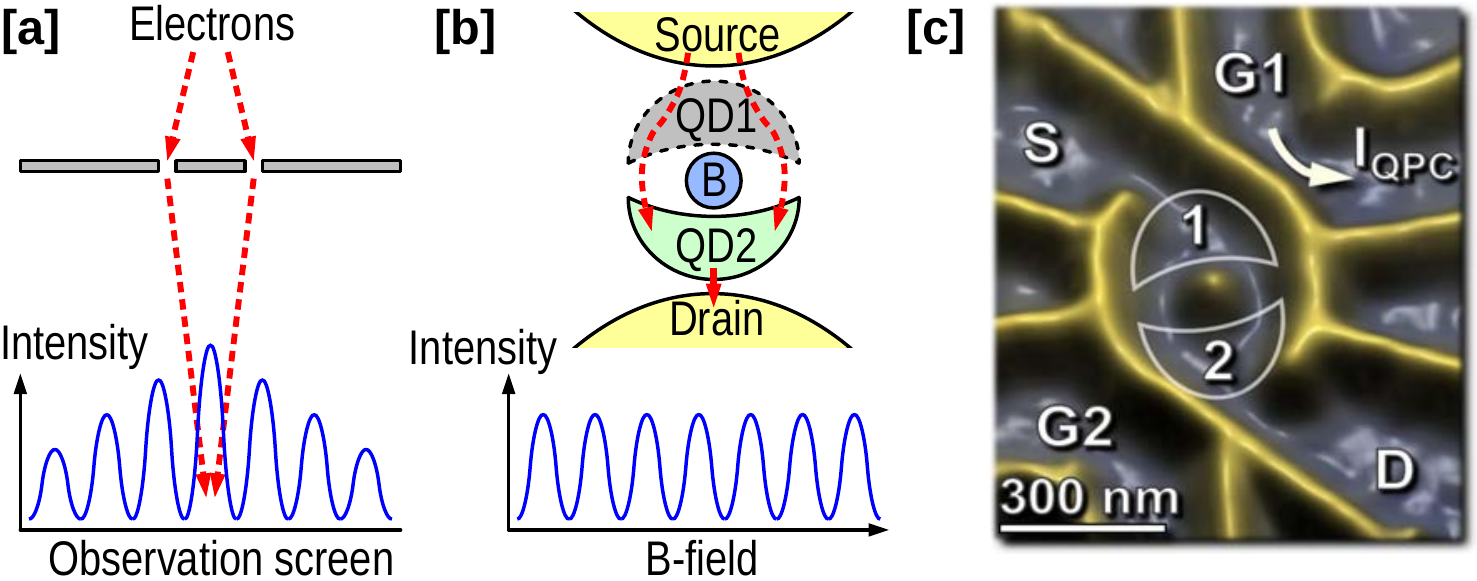}
 \caption{(a) Setup of a traditional double-slit experiment. Electrons passing
 through the two slits give rise to an interference pattern on the observation screen.
 (b) Schematic drawing of the setup used for measuring single-electron Aharonov-Bohm interference.
 Electrons are injected from the source lead, tunnel through QD1 and end up in
 QD2, where they are detected.
 The interference pattern is due to the applied
 B-field, which introduce a phase difference between the left
 and right arm connecting the two quantum dots.
 (c) Double quantum dot used in the experiment. The yellow parts are
 lines written with a scanning force microscope on top of a semiconductor
 heterostructure and represent the potential landscape for the
 electrons. The
 QDs (marked by 1 and 2) are connected by two separate arms,
 allowing partial waves taking different paths to interfere.
 The current in the nearby QPC ($I_\mathrm{QPC}$) is used to monitor the
 electron population in the system. Adapted from Ref. \cite{gustavssonNL:2008}.
 }
\label{fig:AB_setup}
\end{figure}

One of the cornerstone concepts of quantum mechanics is the
superposition principle as demonstrated in the double-slit
experiment \cite{young:1804}.
The partial waves of individual particles passing a double slit
interfere with each other. The ensemble average of many particles
detected on a screen agrees with the interference pattern calculated
using propagating waves [Fig.~\ref{fig:AB_setup}(a)]. This has been
demonstrated for photons, electrons in vacuum \cite{jonsson:1961,
tonomura:1989} as well as for more massive objects like
$C_{60}$-molecules \cite{arndt:1999}.
The Aharonov-Bohm (AB) geometry provides an analogous experiment in
solid-state systems  \cite{aharonov:1959}. Partial waves passing the
arms of a ring acquire a phase difference due to a magnetic flux,
enclosed by the two paths [Fig.~\ref{fig:AB_setup}(b)].
Here, we set out to perform the interference experiment by using a quantum point contact to detect single-electron tunneling in real-time.

We first discuss the experimental conditions necessary for observing
single-electron AB interference. We make use of a geometry
containing two quantum dots within the AB-ring.
Figure~\ref{fig:AB_setup}(c) shows the structure, with the two QDs
(marked by 1 and 2) tunnel-coupled by two separate barriers. It is
the same structure as investigated in chapters~\ref{sec:DQ_main} and
\ref{sec:SP_main}, but this time tuned to a regime where both
barriers connecting the QDs are kept open. Following the sketch in
Fig.~\ref{fig:AB_setup}(b), electrons are provided from the source
lead, tunnel into QD1 and pass on to QD2 through either one of the
two arms. Upon arriving in QD2, the electrons are detected in
real-time by monitoring the conductance of the nearby QPC
\cite{field:1993, vandersypen:2004, schleser:2004, fujisawa:2004}.
Coulomb blockade prohibits more than one excess electron to populate
the structure, implying that the first electron must leave to the
drain before a new one can enter. This enables time-resolved
operation of the charge detector and ensures that we measure
interference due to individual electrons.

To avoid dephasing, the electrons should spend a time as short as
possible on their way from source to QD2. This is achieved by
raising the electrochemical potential of QD1 so that electrons in
the source lead lack an energy $\delta$ required for entering QD1
[see Fig.~\ref{fig:AB_count}(b)]. The time-energy uncertainty
principle still allows electrons to tunnel from source to QD2 by
means of second order processes. The tunneling process is then
limited to a short time scale set by the uncertainty relation, with
$t = \hbar/\delta$.

\subsection{Experimental realization}
\begin{figure}[tb]
 \includegraphics[width=\linewidth]{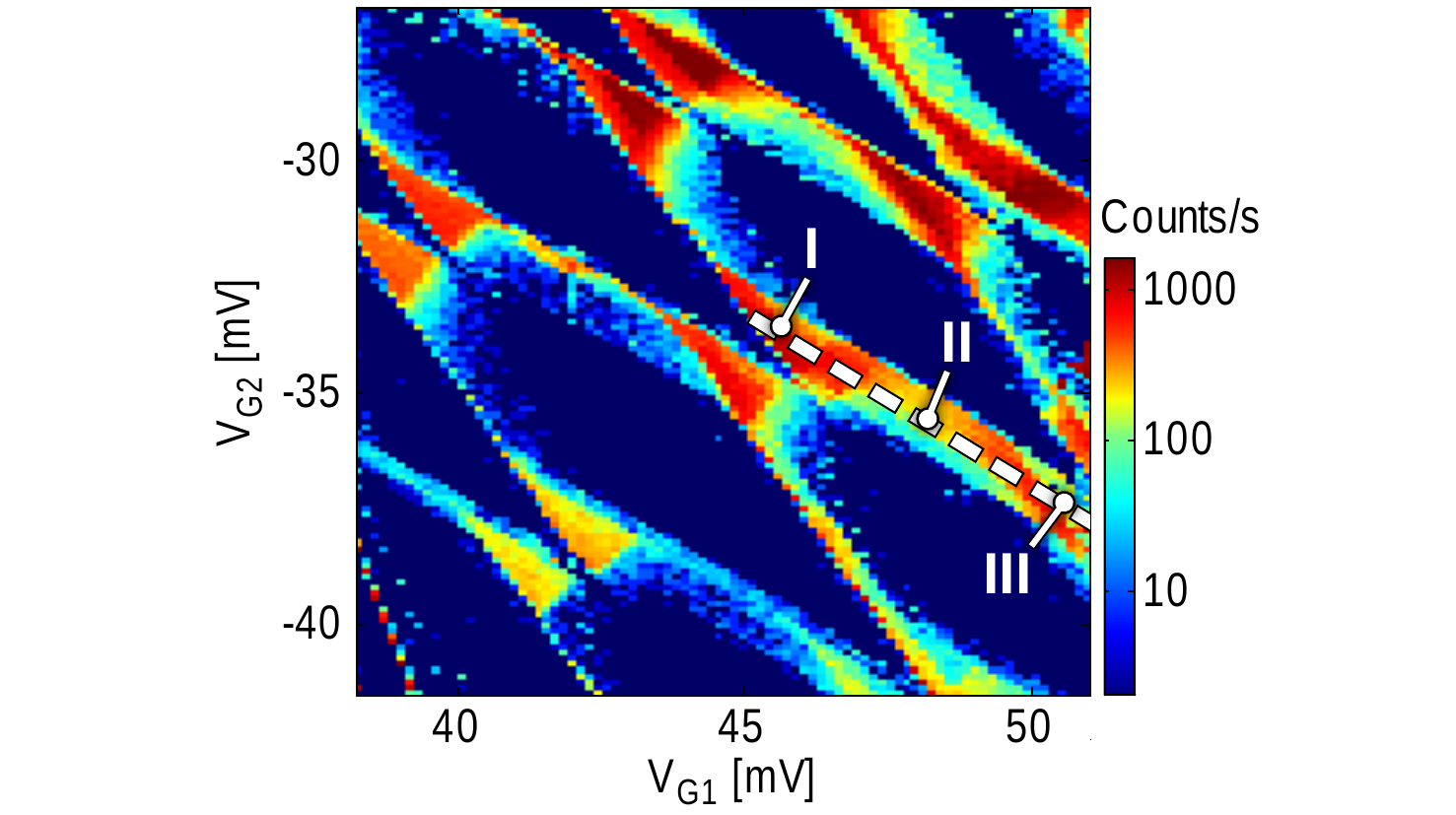}
 \caption{Charge stability diagram of the double quantum dot,
 recorded by counting electrons entering and leaving the structure.
 The data was taken at DQD bias voltage $V_\mathrm{DQD-SD}=600~\mu\mathrm{V}$ and  $B=0~\mathrm{T}$.
 The dashed line marks the region of cotunneling used for measuring
 single-electron Aharonov-Bohm interference.
 }
\label{fig:AB_hexagon}
\end{figure}

In the experiment, we apply appropriate gate voltages to tune the
tunneling rates between the double quantum dot (DQD) and the source
and drain leads to values below $15~\mathrm{kHz}$.
The tunneling coupling between the QDs is set to a few GHz, as
determined from charge localization measurements (see section
\ref{sec:DQ_tunnelCoupling}).
Figure~\ref{fig:AB_hexagon} shows the charge stability diagram of the
DQD systems, measured by counting electrons entering and leaving the
DQD within a fixed period of time. The data was taken with $600~\mu
V$ bias applied between source and drain. The hexagon pattern
together with the triangles of electron transport appearing due to
the applied bias are well-known characteristics of DQD systems
(see chapter~\ref{sec:DQ_main}).
Between the triangles, there are broad, band-shaped regions with low
but non-zero count rates where sequential transport is suppressed
due to Coulomb blockade. The finite count rate in this region is
attributed to electron tunneling involving virtual processes, as
described in section~\ref{sec:DQ_cotunneling}. In the following
paragraph we quickly repeat the main results from that section.

\begin{figure*}[tb] \centering
 \includegraphics[width=0.8\linewidth]{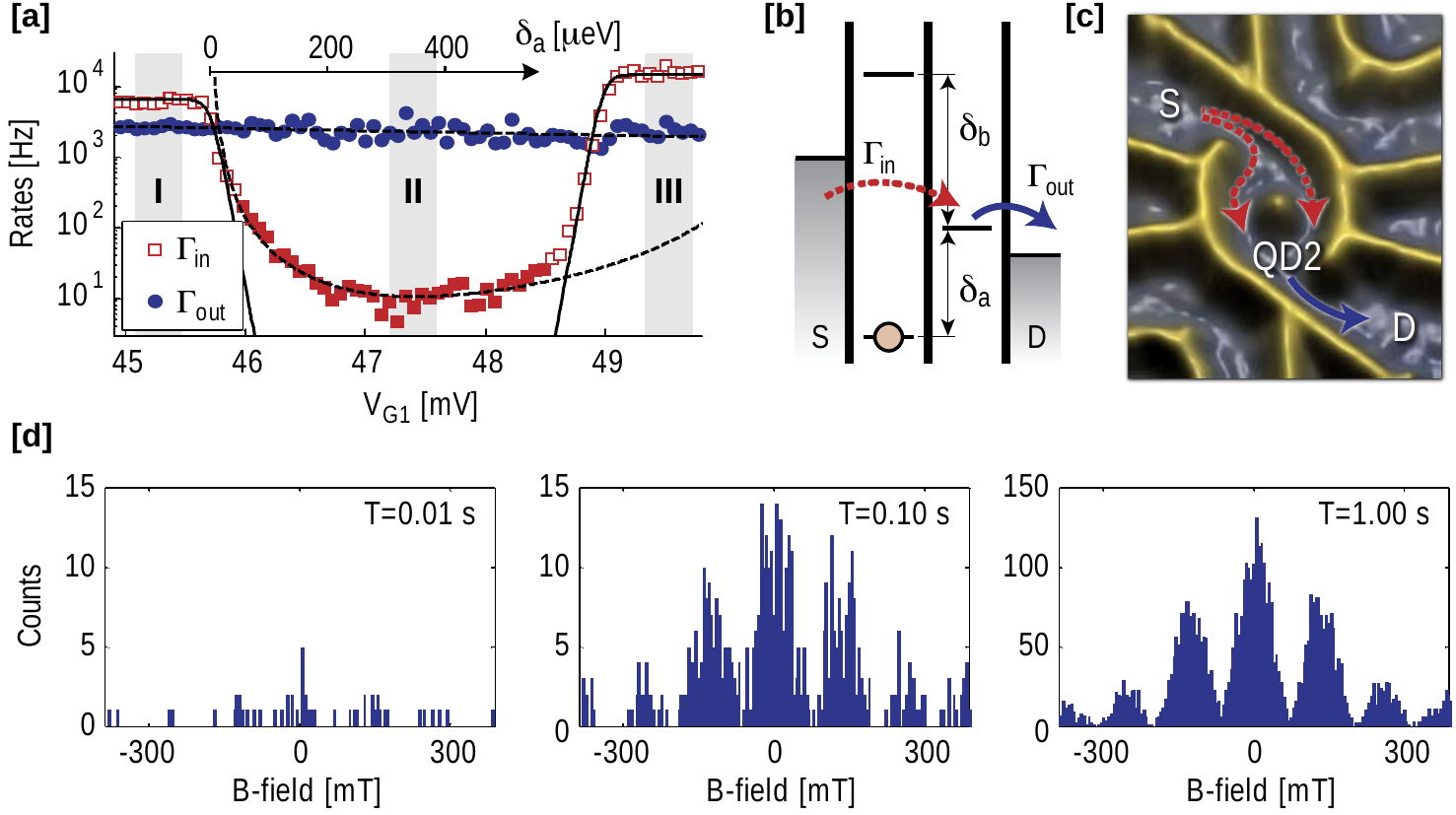}
 \caption{
 (a) Tunneling rates for electrons entering (red) and leaving (blue) the DQD,
 measured along the dashed white line of cotunneling in \FigRef{fig:AB_hexagon}(a).
 The upper x-axis shows $\delta$, the potential
 difference between the state in QD2 and the occupied state of QD1. The solid
 lines are tunneling rates expected from sequential tunneling,
 while the dashed line is a fit to the cotunneling model of Eq.~(\ref{eq:cotunneling}) with parameters $\Gamma_\mathrm{Sa} = 6.4~\mathrm{kHz}$, $\Gamma_\mathrm{Sb} = 14~\mathrm{kHz}$, $t_a=8.3~\mathrm{\mu eV}$ and $t_b=13~\mathrm{\mu
eV}$. The data was taken with $B=340~\mathrm{mT}$.
 (b) Schematic drawings of the energy levels of the DQD at position II in (a). The energy levels of QD1 are shifted so that the electron in QD1 is
 trapped due to Coulomb blockade. Electron transport from source to
 QD2 is still possible through virtual processes.
 (c) The tunneling processes depicted in the double quantum dot structure. When both
 barriers between the QDs are kept open, the cotunneling electron may take any arm when
 going from source to QD2.
 (d) Number of electrons arriving at QD2 within the fixed period of
 time indicated in the upper-right corner, measured as a function of magnetic field. The data was taken at point II in
 (a). The count rate shows an oscillatory pattern with a visibility higher than $90\%$.   Adapted from Ref. \cite{gustavssonNL:2008}.
 }
\label{fig:AB_count}
\end{figure*}

Figure~\ref{fig:AB_count}(a) shows the rates for electrons tunneling
into and out of the DQD measured along the dashed line in
Fig.~\ref{fig:AB_hexagon}. Going along the dashed line corresponds
to lowering the electrochemical potential of QD1 while keeping the
potential of QD2 constant. In the region marked by I, electrons
tunnel sequentially from the source into QD1, continue from QD1 to
QD2 and finally leave QD2 to the drain lead. Proceeding to point II
in \FigRef{fig:AB_count}(a), the electrochemical potential of QD1 is
lowered and an electron eventually gets trapped in QD1 [see sketch
in Fig.~\ref{fig:AB_count}(b)]. At position II, the electron lacks
an energy $\delta_a$ to leave to QD2. Due to the energy-time
uncertainty principle, there is a time-window of length
$\sim\!\hbar/\delta_a$ within which tunneling from QD1 to QD2
followed by tunneling from the source into QD1 is possible without
violating energy conservation. An analogous process is possible
involving the next unoccupied state of QD1, occuring on timescales
$\sim\! \hbar/\delta_b$. This corresponds to electron
\emph{cotunneling} from the source lead to QD2. By continuing to
point III, the unoccupied state of QD1 is shifted into the bias
window and electron transport is again sequential.
The rate for electrons tunneling out of the DQD
[$\Gamma_\mathrm{out}$, blue trace in Fig.~\ref{fig:AB_count}(a)] shows
only slight variations over the region of interest. This is
expected, since the potential of QD2 stays constant along the dashed
line in Fig.~\ref{fig:AB_hexagon}.

Coming back to the sketch of Fig.~\ref{fig:AB_setup}(b), we note
that the cotunneling configuration of case II in
Fig.~\ref{fig:AB_count}(a,b) is ideal for investigating the
Aharonov-Bohm effect for single electrons. Due to the low
probability of the cotunneling process, the source lead provides
low-frequency injection of single electrons into the DQD. The
injected electrons cotunnel through QD1 into QD2 on a timescale $t
\sim \hbar/\delta \sim 1~\mathrm{ps}$ much shorter than the
decoherence time of the system, which is on the order of a few
nanoseconds \cite{folk:2001, eisenberg:2002}. This ensures that
phase coherence is preserved. Finally, the electron stays in QD2 for
a time long enough to be registered by the finite-bandwidth charge
detector. The tunneling processes are sketched in
\FigRef{fig:AB_count}(c).

Next, we tune the system to case II of Fig.~\ref{fig:AB_count}(a)
and count electrons as a function of magnetic field.
Figure~\ref{fig:AB_count}(d) shows snapshots taken at three
different times. The electrons arriving in QD2 build up a
well-pronounced interference pattern with period $130~\mathrm{mT}$.
This corresponds well to one flux quantum $\Phi = h/e$ penetrating
the area enclosed by the two paths.
The visibility of the AB-oscillations is higher than $90\%$, which
is a remarkably large number demonstrating the high degree of phase
coherence in the system. We attribute the high visibility to the
short time available for the cotunneling process \cite{sigrist:2006}
and to strong suppression of electrons being backscattered in the
reverse direction, which is otherwise present in AB-experiments.
Another requirement for the high visibility is that the two tunnel
barriers connecting the QDs are carefully symmetrized. The overall
decay of the maxima of the AB-oscillation with increasing B is
probably due to magnetic field effects on the orbital wavefunctions
in QD1 and QD2.

\begin{figure}[h!]
\centering
 \includegraphics[width=\columnwidth]{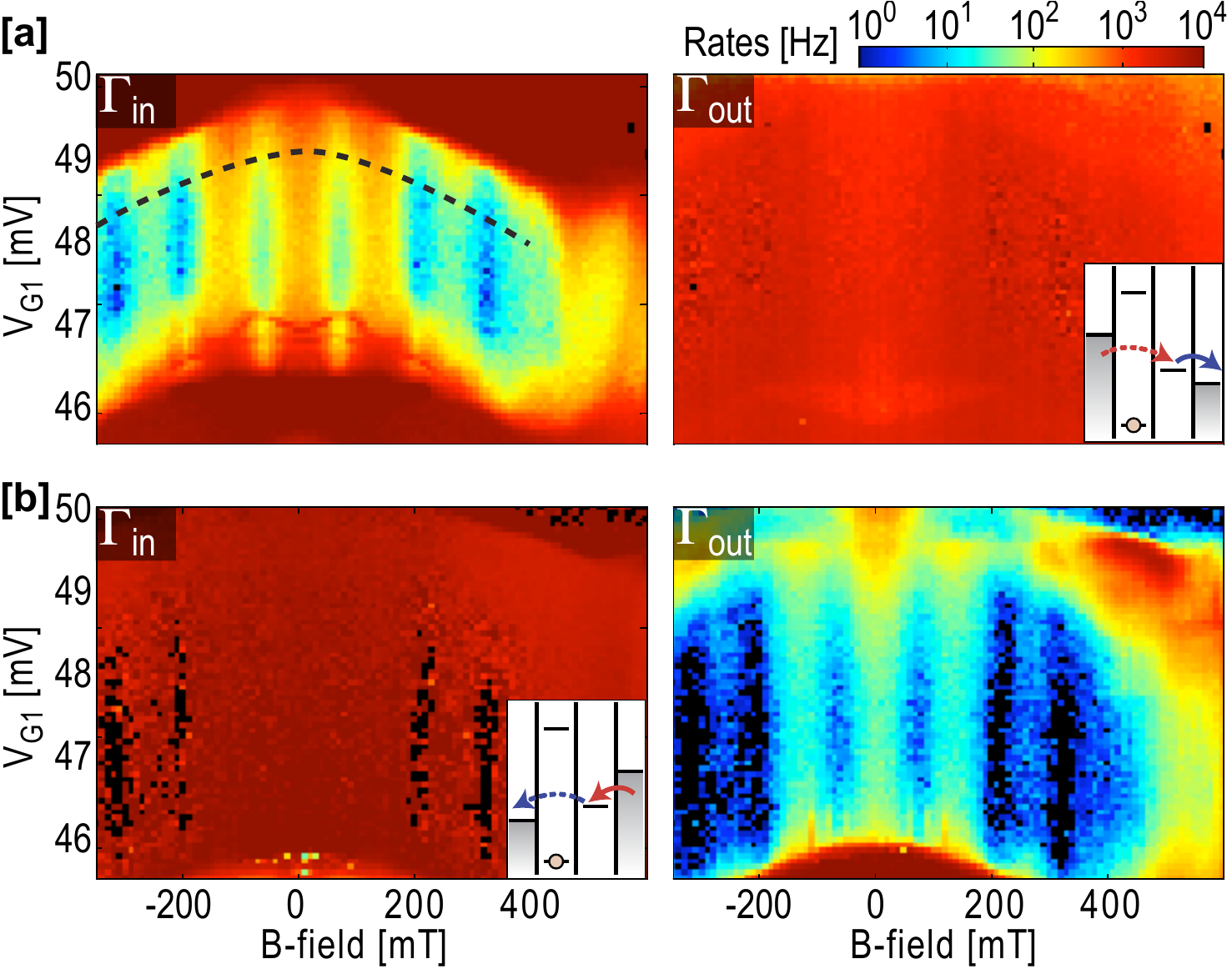}
 \caption{(a) Tunneling rates for electrons entering
 ($\Gamma_\mathrm{in}$) and leaving ($\Gamma_\mathrm{out}$) the DQD,
 measured versus electrochemical potential of QD1 and magnetic field.
 The y-axis corresponds to sweeps along the dashed line in
 Fig.~\ref{fig:AB_hexagon}. Within the cotunneling region,
 $\Gamma_\mathrm{in}$ shows clear B-field periodicity, while
 $\Gamma_\mathrm{out}$ remains constant. This is in agreement with
 the picture where only the electrons tunneling from source to QD2
 encircle the Aharonov-Bohm ring, while electrons leaving to drain
 remain unaffected by the applied B-field. (b) Same as (a), but
 with reverse bias over the DQD. Here, the roles of
$\Gamma_\mathrm{in}$ and $\Gamma_\mathrm{out}$ are inverted. 
 Adapted from Ref. \cite{gustavssonNL:2008}.
 }
\label{fig:AB_gammaInOut}
\end{figure}

In \FigRef{fig:AB_gammaInOut}(a), we investigate the separate rates
for electrons tunneling into and out of the DQD as a function of
magnetic field. The y-axis corresponds to the dashed line in
\FigRef{fig:AB_hexagon}, i.e., to the energy of the states in QD1.
The measurement shows a general shift of the DQD energy with the
applied B-field, which we attribute to changes of the orbital
wavefunctions in the individual QDs. Within the cotunneling region,
$\Gamma_\mathrm{in}$ shows well-defined B-periodic oscillations. At
the same time, $\Gamma_\mathrm{out}$ is essentially independent of
the applied field. This is expected since $\Gamma_\mathrm{out}$
measures the rate at which electrons leave QD2 to the drain, which
occurs independently of the magnetic flux passing through the
AB-ring [see \FigRef{fig:AB_count}(a,c)]. In
\FigRef{fig:AB_gammaInOut}(b), the bias over the DQD is reversed.
This inverts the roles of $\Gamma_\mathrm{in}$ and
$\Gamma_\mathrm{out}$ so that $\Gamma_\mathrm{out}$ corresponds to
the cotunneling process. Here $\Gamma_\mathrm{out}$ shows B-periodic
oscillations while $\Gamma_\mathrm{in}$ remains unaffected. In the
black regions seen in \FigRef{fig:AB_gammaInOut}(b), no counts were
registered within the measurement time of three seconds due to
strong destructive interference for the tunneling-out process. As a
consequence, it was not possible to determine $\Gamma_\mathrm{in}$
in these regions.

In the sequential regime (upper and lower parts of the color maps in
Fig.~\ref{fig:AB_gammaInOut}), one would also expect AB-oscillations
to occur. However, the effect would show up as a modulation of the
coupling between the QDs ($\Gamma_\mathrm{C}$), which involves
timescales of the order $\sim 1/\Gamma_\mathrm{C} \sim
1~\mathrm{ns}$. The detection of single electron motion on such
timescales is presently out of reach due to limited bandwidth of the
detector.

\subsection{Noise in the Aharonov-Bohm regime}
\begin{figure}[tb]
\centering
 \includegraphics[width=\columnwidth]{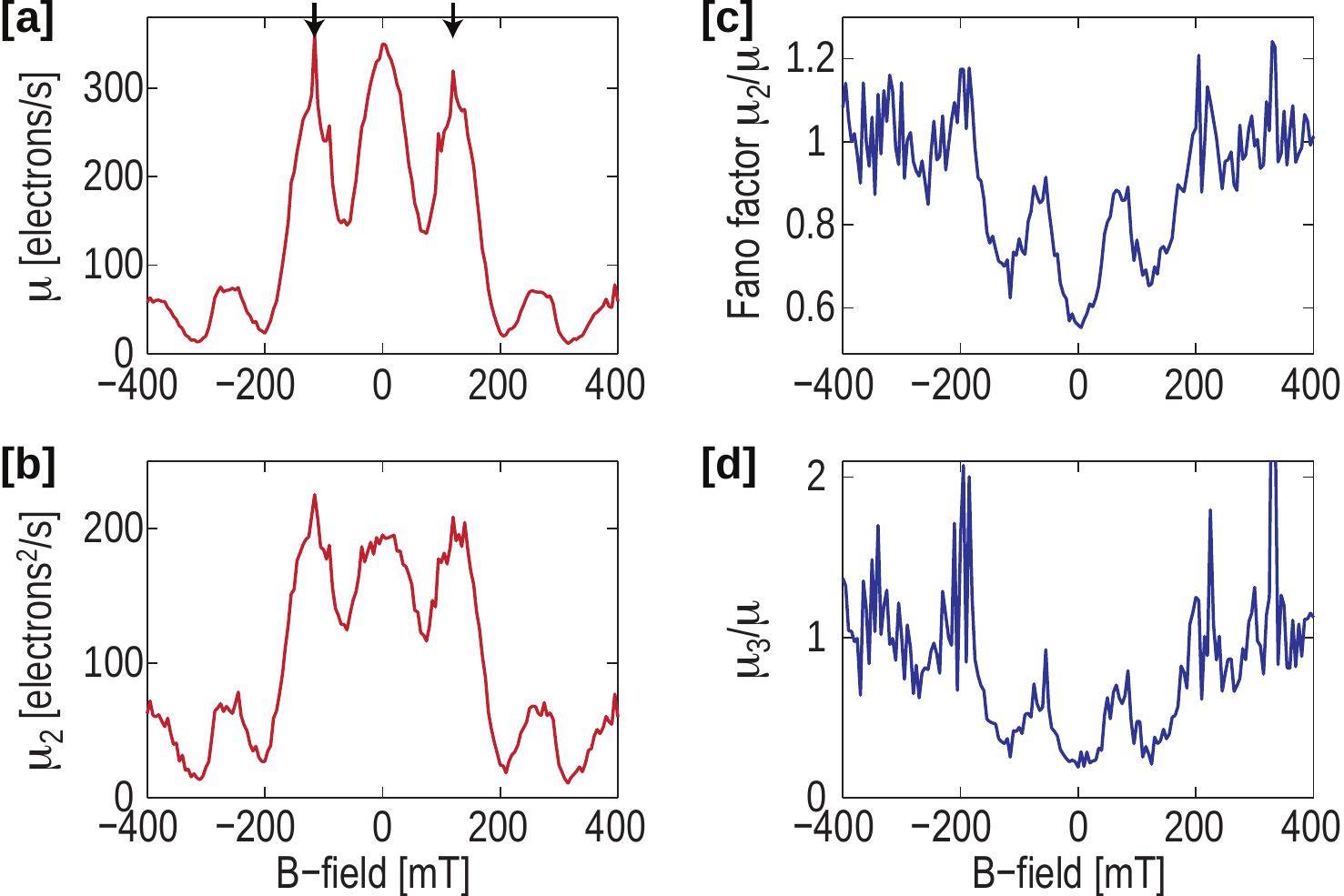}
 \caption{(a) DQD current in the Aharonov-Bohm regime.
 (b) Noise $\mu_2$ of the DQD current. The curve strongly resembles the
 average current shown in (a), with the AB-oscillations clearly visible.
 (c-d) Fano factor $\mu_2/\mu$ and generalized Fano factor for the
 third moment ($\mu_3/\mu$), measured within the same region as the traces
 shown in (a,~b).
 All quantities were extracted from a QPC conductance
 traces of length $T=40\s$, measured with $\Vdsd = 600\uV$ and $\Vqsd =
 300\uV$.
 }
\label{fig:AB_noise}
\end{figure}

In this section we investigate the noise of the current in the
Aharonov-Bohm regime. Using the methods of chapter
\ref{sec:ST_main}, we can extract the noise and the higher moments
of the current distribution directly from the QPC conductance
traces. Figure~\ref{fig:AB_noise}(a) shows a measurement of the
current flowing through the DQD, measured in a regime close to the
upper region of sequential tunneling in
\FigRef{fig:AB_gammaInOut}(a) [$\Vgl = 49\mV$ at $B=0\mT$, dashed
line in \FigRef{fig:AB_gammaInOut}(a)]. When sweeping the magnetic
field, we tune the voltages on gates G1 and G2 to compensate for the
shift of the cotunneling region occurring due to orbital effects in
the QDs. We chose to measure the AB-oscillation at relatively low
DQD detuning; this enhances the cotunneling rates and allows us to
collect more statistics within reasonable measurement times. On the
other hand, it also increases the contribution of sequential
tunneling and photon-assisted tunneling processes, giving the
slightly lower visibility compared to \FigRef{fig:AB_count}(d). The
small spikes seen at $B=\pm 120 \mT$ in \FigRef{fig:AB_noise}(a)
(marked by arrows) are attributed to single-QD excitations.

In \FigRef{fig:AB_noise}(b), we plot the shot noise (second moment
$\mu_2$) of the current distribution, extracted from the same set of
data as used in \FigRef{fig:AB_noise}(a). The noise curve shows
strong similarities to the current trace in (a), with the
AB-oscillations clearly visible. This is reasonable, since we expect
the noise to scale with the magnitude of the current. In
\FigRef{fig:AB_noise}(c), we plot the Fano factor $\mu_2/\mu$,
extracted from the traces in \FigRef{fig:AB_noise}(a,b). Also the
Fano factor displays AB-oscillations, with a minimum occurring at
$B=0\mT$ (with $\mu_2/\mu = 0.55$). We can understand this by
considering the noise calculated for a single QD [see
\EqRef{eq:ST_mu2vsasym} in chapter~\ref{sec:ST_main}]. There, we saw
a reduction of the Fano factor due to Coulomb blockade, with the
lowest noise given in a configuration where the tunneling rates for
entering and leaving the QD were equal.

In the AB-regime, we also measure a current due to two tunneling
rates; one is the cotunneling rate showing strong AB-oscillations
(in this case $\Gin$), while the other ($\Gout$) is a sequential
rate being independent of external magnetic field [compare the rates
$\Gin$ and $\Gout$ in \FigRef{fig:AB_gammaInOut}(a)]. At zero
magnetic field, the cotunneling rate $\Gin$ has a maximum and at
this point it becomes comparable to the sequential rate $\Gout$. The
two tunneling rates are relatively symmetric, giving a reduction of
the Fano factor. For higher magnetic fields, the cotunneling rate
$\Gin$ drops drastically while $\Gout$ stays constant. This results
in a more asymmetric configuration and a Fano factor close to one.

In the region of higher magnetic fields the experimental precision
of the measurement decreases. This is because of the low average
count rate, giving less statistical data for extracting the moments
compared to the region around $B=0\mT$. Finally, in
\FigRef{fig:AB_noise}(d) we plot $\mu_3/\mu$, the generalized Fano
factor for the third moment. This quantity also shows indications of
AB-oscillations, but the experimental uncertainty in the high
B-field range becomes even larger than for the conventional Fano
factor.

\subsection{Temperature effects}
\begin{figure}[tb] \centering
 \includegraphics[width=\linewidth]{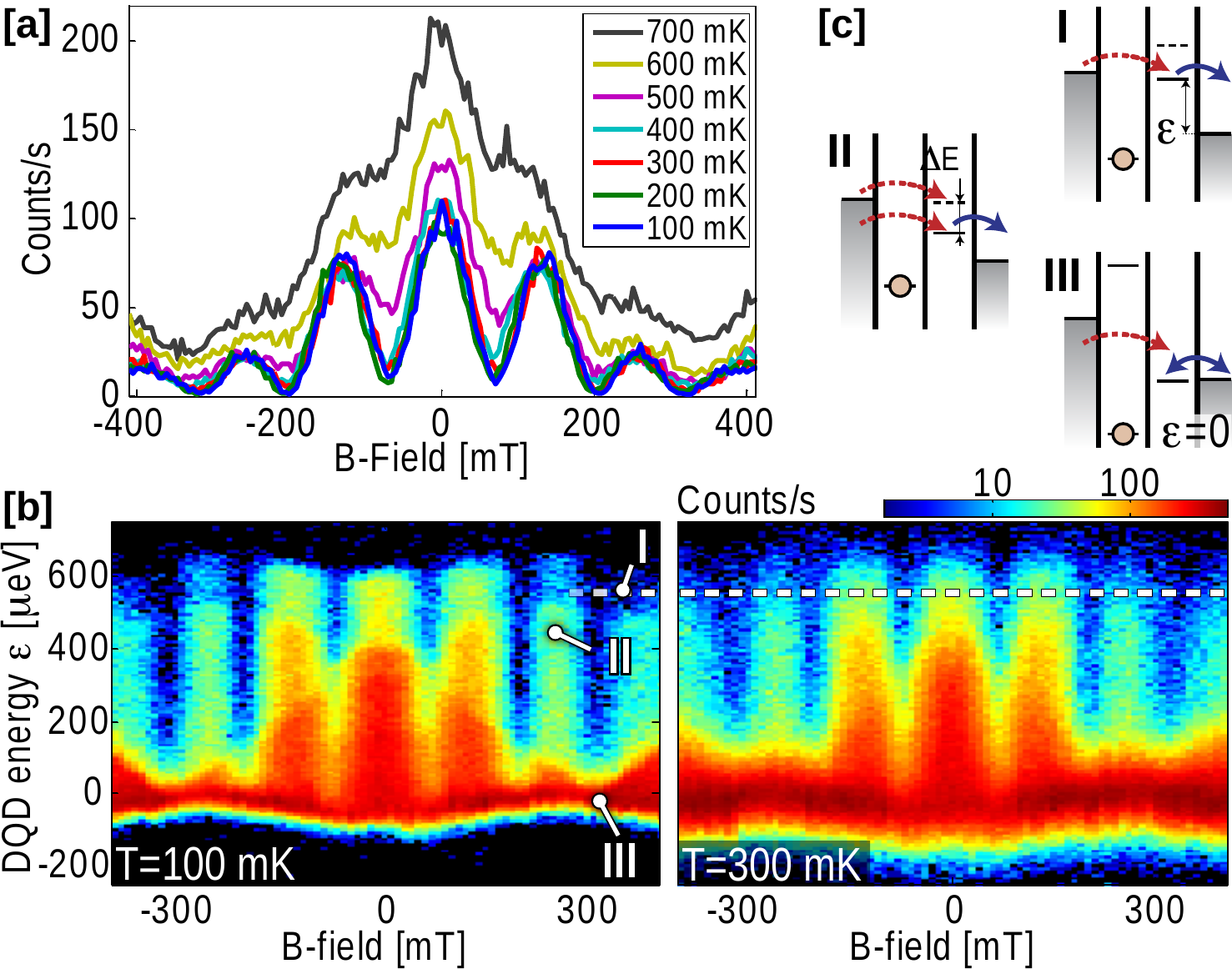}
 \caption{(a) Aharonov-Bohm (AB) oscillations measured at different
 temperatures. At $\sim 400~\mathrm{mK}$, the visibility of the oscillations drops drastically.
 The data was taken along the dashed line in (b).
 (b) Rate of electrons entering QD2, measured versus B-field and
 total energy of the DQD, $\varepsilon$.  The two images show data taken at two
 different temperatures, $T=100~\mathrm{mK}$ and $T=300~\mathrm{mK}$.
 The DQD energy $\varepsilon$
 is taken to be zero when QD2 is aligned with Fermi level of the
 drain. Here, tunneling due to thermal fluctuations between QD2 and the lead gives rise
 to a high count rate (point III). This feature is visibly broadened
 when the temperature is increased.
 In the cotunneling region (point I), the count rate shows clear AB oscillations.
 The elevated temperature only has a slight impact on the AB-visibility. In case II, the
 cotunneling rate goes up compared to case I. We attribute the increase to tunneling into an
 excited state in QD2.
 (c) Diagrams depicting DQD energy levels for the three configurations marked in (b).
 }
\label{fig:AB_temp}
\end{figure}

In Fig.~\ref{fig:AB_count}(a), we investigate how the AB-oscillations
are influenced by elevated temperatures.
The dephasing of open QD systems is thought to be due to
electron-electron interaction \cite{altshuler:1997}, giving
dephasing rates that depend strongly on temperature
\cite{huibers:1998}. Figure~\ref{fig:AB_temp}(a) shows the
temperature dependence of the AB oscillations in our system.
The amplitude of the oscillations remains almost unaffected up to
$\sim\!400~\mathrm{mK}$, indicating that the coherence is not
affected by temperature until the thermal energy becomes comparable
to the single-level spacing of the QDs.

Figure~\ref{fig:AB_temp}(b) shows measurements of the electron count
rate vs magnetic field and the average potential of the DQD, taken
at $T=100~\mathrm{mK}$ and $T=300~\mathrm{mK}$. Contrary to the
measurements presented in Fig.~\ref{fig:AB_count} and
Fig.~\ref{fig:AB_gammaInOut}, the potential difference between QD1
and QD2 is kept constant at $\delta=350\ueV$ while the overall DQD
energy $\varepsilon$ is shifted relative to the leads.
The energy $\varepsilon$ is taken to be zero when the level in QD2
is aligned with the Fermi level of the drain [case III in
Fig.~\ref{fig:AB_temp}(b-c)]. Here, thermal population fluctuations
tunneling between QD2 and the drain lead gives rise to a high count
rate [strong red line in the lower part of
Fig.~\ref{fig:AB_temp}(b)]. The width of the resonant line is set by
the temperature of the electrons in the lead. Indeed, this line is
clearly broader for the $T=300~\mathrm{mK}$ data.

Going to point I in Fig.~\ref{fig:AB_temp}(b,c), the energy of the
DQD is raised compared to the leads and thermal fluctuation are no
longer relevant. Here, electrons can only enter QD2 by cotunneling
from the source lead. The data shows clear Aharanov-Bohm
oscillations at both $T=100~\mathrm{mK}$ and $T=300~\mathrm{mK}$,
with comparable visibility. At the same time, the effect of the
increased temperature is visible in the regime around
$\varepsilon=0$.
As the temperature is further increased, the line of thermal
fluctuations becomes broader and eventually reaches the dashed line
where the AB-oscillations of Fig.~\ref{fig:AB_temp}(a) were
measured. This leads to the sharp decrease of the AB-visibility
demonstrated in Fig.~\ref{fig:AB_temp}(a). We conclude that the
decreased visibility at higher temperatures is due to an increase in
thermal fluctuations of the DQD population.

\subsection{Phase shifts for tunneling involving excited states}
In the following, we investigate the phase of the AB-oscillations
for different states in QD2.
Previous experiments have shown phase shifts of $\pi$ occurring
between consecutive Coulomb resonances in many-electron quantum dots
\cite{schuster:1997,kalish:2005}. To measure AB-oscillations for
consecutive electron fillings requires a relatively large shift of
the gate voltages. Such measurements are difficult to perform in our
setup, since large changes of gate voltages also affect the symmetry
of the left and right arm connecting QD1 and QD2, which may strongly
reduce the visibility of the AB-oscillations. Instead, we look at
excited states of QD2 at fixed electron population
\cite{sigrist:2007b}.

In addition to highlighting temperature effects, the color map in
\FigRef{fig:AB_temp}(b) also shows the existence of excited states
in the QDs. At point II in \FigRef{fig:AB_temp}(b,c), the count rate
is increased compared to case I. We attribute the increase to
cotunneling into an excited state in QD2 (see
section~\ref{sec:inelastic}). Measuring the AB-oscillations at
various DQD energy thus provides a way to investigate relative
phases of the excited states in the QDs. From the data in
\FigRef{fig:AB_temp}(b), we see that the AB-oscillations persist in
regions involving several excited states and that the phase of the
oscillations seems to remain the same in all regions.

\begin{figure}[tb]
\centering
\includegraphics[width=\linewidth]{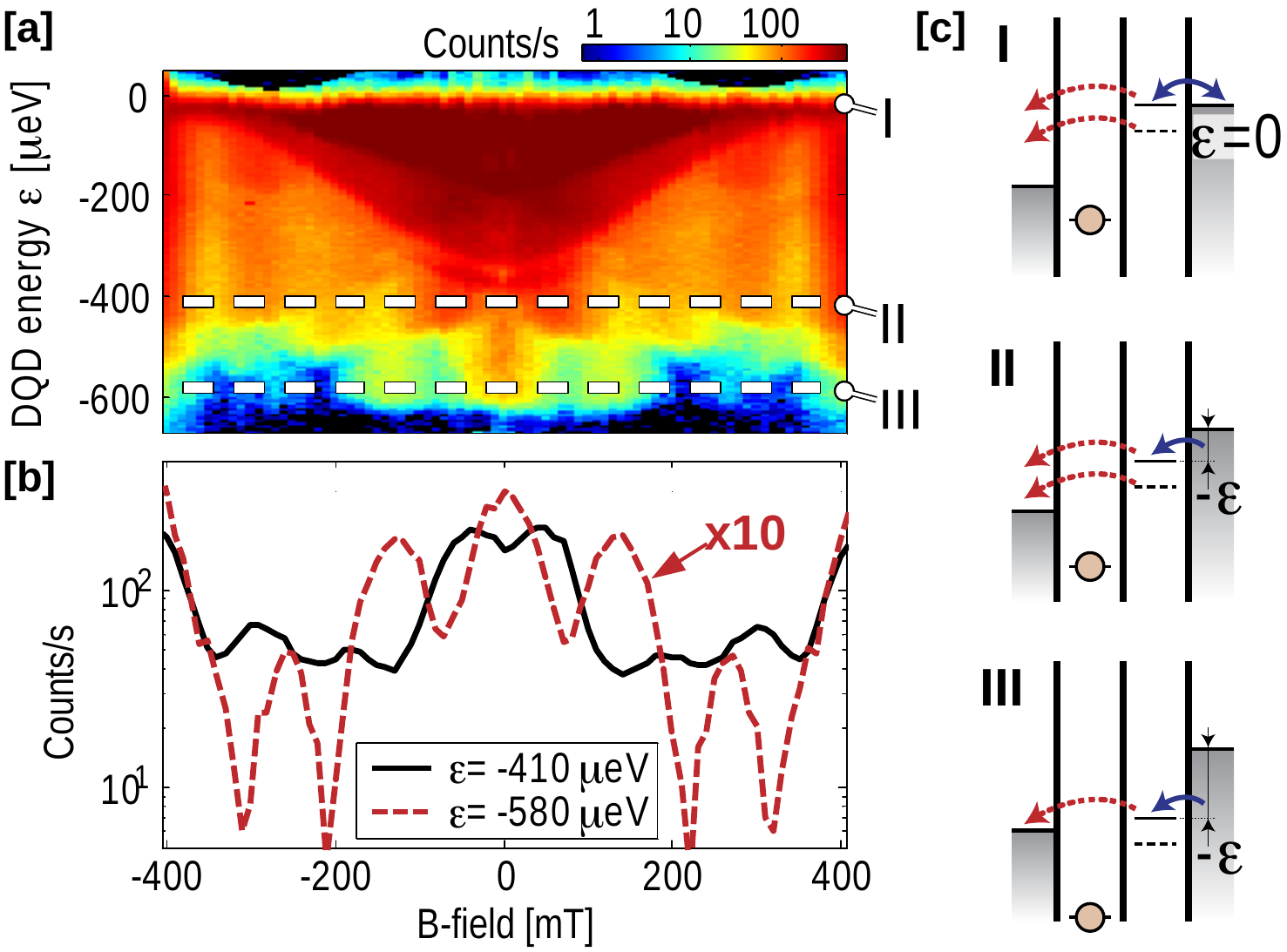}
\caption{(a) Electron count rate, measured versus magnetic field and
total DQD energy relative to the leads, $\varepsilon$. The data was
measured with $V_\mathrm{DQD-SD} = -600\uV$ applied to the DQD. (b)
Count rates measured at the positions marked by the dashed lines in
(a). There is a phase shift of $\sim \! 0.7\pi$ between the two
curves. The trace for $\varepsilon=-580~\mathrm{\mu eV}$ has been
magnified by a factor of ten for better visibility. (c) Energy
diagrams of the DQD for the positions marked by I, II and III in
(a). At point I, the potential of QD2 is lined up with the Fermi
level in the right lead and the tunneling is mainly due to
equilibrium fluctuations between QD2 and the lead. At point II, the
DQD potential is shifted downwards, so that electrons in QD2 may
only leave by cotunneling to the source lead. The energy level
arrangement allows a process involving an excited states of QD2 to
contribute to the cotunneling. Finally, at point III only
cotunneling involving the ground state of QD2 is possible. Adapted from Ref. \cite{gustavssonProcEP2DS:2008}.
}\label{fig:AB_phase}
\end{figure}

Depending on the direction of the applied bias, we can probe
different excited states (see section~\ref{sec:excitedStates}).
For positive bias, electrons cotunnel from source into QD2 and may
thereby put QD2 into either the $\mathrm{(m,n+1)}$-electron ground
state or an $\mathrm{(m,n+1^*)}$-electron excited state [see case II
in \FigRef{fig:AB_temp}(c)]. For negative DQD bias, the cotunneling
involves an electron leaving from QD2 to the source contact. This
involves transitions taking the QD2 into either its
$\mathrm{(m,n)}$-electron ground state or into an
$\mathrm{(m,n^*)}$-electron excited state. Since the energy
difference $E[\mathrm{(n,m^*)}]-E[\mathrm{(n,m+1)}]$ is smaller than
$E[\mathrm{(n,m)}]-E[\mathrm{(n,m+1)}]$, the transition involving
the excited state [$(m,n^*)$] occurs at an energy $\Delta E$
\emph{below} the ground state transition [see case II in
\FigRef{fig:AB_phase}(c)].

Figure~\ref{fig:AB_phase}(a) shows a measurement
of the electron count rate versus magnetic field and DQD energy
$\varepsilon$ for negative DQD bias.
Again, we define $\varepsilon=0$ when the electrochemical potential
of QD2 is aligned with the Fermi level of the drain lead [see case I
in \FigRef{fig:AB_phase}(c)]. Here, the tunneling is mainly due to
equilibrium fluctuations between QD2 and the drain.
As $\varepsilon$ is reduced, the equilibrium fluctuations between
QD2 and drain are no longer possible and electrons can only leave
QD2 by cotunneling to the source. The cotunneling region shows AB-oscillations, but the
oscillations are less uniform compared to the results for positive bias [\FigRef{fig:AB_temp}(b)].
Between the position marked by II and III in
\FigRef{fig:AB_phase}(a), both the intensity and the behavior of the
count rate changes drastically. In \FigRef{fig:AB_phase}(b), we plot
two cross sections from Fig.~\ref{fig:AB_phase}(a), taken at the
positions of the dashed lines. Both traces show AB-oscillations, and
both curves are symmetric around $B=0~\mathrm{T}$ as expected from
the Onsager relations. However, by comparing the positions of the
maxima for $B>0~\mathrm{T}$ we see that the phase is shifted by
$0.7\pi$ between the two curves. The reason for the apparent lack of
phase rigidity is not understood, further measurements are needed
for a more complete understanding of the phenomena.

Starting at point III in Fig.~\ref{fig:AB_phase}(a,c), the
transition involving the [$(m,n^*)$]-electron excited state is below
the Fermi level of the source so that only cotunneling through the
ground state is possible. The trace in Fig.~\ref{fig:AB_phase}(b)
belonging to point III is qualitatively similar to the data shown in
Fig.~\ref{fig:AB_temp}(a), with both curves having a maximum
appearing at $B=0~\mathrm{T}$. The similarity is expected, since
both measurements involve cotunneling through the ground state of
QD2.
Moving to point II, the energy of the DQD is raised and the
transition involving the excited state may also contribute to
transport.
The cotunneling rate measured in this regime is a sum of the
processes involving the ground state and the excited state.
However, since the rates at point II are almost an order of
magnitude larger compared to point III, the behavior is to a large
extent dominated by cotunneling from the excited state.

From this, we conclude that there is a phase shift occurring in the
Aharonov-Bohm signal between tunneling involving the
$\mathrm{(m,n)}$-electron ground state and a
$\mathrm{(m,n^*)}$-electron excited state of QD2. Our findings are
in agreement with previously reported results
\cite{schuster:1997,kalish:2005,sigrist:2007b}, but more
measurements are needed to map out the complete phase behavior of
the QD spectrum.

\subsection{Decoherence due to the quantum point contact}
\begin{figure}[tb] \centering
 \includegraphics[width=\linewidth]{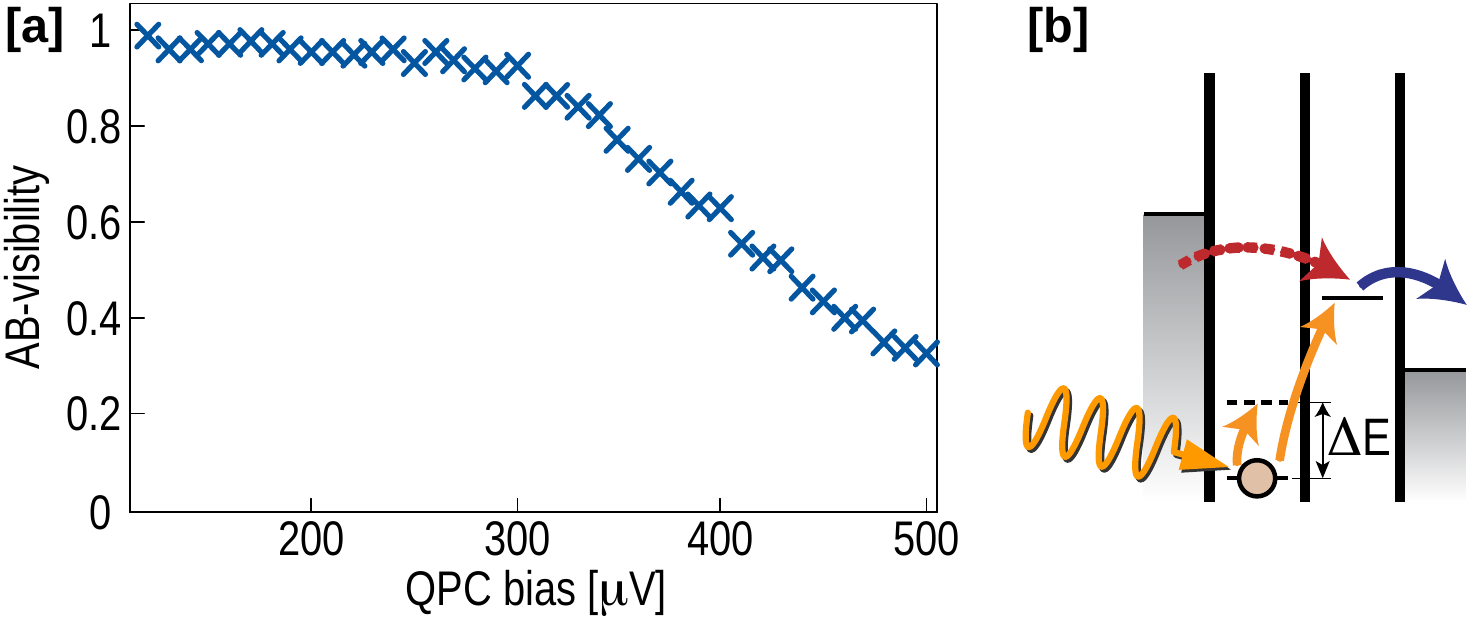}
 \caption{(a) Visibility of the AB-oscillations measured at different QPC bias. The visibility
 stays roughly constant up to $V_\mathrm{QPC} =300\uV$ and then decreases drastically
 with increasing bias voltage.
 We attribute the reduction in visibility to an increase in photon-assisted tunneling.
 (b) Energy level diagram of the DQD in the cotunneling configuration. At high QPC bias, both
 intradot and interdot photon absorption processes become possible.  Adapted from Ref. \cite{gustavssonNL:2008}.
 }
\label{fig:AB_QPC}
\end{figure}

In the experiment, we use the current in the QPC to detect the
charge distribution in the DQD. In principle, the QPC could also
determine whether an electron passed through the left or the right
arm of the ring, thus acting as a which-path detector
\cite{buks:1998, nederPRL:2007}. If the QPC were to detect the
electron passing in one of the arms, the interference pattern should disappear.
In Fig.~\ref{fig:AB_QPC}(a), we show the visibility of the
AB-oscillations as a function of bias on the QPC. The visibility
remains unaffected up to $V_{QPC} \sim \! 250~\mathrm{\mu eV}$, but
drops for higher bias voltages.

We argue that the reduced visibility is not due to which-path
detection. At $V_{QPC} = 400~\mathrm{\mu V}$, the current through
the QPC is $\sim \!\! 10~\mathrm{nA}$. This gives an average time
delay between two electrons passing the QPC of $e/I_{QPC} \sim \!
16~\mathrm{ps}$. Since this time is larger than the typical
cotunneling time, it is unlikely that the electrons in the QPC are
capable of performing an effective which-path measurement.
Instead, we attribute the decrease of the AB-visibility to processes
where the DQD absorbs photons emitted from the QPC. As described in
chapter~\ref{sec:SP_main}, such processes may indeed excite an
electron from one QD to the other, as long as the energy of the
excited state is lower than the energy provided by the QPC bias
\cite{gustavssonPRL:2007}.
The radiation of the QPC may also drive transitions inside the
individual QDs, thus putting one of the dots into an excited state
\cite{onacQD:2006}. A few possible absorption processes are sketched in Fig.~\ref{fig:AB_count}(b).

As long as the QPC bias is lower than both the DQD detuning
($\delta=400~\mathrm{\mu eV}$) and the single-level spacing of the
individual QDs ($\Delta E \sim \!200~\mathrm{\mu eV}$), the AB
visibility in Fig.~\ref{fig:AB_count}(b) is close to unity.
When raising the QPC bias above $\Delta E$, we start exciting the
individual QDs. With increased QPC bias, more states become
available and the absorption process becomes more efficient. This
introduces new virtual paths for the cotunneling process. Since the
different paths may interfere destructively, the interference
pattern is eventually washed out.
In this way, the QPC has a physical back-action on the measurement
which is different from informational back-action
\cite{sukhorukov:2007} and which-path detection previously
investigated \cite{buks:1998,nederPRL:2007}. 

\section{Conclusions}
In conclusion, we have measured current fluctuations in a
semiconductor quantum dot, using a quantum point contact to detect single electron tunneling
through the dot. We show experimentally the reduction of the second
and third moment of the distribution when the quantum dot is symmetrically
coupled to the leads. The setup can be used as a high-precision current meter for measuring ultra-low currents, with resolution several orders of magnitude better than that of conventional current meters.

The quantum point contact does not only serve as a charge detector, but also causes a back-action onto the measured device. Electron scattering in the quantum point contact leads to emission of microwave radiation, which may drive charge transitions in the quantum dot. Turning the perspective around, we show that a double quantum dot can be used as a frequency-selective detector for microwave radiation emitted from mesoscopic structures.

In addition, we demonstrate interference of single electrons in
a solid state environment. Such experiments have previously been limited
to photons or massive particles in a high-vacuum environment in
order to decouple the degrees of freedom as much
as possible from the environment. Our experiments demonstrate the
exquisite control of modern semiconductor nanostructures which
enables interference experiment at the level of single
quasi-particles in a solid state environment.
Once extended to include spin degrees of freedom \cite{loss:2000}
such experiments have the potential to facilitate entanglement
detection \cite{oliver:2002, saraga:2003} or investigate the interference of
particles \cite{nederNature:2007} originating from different
sources.

Financial support from
the Swiss Science Foundation (Schweizerischer Nationalfonds) via
NCCR Nanoscience and from the EU Human Potential Program financed
via the Bundesministerium f\"ur Bildung und Wissenschaft is
gratefully acknowledged.

\appendix
\section{Cumulants or central moments of a distribution}
\label{sec:AP_cumulants} The full distribution function $P_{t_0}(N)$
or the complete set of central moments $\mu_i$ give a complete
description of the current in a system. The moments and the
distribution function contain the same information, making the two
equivalent. Another way to represent the same information is in
terms of the cumulants $C_k$ and the cumulant generating function
$\mathcal{F}(\chi)$. The cumulants are defined as
\cite{mathWorldCumulant:2005}
\begin{equation}\label{eq:ST_cumulants}
   C_k = -(-i)^k \frac{\partial^k}{\partial \chi ^k} \
   \mathcal{F}(\chi) \big|_{\chi=0},
\end{equation}
with the cumulant generating function given by
\begin{equation}\label{eq:ST_cumGenFunc}
   e^{-\mathcal{F}(\chi)} = \sum_N P_{t_0}(N) e^{i N \chi}.
\end{equation}
In terms of the central moments, we have for the first few cumulants
\begin{eqnarray}\label{eq:ST_momToCum}
 \nonumber  && C_1 = \mu_1, ~C_2 = \mu_2, ~ C_3 = \mu_3, \\
  && C_4 = \mu_4-3\mu_2^2, ~C_5 = \mu_5-10 \mu_3 \mu_2.
\end{eqnarray}
The cumulants can be seen as an irreducible representation of the
moments. Again, this means that the knowledge of either the moments
$\mu_i$, the cumulants $C_k$ or the distribution function
$P_{t_0}(N)$ provide the same information.


\end{document}